\documentclass[amsmath,amssymb,floatfix,nofootinbib,prd,showpacs]{revtex4}

\pdfoutput=1
\usepackage{graphicx}
\usepackage{epsfig}
\usepackage{rotate}
\usepackage{amsmath}
\usepackage{amssymb}
\usepackage{amsfonts}
\usepackage{phoenician}
\usepackage{calc}

\usepackage{enumerate}
\usepackage{afterpage}
\usepackage{color}

\usepackage{soul}
\usepackage{ulem}

\usepackage{mathrsfs}

\usepackage{braket}

\usepackage{multirow}

\newcommand{\ud}{\mathrm{d}}
\newcommand{\p}{\partial}
\newcommand{\cH}{\mathcal{H}}
\newcommand{\Perp}{\mathcal{P}}

\newcommand{\e}{{\bf{e}}}
\newcommand{\bx}{{\bf{x}}}
\newcommand{\bq}{{\bf{q}}}
\newcommand{\bk}{{\bf{k}}}
\newcommand{\bp}{{\bf{p}}}
\newcommand{\bn}{{\bf{n}}}
\newcommand{\bm}{{\bf{m}}}

\newcommand{\tbk}{\tilde{\bf{k}}}

\newcommand{\ellvp}{\ell_\bp}
\newcommand{\ellvq}{\ell_\bq}
\newcommand{\mvp}{m_\bp}
\newcommand{\mvq}{m_\bq}

\newcommand{\barell}{\bar{\ell}}
\newcommand{\barm}{\bar{m}}

\newcommand{\tell}{\tilde{\ell}}
\newcommand{\tm}{\tilde{m}}
\newcommand{\ts}{\tilde{s}}

\newcommand{\eS}{\mathcal{ S}}
\newcommand{\R}{\mathcal{ R}}
\newcommand{\G}{\mathcal{ G}}
\newcommand{\M}{\mathcal{M}}

\newcommand{\Q}{\mathcal{Q}}
\newcommand{\F}{\mathcal{ F}}
\newcommand{\W}{\mathcal{ W}}
\newcommand{\T}{\mathcal{ T}}

\newcommand{\I}{\mathcal{ I}}
\newcommand{\K}{\mathcal{ K}}
\newcommand{\J}{\mathcal{ J}}
\newcommand{\WW}{\mathbb{ W}}

\newcommand{\PP}{\mathcal{ P}}

\newcommand{\Omo}{\Omega_{\rm m0}}
\newcommand{\OLo}{\Omega_{\Lambda0}}

\def\be{\begin{equation}}
\def\ee{\end{equation}}
\def\bea{\begin{eqnarray}}
\def\eea{\end{eqnarray}}

\definecolor{darkred}{RGB}{175,0,0}

\newlength{\depthofsumsign}
\newcommand{\nsum}[1][1.8]{
    \mathop{%
        \raisebox
            {-#1\depthofsumsign+1\depthofsumsign}
            {\scalebox
                {#1}
                {$\displaystyle\sum$}%
            }
    }
}

\newcommand{\nsumm}[1][1.3]{
    \mathop{%
        \raisebox
            {-#1\depthofsumsign+1\depthofsumsign}
            {\scalebox
                {#1}
                {$\displaystyle\sum$}%
            }
    }
}

\begin{document}

\title{Relativistic wide-angle galaxy bispectrum on the light cone}

\author{Daniele Bertacca$^a$, Alvise Raccanelli$^\star \let\thefootnote\relax\footnote{$^\star$Marie Sk\l{}odowska-Curie fellow}$$^{b,c}$, Nicola Bartolo$^{d,e,f}$, Michele Liguori$^{d,e,f}$, Sabino Matarrese$^{d,e,f,g}$, Licia Verde$^{b,h}$  \\}

\affiliation{
\vspace{0.25cm}
$^a$Argelander-Institut f\"ur Astronomie, Auf dem H\"ugel 71, D-53121 Bonn, Germany \\ 
$^b$ Institut de Ci\`encies del Cosmos (ICCUB), Universitat de Barcelona (IEEC-UB), Mart\'{i} Franqu\`es 1, E08028 Barcelona, Spain \\
$^c$ Department of Physics and Astronomy, Johns Hopkins University, 3400 N. Charles St., Baltimore, MD 21218, USA \\
$^d$Dipartimento di Fisica Galileo Galilei, Universit\`{a} di Padova,  I-35131 Padova, Italy \\
$^e$INFN Sezione di Padova,  I-35131 Padova, Italy \\
$^f$ INAF-Osservatorio Astronomico di Padova, Vicolo dell Osservatorio 5, I-35122 Padova, Italy\\
$^g$Gran Sasso Science Institute, INFN, I-67100 L'Aquila, Italy\\
$^h$ Instituci\'o Catalana de Recerca i Estudis Avan\c{c}at ICREA, Pg. Lluis Companys 23, 08010 Barcelona, Spain
}

\begin{abstract}
Given the important role that the galaxy bispectrum has recently acquired in cosmology and the scale and precision of forthcoming galaxy clustering observations, it is timely to derive the full expression of the large-scale bispectrum going beyond approximated treatments which neglect integrated terms or higher-order bias terms or use the Limber approximation.
On cosmological scales, relativistic effects that arise from observing the past light cone alter the observed galaxy number counts, therefore leaving their imprints on N-point correlators at all orders.
In this paper we compute for the first time the bispectrum including {\it all} general relativistic, local and integrated, effects at second order, the tracers' bias at second order, geometric effects as well as the 
primordial non-Gaussianity contribution. This is  timely considering that future surveys will probe scales comparable to the horizon where approximations widely used currently may not hold;  neglecting these effects may introduce biases in  estimation of cosmological parameters as well   as primordial non-Gaussianity.
\end{abstract}


\date{\today}

\maketitle


\section{Introduction}

Gravitational instability  drives the evolution of primordial perturbations (as set out by e.g., the inflationary process) into the large-scale structure (LSS) we observe today. In the standard model of cosmology, gravity is described by general relativity (GR). However, for simplicity, to study and describe the formation of LSS,  different approximations are routinely used. For example, we treat large and small scales differently and, on scales well inside the Hubble horizon, we use several aspects of Newtonian gravity. This small vs large-scale splitting has provided an excellent approximation to interpret observations so far, but it may not hold for upcoming LSS surveys, that will probe scales approaching the Hubble horizon, where the Newtonian approximation breaks down. Moreover, observations are performed along the past light cone, which brings in a series of  local and nonlocal (i.e. integrated along the line of sight) corrections, usually called GR projection effects (hereafter they will be abbreviated as {\it GR effects} or {\it corrections}), which are not included in the ``standard"  treatment, where only the local distortion of the radial pattern of the galaxy distribution due to peculiar velocities is considered and the flat-sky limit is assumed (for example see~\cite{Kaiser:1987qv, Hamilton:1997zq})\footnote{In this paper we do not consider ``standard" the magnification due to gravitational lensing which modifies the observed number counts in flux-limited samples (for example, see   \cite{Turner:1980, Turner:1984ch, Moessner:1997qs, Moessner:1997vm, Matsubara:2000pr}) at linear order, and \cite{Schmidt:2008mb} for weak lensing effects on the galaxy three-point correlation function.}.
 
Forthcoming and future surveys will probe very large scales, comparable to the horizon size, where GR effects may not be neglected. Large survey volumes also imply very small statistical errors, enabling high-precision measurements of large-scale structure clustering. Such precision must be matched by high accuracy; hence accurate theoretical modelling is required to correctly interpret the measured  clustering signal.
For this reason, the study of these GR effects on first-order statistics of large scale structure (e.g., the galaxy power spectrum or the two-point correlation function) has received significant attention in recent years, see e.g. \cite{Yoo:2008tj, Yoo:2009au, Yoo:2010ni, Bonvin:2011bg, Challinor:2011bk, Jeong:2011as, Bertacca:2012tp, Yoo:2012se, Yoo:2013tc, Raccanelli:2013multipoli, DiDio:2013bqa, Yoo:2013zga, DiDio:2013sea, Bonvin:2013ogt, Raccanelli:2013gja, Bacon:2014uja, Raccanelli:2015vla, Alonso:2015uua, Montanari:2015rga, Alonso:2015sfa, Fonseca:2015laa, Bonvin:2015kuc, Gaztanaga:2015jrs, Cardona:2016qxn, Raccanelli:2016avd, DiDio:2016ykq}, but also the pioneering work of \cite{Ellis:1971pg}. 
These GR corrections can be implemented in mock galaxy catalogs using the technique recently developed in \cite{Borzyszkowski:2017ayl} (see also \cite{Chisari:2011iq}), while N-body simulations that include dynamical space-time variables in the weak-field approximation have been studied in \citep{Adamek:2013wja, Adamek:2014xba, Adamek:2015eda, Adamek:2016zes, Adamek:2017grt}, and full numerical relativity simulations have been developed in \citep{Bentivegna:2015flc, Giblin:2015vwq, Giblin:2017juu}.

 It is important to note that, for an accurate modeling of first-order statistics, calculations in LSS beyond linear order are needed.
Recently the second-order GR effects 
have been derived
\cite{Bertacca:2014dra, Bertacca:2014wga, Yoo:2014sfa, DiDio:2014lka, Bertacca:2014hwa, Kehagias:2015tda, DiDio:2015bua, DiDio:2016gpd, Umeh:2016nuh, Jolicoeur:2017nyt} and include contributions such as lensing, gravitational Sachs-Wolfe (SW) and integrated Sachs-Wolfe (ISW) effects, primordial non-Gaussianity and bias (see e.g. \cite{Schmidt:2008mb, McDonald:2009dh, Giannantonio:2009ak, Baldauf:2011bh, Baldauf:2012hs, Desjacques:2012eb, Tasinato:2013vna, Biagetti:2014pha, Bertacca:2015mca, Bartolo:2015qva, Tellarini:2015faa}, the recent review \cite{Desjacques:2016bnm} and references therein).
It is interesting to note that, while for cosmic microwave background studies the second-order evolution of perturbations in GR has a longer history \cite{Hu:1993tc,Pyne:1995bs,Mollerach:1997up,Matarrese:1997ay,Bartolo:2005kv, Bartolo:2006fj, Bartolo:2006cu,Fitzpatrick:2009ci, Nitta:2009jp, Boubekeur:2009uk}, only recently LSS observations are reaching a comparable level of accuracy to warrant generalisation of the treatment beyond linear order.

Besides the first-order statistics, higher-order statistics encode highly complementary information, which, while at possibly lower signal-to-noise ratios compared to the power spectrum, is nevertheless of crucial importance.  
In particular, the bispectrum can offer a direct handle on galaxy bias \cite{Fry:1996fg, Matarrese:1997sk, Verde:2001sf, McDonald:2009dh, Baldauf:2012hs, Gil-Marin:2014sta, Desjacques:2016bnm} and on primordial 
non-Gaussianity \cite{Sefusatti:2007ih, Jeong:2009vd, Giannantonio:2009ak, Bartolo:2010ec,Bartolo:2010qu,  Baldauf:2011bh, Baldauf:2012hs, Tasinato:2013vna, Bartolo:2015qva, Tellarini:2015faa}, as well as helping to break other degeneracies among cosmological parameters~\cite{Sefusatti:2006pa}. The expression of the bispectrum at tree level using the Newtonian approximation has been known for a while (see for example \cite{Bernardeau:2001qr}) and subsequent improvements have been developed by \cite{Scoccimarro:2000ee, Smith:2007sb, Matsubara:2011ck, Desjacques:2011mq, GilMarin:2011ik, Schmittfull:2012hq, Yamamoto:2016anp, Hashimoto:2017klo}, always within the Newtonian framework (see also \cite{Matsubara:1997zj, Moscardini:1999ba, Suto:1999id}).

Until very recently, any modeling of the galaxy bispectrum has relied on small-scales approximations, neglecting relativistic effects and radial correlations, given that LSS precise measurements were available only on relatively small scales.
However, galaxy surveys are now at the point in which measurements of LSS are becoming extremely precise, and will be available on very large scales, therefore reaching the regime in which, as said above, a proper GR treatment is necessary and approximations such as the plane-parallel one fail. 
Hence, it is timely to have a full  expression of the galaxy bispectrum in GR, avoiding approximations that are, in principle, inaccurate for correlations on very large scales.
To this aim GR, bias and light-cone effects must be computed beyond linear order, at least to second order in the perturbations \cite{Bertacca:2015mca,DiDio:2015bua, Umeh:2016nuh, Jolicoeur:2017nyt}.

Various investigations on modeling the bispectrum including GR effects on large scales have been made recently. In particular, \cite{DiDio:2014lka, Bertacca:2014hwa, DiDio:2015bua, DiDio:2016gpd} compute the three-point correlation function in configuration space, including only some projection terms, but not the bias and relativistic perturbation solutions at second order (i.e. results obtained in \cite{Matarrese:1997ay, Bartolo:2005kv, Bartolo:2010rw, Bruni:2013qta, Bruni:2014xma, Villa:2015ppa}); 
Refs.~\cite{Umeh:2016nuh, Jolicoeur:2017nyt} analyzed the bispectrum in Fourier space including some bias terms but not nonlocal integrated terms (which \cite{DiDio:2015bua, DiDio:2016gpd} showed can be dominant for long radial correlations). In addition, the flat-sky limit is assumed even at very large scales, where this approximation has been shown to fail for the two-point correlation case~\cite{Szalay:1997cc,  Szapudi:2004gh, Papai:2008bd, Raccanelli:2010hk, Bertacca:2012tp, Raccanelli:2016avd}.
 Of all the possible GR effects, it is well known that, at least  on first-order statistics, magnification bias and wide angle corrections are the dominant ones~\cite{Raccanelli:2010hk, Raccanelli:2013multipoli, Raccanelli:2013gja, Raccanelli:2015vla, Borzyszkowski:2017ayl}. These are also expected to be important  for  higher-order statistics, but other contributions may  turn out to  be comparable; a full numerical investigation for different surveys is needed in order to have more details on this.
Given all these considerations, we extend previous works and derive, for the first time, the full wide-angle GR second-order expression of the bispectrum, i.e. including second-order bias, non-Gaussianity, velocities, Sachs-Wolfe, integrated Sachs-Wolfe and time-delay terms and lensing distortions from convergence and shear.
This work builds on Refs.~\cite{Bertacca:2014wga, Bertacca:2014dra, Bertacca:2014hwa} where SW and lensing contributions to the second-order matter overdensity are presented. 
We envision that the resulting full expression of the bispectrum will be important for accurate estimations of primordial non-Gaussianity and cosmological parameters, taking into account all effects at very large scales, therefore fully exploiting the ultra-large-scale correlations that will be measured for the first time in the next decade.
It will also provide an additional precise test of Einstein GR on those very large scales.
Beside the effects we consider here,  if the probability of observing a galaxy depends on its ellipticity,  there is a selection effect beyond magnification,  which is affected by shear and intrinsic alignment (see, for example, \cite{Hirata:2009qz, Krause:2010tt}).  Here for simplicity we are not considering this orientation-dependent selection effect which will produce  new terms both at first and second order that depend on the large-scale tidal field. The inclusion of this effect is left to future work.

We improve upon previous works in the literature in the sense that for the first time we write down the full expression in Fourier space, and we do so using what we call the ``spherical-Bessel" formalism, which was first developed by ~ \cite{Peebles1973} and, subsequently, in redshift space by \cite{Heavens:1994iq} for the two-point correlation function. (Let us point out another interesting approach studied in \cite{Dai:2012bc, Dai:2012ma}, where they compute bispectra with 
total-angular-momentum waves.)
The expressions are inevitably lengthy and involved but whenever possible we give a physical insight on their meaning and we summarize schematically the structure of the full expression in Section~\ref{sec:summary}.   
The expressions provided here will not be directly relevant to data analyses, their value is in providing a starting point to devise useful approximations, and a quantitative evaluation of the different contributions. The next natural step is to  derive a more compact expression including the dominant terms, for different geometrical configurations ({\it shapes}), which will be presented elsewhere.

The rest of the paper is organized as follows. We begin by presenting the second-order number counts on the light cone in Sec.~\ref{sec:Second-ordernumbercounts}. This is a key ingredient in the expression for the bispectrum which we present in Sec. \ref{sec:bisp}, which expression depends on several kernels that are written down explicitly in Sec.~\ref{sec:kernels} for the first-order, and Sec.~\ref{calM-second_order} at second order.
The second-order bias is derived in Sec.\ref{sec:bias}.
We conclude in Sec.\ref{Sec:Conclusions}.
The appendices report useful relations and expressions needed to follow the derivation of our main results and link our analytical expressions to numerical codes available in the literature\footnote{In Appendix \ref{SONG}, 
we give a possible prescription on how to insert the numerical outputs obtained in SONG \cite{Pettinari:2014vja}.} into our expressions.


\section{Executive summary}
\label{sec:summary}

Here we summarize in a concise way the main result of this work, showing the main expression for the spherical-Bessel bispectrum, including all general relativistic, local and integrated, wide-angle and mode-coupling terms, that we call $\mathcal{S}_3(\{k\ell m\}_1,\{k\ell m\}_2,\{k\ell m\}_3$), where $\{k\ell m\}_i = (k_i,\ell_i,m_i)$ for $i=1,2,3$. 
We write the galaxy bispectrum using the ``spherical-Bessel" formalism first proposed by ~ \cite{Peebles1973} and, subsequently, in redshift space by \cite{Heavens:1994iq} (see also \cite{Binney:1991nv}) for the two-point correlation function, and then by~\cite{Verde:2000xj} for the three-point statistics. The same formalism (for the power spectrum) has been applied to real data in~\cite{Percival:2004fs} and extended to include GR effects in~\cite{Yoo:2013tc}. 
 [See also Ref. \cite{Kitching:2016zkn} where they discuss the limit of  various approximations (e.g. flat-sky,  limber) which are applied to the lensing signal.]
Here we generalize this formalism to include all wide-angle and relativistic terms, and write down the full expression of the galaxy bispectrum, without any approximations  besides second-order perturbative expansion in the perturbations and the bias and without neglecting any contributions (apart from vector and tensor perturbations). 
In this section we report the general expression for $\eS_3(\{k\ell m\}_1,\{k\ell m\}_2,\{k\ell m\}_3)$, and in the rest of the paper we show its derivation and the explicit expression for all the terms. Here and hereafter:
\bea
\label{eq:bispectrum_summ}
\eS_3(\{k\ell m\}_1,\{k\ell m\}_2,\{k\ell m\}_3)= \langle \Delta^g_{\ell_1  m_1} (k_1) \Delta^g_{\ell_2 m_2}(k_2) \Delta^g_{\ell_3 m_3}(k_2)\rangle&=& \frac{1}{2} \sum_{t,u,v=1,2 \atop t+u+v=4} \bigg \langle \Delta^{g(t)}_{\ell_1  m_1} (k_1) \Delta^{g(u)}_{\ell_2 m_2} (k_2) \Delta^{g(v)}_{\ell_3 m_3}(k_3) \bigg\rangle \; , \nonumber\\
\eea
where $\Delta^g$ is the observed galaxy fractional number overdensity (for further details, see Sec.\ \ref{sec:Second-ordernumbercounts}) and  contains the contributions of all the local and integrated terms including bias, and the indexes $t, u, v$ indicate the order, so that in the right-hand side two of them will be $1$ and one will be $2$, cyclically. 
To be precise  $\mathcal{S}_3$ is actually the spherical 3-point correlation function of Fourier-space galaxy overdensity (but, for simplicity, we will refer to it as the bispectrum); in the same way as for the power spectrum, the correlation of the field $\delta$ in Fourier space $\langle\delta (\bk_1)\delta (\bk_2)\rangle$ is related to the power spectrum $P(k)$ via a Dirac delta function $\langle\delta (\bk_1)\delta (\bk_2)\rangle\propto P(k_1)\delta^D(\bk_1+\bk_2)$.

As an example, let us consider below only the first additive term of Eq.\eqref{eq:bispectrum_summ}, where $t=2$ $u,v=1$ (the other terms will be just permutations of this).
Since each $\Delta^{g(t)}_{\ell  m} (k)$ is a sum of several components, it will be a sum of several contributions which we indicate by the running indices $a,b,c$. Each of these indices will model the effect of all the physical quantities we consider:
\begin{equation}
\label{eq:deltas}
\Delta_{g} = \Delta_{\delta} + \Delta_{\rm rsd} + \Delta_{\rm v} + \Delta_{\rm L} + \Delta_{\rm pot} \, ,
\end{equation}
where $\delta$ refers to the matter overdensity (intrinsic clustering), ``${\rm rsd}$" and ``${\rm v}$" are velocity (peculiar velocities (RSD) and doppler) terms,
$\rm L$ contains lensing terms and ``${\rm pot}$" gravitational potentials (ISW and STD).
Each of  these quantities enter (alternatively) at the first and second order.
Within our formalism, in this specific example, the index $a$ is associated with a second-order quantity and refers to the terms expanded in Eq.\ (\ref{Deltag-2}), while the indices $b,c$ are associated with first-order quantities and refer to the terms that we will make explicit in Eq.\ (\ref{Deltag1}).
Therefore, we have
\bea \label{Gamma2}
 \bigg\langle \frac{1}{2}\Delta^{a(2)}_{\ell_1  m_1}(k_1)  \Delta^{b(1)}_{\ell_2 m_2} (k_2)   \Delta^{c(1)}_{\ell_3 m_3}(k_3) \bigg\rangle
 = 2  \Upsilon^{abc}_{\ell_1 m_1 \ell_2 m_2 \ell_3 m_3}(k_1,k_2,k_3) \, .
\eea
Explicitly (for derivation details, see Sec. \ref{sec:Bispectr-ScalarCase} and Appendix \ref{Sec:Upsilon}) the bispectrum building blocks $\Upsilon^{ abc}_{\ell_1 m_1 \ell_2 m_2 \ell_3 m_3}$ are functions that contain all of the information needed to express the {\it observed} bispectrum:
\bea
\label{Upsilon}
\Upsilon^{abc}_{\ell_1 m_1 \ell_2 m_2 \ell_3 m_3}(k_1,k_2,k_3) &=& \sum_{{{\ellvp}_1 {\mvp}_1 {\ellvq}_1 {\mvq}_1 {\barell}_1 {\barm}_1}}  \digamma_{m_1 m_2 m_3 {\mvp}_1  {\mvq}_1 {\barm}_1}^{~~~~\ell_2 ~\ell_3 ~{\ellvp}_1 ~ {\ellvq}_1 ~{\barell}_1} ~~\Bigg\{
  \int \frac{q_2^2\ud q_2}{(2\pi)^3}   \frac{q_3^2\ud q_3}{(2\pi)^3} \nonumber \\
&  & \times \left[
   \M_{\ell_1 m_1 {{\ellvp}_1 {\mvp}_1 {\ellvq}_1 {\mvq}_1 {\barell}_1 {\barm}_1 } }^{a(2)}(k_1; q_2, q_3) ~ \M_{\ell_2}^{b(1)}(k_2, q_2)  ~\M_{\ell_3}^{c(1)}(k_3, q_3)\right] P_\Phi(q_2) P_\Phi(q_3)\Bigg\}\;, \nonumber \\
\eea
where the sum is over ${{\ellvp}_1 {\mvp}_1 {\ellvq}_1 {\mvq}_1 {\barell}_1 {\barm}_1 }$, and
\[\digamma_{m_1 m_2 m_3 {\mvp}_1  {\mvq}_1 {\barm}_1}^{~~~~\ell_2 ~\ell_3 ~{\ellvp}_1 ~ {\ellvq}_1 ~{\barell}_1}=(-1)^{\ell_2 +\ell_3-m_2-m_3+{\mvp}_1+{\mvq}_1+{\barm}_1} \G^{{\ellvp}_1 {\barell}_1 \ell_{2} }_{-{\mvp}_1 {\barm}_1 -m_2 }\G^{{\ellvq}_1 {\barell}_1 \ell_{3} }_{{\mvq}_1 {\barm}_1 m_{3}}\;.\]
 Here we have used the Gaunt integral
\be
\label{gaunt}
 \G^{\ell_1 \ell_2 \ell_{3} }_{m_1 m_2 m_3}= \int \ud^2 \hat \bn ~ Y_{\ell_1 m_1} (\hat \bn)  Y_{\ell_2 m_2} (\hat \bn) Y_{\ell_3 m_3} (\hat \bn) = \sqrt{\frac{(2\ell_1+1)(2\ell_2+1)(2\ell_3+1)}{4\pi}} \left(
\begin{array}{ccc}
\ell_1 & \ell_2 & \ell_3 \\
0 & 0 & 0 
\end{array}
\right)
\left(
\begin{array}{ccc}
\ell_1 &\ell_2& \ell_3 \\
m_1 & m_2 &m_3
\end{array}
\right)\;.
\ee
The spherical multipole functions $\M$ contain all the contributions (at first--indicated by superscript (1)--and second--superscript (2)--order) for the density, velocity, lensing and gravitational potentials, and their combinations; their explicit expressions will be presented below. 
Here $P_{\Phi}(k)$ is the primordial linear power spectrum of the Bardeen gravitational potential; it will be clear below why it is advantageous to express the second-order GR bispectrum for number counts in terms of a primordial quantity, but we will anticipate it briefly here. The primordial linear Bardeen potential is the only relevant spatial field that has no gauge issue, that--by construction--remains invariant in time, is statistically homogeneous and isotropic and thus for which it makes sense to perform a three-dimensional Fourier transform. In fact, for any quantity that evolves along the line of sight or may be affected by projection effects such as lensing or ISW, or for which the flat-sky approximation does not hold, there is no unambiguous three-dimensional Fourier transform.

It is also useful to provide a relation between the spherical Bessel bispectrum $\eS$ and the  bispectrum in Fourier angular space, $B_{\ell_1\ell_2 \ell_3 }(k_1,k_2,k_3)$:
\bea
\eS_3(\{k\ell m\}_1,\{k\ell m\}_2,\{k\ell m\}_3) =
\left(
\begin{array}{ccc}
\ell_1 &\ell_2& \ell_3 \\
m_1 & m_2 &m_3
\end{array}
\right)
B_{\ell_1\ell_2 \ell_3 }(k_1,k_2,k_3) \, ,
\eea
and between the 3-point function and the bispectrum in Fourier angular space:
\bea
\label{eq:bispectrum_fourier_final}
\langle \Delta_g (\bk_1) \Delta_g (\bk_2) \Delta_g(\bk_3)\rangle  &=& \sum_{\ell_i  m_i=1}^{3}
 \frac{(4\pi)^3 ~ (-i)^{\ell_1+\ell_2+\ell_3}}{k_1k_2k_3\left(2/\pi\right)^{3/2}} Y_{\ell_1 m_1} ({\hat \bk}_1) Y_{\ell_2 m_2} ({\hat \bk}_2) Y_{\ell_3 m_3} ({\hat \bk}_3)  \nonumber\\
&& \times \left(
\begin{array}{ccc}
\ell_1 &\ell_2& \ell_3 \\
m_1 & m_2 &m_3
\end{array}
\right)
B_{\ell_1\ell_2 \ell_3 }(k_1,k_2,k_3) \, , 
\eea
after resumming over\footnote{The denominator of the first term in Eq.~(\ref{eq:bispectrum_fourier_final}) reads $\aleph_\ell^*(k_1)\aleph_\ell^*(k_2)\aleph_\ell^*(k_3)$ [see Eqs (\ref{Bispectrum}) and (\ref{Bispectrum-2})], with the coefficients being $\aleph_\ell = k\left(2/\pi\right)^{1/2}$ in the formalism of~\citep{Verde:2000xj, Yoo:2013tc}, while they become $\aleph_\ell(k)=4\pi i^\ell$ in the total angular momentum formalism of~\citep{Dai:2012bc, Dai:2012ma} (for details, see Appendix  \ref{klm-basis}).
$\{\ell_1  m_1, \ell_2 m_2, \ell_3 m_3\}$. Here and hereafter $\hat{\bf k}$ denotes the angular  position  on the unit sphere of the corresponding unit vector. }  

As usual, $Y_{\ell m}$  here denote  the spherical harmonics and $\left(
\begin{array}{ccc}
\ell_1 &\ell_2& \ell_3 \\
m_1 & m_2 &m_3
\end{array}
\right)$ the Wigner 3-j symbols. 
In our case, the bispectrum in Fourier angular space is written as
\bea
B_{\ell_1\ell_2 \ell_3 }(k_1,k_2,k_3) &=& \nsum_{abc} \sqrt{\frac{(2\ell_1+1)(2\ell_2+1)(2\ell_3+1)}{4\pi}} 
 \sum_{{\ellvp}_1  {\ellvq}_1 {\barell}_1} (-1)^{-({\barell}_1 +  {\ellvq}_1 +{\ellvp}_1)}  \frac{(2{\barell}_1+1)(2{\ellvq}_1+1)(2{\ellvp}_1+1)}{4\pi} \nonumber \\
&&\times   \left(
\begin{array}{ccc}
{\ellvq}_1 & \ell_1 & {\ellvp}_1 \\
0 & 0 & 0
\end{array}
\right)
\left(
\begin{array}{ccc}
{\ellvp}_1 & \ell_2 & {\barell}_1 \\
0 & 0 & 0
\end{array}
\right)
\left(
\begin{array}{ccc}
 {\barell}_1 & \ell_3 & {\ellvq}_1 \\
0 & 0 & 0
\end{array}
\right)
  \left\{
\begin{array}{ccc}
\ell_1 &\ell_2& \ell_3 \\
{\barell}_1 &  {\ellvq}_1 &{\ellvp}_1 
\end{array}
\right\} \nonumber\\
&&\times  (-1)^{\ell_2 +\ell_3} \int \frac{q_2^2\ud q_2}{(2\pi)^3}   \frac{q_3^2\ud q_3}{(2\pi)^3}\left[ \K_{\ell_1 {\ellvp}_1  {\ellvq}_1 {\barell}_1 }^{a(2)}(k_1; q_2, q_3)  ~ \M_{\ell_2}^{b(1)}(k_2, q_2)  ~\M_{\ell_3}^{c(1)}(k_3, q_3)\right] P_\Phi(q_2) P_\Phi(q_3) \nonumber \\
&+& ~{\rm cyc}\, ,
\eea
where
$(-1)^{m_1}
 \G^{\ell_1 {\ellvp}_1 {\ellvq}_1 }_{-m_1  {\mvp}_1 {\mvq}_1}  \K_{\ell_1 {\ellvp}_1  {\ellvq}_1 {\barell}_1 }^{a(2)}(k_1; q_2, q_3) = \M_{\ell_1 m_1 {\ellvp}_1 {\mvp}_1 {\ellvq}_1 {\mvq}_1 {\barell}_1 {\barm}_1}^{a(2)}(k_1; q_2, q_3)$.

Thus all local and integrated effects due to, for example, the bias (see Sec.\ref{sec:bias}), lensing, SW,  gravitational evolution,  non-Gaussianity  etc., are enclosed  in the spherical multipole functions  ${\cal M}$. [Precisely, the generating functions are in  Eqs. (\ref{delta_g^(2)}-\ref{v^(2)}), (\ref{MdeltaPhi(2)}-\ref{M-TQ}), (\ref{vphi}-\ref{vv}), (\ref{Ixall}-\ref{nablaperpTxall}),  (\ref{SperpxAll}-\ref{SperpSperp}),  (\ref{T^(2)}-\ref{I^(2)}),  (\ref{kappa^(2)}), (\ref{phiperpS}-\ref{phinablaT}), (\ref{M-Deltag2-gamma-sq}), (\ref{M-gamma-Post-Born}), (\ref{M-T^(2)-2}-\ref{M-I^(2)-2}) and (\ref{M-kappa^(2)-2}).]

The rest of the paper is devoted to  compute the full expression of these functions (see Secs.\ \ref{sec:kernels} and   \ref{calM-second_order} ), 
their interplay in the calculation and some subtle cancellations of the bispectrum building blocks $\Upsilon$, see Appendix \ref{Sec:Upsilon}.
Finally,  in Sec.\ref{sec:bias}, we explicitly show how to include the second order galaxy bias in the expression for the overdensity  (this is  also a contribution to the functions ${\cal M}$).
Throughout the paper we assume the following conventions: units, $c=G = 1$; signature $(-, +, +, +)$; Greek indices run over $0, 1, 2, 3$, and Latin ones over $1, 2, 3$.

second-order


\section{Second-order number counts on the light cone}
\label{sec:Second-ordernumbercounts}
In this section we start presenting all GR effects that have been computed previously by\footnote{This result was obtained by using  
the ``cosmic rulers" approach developed in \cite{Jeong:2011as, Schmidt:2012ne}, and generalized in \cite{Bertacca:2014dra, Bertacca:2014wga} at second order.} \cite{Bertacca:2014hwa}.
We begin by assuming a concordance background model, and at first-order we neglect anisotropic stress,
vector and tensor perturbations. 
In the Poisson gauge, the metric and peculiar velocity   expanded to second order are \cite{Matarrese:1997ay}
 \begin{eqnarray} 
 \label{Poiss-metric}
 \ud s^2 &=& a(\eta)^2\left\{-\left(1 + 2\Phi +\Phi^{(2)}\right)\ud\eta^2+2\omega_{i}^{(2)}\ud\eta \, \ud x^i+\left[\delta_{ij} \left(1 -2\Phi -\Psi^{(2)}\right)+\frac{1}{2}\hat h_{ij}^{(2)}\right]\ud x^i\ud x^j\right\} ,\\
v^{i  }&=& \p^i v +\frac{1}{2}v^{i (2)},~~ v^{i (2)}= \p^i v^{(2)}+ \hat v^{i (2)}, 
\end{eqnarray}
where  $\eta$ is the conformal time, $a$ is the scale-factor, and we have omitted the superscript ${(1)}$  indicating terms of the first-order expansion on familiar quantities such as the metric perturbation  $\Phi$, Newtonian potential $\Psi$ and  the galaxy peculiar velocity $\p^iv$.  Here and hereafter, second-order terms and indicated by the superscript (2).
 Here $ \hat \omega_i^{(n)}$ is a solenoidal vector, i.e. $\p^i\hat \omega_i^{(n)}=0$, and $\hat h^{(n)}_{ij}$ the tensor perturbation, i.e. $\p^i \hat h^{(n)}_{ij}=\hat h_i^{i(n)}=0$, see \cite{Bardeen:1985tr, Kodama:1985bj, Bertschinger:1993xt, Ma:1995ey, Matarrese:1997ay, Mukhanov:2005sc, Malik:2008im}.

Redshift-space or {\it redshift-frame} is the ``cosmic laboratory'' where we probe the observations. In redshift-space we use coordinates which effectively flatten our past light-cone so that the photon geodesic from an observed galaxy has the following conformal space-time coordinates \cite{Jeong:2011as, Schmidt:2012ne, Bertacca:2014wga}:
\begin{equation}
\bar{x}^\mu=(\eta,\; \bar {\bf x})=(\eta_0-\bar \chi, \; \bar \chi \, \hat {\bf n}).
\end{equation}
Here $\bar \chi(z)$ is the comoving distance to the observed redshift, calculated in the background  (i.e., a  redshift-space quantity), $\hat {\bf n}$ is the observed direction to the  galaxy, i.e. $\hat n^i=\bar x^i/\bar \chi=\delta^{ij} (\p \bar \chi/\p \bar x^j)$ . Using $\bar \chi$ as an affine parameter in the redshift frame (at zeroth order), the total derivative along the past light-cone is $\ud / \ud \bar \chi = - \p/ \p \eta + \hat n^i \p/\p \bar x^i$.  
In our analysis we use only the observed redshift $z$ rather than the background (Hubble flow) redshift. In particular all background quantities are not evaluated at the background, redshift (i.e. the redshift that would have been observed for the same source without any perturbations along the line of sight), they are instead evaluated at the observed redshift which include real world effects such as peculiar velocities.

Defining $x^\mu (\chi)$ as the coordinates in the {\it physical frame} [see Eq.~(\ref{Poiss-metric})], where $\chi$ is the physical comoving distance of the source,  we can set up a mapping between redshift space and real space (the ``physical frame")  up to second order in the following way: $x^\mu(\chi)=\bar{x}^\mu(\bar\chi)+\Delta x^{\mu (1)} (\bar \chi)+\Delta x^{\mu (2)} (\bar \chi)/2$. 

In this work, $\perp$ denotes projection into the screen space (with projector $\Perp^{ij}= \delta^{ij} -  \hat n^i \hat n^j$), $\|$ indicates projection along the unit line-of-sight vector $\hat n^i$, and we define the derivatives 
\begin{eqnarray}
&&\partial_\parallel=n^j \partial_j,
~~ \partial_\perp^i = \Perp^{ij}\partial_j= \partial^i -  n^i \partial_\parallel
,~~ \nabla^2_\perp = \partial_{\perp i}\partial_\perp^i =\frac{1}{\bar \chi^2}  \triangle_{\hat \bn} =\nabla^2
- \partial_\parallel^2 - \frac{2}{\bar \chi}\partial_\parallel.
\end{eqnarray}

Now  we want to study the physical number density  of galaxies $n_g$ as a function of the physical comoving coordinates $x^\mu$ and the magnification $\M$.
In particular we consider the cumulative physical number density sample with a flux larger than a observed limit $\bar \F$ which can be translated  in terms of a the inferred threshold luminosity $\bar L(z)$  ($n_g=N$  in \cite{Challinor:2011bk, DiDio:2013bqa}). 
The  physical number  density  contained within a volume $\mathcal{V}$ is given by
\begin{eqnarray}
 \label{N}
\mathcal{N}=
&&\int_{ \mathcal{\bar V}} \sqrt{-\hat g(x^\alpha)}\: a^3(x^0) \: n_g(x^\alpha,  \M)\: \ud \mathcal{V}\;,
\end{eqnarray}
where $ n_g(x^\alpha,  \M)$ is the physical number density  which occupies the comoving physical volume $\ud \mathcal{V}$, $a$ is  the scale factor, $\sqrt{- \hat g}=\sqrt{- g}/ a^4$, $\hat g^{\mu \nu}$ is the comoving metric.
In the redshift frame is, by definition,
\begin{equation}
\label{N2}
\mathcal{N} =
\int_{ \mathcal{\bar V}} {\bar a^3(\bar x^0) \,  n_g\left(\bar x^0,\bar {\bf x}, \bar L \right)} \, \ud^3\bar {\bf x} \;,
\end{equation}
where the observed comoving volume is $\ud^3\bar {\bf x} = \ud \mathcal{ \bar V}$.
Then,  relating the observed galaxy number density with the physical one, i.e. Eqs.\ (\ref{N}), (\ref{N2}),  we  obtain the observed fractional number overdensity
\begin{equation}
\label{Delta_g}
\Delta_g=\frac{ n_g (\bar x^0, \bar{\bf x}, \bar L) - \bar n_g(\bar x^0, \bar L)}{ \bar n_g(\bar x^0, \bar L)} = \Delta_g^{(1)} + \frac{1}{2}\Delta_g^{(2)}\;,
\end{equation}
where
\begin{eqnarray}
\label{Deltag-1}
\Delta_g^{(1)} &=& \frac{\Delta n_g^{(1)}}{\bar n_g}+ 3 \Delta \ln a^{(1)} + \Delta \sqrt{-g}^{(1)}+\Delta V^{(1)}
\end{eqnarray}
and 
\begin{eqnarray}
\label{Deltag-2}
&&\Delta_g^{(2)} = \frac{\Delta n_g^{(2)}}{\bar n_g}+ 3 \Delta \ln a^{(2)} + \Delta \sqrt{-g}^{(2)}+\Delta V^{(2)} + 6 \left( \frac{\Delta n_g^{(1)}}{\bar n_g}+ \Delta \sqrt{-g}^{(1)}+\Delta V^{(1)}\right) \, \Delta \ln a^{(1)}+ 6 \left( \Delta \ln a^{(1)}\right)^2\nonumber \\
&&+ 2  \frac{\Delta n_g^{(1)}}{\bar n_g}  \Delta V^{(1)} + 2 \Delta \sqrt{-g}^{(1)} \Delta V^{(1)}+ 2  \frac{\Delta n_g^{(1)}}{\bar n_g}  \Delta \sqrt{-g}^{(1)}  \;.
\end{eqnarray}
Here we have considered the corrections up to second order of the volume:
\be 
\bigg|{\ud \mathcal{V} \over  \ud  \mathcal{ \bar V} }\bigg| = 1 + \Delta V^{(1)} + \frac{1}{2}\Delta V^{(2)}\;,
\ee
the scalar factor:
\begin{equation}
\label{a}
a(x^0(\chi))=\bar a \left(1 +\Delta \ln a^{(1)}+\frac{1}{2} \Delta \ln a^{(2)}\right)\;,
\end{equation}
where $\bar a = a(\bar x^0(\bar \chi))=1/(1+z)$, and
\begin{eqnarray}
\label{deltaSqrtg}
\sqrt{-\hat g(x^\alpha)}=1+\Delta\sqrt{-\hat g (\bar x^\alpha)}^{\,(1)}+\frac{1}{2}\Delta\sqrt{-\hat  g (\bar x^\alpha)}^{\,(2)}\;.
\end{eqnarray}
Finally, we used
\begin{equation}
\label{Deltang}
n_g(x^\alpha,\M)=\bar n_g(\bar x^0, \bar L) +\Delta n_g(\bar x^\alpha, \bar L)^{(1)}+\frac{1}{2}\Delta n_g(\bar x^\alpha, \bar L)^{(2)}\;,
\end{equation}
where  $\Delta n_g(\bar x^\alpha, \bar L)^{(1)}$ and $\Delta n_g(\bar x^\alpha, \bar L)^{(2)}$ contain also the fluctuation of the luminosity distance at the observed redshift $z$  (e.g. see \cite{Bertacca:2014wga}). In order to write explicitly all above relations, first of all, let us define at first-order the following quantities: 
\begin{itemize}
\item  the scale factor correction\footnote{From now on, in the expressions we have removed all the terms that are evaluated by the observer, as they represent unobservable monopole-like terms.}
\be
\label{Poiss-Deltalna-1}
\Delta \ln a^{(1)}  = - \Phi + \p_\| v + 2I^{(1)}  \;, 
\ee
where $I^{(1)}$ is the integrated Sachs-Wolfe (ISW) effect at first-order, i.e.
\be \label{Poiss-iota}
I^{(1)}   =  - \int_0^{\bar \chi} \ud \tilde \chi \, \Phi {'}  \;,
\ee
where, here and hereafter, the prime $'$ denotes $\p/\p\eta$;
 
 \item the weak-lensing convergence term
\be
 \kappa^{(1)} = \int_0^{\bar \chi} \ud \tilde \chi  \left(\bar \chi-\tilde \chi\right) \frac{\tilde \chi}{ \bar \chi} \,   \tilde \nabla^2_\perp \Phi\;; 
 \ee

\item  the weak-lensing shear $\gamma_{ij}^{(1)}$ and rotation $\vartheta_{ij}$ terms, where

\bea
\gamma_{ij}^{(1)} &=&  2 \int_0^{\bar \chi} \ud \tilde \chi \left[\left(\bar \chi-\tilde \chi\right) \frac{\tilde \chi}{ \bar \chi}  \tilde \p_{\perp (i}  \tilde \p_{\perp j)} \Phi \right]-\Perp_{ij} \kappa^{(1)}\;,\label{gamp}  \\
\vartheta_{ij}^{(1)}\vartheta^{ij(1)} &=&   +\frac{2}{\bar \chi^2}  \int_0^{\bar \chi} \ud \tilde \chi\bigg[ \left(\bar \chi-\tilde \chi\right)  \tilde \p_{\perp i}  \Phi \bigg]  \times \int_0^{\bar \chi} \ud \tilde \chi\bigg[ \left(\bar \chi-\tilde \chi\right)  \tilde \p^i_{\perp}  \Phi \bigg] \;,\label{varp} 
\eea
where $\tilde \p_i= \p /\p \tilde x^i$;

\item the radial displacement at first order that corresponds to the usual (Shapiro) time-delay (STD) term \cite{Challinor:2011bk} 
\be
\label{Poiss-s-1}
 T^{(1)} =- 2 \int_0^{\bar \chi} \ud \tilde \chi \Phi \;;
\ee

\item we define the following 3D nonlocal vector \cite{Bertacca:2014dra}
 \be
S^{i (1)} = -  \int_0^{\bar \chi} \ud \tilde \chi \left( \tilde\p^i \Phi   -\frac{1}{\tilde \chi} \hat n^i\Phi  \right)
\label{Poiss-varsigma}
\ee
which can be split as
\be
S_{\perp}^{i (1)} =\Perp^i_j S_{\perp}^{j (1)}= - \int_0^{\bar \chi} \ud \tilde \chi \, \tilde\p^i_\perp \Phi   \;, \quad {\rm and} \quad S_{\|}^{(1)}  = \hat n_i S^{i (1)} = -  \Phi + I^{(1)}  + \int_0^{\bar \chi} \ud \tilde \chi \frac{\Phi}{\tilde \chi} \;.
\ee
The physical meaning of this vector can be understood by noting that the transverse part, i.e. $S_{\perp}^{i (1)}$, can be related with $ T^{(1)}$ and $ \kappa^{(1)}$ in the following way
$$-2 \kappa^{(1)} =2 \p_{\perp i} S^{i (1)} - \p_{\perp i} T^{(1)}\;.$$
\item Finally the overdensity
\begin{equation}
\delta_g^{(1)}=\delta_g(\bar x^\alpha, \bar L)^{(1)}=\frac{{n_g(\bar x^\alpha, \bar L)}^{(1)}}{{n_g(\bar x^0, \bar L)}^{(0)}}\;, 
\label{eq:deltagfirstorder}
\end{equation}
where
$\bar n_g={n_g(\bar x^0, \bar L)}^{(0)}$ is the background number density of sources with luminosity exceeding $\bar L$ and  $n_g^{(1)}={n_g(\bar x^\alpha, \bar L)}^{(1)}$. 
\end{itemize}

Expanding the galaxy fractional number overdensity, at linear order we find  \cite{Challinor:2011bk, Jeong:2011as,Bertacca:2014dra, Bertacca:2014wga, Bertacca:2014hwa}  
\begin{eqnarray}
\label{Deltag1}
\Delta_g^{(1)} &=&\delta_g^{(1)} +\left[ b_e  - \frac{\cH'}{\cH^2} - 2 \Q  - 2 \frac{\left(1- \Q \right)}{\bar \chi \cH} \right]  \Delta \ln a^{(1)}  +\left(-1+2 \Q \right)  \Phi - \frac{1}{\cH}  \p_\|^2 v  + \frac{1}{\cH} \Phi {'} -2 \frac{\left(1- \Q \right)}{\bar \chi}  T^{(1)}  \nonumber\\
&& - 2 \left(1- \Q \right) \kappa^{(1)} \; ,
\end{eqnarray}
where
\begin{eqnarray}
\label{be}
b_e =\frac{\p \ln \bar n_g (\bar a , \bar L)}{\p \ln \bar a}+3
\end{eqnarray}
is the evolution bias term related to the background number density,
and $\cH=a'/a$ is the conformal Hubble factor and we have defined the background magnification bias  as
\begin{equation}
\Q(\bar x^0, \bar L)=-\frac{\p \ln \bar n_g}{\p \ln \bar L}\bigg|_{\bar a}\;.
\end{equation}

All the quantities defined at linear order above, at second order become~\cite{Bertacca:2014dra, Bertacca:2014wga, Bertacca:2014hwa}
\begin{eqnarray}
\label{Poiss-Deltalna-2}
 && \Delta\ln a^{(2)} =- \Phi^{(2)}+ \p_\| v^{(2)}+ \hat v^{(2)}_\| +  3  {\Phi}^2  -  \left( \p_\| v \right)^2+ \p_{\perp i} v \,  \p^i_{\perp} v  -2  \p_\| v \,  \Phi  - \frac{2}{\cH}\left( \Phi  -  \p_\| v \right) \left(  \Phi {'} - \p_\|^2 v  \right)  - 4 \bigg(  \Phi   +  \frac{1}{\cH}  \p_\|^2 v  \nonumber \\
&&   - \frac{1}{\cH}  \Phi {'} \bigg) I^{(1)}   +2  \;   \left(2\Phi{'} +\p_\| \Phi -  \p_\|^2 v \right) T^{(1)}  +   4 \bar \chi \p_{\perp i}\left(-\Phi  +   \p_\| v  \right)   S_{\perp}^{i (1)}   +2 \bigg[\bar \chi \p_{\perp i}\left(\Phi  -   \p_\| v  \right)   + \p_{\perp i} v \bigg]  \p_{\perp}^i T^{(1)}+ 8\Phi  \kappa^{(1)}   \nonumber \\
&&    + 2I^{(2)} + 8  \left(I^{(1)}\right)^2    +4 \int_0^{\bar \chi}  \ud \tilde{\chi} \Bigg[   \Phi{''}  T^{(1)}  + 2 \Phi \Phi{'}  +  2 \Phi{'}  I^{(1)}    + 2 \Phi    \tilde \p_{\perp j}S_{\perp}^{j(1)}    - 2  \tilde \chi  \tilde \p_{\perp i} \Phi{'}    S_\perp^{i(1)}   - 2 \bigg( \frac{\ud \Phi}{\ud \tilde \chi}     -  \frac{1}{\tilde\chi}  \Phi  \bigg) \kappa^{(1)}  \nonumber\\
    &&       + \tilde \chi  \tilde \p_{\perp i} \Phi{'}  \tilde \p_\perp^i T^{(1)} \Bigg]\;,
\end{eqnarray}

\begin{eqnarray}
 \label{Poiss-kappa-2}
 &&\kappa^{(2)}= \frac{1}{2}  \int_0^{\bar \chi} \ud \tilde \chi  \left(\bar \chi-\tilde \chi\right) \frac{\tilde \chi}{ \bar \chi}   \tilde \nabla^2_\perp \left( \Phi^{(2)} + 2 \omega^{(2)}_{\| }+\Psi^{(2)} - \frac{1}{2} \hat h^{(2)}_{\| } \right)      + \frac{1}{2}  \int_0^{\bar \chi} \ud \tilde \chi \bigg(-2  \tilde \p_\perp^i \omega_i^{ (2)} + \frac{4}{\tilde\chi} \omega_\|^{ (2)} + \Perp^{ij} n^k  \tilde \p_i  \hat h_{jk}^{ (2)}  - \frac{3}{\tilde \chi}  \hat h_\|^{ (2)}\bigg)    \nonumber \\
  &&    - \frac{2}{ \bar \chi}\bigg( 2 \bar\chi I^{(1)} +2\int_0^{\bar \chi} \ud \tilde \chi \tilde \chi \Phi {'}  +T^{(1)}   +  \frac{1}{\cH}\Delta \ln a  \bigg)   \left(\kappa^{(1)}  - \frac{\bar \chi}{2}\nabla^2_\perp T^{(1)}\right)   +2S_{\perp}^{i }   \bigg(  \p_{\perp i } T^{(1)} +  \frac{1}{\cH} \p_{\perp i}  \Delta \ln a^{(1)} 
  + 2\bar \chi\p_{\perp i}  I^{(1)}      \nonumber \\
    &&  + 2\p_{\perp i}  \int_0^{\bar \chi} \ud \tilde \chi   \tilde \chi  \Phi{'}    \bigg) +2 \int_0^{\bar \chi} \ud \tilde{\chi}  \frac{ \tilde \chi }{\bar \chi} \bigg[  + 2\tilde \chi   \tilde \nabla^2_{\perp} \Phi I^{(1)}  + 2  \tilde \nabla^2_{\perp} \Phi \int_0^{\tilde \chi} \ud \tilde{\tilde \chi} \tilde{\tilde \chi}  \Phi {'}   + 2  \tilde \chi \p_{\perp i}  \Phi   \tilde \p_\perp^i I^{(1)} + 2 \p_{\perp i}  \Phi   \tilde \p_{\perp i} \int_0^{\tilde \chi} \ud \tilde{\tilde \chi} \tilde{\tilde \chi}  \Phi {'}   + \frac{4}{\tilde \chi}   \Phi  S_{\|}^{(1)}   \nonumber \\ 
&& - 2 \Phi   \tilde \p_{\perp m} S^{m (1)}   -  \tilde \p_{\perp i} \Phi  \p_\perp^i T^{(1)}  - \frac{2}{\tilde \chi}\Phi  \kappa^{(1)}\bigg]  + 2\int_0^{\bar \chi} \ud \tilde \chi \left(\bar \chi-\tilde \chi\right)\frac{\tilde \chi}{ \bar \chi}  \Bigg[ - 2\tilde\p^i_\perp  \Phi   \tilde \p_{\perp i} \Phi + 2 \tilde\p^i_\perp  \Phi  \tilde \p_{\perp i} I^{(1)}   - 2 \Phi  \tilde \nabla^2_\perp \Phi + 2  \tilde \nabla^2_\perp \Phi I^{(1)}    \nonumber\\
&&   - \tilde \nabla^2_{\perp} \Phi  ~ T^{(1)}    -   \tilde \p_{\perp }^i    \Phi{'}  \tilde \p_{\perp i} T^{(1)}   +  \frac{2}{\tilde\chi}  \bigg( -     \frac{1}{\tilde\chi}\Phi +  \frac{\ud}{\ud \tilde \chi} \Phi\bigg)  \kappa^{(1)}  
 +  \frac{1}{\tilde\chi} \tilde \p_{\perp i}  \Phi ~ S_{\perp}^{i(1)}    - \frac{3}{2 \tilde \chi}  \tilde \p_{\perp i} \Phi   \p_\perp^i T^{(1)}    + \tilde \chi \bigg(  \tilde \p_{\perp i} \tilde \nabla^2_{\perp} \Phi + \frac{1}{\tilde\chi} \tilde \p_{\perp i}  \Phi{'}    \bigg) \nonumber\\
&& \times \left( 2   S_\perp^{i(1)} -  \p_\perp^i T^{(1)}\right)+  \tilde\chi \tilde \p^{(j}_{\perp} \tilde \p^{m)}_{\perp} \Phi \left( 2 \tilde \p_{\perp(m}  S_{\perp j)}^{(1)} -  \tilde \p_{\perp (m} \tilde  \p_{\perp j)}  T^{(1)}\right) +  2\Phi{'} \tilde \p_{\perp m} S_{\perp}^{m(1)} - \left( \Phi{'}   + \frac{1}{\tilde \chi} \Phi  \right) \tilde \nabla_\perp^2 T^{(1)} \Bigg] \;,
  \end{eqnarray}
  where $ A_{(i} B_{j)} = \left(A_{i} B_{j}+ B_{i}A_{j}\right)/2$, 
 and
\begin{eqnarray}
\label{Poiss-iota2}
I^{(2)}& =& -\frac{1}{2} \int_0^{\bar \chi} \ud \tilde \chi \left(\Phi^{(2)}{'} +2  \omega^{(2)}_{\| }{'} + \Psi^{(2)}{'}- \frac{1}{2} \hat h^{(2)}_{\| }{'} \right),
\\
\label{Poiss-varsigma2}
S_{\perp}^{i(2)} &=& -\frac{1}{2} \int_0^{\bar \chi} \ud \tilde \chi \left[ \tilde\p^i_\perp \left( \Phi^{(2)} +2  \omega^{(2)}_{\| } + \Psi^{(2)}- \frac{1}{2} \hat h^{(2)}_{\| }\right) + \frac{1}{\tilde \chi} \left(-2 \omega^{i (2)}_{\perp }+  n^k \hat h_{kj}^{(2)} \Perp^{ij}  \right)\right], \\
\label{Poiss-s-1}
 T^{(2)} &=& - \int_0^{\bar \chi} \ud \tilde \chi \left(\Phi^{(2)} +2 \omega^{(2)}_{\| }+ \Psi^{(2)}- \frac{1}{2}h^{(2)}_{\| }\right) .
\end{eqnarray}
Finally, at second order the galaxy fractional overdensity is \cite{Bertacca:2014dra, Bertacca:2014wga, Bertacca:2014hwa}
\begin{eqnarray}
\label{Deltag-2}
&&\Delta_g^{(2)} = \delta_g^{(2)}  +\left[ b_e-2 \Q  -   \frac{\cH'}{\cH^2} - \left(1 - \Q\right) \frac{2}{\bar \chi \cH} \right] \Delta \ln a^{(2)}  - \left(1 - \Q\right) \left( 2  \Psi^{(2)} +\frac{1}{2}\hat h_{\|}^{(2)} \right)    - \left(1- \Q \right) \frac{2}{\bar \chi} T^{(2)} - 2 \left(1- \Q \right) \kappa^{(2)} \nonumber \\  
&&  + \Phi^{(2)}+  \frac{1}{ \cH} \Psi^{(2)}{'}-  \frac{1}{2 \cH} \hat h^{(2)}_{\| }{'} -\frac{1}{ \cH }\p_\|^2 v^{(2)} -\frac{1}{ \cH }   \p_\|\hat v^{(2)}_\| +2\left(-1+2\Q\right)\Phi \delta_g^{(1)} - \frac{2}{\cH} \delta_g^{(1)} \p_\|^2 v  +\frac{2}{\cH}\delta_g^{(1)}  \Phi {'} + \left( \p_\| v  \right)^2 \nonumber \\ 
&&  + \frac{2}{\cH}\left(2\Q+\frac{\cH'}{\cH^2}\right)\Phi \Phi {'} + \left(-5 + 4\Q  +4 \Q^2 - 4 \frac{\p \Q}{\p \ln \bar L} \right){\Phi}^2    - \frac{2}{\cH}\left(1+ 2\Q  +\frac{\cH'}{\cH^2}\right)\Phi \p_\|^2 v + \frac{2}{\cH^2}\left( \Phi {'}  \right)^2 + \frac{2}{\cH^2}\left(\p_\|^2 v  \right)^2   \nonumber \\
 &&  +\frac{4}{ \cH }\p_\| v  \p_\| \Phi   - \frac{2}{\cH^2} \Phi \p_\|^3 v   -\frac{2}{\cH}\Phi \p_\| \Phi    + \frac{2}{\cH^2}\Phi \frac{\ud \Phi {'} }{\ud \bar \chi}- \frac{2}{\cH^2} \p_\| v \frac{\ud \Phi {'} }{\ud \bar \chi}   - \frac{2}{\cH^2}\Phi  \p_\|^2 \Phi   -  \frac{4}{\cH^2}\p_\|^2 v   \Phi {'}     +\frac{2}{\cH} \left(1+\frac{\cH'}{\cH^2} \right) \p_\| v \p_\|^2 v \nonumber \\
 && + \frac{2}{\cH^2} \p_\| v  \p_\|^2 \Phi  + \frac{2}{\cH} \left(1 -\frac{\cH'}{\cH^2} \right)  \p_\| v  \Phi {'}+\frac{2}{\cH}\p_{\perp i} v \p^i_\perp \Phi   -\frac{4}{ \cH } \p_{\perp i} v   \p_{\perp}^i \p_\|  v   +\left(-1  +\frac{4}{ \bar \chi \cH } \right) \p_{\perp i} v    \p_{\perp}^i v +  \frac{2}{\cH^2}\p_\| v \p_\|^3v   \nonumber \\
  && + \Bigg\{ \bigg[  -2 b_e - 4 \Q   + 4 b_e \Q  - 8 \Q^2  + 8 \frac{\p \Q}{\p \ln \bar L} +4 \frac{\p \Q}{\p \ln \bar a}   + 2  \frac{\cH'}{\cH^2} \left(1 -2 \Q\right)   + \frac{4}{\bar \chi \cH}\bigg(-1+\Q +2\Q^2 - 2\frac{\p \Q}{\p \ln \bar L} \bigg) \bigg] \Phi \nonumber \\
  &&    +2 \left[ b_e - 2\Q   -  \frac{\cH'}{\cH^2}  - \frac{2}{\bar \chi \cH} \left(1 - \Q\right)  \right] \delta_g^{(1)}   - \frac{2}{\cH}  \frac{\ud  \delta_g^{(1)} }{\ud \bar \chi} +  \frac{2}{\cH}  \left[ -   b_e   + 2 \Q     +  \frac{\cH' }{\cH^2}    +  \frac{2}{\bar \chi \cH}  \left(1 - \Q\right)\right]  \p_\|^2 v  - \frac{4}{\cH} \Q \p_\|  \Phi  \nonumber \\    
&&    + \frac{2}{\cH}\left[  -2  +  b_e  - \frac{\cH' }{\cH^2}  -   \frac{2}{\bar \chi \cH} \left(1 - \Q\right)\right]  \Phi {'}     + 4\bigg[  - \bigg(b_e-b_e\Q +2 \Q^2 -2\frac{\p \Q}{\p \ln \bar L}-\frac{\p \Q}{\p \ln \bar a} \bigg) +  \frac{\cH'}{\cH^2} \left(1 - \Q\right)  \nonumber \\    
 &&     + \frac{1}{\bar \chi \cH}  \left(1-\Q +2\Q^2  -2\frac{\p \Q}{\p \ln \bar L}\right) \bigg] \left(\frac{T^{(1)}}{\bar \chi}+\kappa^{(1)} \right)\Bigg\}\Delta \ln a^{(1)}  + \Bigg\{-b_e +b_e^2+ \frac{\p b_e}{\p \ln \bar a}+6\Q-4 \Q b_e+4 \Q^2  -4\frac{\p \Q}{\p \ln \bar L} \nonumber \\    
&&   - 4\frac{\p \Q}{\p \ln \bar a} + \left(1-2 b_e + 4\Q \right) \frac{\cH' }{\cH^2} -\frac{\cH'' }{\cH^3} +3\left( \frac{\cH' }{\cH^2} \right)^2 + \frac{6}{\bar \chi} \frac{\cH' }{\cH^3} \left(1 - \Q\right) +  \frac{2}{\bar \chi^2 \cH^2} \bigg(1-\Q +2\Q^2  -2\frac{\p \Q}{\p \ln \bar L}\bigg) \nonumber \\
&&      + \frac{2}{\bar \chi \cH} \left[ 1 - 2b_e - \Q  + 2b_e \Q  - 4 \Q^2 +4\frac{\p \Q}{\p \ln \bar L} +2 \frac{\p \Q}{\p \ln \bar a} \right]    \Bigg\} \left(\Delta \ln a^{(1)}\right)^2 + 4  \bigg[  +\frac{1}{\cH} \left(1 -\frac{\cH' }{\cH^2}\right)\Phi {'} + \frac{1}{\cH}   \p_\| \Phi   \nonumber \\
&&  +  \frac{1}{\cH} \left(1 +  \frac{\cH' }{\cH^2} \right)\p_\|^2 v    +  \frac{1}{\cH^2} \p_\|^2 \Phi  + \frac{1}{\cH^2} \p_\|^3 v   - \frac{1}{\cH^2}\frac{\ud \Phi {'} }{\ud \bar \chi}  \bigg] I^{(1)} + \bigg[ - \frac{4}{\bar \chi}  \left(1- \Q \right) \delta_g^{(1)}   - 2   \p_{\|}\delta_g^{(1)}   - \frac{4}{\bar \chi \cH} \left(1  - \Q \right)  \Phi {'}  \nonumber \\
 &&  + \frac{4}{\bar \chi} \left( -1 + \Q+2 \Q^2- 2\frac{\p \Q}{\p \ln \bar L} \right)  \Phi +2\left(1-2 \Q\right) \p_{\|}\Phi  + \frac{4}{\bar \chi \cH} \left(1- \Q \right) \p_\|^2 v  +\frac{2}{\cH} \p_\|^3 v.   -\frac{2}{\cH} \p_\| \Phi {'}    \bigg] T^{(1)}   \nonumber \\ 
 &&   +\bigg(1-\Q +2\Q^2   -2\frac{\p \Q}{\p \ln \bar L}\bigg) \left[  \frac{2}{\bar \chi^2} \left(  T^{(1)} \right)^2 + \frac{4}{\bar \chi} T^{(1)}   \kappa^{(1)}  \right] +4 \bigg[ - \left(1-\Q -2\Q^2 + 2\frac{\p \Q}{\p \ln \bar L}\right) \Phi  + \frac{1}{\cH} \left(1- \Q \right) \p_\|^2 v  \nonumber \\ 
 &&    -\frac{1}{\cH}\left(1- \Q \right) \Phi {'}   -  \left(1- \Q \right) \delta_g^{(1)} \bigg]\kappa^{(1)} +  \left(1- \Q \right)  \vartheta_{ij}^{(1) }\vartheta^{ij(1)}  + 2\left(1-\Q +2\Q^2- 2\frac{\p \Q}{\p \ln \bar L}\right) \left(\kappa^{(1)}\right)^2  -2  \left(1- \Q \right)\big|\gamma^{(1)}\big|^2 \nonumber \\
&&       + 4 \bigg[  \frac{\bar \chi}{\cH}\bigg(  \p_{\perp i}\Phi {'} -  \p_{\perp i} \p_\|^2 v  \bigg)   +   \bar \chi  \p_{\perp i} \delta_g^{(1)}   +  \bar \chi  \p_{\perp i} \Phi  -2\bar \chi \left(1  -  \Q \right)\p_{\perp i} \Phi +  \frac{1}{\cH}\left(1- \Q \right)   \p_{\perp i}  \Delta \ln a^{(1)} \bigg] S_{\perp}^{i (1)}  \nonumber \\
 &&      -4 (1-\Q)S_{\perp }^{i (1)}S_{\perp }^{j (1)} \delta_{ij}   +2 \bigg[  \frac{2}{\bar \chi \cH} \p_{\perp i} v   -  \frac{\bar \chi}{\cH} \p_{\perp i} \Phi {'} + \frac{\bar \chi}{\cH} \p_{\perp i} \p_\|^2 v   -\frac{2}{\cH}\p_{\perp i}\p_\| v  \   -   \bar \chi  \p_{\perp i} \delta_g^{(1)}   +   \bar \chi   \left(1-  2 \Q \right) \p_{\perp i} \Phi     \bigg]  \p_\perp^i T^{(1)}  \nonumber \\
&&     +4\Q^{(1)} \left[ \Phi  - \left( 1 - \frac{1}{\bar \chi \cH} \right) \Delta \ln a^{(1)}  +\frac{1}{\bar \chi} T^{(1)} +  \kappa^{(1)}\right]   + 8(1-\Q)\Bigg\{  \int_0^{\bar \chi}  \ud \tilde{\chi}\bigg[       -  \Phi  \tilde \p_{\perp m}S_{\perp}^{m(1)}  +  \left( \frac{\ud \Phi}{\ud \tilde \chi}   -  \frac{1}{\tilde\chi} \Phi  \right) \kappa^{(1)}  \bigg]   \nonumber \\
 && - \frac{1}{\bar \chi}\int_0^{\bar \chi}  \ud \tilde{\chi} \bigg(    { \Phi}^2  +  \Phi{'} T^{(1)} +2 \Phi \kappa^{(1)}   + \tilde \chi \tilde \p_{\perp i}\Phi  \tilde \p_\perp^i T^{(1)}   \bigg) +  \frac{1}{\bar \chi} \int_0^{\bar \chi}  \ud \tilde{\chi} ~ (\bar \chi - \tilde \chi) \bigg[  -  2 \Phi   \tilde \p_{\perp m}S_{\perp}^{m(1)}     
 + 2  \left( \frac{\ud  \Phi}{\ud \tilde \chi}   -  \frac{1}{\tilde\chi} \Phi  \right) \kappa^{(1)}   \bigg] \Bigg\}\;.   
 \nonumber \\
   \end{eqnarray}
 where  the density contrast at second order has been defined as
\begin{equation}
 \delta_g^{(2)}= \frac{{n_g(\bar x^\alpha, \bar L)}^{(2)}}{{n_g(\bar x^0, \bar L)}^{(0)}}\;,
\end{equation}
and for convenience of notation we sometimes use 
 $n_g^{(2)}={n_g(\bar x^\alpha, \bar L)}^{(2)}$.
 Here  we have defined the first-order magnification bias in the following way
\begin{equation}
\Q^{(1)}(\bar x^\alpha, \bar L)
=-\frac{\p  \delta_g^{(1)}}{\p \ln \bar L}\Bigg|_{\bar a}\;.
\end{equation}
Let us point out that, using the result from Ref.~\cite{Bertacca:2014hwa}, we have indirectly assumed that galaxy velocities follow the matter velocity field, i.e. no velocity bias, and we have used the velocity equation both at first and second order (for example see Appendix \ref{appendix:A}).

The  readers interested in the primordial non-Gaussianity contribution (or $f_{\rm NL}$ for short) should keep in mind that  it is implicitly enclosed in second-order primordial quantities such as $\Phi^{(2)}$; see e.g., Sec.\ \ref{LS}.

Now, replacing Eqs.~(\ref{Poiss-Deltalna-1}) in Eq. (\ref{Deltag1}) we have the galaxy density contrast at first-order that it is given by 8 terms involving $\delta_g$, first and second derivatives of the velocity, potential, convergence, ISW and STD:
 \begin{eqnarray}
\label{Deltag1-esplicito}
\Delta_g^{(1)} &=&\delta_g^{(1)} - \frac{1}{\cH}  \p_\|^2 v  + \left[ b_e  - \frac{\cH'}{\cH^2} - 2 \Q  - 2 \frac{\left(1- \Q \right)}{\bar \chi \cH} \right] \p_\| v - \left[ b_e  - \frac{\cH'}{\cH^2} - 4 \Q +1 - 2 \frac{\left(1- \Q \right)}{\bar \chi \cH} \right]  \Phi - 2 \left(1- \Q \right) \kappa^{(1)} \nonumber\\
&&+ \frac{1}{\cH} \Phi {'} + 2\left[ b_e  - \frac{\cH'}{\cH^2} - 2 \Q  - 2 \frac{\left(1- \Q \right)}{\bar \chi \cH} \right] I^{(1)}    -2 \frac{\left(1- \Q \right)}{\bar \chi}  T^{(1)} =\sum_{a=1}^8 \Delta_g^{a(1)} \;.
\end{eqnarray}

At second order, replacing Eqs.~(\ref{Poiss-Deltalna-1}),  (\ref{Poiss-Deltalna-2})  and (\ref{Poiss-kappa-2})  in Eq.~(\ref{Deltag-2}), we have
\begin{eqnarray}
\label{Deltag2-explicito}
 \Delta_g^{(2)} =\Delta_{g\, {\rm loc}}^{(2)}+\Delta_{g\,{\rm int}}^{(2)}\;,
 \end{eqnarray}
where we have separated the second-order contribution to $\Delta_g^{(2)}$ in local (subscript loc) and integrated (subscript int) terms.
Here the local contribution is the sum of three terms
 \begin{equation}
 \label{Deltag2-loc}
 \Delta_{g\, {\rm loc}}^{(2)} =\sum_{i=1}^3  \Delta_{g\, {\rm loc-}i}^{(2)}\;,
 \end{equation}
where 
\begin{eqnarray}
\label{Deltag2-loc-1}
&&\Delta_{g\, {\rm loc}-1}^{(2)} = \delta_g^{(2)}   -2  \left(1 - \Q\right) \Psi^{(2)} - \frac{1}{2}\left(1 - \Q\right) \hat h_{\|}^{(2)}   -\left[ b_e-2 \Q -1 -   \frac{\cH'}{\cH^2} - \left(1 - \Q\right) \frac{2}{\bar \chi \cH} \right]  \Phi^{(2)} +  \frac{1}{ \cH} \Psi^{(2)}{'} -  \frac{1}{2 \cH} \hat h^{(2)}_{\| }{'}\nonumber \\
&&  -\frac{1}{ \cH }\p_\|^2 v^{(2)} -\frac{1}{ \cH }   \p_\|\hat v^{(2)}_\| +\left[ b_e-2 \Q  -   \frac{\cH'}{\cH^2} - \left(1 - \Q\right) \frac{2}{\bar \chi \cH} \right] \left( \p_\| v^{(2)}+ \hat v^{(2)}_\|  \right) \
\end{eqnarray}
contains all local contributions with second-order perturbation terms,
\begin{eqnarray}
\label{Deltag2-loc-2}
&&\Delta_{g\, {\rm loc}-2}^{(2)} = 2 \left[- b_e -1+ 4 \Q   +  \frac{\cH'}{\cH^2}  + \frac{2}{\bar \chi \cH} \left(1 - \Q\right)  \right]  \delta_g^{(1)} \Phi +2 \left[ b_e - 2\Q   -  \frac{\cH'}{\cH^2}  - \frac{2}{\bar \chi \cH} \left(1 - \Q\right)  \right]  \delta_g^{(1)} \p_\| v   - \frac{2}{\cH} \delta_g^{(1)} \p_\|^2 v  \nonumber \\
&&   +\frac{2}{\cH}\delta_g^{(1)}  \Phi {'}  +  \frac{2}{\cH}  \frac{\ud  \delta_g^{(1)} }{\ud \bar \chi} \Phi -  \frac{2}{\cH}  \frac{\ud  \delta_g^{(1)} }{\ud \bar \chi} \p_\| v +   \Bigg[ -5 + 4b_e    +b_e^2  + \frac{\p b_e}{\p \ln \bar a} +8 \Q  +16 \Q^2    - 8 b_e \Q   - 16 \frac{\p \Q}{\p \ln \bar L}      -8 \frac{\p \Q}{\p \ln \bar a}   \nonumber \\
&&      +2 \left( -2 -  b_e  + 4 \Q \right) \frac{\cH'}{\cH^2} -\frac{\cH'' }{\cH^3} +3\left( \frac{\cH' }{\cH^2} \right)^2 + \frac{6}{\bar \chi} \frac{\cH' }{\cH^3} \left(1 - \Q\right)    + \frac{4}{\bar \chi \cH} \left(  - 4\Q^2  - b_e  + b_e \Q    + 4\frac{\p \Q}{\p \ln \bar L}   + \frac{\p \Q}{\p \ln \bar a} \right)  \nonumber \\
     &&    +  \frac{2}{\bar \chi^2 \cH^2} \bigg(1-\Q +2\Q^2  -2\frac{\p \Q}{\p \ln \bar L}\bigg)     \Bigg] \Phi^2 +2 \Bigg[ - b_e  - 6 \Q  - b_e^2  + 6 \Q b_e - 8 \Q^2   + 8 \frac{\p \Q}{\p \ln \bar L}   +6 \frac{\p \Q}{\p \ln \bar a}     - \frac{\p b_e}{\p \ln \bar a}  \nonumber \\
&&+ \left(1+ 2b_e -6 \Q \right) \frac{\cH' }{\cH^2}+\frac{\cH'' }{\cH^3} -3\left( \frac{\cH' }{\cH^2} \right)^2  - \frac{6}{\bar \chi} \frac{\cH' }{\cH^3} \left(1 - \Q\right) +  \frac{2}{\bar \chi \cH}  \left(-1+ 2b_e  + \Q +6\Q^2    - 2b_e \Q    - 6\frac{\p \Q}{\p \ln \bar L}   - 2 \frac{\p \Q}{\p \ln \bar a} \right) \nonumber \\
   &&  +  \frac{2}{\bar \chi^2 \cH^2} \bigg(- 1 +\Q - 2\Q^2  + 2\frac{\p \Q}{\p \ln \bar L}\bigg) \Bigg] \Phi \p_\| v +   \frac{2}{\cH}  \bigg[ -1- 5 \Q   +   2b_e  -  3 \frac{\cH'}{\cH^2} -  \frac{4 \left(1 - \Q\right) }{\bar \chi \cH}  \bigg] \Phi \p_\|^2 v     - \frac{2}{\cH^2} \Phi \p_\|^3 v   \nonumber \\
 &&    + \frac{2}{\cH}\bigg(2  - 2b_e +4 \Q   + 3\frac{\cH' }{\cH^2}    +   \frac{4 \left(1 - \Q\right) }{\bar \chi \cH}   \bigg)\Phi \Phi {'} + \frac{2}{\cH} \left(-1 +2 \Q \right)\Phi \p_\|  \Phi   - \frac{2}{\cH^2}\Phi  \p_\|^2 \Phi   + \frac{2}{\cH^2}\Phi \frac{\ud \Phi {'} }{\ud \bar \chi}       + \frac{2}{\cH^2}\left( \Phi {'}  \right)^2+ \frac{2}{\cH^2}\left(\p_\|^2 v  \right)^2  \nonumber \\
  && + \Bigg\{ 1  -2 b_e +b_e^2+ \frac{\p b_e}{\p \ln \bar a} + 8 \Q  -4 \Q b_e+4 \Q^2  -4\frac{\p \Q}{\p \ln \bar L}-4\frac{\p \Q}{\p \ln \bar a} + 2\left(1- b_e + 2\Q \right) \frac{\cH' }{\cH^2}  -\frac{\cH'' }{\cH^3}  +3\left( \frac{\cH' }{\cH^2} \right)^2 \nonumber \\
&&  + \frac{6}{\bar \chi} \frac{\cH' }{\cH^3} \left(1 - \Q\right) + \frac{2}{\bar \chi \cH} \left[ 2 - 2b_e -2 \Q  + 2b_e \Q  - 4 \Q^2 +4\frac{\p \Q}{\p \ln \bar L} +2 \frac{\p \Q}{\p \ln \bar a} \right]  +  \frac{2}{\bar \chi^2 \cH^2} \bigg(1-\Q +2\Q^2  -2\frac{\p \Q}{\p \ln \bar L}\bigg)     \Bigg\}  \left(\p_\| v \right)^2 \nonumber \\
&&   + \frac{2}{\cH} \left[-1+  2b_e - 2 \Q   -  3 \frac{\cH'}{\cH^2}   -\frac{4  \left(1 - \Q\right) }{\bar \chi \cH}  \right]  \Phi {'}  \p_\| v    + \frac{2}{\cH} \left[ 1 - 2 b_e +4 \Q  +  3 \frac{\cH'}{\cH^2} + \frac{4  \left(1 - \Q\right)}{\bar \chi \cH} \right] \p_\| v  \p_\|^2 v \nonumber \\  
 &&   +4\frac{\left(1-\Q\right)}{ \cH } \p_\| \Phi \p_\| v    + \frac{2}{\cH^2} \p_\| v  \p_\|^2 \Phi - \frac{2}{\cH^2} \p_\| v \frac{\ud \Phi {'} }{\ud \bar \chi}   +  \frac{2}{\cH^2}\p_\| v \p_\|^3v  -  \frac{4}{\cH^2}  \Phi {'}  \p_\|^2 v      +4\Q^{(1)} \left[   \left( 2 - \frac{1}{\bar \chi \cH} \right) \Phi -\left( 1 - \frac{1}{\bar \chi \cH} \right) \p_\| v \right]  \nonumber \\
\end{eqnarray}
contains all local projection terms with time and partial derivatives along the line-of-sight direction,
and
 \begin{eqnarray}
\label{Deltag2-loc-3}
&&\Delta_{g\, {\rm loc}-3}^{(2)} =\frac{2}{\cH}\p_{\perp i} v \p^i_\perp \Phi    -\frac{4}{ \cH } \p_{\perp i} v   \p_{\perp}^i \p_\|  v+ \left( -1 + b_e -   \frac{\cH'}{\cH^2} -2  \Q  \right) \p_{\perp i} v \,  \p^i_{\perp} v \;
\end{eqnarray}
incorporates all local projection terms with transverse partial derivatives.

The integrated contribution can be divided into a sum of five terms
\begin{eqnarray}
\label{Deltag2-int}
&& \Delta_{g\, {\rm int}}^{(2)}=\sum_{i=1}^5 \Delta_{g\, {\rm int-}i}^{(2)}\;.
\end{eqnarray}
In particular,
\begin{eqnarray}
\label{Deltag2-int-1}
&& \Delta_{g\, {\rm int-1}}^{(2)}= + 4\bigg\{\bigg[- 2b_e + 2 b_e \Q  - 4 \Q^2    + 4 \frac{\p \Q}{\p \ln \bar L} +2 \frac{\p \Q}{\p \ln \bar a} +   2 \frac{\cH'}{\cH^2} \left(1 - \Q\right)  + \frac{2}{\bar \chi \cH}\bigg(2\Q^2- 2\frac{\p \Q}{\p \ln \bar L} \bigg)   \bigg] \Phi   \nonumber \\
&&  +  \frac{1}{\cH} \left[  1 -2 b_e +4 \Q + 3 \frac{\cH' }{\cH^2}      +  \frac{4\left(1 - \Q\right)}{\bar \chi \cH}  \right]  \p_\|^2 v     + \frac{1}{\cH}\left[  -1  + 2 b_e  - 2 \Q - 3\frac{\cH' }{\cH^2}    -   \frac{4 \left(1 - \Q\right) }{\bar \chi \cH}  \right]   \Phi {'} + \frac{1}{\cH}  \left( 1 - 2 \Q  \right)\p_\|  \Phi\nonumber \\
&& +  \frac{1}{\cH^2} \p_\|^2 \Phi  + \frac{1}{\cH^2} \p_\|^3 v  - \frac{1}{\cH^2}\frac{\ud \Phi {'} }{\ud \bar \chi}      + \left[ b_e - 2\Q   -  \frac{\cH'}{\cH^2}  - \frac{2}{\bar \chi \cH} \left(1 - \Q\right)  \right] \delta_g^{(1)}  - \frac{1}{\cH}  \frac{\ud  \delta_g^{(1)} }{\ud \bar \chi}  \Bigg\} I^{(1)} +4 \Bigg\{b_e+b_e^2+ \frac{\p b_e}{\p \ln \bar a} \nonumber \\
&&  +2\Q-4 \Q b_e+4 \Q^2  -4\frac{\p \Q}{\p \ln \bar L}-4\frac{\p \Q}{\p \ln \bar a}  
+ \left(-1-2 b_e + 4\Q \right) \frac{\cH' }{\cH^2} -\frac{\cH'' }{\cH^3} +3\left( \frac{\cH' }{\cH^2} \right)^2 + \frac{6}{\bar \chi} \frac{\cH' }{\cH^3} \left(1 - \Q\right)  \nonumber \\
&&   + \frac{2}{\bar \chi \cH} \left[-1  - 2b_e + \Q - 4 \Q^2   + 2b_e \Q  +4\frac{\p \Q}{\p \ln \bar L} +2 \frac{\p \Q}{\p \ln \bar a} \right]  +  \frac{2}{\bar \chi^2 \cH^2} \bigg(1-\Q +2\Q^2  -2\frac{\p \Q}{\p \ln \bar L}\bigg)    \Bigg\} \left(I^{(1)}\right)^2  \nonumber \\
&&   + \frac{4}{\bar \chi}\bigg[  - \left(b_e-b_e\Q+2 \Q^2 -2\frac{\p \Q}{\p \ln \bar L}-\frac{\p \Q}{\p \ln \bar a} \right) +  \frac{\cH'}{\cH^2} \left(1 - \Q\right)  \ + \frac{1}{\bar \chi \cH}  \left(1-\Q +2\Q^2  -2\frac{\p \Q}{\p \ln \bar L}\right) \bigg]  I^{(1)} T^{(1)}\nonumber \\
&&     +   \Bigg\{   - \frac{4}{\bar \chi}  \left(1- \Q \right) \delta_g^{(1)} - 2   \p_{\|}\delta_g^{(1)}  + \frac{4}{\bar \chi} \left( -1 + \Q+2 \Q^2- 2\frac{\p \Q}{\p \ln \bar L} \right)  \Phi   +  2\left[ - b_e + 2 \Q  +   \frac{\cH'}{\cH^2} +  \frac{4 \left(1 - \Q\right)}{\bar \chi \cH} \right]  \p_\|^2 v  \nonumber \\
&&   +4\left[ b_e-2 \Q  -   \frac{\cH'}{\cH^2}  -\frac{3\left(1  - \Q \right)}{\bar \chi \cH}  \right] \Phi{'} +2\left[1+b_e - 4 \Q - \frac{\cH'}{\cH^2} - \left(1 - \Q\right) \frac{2}{\bar \chi \cH} \right]  \p_\| \Phi    +\frac{2}{\cH} \p_\|^3 v  -\frac{2}{\cH} \p_\| \Phi {'}    \Bigg\} T^{(1)} \nonumber \\
&&  + \frac{2}{\bar \chi^2} \bigg(1-\Q +2\Q^2    -2\frac{\p \Q}{\p \ln \bar L}\bigg) \left(  T^{(1)} \right)^2+ 4 \bigg\{  \left[ -1 - 3 \Q  + 2\Q^2 - 2\frac{\p \Q}{\p \ln \bar L}+2b_e -   2\frac{\cH'}{\cH^2} -  \frac{5\left(1 - \Q\right)}{\bar \chi \cH}    \right] \Phi \nonumber \\ 
&&   + \frac{2 \left(1- \Q \right) }{\bar \chi}\int_0^{\bar \chi} \ud \tilde \chi \tilde \chi \Phi {'}   + \frac{\left(1- \Q \right)}{\bar \chi\cH}\p_\| v     + \frac{1}{\cH} \left(1- \Q \right) \p_\|^2 v   -\frac{1}{\cH}\left(1- \Q \right) \Phi {'}     -  \left(1- \Q \right) \delta_g^{(1)} \bigg\}\kappa^{(1)} \nonumber\\
 && + \frac{8}{\bar \chi} \bigg(1-\Q +\Q^2    -\frac{\p \Q}{\p \ln \bar L}\bigg) T^{(1)}   \kappa^{(1)}     + 2\left(1-\Q +2\Q^2- 2\frac{\p \Q}{\p \ln \bar L}\right) \left(\kappa^{(1)}\right)^2  \nonumber \\
 && + 4\bigg[ + 2 - b_e + b_e\Q - 2\Q   - 2 \Q^2 + 2\frac{\p \Q}{\p \ln \bar L} + \frac{\p \Q}{\p \ln \bar a} +  \frac{\cH'}{\cH^2} \left(1 - \Q\right) 
  + \frac{1}{\bar \chi \cH}  \left(3-3\Q +2\Q^2  -2\frac{\p \Q}{\p \ln \bar L}\right) \bigg] I^{(1)}  \kappa^{(1)}  \nonumber \\
  && +\bar \chi  \left(1- \Q \right)  \bigg[- 4\left( 1 + \frac{1}{\bar \chi\cH}\right)I^{(1)} - \frac{4}{\bar \chi}\int_0^{\bar \chi} \ud \tilde \chi \tilde \chi \Phi {'}  -\frac{2}{\bar \chi}T^{(1)}    + \frac{2}{\bar \chi\cH}\Phi -  \frac{2}{\bar \chi\cH}\p_\| v \bigg] \nabla^2_\perp T^{(1)}       \nonumber \\
  &&   +4\Q^{(1)} \left[  - 2\left( 1 - \frac{1}{\bar \chi \cH} \right)  I^{(1)}  +\frac{1}{\bar \chi} T^{(1)} +  \kappa^{(1)}\right] \;,
 \end{eqnarray}
contains convergence, ISW and STD terms. 

In
\begin{eqnarray}
\label{Deltag2-int-2}
&& \Delta_{g\, {\rm int-2}}^{(2)}=+ 4 \bigg\{     \bar \chi  \p_{\perp i} \delta_g^{(1)}   -  \bar \chi \left[ 1 + b_e-4 \Q  -   \frac{\cH'}{\cH^2} -  2\frac{\left(1 - \Q\right)}{\bar \chi \cH} \right]   \p_{\perp i} \Phi   +   \bar \chi \left[ b_e-2 \Q  -   \frac{\cH'}{\cH^2} - 2\frac{ \left(1 - \Q\right)}{\bar \chi \cH} \right]  \p_{\perp i} \p_\| v  \nonumber \\
&&    + \frac{\bar \chi}{\cH} \p_{\perp i}\Phi {'}   -  \frac{\bar \chi}{\cH} \p_{\perp i} \p_\|^2 v    \Bigg\} S_{\perp}^{i (1)} +2 \Bigg\{    -   \bar \chi  \p_{\perp i} \delta_g^{(1)} -  \frac{\bar \chi}{\cH} \p_{\perp i} \Phi {'}     - \bar \chi\left[ b_e-2 \Q  -   \frac{\cH'}{\cH^2}  +\frac{2\Q}{\bar \chi \cH}   \right]  \p_{\perp i}  \p_\| v       \nonumber \\
 &&     + \bar \chi  \left[ 2 + b_e - 4 \Q  -   \frac{\cH'}{\cH^2} - \left(1 - \Q\right) \frac{2}{\bar \chi \cH} \right] \p_{\perp i} \Phi 
 + \left[ b_e-2 \Q  -   \frac{\cH'}{\cH^2} +  \frac{2\Q}{\bar \chi \cH}  \right] \p_{\perp i} v       + \frac{\bar \chi}{\cH} \p_{\perp i} \p_\|^2 v\Bigg\}  \p_{\perp}^i T^{(1)}  \nonumber \\
   &&   -4 \left(1- \Q \right) S_{\perp}^{i }     \p_{\perp i } T^{(1)}  - 8\left(1- \Q \right) \bar \chi\p_{\perp i}  I^{(1)}   S_{\perp}^{i (1)}    - 8\left(1- \Q \right) S_{\perp}^{i (1)}  \p_{\perp i}  \int_0^{\bar \chi} \ud \tilde \chi   \tilde \chi  \Phi{'}        +  \left(1- \Q \right)  \vartheta_{ij}^{(1) }\vartheta^{ij(1)}  \nonumber \\
   &&    -4 (1-\Q)S_{\perp }^{i (1)}S_{\perp }^{j (1)} \delta_{ij} \;,  
 \end{eqnarray}
 we have terms like $\p_{\perp i} A^{(1)} \p_{\perp}^{i}B^{(1)}$, where $A^{(1)}$ or $B^{(1)}$ can be a local or an integrated term at first-order. 
 
In
\begin{eqnarray}
\label{Deltag2-int-3}
&& \Delta_{g\, {\rm int-3}}^{(2)}= - \left(1- \Q \right) \frac{2}{\bar \chi} T^{(2)}   + 2\left[ b_e-2 \Q  -   \frac{\cH'}{\cH^2} - \left(1 - \Q\right) \frac{2}{\bar \chi \cH} \right] I^{(2)} + 2 \left(1- \Q \right) \bar \chi \p_{\perp i}  S^{i(2)} - \bar \chi \left(1- \Q \right) \nabla_{\perp}^2  T^{(2)}\;,  \nonumber \\
 \end{eqnarray}
we find integrated terms such as, for example, ISW and STD at second order.

In the fourth term,
\begin{eqnarray}
\label{Deltag2-int-4}
&& \Delta_{g\, {\rm int-4}}^{(2)}= + 8(1-\Q)  \int_0^{\bar \chi}  \ud \tilde{\chi}\bigg[       -  \Phi  \tilde \p_{\perp m}S_{\perp}^{m(1)}   +  \left( \frac{\ud \Phi}{\ud \tilde \chi}   -  \frac{1}{\tilde\chi} \Phi  \right) \kappa^{(1)}  \bigg] \nonumber\\
&& - \frac{8(1-\Q)}{\bar \chi}\int_0^{\bar \chi}  \ud \tilde{\chi} \bigg(    { \Phi}^2  +  \Phi{'} T^{(1)} +2 \Phi \kappa^{(1)}   + \tilde \chi \tilde \p_{\perp i}\Phi \tilde \p_\perp^i T^{(1)}   \bigg)  +4 \left[ b_e-2 \Q  -   \frac{\cH'}{\cH^2} - \left(1 - \Q\right) \frac{2}{\bar \chi \cH} \right]  \nonumber \\
  && \times \int_0^{\bar \chi}  \ud \tilde{\chi}  \bigg[   \Phi{''}  T^{(1)} + 2 \Phi \Phi{'}  +  2 \Phi{'}  I^{(1)}    + 2 \Phi    \tilde \p_{\perp j}S_{\perp}^{j(1)}    - 2  \tilde \chi  \tilde \p_{\perp i} \Phi{'}    S_\perp^{i(1)} - 2 \bigg( \frac{\ud \Phi}{\ud \tilde \chi}     -  \frac{1}{\tilde\chi}  \Phi  \bigg) \kappa^{(1)} + \tilde \chi  \tilde \p_{\perp i} \Phi{'} \tilde  \p_\perp^i T^{(1)} \bigg] \nonumber \\
 &&+  \frac{8(1-\Q)}{\bar \chi} \int_0^{\bar \chi}  \ud \tilde{\chi} ~ (\bar \chi - \tilde \chi) \Bigg[  -  2 \Phi   \tilde \p_{\perp m}S_{\perp}^{m(1)}     
 + 2  \left( \frac{\ud  \Phi}{\ud \tilde \chi}   -  \frac{1}{\tilde\chi} \Phi  \right) \kappa^{(1)}  \nonumber \Bigg] -4 \left(1- \Q \right) \int_0^{\bar \chi} \ud \tilde{\chi}  \frac{ \tilde \chi }{\bar \chi} \Bigg[  + 2\tilde \chi   \tilde \nabla^2_{\perp} \Phi I^{(1)}\nonumber \\
  &&   + 2  \tilde \nabla^2_{\perp} \Phi \int_0^{\tilde \chi} \ud \tilde{\tilde \chi} \tilde{\tilde \chi}  \Phi {'}   + 2  \tilde \chi \p_{\perp i}  \Phi   \tilde \p_\perp^i I^{(1)} + 2 \tilde  \p_{\perp i}  \Phi  ~  \tilde \p_{\perp i} \int_0^{\tilde \chi} \ud \tilde{\tilde \chi} \tilde{\tilde \chi}  \Phi {'}     - 2 \Phi   \tilde \p_{\perp m} S_{\perp}^{m (1)}   -  \tilde \p_{\perp i} \Phi \tilde \p_\perp^i T^{(1)}   - \frac{2}{\tilde \chi}\Phi  \kappa^{(1)}\Bigg]  \nonumber \\ 
&&   -4 \left(1- \Q \right) \int_0^{\bar \chi} \ud \tilde \chi \left(\bar \chi-\tilde \chi\right)\frac{\tilde \chi}{ \bar \chi}  \Bigg[ - 2\tilde\p^i_\perp  \Phi   \tilde \p_{\perp i} \Phi  + 2 \tilde\p^i_\perp  \Phi  \tilde \p_{\perp i} I^{(1)}   - 2 \Phi  \tilde \nabla^2_\perp \Phi + 2  \tilde \nabla^2_\perp \Phi I^{(1)}     - \tilde \nabla^2_{\perp} \Phi  ~ T^{(1)}    -   \tilde \p_{\perp }^i    \Phi{'}  \tilde \p_{\perp i} T^{(1)} \nonumber\\
&&     +  \frac{2}{\tilde\chi}  \bigg( -     \frac{1}{\tilde\chi}\Phi +  \frac{\ud}{\ud \tilde \chi} \Phi\bigg)  \kappa^{(1)}  
 +  \frac{1}{\tilde\chi} \tilde \p_{\perp i}  \Phi ~ S_{\perp}^{i(1)}    - \frac{3}{2 \tilde \chi}  \tilde \p_{\perp i} \Phi \tilde   \p_\perp^i T^{(1)}    + \tilde \chi \bigg(  \tilde \p_{\perp i} \tilde \nabla^2_{\perp} \Phi + \frac{1}{\tilde\chi} \tilde \p_{\perp i}  \Phi{'}    \bigg) \left( 2   S_\perp^{i(1)} -  \tilde\p_\perp^i T^{(1)}\right) \nonumber\\
&&  + 2 \tilde \nabla^2_{\perp} \Phi~ \kappa  +  2\Phi{'} \tilde \p_{\perp m} S_{\perp}^{m(1)} - \left( \Phi{'}   + \frac{1}{\tilde \chi} \Phi  \right) \tilde \nabla_\perp^2 T^{(1)} \Bigg] 
\end{eqnarray}
we identify contributions in which the product between a local and integrated term (or two local terms) is within another integral along the line of sight.
Finally in the fifth term 
 \begin{eqnarray}
 \label{Deltag2-int-5}
 && \Delta_{g\, {\rm int-5}}^{(2)}=  -2  \left(1- \Q \right)\big|\gamma^{(1)}\big|^2  + 8\left(1- \Q \right) \int_0^{\bar \chi} \ud \tilde \chi (\bar \chi- \tilde \chi) \frac{\tilde \chi}{\bar \chi}\Bigg[\left( \tilde \p^{(j}_{\perp} \tilde \p^{m)}_{\perp} - \frac{1}{2} \Perp^{jm}  \tilde \nabla^2_{\perp} \right)\int_0^{\tilde \chi} \ud \tilde{\tilde \chi} 
  ( \tilde \chi -  \tilde{\tilde \chi}) \frac{\tilde \chi}{ \tilde{\tilde \chi}}\Phi \nonumber \\
  &&  \times \left( \tilde \p^{(j}_{\perp} \tilde \p^{m)}_{\perp} - \frac{1}{2} \Perp^{jm}  \tilde \nabla^2_{\perp} \right)\Phi\Bigg]   \;.
  \end{eqnarray}
we find all symmetric trace-free terms with orthogonal partial derivatives.
Obviously,  Eqs (\ref{Deltag2-loc}) and (\ref{Deltag2-int}) are written according to the properties of the various local and integral terms.
 Hereafter for simplicity we will compress 
 all these equations by writing
\begin{equation}
\Delta^{(2)}_g=\sum_a \Delta_g^{a(2)}\,,
\label{eq:sumsecondorder}
\end{equation}
where Eq.(\ref{eq:sumsecondorder}) is given by the sum of all terms contained in Eqs. (\ref{Deltag2-loc-1}), (\ref{Deltag2-loc-2}), (\ref{Deltag2-loc-3}), (\ref{Deltag2-int-1}), (\ref{Deltag2-int-2}), (\ref{Deltag2-int-3}), (\ref{Deltag2-int-4}) and (\ref{Deltag2-int-5}).


\section{Bispectrum}
\label{sec:bisp}
The spherical Bessel representation uses a complete set of orthogonal basis functions $|k\ell m \rangle$ in a spherical Fourier space (for a very brief review of this formalism, see Appendix~\ref{klm-basis}). A scalar field like $\Delta_g$ in configuration space can be decomposed in the following way:
\be
\Delta_{\ell m}^{g}(k)=\langle k \ell m |\Delta_{g} \rangle= \int \ud^3\bx ~ \langle k \ell m | \bx \rangle \langle \bx |\Delta_{g} \rangle\;.
\ee

It is important to note that by definition the monopole $\Delta_{00}^{g}$ can be removed and set to zero for $k \to 0$. In general, we can discard $\Delta_{00}^{g}$ because only the mean $\bar n_g(z)$  contributes to the monopole.
Therefore we will compute the spherical power spectrum only for $\ell >0$. 

Here we introduce a more realistic definition of $\langle \bx |\Delta_{g} \rangle = \Delta_{g} (\bx)$ which inÊ a realistic case becomes:
\be 
\langle \bx |\Delta_{g} \rangle \longrightarrow  \W (\bar \chi)  \Delta_{g} (\bx)\;,
\ee
where we have included the radial selection function $ \W (\bar \chi) $.

Then at first and second order we find
\be\label{sumklm}
\Delta_{\ell m}^{g}(k) =\Delta_{\ell m}^{g(1)}(k) +\frac{1}{2} \Delta_{\ell m}^{g(2)}(k) + ...  = \sum_b  \Delta_{\ell m}^{b(1)}(k) +\frac{1}{2}\sum_a  \Delta_{\ell m}^{a(2)}(k) + ... \;, 
\ee
where the index $b$ runs over the terms defined in Eq. (\ref{Deltag1-esplicito}) and $a$
represents the index of summation over all additive terms at second order, see Eq. (\ref{eq:sumsecondorder}).
Finally, $\Delta_{\ell m}^{g(1)}(k)$ is the spherical Bessel transform of  Eq.~(\ref{Deltag1-esplicito}) and the second-order terms are the spherical Bessel transforms of Eq. (\ref{eq:sumsecondorder}).

\subsection{First-order terms} \label{first_order-klm}

Considering each term of the first summation on the right-hand side  of Eq. (\ref{sumklm}), we have 
\be
\Delta_{\ell m}^{b(1)}(k)=\langle k \ell m |\Delta^{b(1)} \rangle= \int \ud^3 \bx ~ \W (\bar \chi) ~   \langle k \ell m | \bx \rangle
 \Delta_{g}^{b(1)} (\bx) \;,
\ee
where we have already included the radial selection function.
Usually, we can define a generic first-order perturbation as follows
\bea
 \Delta_{g}^{b(1)} (\bx)=  \int_0^{\bar \chi} \ud \tilde \chi ~\WW^{b} \left(\bar \chi, \tilde \chi, \eta, \tilde \eta, \frac{\p}{\p \tilde \chi}, \frac{\p}{\p \tilde \eta}, \triangle_{\hat \bn}\right)\left[ \int \frac{\ud^3\bk}{(2\pi)^3}  \T^{b}(\bk,\tilde \eta) \Phi_{\rm p}(\bk) e^{i \bk \tilde \bx}\right]\;,
 \label{eq:defWW}
\eea
where $\Phi_{\rm p}(\bk)$ is the primordial potential set during the inflation epoch, $\T^{b}( \bk, \eta)$ is a generalized transfer function which relates the linear primordial potential with a generic perturbation term (here labeled with $b$).
Here $\bk$ is a Fourier space vector, $\hat \bn$ is the observed galaxy unity vector on the sky,
 and we have used the following relations
$$ \int \ud^3 \bx  = \int \ud \bar \chi~\bar \chi^2 \int \ud^2\hat \bn\;$$
and 
\[\langle \bx| k \ell m  \rangle=\aleph_\ell(k) j_\ell (k  \chi)  Y_{\ell m}(\hat \bn)\;,\]
where $\aleph_\ell(k)$ is a function that depends on $\ell$ and $k$  (see Appendix \ref{klm-basis})  which specific expression depends on the conventions adopted in the expansion.

For each contribution in Eq. (\ref{Deltag1-esplicito}), we have defined a weight function $ \WW^{b }$ which is a generic operator that depends on $\bar \chi, \eta,  \p/\p \bar \chi,\p/\p  \eta$  and $\triangle_{\hat \bn}$. This operator encloses the physical effects due to the fact that in any observation we collect  the photons emitted from a source after they have traveled through the  past light cone of the observer. It is this  ``projection effect'' on the perturbations  that is captured by $\WW$; $\WW$ takes slightly different explicit expressions depending on what perturbation is being considered, as labeled by its superscript index. Applying it to the spherical harmonics, we find
 \be \label{WW}
 \WW^{b } \left(\bar \chi, \tilde \chi, \eta, \tilde \eta, \frac{\p}{\p \tilde \chi}, \frac{\p}{\p \tilde \eta}, \triangle_{\hat  \bn}\right) Y_{\ell m}(\hat \bn)=  \WW_\ell^{b}  \left(\bar \chi, \tilde \chi, \eta, \tilde \eta, \frac{\p}{\p \tilde \chi}, \frac{\p}{\p \tilde \eta}\right) Y_{\ell m}(\hat \bn)\;,
 \ee
where, in $\WW_\ell^{b(1)}$, through the relation
$$ \triangle_{ \hat \bn} Y_{\ell m}(\hat {\bn}) =- \ell(\ell+1) Y_{\ell m}(\hat {\bn}) \, ,$$ 
 we have removed its angular dependence.
 
 Using Eq. (\ref{WW}), we have
\be
\WW^{b} \left(\bar \chi, \tilde \chi, \eta, \tilde \eta, \frac{\p}{\p \tilde \chi}, \frac{\p}{\p \tilde \eta}, \triangle_{\hat  \bn}\right) \left[ \T^{b}(\bk,\tilde \eta) e^{i \bk \tilde \bx}\right] = \sum_{\ell m} 4\pi i^\ell \left[\WW_\ell^{b} \left(\bar \chi, \tilde \chi, \eta, \tilde \eta, \frac{\p}{\p \tilde \chi}, \frac{\p}{\p \tilde \eta}\right)  \T^{b}(\bk,\tilde \eta) j_\ell (k \tilde \chi) \right] Y_{\ell m}^*(\hat \bk) Y_{\ell m}(\hat \bn)\;.
\ee
Then, taking into account that
\be
\label{ortang}
\int \ud^2 \hat \bn ~ Y_{\ell_1 m_1} (\hat \bn)  Y^*_{\ell_2 m_2} (\hat \bn) =  \delta^K_{\ell_1 \ell_2} \delta^K_{m_1 m_2} \;,
\ee
 where  $\delta^K$ denotes the Kronecker delta and * denotes complex conjugate, 
 we find
\be \label{Delta_ellk(k)^1}
\Delta_{\ell m}^{b(1)}(k)= \int \frac{\ud^3\tilde \bk}{(2\pi)^3} \M_\ell^{b(1)}(k,\tilde k) Y_{\ell m}^*({\hat  \tbk})  \Phi_{\rm p}( \tbk)\;,
\ee
where
\be
\M_\ell^{b(1)}(k,\tilde k) =\int \ud \bar \chi~\bar \chi^2  \W (\bar \chi) [4\pi i^\ell \aleph_\ell^*(k)] j_\ell (k \bar \chi)\int_0^{\bar \chi} \ud \tilde \chi ~\left[\WW_\ell^{b} \left(\bar \chi, \tilde \chi, \eta, \tilde \eta, \frac{\p}{\p \tilde \chi}, \frac{\p}{\p \tilde \eta}\right)  \T^{b}(\tilde \bk,\tilde \eta) j_\ell (\tilde k \tilde \chi)\right]\;.
\ee

Before concluding this subsection, it is useful to calculate the spherical power spectrum (see also \cite{Yoo:2013tc}), i.e.
\be
\langle \Delta_{\ell m}^{b(1)*}(k) ~ \Delta_{\ell' m'}^{c(1)}(k')\rangle = \delta^K_{\ell \ell'} \delta^K_{m m'} \int \frac{\tilde k^2 \ud \tilde k}{(2\pi)^3} ~ \M_\ell^{b(1) *} (k,\tilde k)~\M_\ell^{c(1)}(k',\tilde k)~ P_\Phi(\tilde k)\;,
\ee
where $P_\Phi(\tilde k)$ is the power spectrum of the potential at initial epoch 
\[\langle \Phi_{\rm p}^*( \tbk) ~ \Phi_{\rm p}( \tbk')\rangle = (2\pi)^3  \delta^D (\tbk-\tbk') P_\Phi(\tilde k)  \;.\]

\subsection{Second order (scalar case)}\label{calM-second_order-Master_eq}

In the same way, at second order, the spherical Bessel transform of each term contained in the last summation of Eq. (\ref{sumklm}) can be written as
\be
\label{Delta-2order-klm}
\frac{1}{2}\Delta_{\ell m}^{a(2)}(k)= \int \ud^3 \bx ~ \W (\bar \chi) ~\langle k \ell m | \bx \rangle ~ \frac{1}{2}\Delta_g^{a(2)}(\bx)\;.
\ee
In general (for the scalar case), similarly to what was done at first-order,  let us define the spherical multipole functions  $ \widetilde \M^{(2)}$, at second order,
 in the following way:
 \be
\label{Delta_ellk(k)^2-1}
\frac{1}{2}\Delta_{\ell m}^{a(2)}(k)=\sum_{\ellvp \mvp \ellvq \mvq} \int \frac{\ud^3 \bp}{(2\pi)^3} \frac{\ud^3 \bq}{(2\pi)^3} \widetilde{\M}_{\ell m \ellvp \mvp \ellvq \mvq}^{a(2)}(k; \bp, \bq) Y_{\ellvp \mvp}^*(\hat \bp)Y_{\ellvq \mvq}^*(\hat \bq)  \Phi_{\rm p}(\bp) \Phi_{\rm p}(\bq)\;.
\ee
In addition, as we will show explicitly in Sec.\ \ref{calM-second_order} (see also Appendix \ref{NewcalM2}), 
$\widetilde{\M}$ can be expanded as
\be
\widetilde{\M}_{\ell m \ellvp \mvp \ellvq \mvq}^{a(2)}(k; \bp, \bq) = \sum_{\barell\barm}\M_{\ell m \ellvp \mvp \ellvq \mvq \barell \barm}^{a(2)}(k; p, q)  Y_{\barell \barm} (\hat \bp)Y_{\barell \barm}^*(\hat \bq)\;,
\ee
where we used the following theorem that relates Legendre polynomials $\PP_{\barell} $ to spherical harmonics
\be
\PP_{\barell} (\hat \bp \cdot \hat \bq) = \frac{4 \pi}{2 \barell +1} \sum_{\barm=-\barell}^{\barell}   Y_{\barell \barm} (\hat \bp) Y_{\barell \barm}^*(\hat \bq)\;,
\ee
and we obtain
\be
\label{Delta_ellk(k)^2}
\frac{1}{2}\Delta_{\ell m}^{a(2)}(k)=\sum_{\ellvp \mvp \ellvq \mvq \barell\barm} \int \frac{\ud^3 \bp}{(2\pi)^3} \frac{\ud^3 \bq}{(2\pi)^3} \M_{\ell m \ellvp \mvp \ellvq \mvq \barell \barm}^{a(2)}(k; p, q)   Y_{\ellvp \mvp}^*(\hat \bp)  Y_{\barell \barm} (\hat \bp) Y_{\ellvq \mvq}^*(\hat \bq) Y_{\barell \barm}^*(\hat \bq) \Phi_{\rm p}(\bp) \Phi_{\rm p}(\bq)\;.
\ee

As we will see in Sec.\ \ref{calM-second_order} and Appendix \ref{NewcalM2}, $\M_{\ell m \ellvp \mvp \ellvq \mvq \barell \barm}^{a(2)}(k; p, q) $
 will be computed explicitly for all terms contained in Eq. (\ref{eq:sumsecondorder}). 
Finally, using Eqs. (\ref{Delta_ellk(k)^1}) and (\ref{Delta_ellk(k)^2}), in the next section we derive the expression for the spherical Bessel bispectrum.

\subsection{Master relation for bispectrum (scalar case)}\label{sec:Bispectr-ScalarCase}

In order to compute the projected bispectrum in spherical Fourier space, let us start with the following relation\footnote{
In general, one might think that
$\langle \Delta_g (\bk_1) \Delta_g (\bk_2) \Delta_g(\bk_3)\rangle = (2\pi)^3 \delta^D(\bk_1+ \bk_2+\bk_3)  B(\bk_1, \bk_2, \bk_3).$
However, we cannot {\it a priori} assume that the bispectrum in redshift space preserves the property of homogeneity and isotropy; therefore, we need to proceed as in Eq.~\eqref{Bispectrum}.}

\bea \label{Bispectrum}
\langle \Delta_g (\bk_1) \Delta_g (\bk_2) \Delta_g(\bk_3)\rangle&=&\sum_{\ell_1  m_1\ell_2 m_2\ell_3 m_3}
 \frac{(4\pi)^3 ~ (-i)^{\ell_1+\ell_2+\ell_3}}{ \aleph^*_{\ell_1}(k_1) \aleph^*_{\ell_2}(k_2) \aleph^*_{\ell_3}(k_3)} Y_{\ell_1 m_1} ({\hat \bk}_1) Y_{\ell_2 m_2} ({\hat \bk}_2) Y_{\ell_3 m_3} ({\hat \bk}_3) \nonumber\\
&& \times  \langle \Delta^g_{\ell_1  m_1} (k_1) \Delta^g_{\ell_2 m_2}(k_2) \Delta^g_{\ell_3 m_3}(k_2)\rangle\;,
\eea
where we used 
 \[
  \Delta_g(\bk)= \sum_{\ell m} \frac{4\pi (-i)^\ell}{ \aleph^*_\ell(k)} Y_{\ell m} ({\hat \bk})  \Delta^g_{\ell m}(k)\;.
 \]
See also  Eq.(\ref{phi-klm_to_phi-veck}).
In particular
\bea
\label{Bispectrum-general}
\langle \Delta^g_{\ell_1  m_1} (k_1) \Delta^g_{\ell_2 m_2}(k_2) \Delta^g_{\ell_3 m_3}(k_2)\rangle&=& \bigg\langle \frac{1}{2}\Delta^{g(2)}_{\ell_1  m_1} (k_1) \Delta^{g(1)}_{\ell_2 m_2} (k_2) \Delta^{g(1)}_{\ell_3 m_3}(k_3) \bigg\rangle +  \bigg\langle  \Delta^{g(1)}_{\ell_1  m_1} (k_1)   \frac{1}{2}\Delta^{g(2)}_{\ell_2 m_2} (k_2) \Delta^{g(1)}_{\ell_3 m_3}(k_3) \bigg\rangle  \nonumber \\ 
&+&  \bigg\langle  \Delta^{g(1)}_{\ell_1  m_1} (k_1) \Delta^{g(1)}_{\ell_2 m_2} (k_2)  \frac{1}{2}\Delta^{g(2)}_{\ell_3 m_3}(k_3) \bigg\rangle \;.
\eea
For simplicity, let us consider only the first additive term of Eq.(\ref{Bispectrum-general})
\be\label{Bispectrum-general-2-1-1}
 \bigg\langle \frac{1}{2}\Delta^{g(2)}_{\ell_1  m_1} (k_1) \Delta^{g(1)}_{\ell_2 m_2} (k_2) \Delta^{g(1)}_{\ell_3 m_3}(k_3) \bigg\rangle =   \sum_{abc}  \bigg\langle \frac{1}{2}\Delta^{a(2)}_{\ell_1  m_1}(k_1)  \Delta^b_{\ell_2 m_2} (k_2)   \Delta^c_{\ell_3 m_3}(k_3) \bigg\rangle
\ee
where the index $a$ identifies all the terms at second order in Eq.\  (\ref{eq:sumsecondorder}), and the indexes $b$ and $c$ identify all the terms in Eq.\ (\ref{Deltag1-esplicito}). 
For $a$, $b$ and $c$ fixed we have
\bea
 \bigg\langle \frac{1}{2}\Delta^{a(2)}_{\ell_1  m_1}(k_1)  \Delta^b_{\ell_2 m_2} (k_2)   \Delta^c_{\ell_3 m_3}(k_3) \bigg\rangle & =&\sum_{{\ellvp}_1 {\mvp}_1 {\ellvq}_1 {\mvq}_1 {\barell}_1 {\barm}_1} \int \frac{p_1^2\ud p_1}{(2\pi)^3} \frac{q_1^2 \ud q_1}{(2\pi)^3}  \frac{q_2^2\ud q_2}{(2\pi)^3} \frac{q_3^2\ud q_3}{(2\pi)^3}  \quad \quad \quad \quad \quad \quad\quad \quad \quad \quad  \quad \quad \quad \quad \quad \nonumber \\
 && \times  ~ \left[\M_{\ell_1 m_1 {\ellvp}_1 {\mvp}_1 {\ellvq}_1 {\mvq}_1 {\barell}_1 {\barm}_1}^{a(2)}(k_1; p_1, q_1) ~ \M_{\ell_2}^{b(1)}(k_2, q_2)  ~\M_{\ell_3}^{c(1)}(k_3, q_3)\right] \nonumber \\
&& \times  \int \ud^2 \hat \bp_1  \ud^2 \hat \bq_1 \ud^2 \hat \bq_2 \ud^2 \hat \bq_3 ~
  Y_{{\ellvp}_1 {\mvp}_1}^*(\hat \bp_1)  Y_{\barell_1 \barm_1} (\hat \bp_1) Y_{{\ellvq}_1 {\mvq}_1}^*(\hat \bq_1) Y_{\barell_1 \barm_1}^*(\hat \bq_1)   \nonumber  \\
&& \times  ~ Y_{\ell_2 m_2}^*(\hat \bq_2)Y_{\ell_3 m_3}^*(\hat \bq_3)  ~ \big\langle\Phi_{\rm p}(\bp_1) \Phi_{\rm p}(\bq_1)\Phi_{\rm p}(\bq_2)\Phi_{\rm p}(\bq_3) \big\rangle\;. 
\eea
Taking into account that
 \bea && \big\langle\Phi_{\rm p}(\bp_1) \Phi_{\rm p}(\bq_1)\Phi_{\rm p}(\bq_2)\Phi_{\rm p}(\bq_3) \big\rangle = (2\pi)^6\left[\delta^D (\bp_1+\bq_1)  P_\Phi(p_1)\delta^D (\bq_2+\bq_3) P_\Phi(q_2)  \right] \nonumber  \\ 
&& +   (2\pi)^6\left[\delta^D (\bp_1+\bq_2)  P_\Phi(q_2) \delta^D (\bq_1+\bq_3) P_\Phi(q_3)  \right] +  (2\pi)^6\left[ \delta^D (\bp_1+\bq_3) P_\Phi(q_3) \delta^D (\bq_1+\bq_2) P_\Phi(q_2)  \right]
\eea 
and using
\[(2\pi)^3\delta^D (\bq_i+\bq_j) =\frac{\pi}{2 q_i^2} \delta^D(q_i - q_j) \sum_{\ell_{ij}' m_{ij}'}  (4 \pi)^2 (-1)^{\ell_{ij}' + m_{ij}' }Y_{\ell_{ij}' m_{ij}'} (\hat \bq_i) Y_{\ell_{ij}' -m_{ij}'}(\hat \bq_j)\;, \]
we find 
\be
\label{Gammas}
 \bigg\langle \frac{1}{2}\Delta^{a(2)}_{\ell_1  m_1}(k_1)  \Delta^b_{\ell_2 m_2} (k_2)   \Delta^c_{\ell_3 m_3}(k_3) \bigg\rangle =
 ~ \Upsilon^{[1]abc}_{\ell_1 m_1 \ell_2 m_2 \ell_3 m_3}(k_1,k_2,k_3) + \Upsilon^{[2]abc}_{\ell_1 m_1 \ell_2 m_2 \ell_3 m_3}(k_1,k_2,k_3) + \Upsilon^{[3]abc}_{\ell_1 m_1 \ell_2 m_2 \ell_3 m_3}(k_1,k_2,k_3)
 \ee
 where
 $ \Upsilon^{[j]abc}_{\ell_1 m_1 \ell_2 m_2 \ell_3 m_3}$, for $j=1,2,3$, are the bispectrum building blocks and contain all the information on the bispectrum in redshift space. Explicitly, we have 
\bea
\label{Upsilon[1]}
\Upsilon^{[1]abc}_{\ell_1 m_1 \ell_2 m_2 \ell_3 m_3}(k_1,k_2,k_3) = \sum_{{\ellvp}_1 {\mvp}_1 {\ellvq}_1 {\mvq}_1 {\barell}_1 {\barm}_1 \ell_{1}' m_{1}'} \delta^K_{m_2-m_3}  \delta^K_{\ell_2 \ell_3}  (-1)^{\ell_{1}' +\ell_2+m_2+{\mvp}_1} \G^{{\ellvp}_1 {\barell}_1 \ell_{1}' }_{-{\mvp}_1 {\barm}_1 m_{1}' }\G^{{\ellvq}_1 {\barell}_1 \ell_{1}' }_{{\mvq}_1 {\barm}_1 m_{1}' } \nonumber \\
  \int \frac{q_1^2\ud q_1}{(2\pi)^3}   \frac{q_2^2\ud q_2}{(2\pi)^3}\left[\M_{\ell_1 m_1 {\ellvp}_1 {\mvp}_1 {\ellvq}_1 {\mvq}_1 {\barell}_1 {\barm}_1}^{a(2)}(k_1; q_1, q_1) ~ \M_{\ell_2}^{b(1)}(k_2, q_2)  ~\M_{\ell_3}^{c(1)}(k_3, q_2)\right] P_\Phi(q_1) P_\Phi(q_2)\;,
\eea
\bea
\label{Upsilon[2]}
\Upsilon^{[2]abc}_{\ell_1 m_1 \ell_2 m_2 \ell_3 m_3}(k_1,k_2,k_3) = \sum_{{\ellvp}_1 {\mvp}_1 {\ellvq}_1 {\mvq}_1 {\barell}_1 {\barm}_1}  (-1)^{\ell_2 +\ell_3-m_2-m_3+{\mvp}_1+{\mvq}_1+{\barm}_1} \G^{{\ellvp}_1 {\barell}_1 \ell_{2} }_{-{\mvp}_1 {\barm}_1 -m_2 }\G^{{\ellvq}_1 {\barell}_1 \ell_{3} }_{{\mvq}_1 {\barm}_1 m_{3}}~~~~~~~ \nonumber \\
  \int \frac{q_2^2\ud q_2}{(2\pi)^3}   \frac{q_3^2\ud q_3}{(2\pi)^3}\left[\M_{\ell_1 m_1 {\ellvp}_1 {\mvp}_1 {\ellvq}_1 {\mvq}_1 {\barell}_1 {\barm}_1}^{a(2)}(k_1; q_2, q_3) ~ \M_{\ell_2}^{b(1)}(k_2, q_2)  ~\M_{\ell_3}^{c(1)}(k_3, q_3)\right] P_\Phi(q_2) P_\Phi(q_3)\;,
\eea
and  $\Upsilon^{[3]abc}_{\ell_1 m_1 \ell_2 m_2 \ell_3 m_3}(k_1,k_2,k_3) $ is obtained from $\Upsilon^{[2]abc}_{\ell_1 m_1 \ell_2 m_2 \ell_3 m_3}(k_1,k_2,k_3)$ if we note that $ \M_{\ell m \ellvp \mvp \ellvq \mvq \barell \barm}^{ a(2)}(k; p, q) = \M_{\ell m \ellvq \mvq \ellvp \mvp \barell \barm}^{ a(2)}(k;q, p)$.
Here  ${\cal G}$ is the usual Gaunt integral [see Eq.(\ref{gaunt})].
For symmetry reasons one may expect that \[\Upsilon^{[2]abc}_{\ell_1 m_1 \ell_2 m_2 \ell_3 m_3}(k_1,k_2,k_3)=\Upsilon^{[3]abc}_{\ell_1 m_1 \ell_2 m_2 \ell_3 m_3}(k_1,k_2,k_3) \;.\] In Appendix \ref{Sec:Upsilon} will show  explicitly that this is the case. The explicit proof offers a consistency check on our calculations and expressions.
In Appendix \ref{Sec:Upsilon} we will  also prove that \[ \Upsilon^{[1]abc}_{\ell_1 m_1 \ell_2 m_2 \ell_3 m_3}(k_1,k_2,k_3) \sim \delta^K_{\ell_1 0}\;.\]
This implies that we can discard $\Upsilon^{[1]abc}$ because here we consider only terms with $\ell > 0$. 
Finally we will demonstrate that the three-point function of the coefficients defined in Eq.(\ref{Bispectrum-general}) may be factorized by 
isotropy\footnote{Precisely, here the term isotropy means that $B_{\ell_1\ell_2 \ell_3 }(k_1,k_2,k_3)$ is rotationally invariant for rotations around the line of sight passing through the circumcenter.} into
\be
\langle \Delta^g_{\ell_1  m_1} (k_1) \Delta^g_{\ell_2 m_2}(k_2) \Delta^g_{\ell_3 m_3}(k_2)\rangle = \left(
\begin{array}{ccc}
\ell_1 &\ell_2& \ell_3 \\
m_1 & m_2 &m_3
\end{array}
\right)
B_{\ell_1\ell_2 \ell_3 }(k_1,k_2,k_3)\;,
\ee
where $\ell_1+ \ell_2 + \ell_3 =$even, $m_1 + m_2+m_3=0$ and $\ell_1,\; \ell_2,\; \ell_3 $ satisfy the triangle rule.

From the above relation and Eq.(\ref{Bispectrum}) we can write one of the main results of this paper, i.e.
\bea \label{Bispectrum-2}
\langle \Delta_g (\bk_1) \Delta_g (\bk_2) \Delta_g(\bk_3)\rangle&=&\sum_{\ell_1  m_1\ell_2 m_2\ell_3 m_3}
 \frac{(4\pi)^3 ~ (-i)^{\ell_1+\ell_2+\ell_3}}{ \aleph^*_{\ell_1}(k_1) \aleph^*_{\ell_2}(k_2) \aleph^*_{\ell_3}(k_3)} Y_{\ell_1 m_1} ({\hat \bk}_1) Y_{\ell_2 m_2} ({\hat \bk}_2) Y_{\ell_3 m_3} ({\hat \bk}_3) \nonumber\\
&& \times   \left(
\begin{array}{ccc}
\ell_1 &\ell_2& \ell_3 \\
m_1 & m_2 &m_3
\end{array}
\right)
B_{\ell_1\ell_2 \ell_3 }(k_1,k_2,k_3)\;.
\eea

The rest of paper is devoted to computing explicitly all $\M_{\ell m \ellvp \mvp \ellvq \mvq \barell \barm}^{a(2)}(k; p, q)$ in  Eq. (\ref{eq:sumsecondorder}) and in Sec.\ \ref{sec:bias} we give a prescription to incorporate 
consistently the bias at second order within $\delta_g^{(2)}$.

\section{Spherical multipole functions at first-order}
\label{sec:kernels}
Using the prescription in \ref{first_order-klm}, it is easy to rewrite the matter overdensity at first order [Eq.(\ref{Deltag1-esplicito})]  in terms of $\M_\ell^{b(1)}(k,\tilde k)$.
Specifically, we obtain the following relations (see also \cite{Yoo:2013tc}):
\bea
\begin{array}{c}
\M_\ell^{\delta_g (1)}(k,\tilde k)=\int \ud \bar \chi~\bar \chi^2  \W (\bar \chi) [4\pi i^\ell \aleph_\ell^*(k)] j_\ell (k \bar \chi)   j_\ell (\tilde k \bar \chi) \T^{\delta_g}(\tilde \bk, \eta)
 \; , \\
\\
\M_\ell^{ \p_\|^2 v (1)}(k,\tilde k)=\int \ud \bar \chi~\bar \chi^2  \W (\bar \chi) [4\pi i^\ell \aleph_\ell^*(k)] \left[- \frac{1}{\cH (\eta)} \right] j_\ell (k \bar \chi)\left[ \frac{\p^2}{\p \bar \chi^2}  j_\ell (\tilde k \bar \chi) \right]  \T^{v }(\tilde \bk, \eta) \;,  \\
\\
\M_\ell^{\p_\| v (1)}(k,\tilde k)=\int \ud \bar \chi~\bar \chi^2  \W (\bar \chi) [4\pi i^\ell \aleph_\ell^*(k)]\left[b_e(\eta)  - \frac{\cH'(\eta)}{\cH^2(\eta)} - 2 \Q(\eta)  - 2 \frac{\left(1- \Q(\eta) \right)}{\bar \chi \cH(\eta)}  \right] j_\ell (k \bar \chi)\left[ \frac{\p}{\p \bar \chi}  j_\ell (\tilde k \bar \chi) \right]  \T^{v }(\tilde \bk, \eta)\;,   \\
\\
\M_\ell^{\Phi  (1)}(k,\tilde k)=- \int \ud \bar \chi~\bar \chi^2  \W (\bar \chi) [4\pi i^\ell \aleph_\ell^*(k)]   \left[ b_e(\eta)  - \frac{\cH'(\eta)}{\cH^2(\eta)} - 4 \Q(\eta) +1 - 2 \frac{\left(1- \Q (\eta) \right)}{\bar \chi \cH (\eta)} \right]  j_\ell (k \bar \chi)   j_\ell (\tilde k \bar \chi) \T^{\Phi}(\tilde \bk, \eta) \;,  \\
\\
\M_\ell^{ \Phi {'} (1)}(k,\tilde k)=\int \ud \bar \chi~\bar \chi^2  \W (\bar \chi) [4\pi i^\ell \aleph_\ell^*(k)] \left[\frac{1}{\cH (\eta)} \right] j_\ell (k \bar \chi) j_\ell (\tilde k \bar \chi)  \frac{\p}{\p \eta}  \T^{\Phi}(\tilde \bk, \eta) \;,  \\
\\
\M_\ell^{\kappa(1)}(k,\tilde k) =\int \ud \bar \chi~\bar \chi^2  \W (\bar \chi) [8\pi \ell (\ell+1) i^\ell \aleph_\ell^*(k)] [1-\Q(\eta)] j_\ell (k \bar \chi)\int_0^{\bar \chi} \ud \tilde \chi ~\left[ \frac{ \left(\bar \chi-\tilde \chi\right)}{ \bar \chi \tilde \chi}   j_\ell (\tilde k \tilde \chi)  \T^{\Phi}(\tilde \bk,\tilde \eta) \right]
  \;,  \\
\\
\M_\ell^{I (1)}(k,\tilde k)=-\int \ud \bar \chi~\bar \chi^2  \W (\bar \chi) [8\pi i^\ell \aleph_\ell^*(k)]  \left[ b_e(\eta)  - \frac{\cH'(\eta)}{\cH^2(\eta)} - 2 \Q(\eta)  - 2 \frac{\left(1- \Q(\eta) \right)}{\bar \chi \cH(\eta)} \right]  j_\ell (k \bar \chi)\int_0^{\bar \chi} \ud \tilde \chi ~\left[  j_\ell (\tilde k \tilde \chi)  \frac{\p}{\p \tilde \eta}  \T^{\Phi}(\tilde \bk,\tilde \eta) \right]
 \;,  \\
\\
\M_\ell^{T(1)}(k,\tilde k)=\int \ud \bar \chi~\bar \chi^2  \W (\bar \chi) [16\pi i^\ell \aleph_\ell^*(k)] \left[\frac{1-\Q(\eta)}{\bar \chi}\right] j_\ell (k \bar \chi)\int_0^{\bar \chi} \ud \tilde \chi ~\left[  j_\ell (\tilde k \tilde \chi)   \T^{\Phi}(\tilde \bk,\tilde \eta) \right] \;.  
\end{array}
\label{M-1} 
\eea

Note that here and hereafter for clarity we have made explicit what term the  superscript index $b$ of ${\cal M}$ refers to, by writing in its place the physical quantity involved e.g. $\delta_g$ for galaxy over density, $\kappa$ for convergence etc.
Here we used $$\nabla_\perp^2 Y_{\ell m}(\hat {\bn}) =\frac{1}{\bar \chi^2}   \triangle_{ \hat \bn} Y_{\ell m}(\hat {\bn}) =-\frac{1}{\bar \chi^2} \ell(\ell+1) Y_{\ell m}(\hat {\bn}) \, ,$$ and the following definitions:

 \be \label{T}
\delta_g^{(1)} ( \bk, \eta) =\T^{\delta_g}( \bk, \eta) \Phi_{\rm p}(\bk)\;, \quad \quad v^{(1)} ( \bk, \eta) =\T^{v }( \bk, \eta) \Phi_{\rm p}(\bk)\;, \quad \quad {\rm and } \quad \quad \Phi^{(1)} ( \bk, \eta) =\T^{\Phi }( \bk, \eta) \Phi_{\rm p}(\bk)\;. \\
 \ee

\section{Spherical multipole functions at second  order} 
\label{calM-second_order}

In this section we will apply the instructions contained in Sec.\ \ref{calM-second_order-Master_eq} where we proved that a generic (scalar) term at second order can be expressed in terms of  $\M_{\ell m \ellvp \mvp \ellvq \mvq \barell \barm}^{a(2)}(k; p, q)$. In the next subsections we will consider separately each additive term contained in Eqs. (\ref{Deltag2-loc}) and (\ref{Deltag2-int}), i.e. $ \Delta_{g\, {\rm loc-}i}^{(2)}$ for $i$ runs from $1$ to $3$, and $ \Delta_{g\, {\rm int-}j}^{(2)}$ where $j$ runs  from $1$ to $5$.

\subsection{ Terms from $ \Delta_{g\, {\rm loc-}1}^{(2)}$ [see Eq.\ (\ref{Deltag2-loc-1})]}
\label{Sec:Deltag2-loc-1}
Equation (\ref{Deltag2-loc-1}) contains all local contributions with second-order perturbation terms.
These additive terms can be written in the following way:
\bea
\label{Deltag2-loc-1-general_relation}
 \frac{1}{2}\Delta^{\alpha[b](2)}_{\ell m}(k) =  \int \ud^3 \bx  ~   \W (\bar \chi) \aleph^*_\ell(k)  j_\ell(k\chi)  Y^*_{\ell m} ({\hat \bn})    ~\WW^{\alpha} \left(\bar \chi, \eta, \frac{\p}{\p \bar \chi}, \frac{\p}{\p  \eta} \right)\frac{1}{2} \Delta^{b (2)}(\bx, \eta)\;,
 \eea
where $\WW^{\alpha} $ 
  has been introduced in  Eq. (\ref{eq:defWW}), 
 and the superscript $\alpha[b]$ means that we are taking combinations of different explicit   ``projection effect''  expressions $\WW$, labeled by the superscript index $\alpha$, with $\Delta$ labeled by the superscript index $b$.  
  
Here $ \Delta^{b (2)}=\delta_g^{(2)},\Phi^{(2)},\Psi^{(2)},v^{(2)}$, respectively, for $b=\delta_g^{(2)},\Phi^{(2)},\Psi^{(2)},v^{(2)}$ and 
\bea\label{F2-b}
\frac{1}{2} \Delta^{b (2)}(\tilde \bk, \eta) &= & \int \frac{\ud^3 \bp}{(2\pi)^3} \frac{\ud^3 \bq}{(2\pi)^3} ~(2\pi)^3 \delta^D \left(\bp+\bq-\tilde \bk \right)~  \frac{1}{2} F^{b (2)}(\bp, \bq, \tilde k; \eta) \T^{\Phi}(\bp, \eta)\T^{\Phi}(\bq, \eta)  \Phi_{\rm p}(\bp) \Phi_{\rm p}(\bq) \;,
\eea
where $F^{b (2)}(\bp, \bq, \tilde k; \eta)$ are specific kernel functions (fully symmetric functions) and depend also on the  parameters related to primordial non-Gaussianity. Explicit expressions for  $F^{b (2)}(\bp, \bq, \tilde k; \eta)$ are reported in Appendix \ref{F_2}.

Using Eq.\ (\ref{Deltag2-loc-1-general_relation}) we find
\bea
\label{Fourier_Deltag2-loc-1}
&&\frac{1}{2} \WW^{\alpha} \left(\bar \chi, \eta, \frac{\p}{\p \bar \chi}, \frac{\p}{\p  \eta} \right) \Delta^{b (2)}(\bx, \eta) =  \int \frac{\ud^3\tilde \bk}{(2\pi)^3} ~ \frac{1}{2} \WW^{\alpha} \left(\bar \chi, \eta, \frac{\p}{\p \bar \chi}, \frac{\p}{\p  \eta} \right)\left[\Delta^{b (2)}(\tilde \bk, \eta) e^{i \tilde\bk  \bx}\right]\nonumber\\
&& = \sum_{\ellvp \mvp \ellvq \mvq}  (4\pi)^2 i^{\ellvp+\ellvq}\Bigg\{  \int \frac{\ud^3 \bp}{(2\pi)^3} \frac{\ud^3 \bq}{(2\pi)^3}  \WW^{\alpha} \left(\bar \chi, \eta, \frac{\p}{\p \bar \chi}, \frac{\p}{\p  \eta} \right)\left[\frac{1}{2} F^{b (2)}(\bp, \bq, \tilde k; \eta) j_{\ellvp} (p \bar \chi) j_{\ellvq} (q \bar \chi) \T^{\Phi}( \bp, \eta)\T^{\Phi}( \bq, \eta)\right] \nonumber\\
&&~\times Y_{\ellvp \mvp}^*(\hat \bp)Y_{\ellvq \mvq}^*(\hat \bq)  \Phi_{\rm p}(\bp) \Phi_{\rm p}(\bq) \Bigg\} Y_{\ellvp \mvp}(\hat \bn)Y_{\ellvq \mvq}(\hat \bn)\;.
\eea
Here, by construction, we note that, in these kernels, $\tilde k$ is related to $p,q,$ and the angle between them, $\cos(\theta_{\rm p q})$ via 
 $$\tilde k[p,q,\cos(\theta_{\rm p q})] = \sqrt{p^2+q^2+2pq\cos(\theta_{\rm p q})}\;,$$ where $\cos(\theta_{\rm p q})= \hat \bp \cdot \hat \bq$.
Then $F^{b (2)}(\bp, \bq, \tilde k; \eta) =F^{b (2)} [p,q,\cos(\theta_{\rm p q}), \eta]$ and in general we can expand in Legendre polynomials the dependence of the kernel on $\theta_{\rm pq}$ using 
\be\label{projLegendre}
 F_{\bar \ell}^{b (2)}(p, q; \eta) =\frac{2\bar \ell +1}{2}\int_{-1}^{1} \ud\cos(\theta_{\rm p q}) ~ \PP_{\barell} [\cos(\theta_{\rm p q})] F^{b (2)} [p,q,\cos(\theta_{\rm p q}); \eta]\,.
\ee
 Hence the second-order kernels can be expanded as 
\be
\frac{1}{2} F^{b (2)}(\bp, \bq, \tilde k; \eta) =\sum_{\bar \ell} \frac{1}{2} F_{\bar \ell}^{b (2)}(p, q; \eta) ~\PP_{\barell} [\cos(\theta_{\rm p q})]=
\sum_{\bar \ell \bar m} \frac{4\pi}{2\bar \ell +1} ~ \frac{1}{2}F_{\bar \ell}^{b (2)}(p, q; \eta)    Y_{\barell \barm} (\hat \bp) Y_{\barell \barm}^*(\hat \bq)\;.
\ee

In Appendix \ref{NewcalM2} we present an alternative way to compute these terms. This new method might be important if $ F^{b (2)}(\bp, \bq, \tilde k; \eta)$ is function of ${\tilde k}^n$, where $n<0$.

Finally we find
 \bea
 \M_{\ell m \ellvp \mvp \ellvq \mvq \barell \barm}^{\alpha[b](2)}(k; p, q) &=& (4\pi)^3 \aleph^*_\ell(k)  (-1)^m i^{\ellvp+\ellvq} (2\bar \ell +1)^{-1} \G^{\ell \ellvp \ellvq }_{-m  \mvp \mvq} \int \ud \bar \chi~\bar \chi^2  \W (\bar \chi)
\nonumber\\ 
&\times & \Bigg\{ \WW^{\alpha} \left(\bar \chi,  \eta, \frac{\p}{\p \bar \chi}, \frac{\p}{\p  \eta} \right)  \left[\frac{1}{2}  F_{\bar \ell}^{b (2)}(p, q; \eta) j_{\ellvp} (p \bar \chi) j_{\ellvq} (q \bar \chi) \T^{\Phi}(\bp, \eta)\T^{\Phi}(\bq, \eta)\right]  \Bigg\} j_\ell (k  \bar \chi)\;.\nonumber\\
\eea
Below we show explicitly all the terms in Eq.\ (\ref{Deltag2-loc-1}):
 \bea\label{delta_g^(2)}
 \M_{\ell m \ellvp \mvp \ellvq \mvq \barell \barm}^{ \delta_g^{(2)}(2)}(k; p, q) &=& (4\pi)^3 \aleph^*_\ell(k)  (-1)^m i^{\ellvp+\ellvq} (2\bar \ell +1)^{-1} \G^{\ell \ellvp \ellvq }_{-m  \mvp \mvq} \int \ud \bar \chi~\bar \chi^2  \W (\bar \chi)
\nonumber\\ 
&\times &   \left[\frac{1}{2}  F_{\bar \ell}^{ \delta_g (2)}(p, q; \eta) j_{\ellvp} (p \bar \chi) j_{\ellvq} (q \bar \chi) \T^{\Phi}(\bp, \eta)\T^{\Phi}(\bq, \eta) \right]  j_\ell (k  \bar \chi)\;,
\eea

 \bea\label{Psi^(2)}
 \M_{\ell m \ellvp \mvp \ellvq \mvq \barell \barm}^{\Psi^{(2)}(2)}(k; p, q) &=& (4\pi)^3 \aleph^*_\ell(k)  (-1)^m i^{\ellvp+\ellvq} (2\bar \ell +1)^{-1} \G^{\ell \ellvp \ellvq }_{-m  \mvp \mvq} \int \ud \bar \chi~\bar \chi^2  \W (\bar \chi)\left[  -2  \left(1 - \Q\right)\right]
\nonumber\\ 
&\times &  \left[\frac{1}{2}  F_{\bar \ell}^{\Psi (2)}(p, q; \eta) j_{\ellvp} (p \bar \chi) j_{\ellvq} (q \bar \chi) \T^{\Phi}(\bp, \eta)\T^{\Phi}(\bq, \eta) \right]  j_\ell (k  \bar \chi)\;,
\eea

 \bea
 \M_{\ell m \ellvp \mvp \ellvq \mvq \barell \barm}^{\Phi^{(2)}(2)}(k; p, q) &=& (4\pi)^3 \aleph^*_\ell(k)  (-1)^{m+1} i^{\ellvp+\ellvq} (2\bar \ell +1)^{-1} \G^{\ell \ellvp \ellvq }_{-m  \mvp \mvq} \int \ud \bar \chi~\bar \chi^2  \W (\bar \chi) \bigg[ b_e-2 \Q -1 -   \frac{\cH'}{\cH^2} 
\nonumber\\ 
& &-  2\frac{\left(1 - \Q\right)}{\bar \chi \cH} \bigg]\times \left[\frac{1}{2}  F_{\bar \ell}^{\Phi (2)}(p, q; \eta) j_{\ellvp} (p \bar \chi) j_{\ellvq} (q \bar \chi)  \T^{\Phi}(\bp, \eta)\T^{\Phi}(\bq, \eta) \right]  j_\ell (k  \bar \chi)\;,
\eea

 \bea
 \M_{\ell m \ellvp \mvp \ellvq \mvq \barell \barm}^{\Psi^{(2)}{'} (2)}(k; p, q) &=& (4\pi)^3 \aleph^*_\ell(k)  (-1)^m i^{\ellvp+\ellvq} (2\bar \ell +1)^{-1} \G^{\ell \ellvp \ellvq }_{-m  \mvp \mvq} \int \ud \bar \chi~\bar \chi^2  \W (\bar \chi) \left(\frac{1}{ \cH}\right)
\nonumber\\ 
&\times & \frac{\p}{\p  \eta} \left[\frac{1}{2}  F_{\bar \ell}^{\Psi (2)}(p, q; \eta)  \T^{\Phi}(\bp, \eta)\T^{\Phi}(\bq, \eta) \right]  j_{\ellvp} (p \bar \chi) j_{\ellvq} (q \bar \chi)  j_\ell (k  \bar \chi)\;,
\eea 

 \bea
 \M_{\ell m \ellvp \mvp \ellvq \mvq \barell \barm}^{\p_\| v^{(2)}  (2)}(k; p, q) &=& (4\pi)^3 \aleph^*_\ell(k)  (-1)^m i^{\ellvp+\ellvq} (2\bar \ell +1)^{-1} \G^{\ell \ellvp \ellvq }_{-m  \mvp \mvq} \int \ud \bar \chi~\bar \chi^2  \W (\bar \chi)\left[ b_e-2 \Q  -   \frac{\cH'}{\cH^2} -2  \frac{\left(1 - \Q\right)}{\bar \chi \cH} \right]
\nonumber\\ 
&\times &\frac{1}{2}  F_{\bar \ell}^{v (2)}(p, q; \eta)  \frac{\p}{\p \bar \chi}  \left[ j_{\ellvp} (p \bar \chi) j_{\ellvq} (q \bar \chi) \right]  ~ j_\ell (k  \bar \chi) ~\T^{\Phi}(\bp, \eta)\T^{\Phi}(\bq, \eta) \;,
\eea

 \bea\label{v^(2)}
 \M_{\ell m \ellvp \mvp \ellvq \mvq \barell \barm}^{\p_\|^2 v^{(2)} (2)}(k; p, q) &=& (4\pi)^3 \aleph^*_\ell(k)  (-1)^{m+1} i^{\ellvp+\ellvq} (2\bar \ell +1)^{-1} \G^{\ell \ellvp \ellvq }_{-m  \mvp \mvq} \int \ud \bar \chi~\bar \chi^2  \W (\bar \chi)\left(\frac{1}{ \cH }\right)
\nonumber\\ 
&\times &\frac{1}{2}  F_{\bar \ell}^{v (2)}(p, q; \eta)  \frac{\p^2}{\p \bar \chi^2}  \left[ j_{\ellvp} (p \bar \chi) j_{\ellvq} (q \bar \chi) \right]  ~ j_\ell (k  \bar \chi) ~\T^{\Phi}(\bp, \eta)\T^{\Phi}(\bq, \eta) \;.
\eea

\subsection{Terms from $\Delta_{g\, {\rm loc}-2}^{(2)}$ [see Eq.\ (\ref{Deltag2-loc-2})]}

All local projection terms with time and partial space derivatives along the line-of-sight direction of $\Delta_{g\, {\rm loc}-2}^{(2)}$ in Eq.\ (\ref{Deltag2-loc-2}) can be written in the following way:
\begin{equation}
{1 \over 2} \Delta^{ij(2)}(\bx, \eta) = \left[\WW^{i}\left(\bar \chi,  \eta, \frac{\p}{\p  \bar \chi}, \frac{\p}{\p  \eta}, \triangle_{\hat  \bn}\right) \Delta^{i(1)}(\bx, \eta) \right]~ \left[ \WW^{j}\left(\bar \chi,  \eta, \frac{\p}{\p \bar \chi}, \frac{\p}{\p  \eta}, \triangle_{\hat  \bn}\right)  \Delta^{j(1)}(\bx, \eta) \right]\;,
\end{equation}
where the superscript $ij$ means that we are taking combinations of different $\WW^b \Delta^{b(1)}$   labeled by the superscript indices $b=i$ or $j$,  where $\Delta^{i(1)}(\bx, \eta)$ can be $\Phi,v,\delta_g^{(1)}$. [Note that in Eq.~(\ref{sumklm}) the index $a$  also runs through all  combinations of $i,j$.] Applying the weight function $ \WW^{j}$ on $\Delta^{j(1)}$ we have
\bea
 \WW^{j}\left(\bar \chi,  \eta, \frac{\p}{\p  \bar \chi}, \frac{\p}{\p  \eta}, \triangle_{\hat  \bn}\right) \Delta^{j(1)}(\bx, \eta) = \int \frac{\ud^3\bk}{(2\pi)^3}   \WW^{j}\left(\bar \chi,  \eta, \frac{\p}{\p \bar \chi}, \frac{\p}{\p  \eta}, \triangle_{\hat  \bn}\right)   \T^{j}(\bk, \eta) \Phi_{\rm p}(\bk) e^{i \bk \bx} \\
= \sum_{\ell m} 4\pi i^\ell  Y_{\ell m}(\hat \bn) \int \frac{\ud^3\bk}{(2\pi)^3}  \left[\WW_\ell^{j} \left(\bar \chi,  \eta, \frac{\p}{\p \bar \chi}, \frac{\p}{\p  \eta} \right)   \T^{j}(\bk, \eta) j_\ell (k \bar \chi) \right] Y_{\ell m}^*(\hat \bk) \Phi_{\rm p}(\bk)\;,
\eea
where we have used Eq. (\ref{WW}).
Starting from Eq.(\ref{Delta-2order-klm}) and using Eq.(\ref{gaunt}), we find
\bea
\frac{1}{2}\Delta_{\ell m}^{ij(2)}(k)&=&\sum_{\ellvp \mvp \ellvq \mvq} \int \frac{\ud^3 \bp}{(2\pi)^3} \frac{\ud^3 \bq}{(2\pi)^3} \Bigg\{ (4\pi)^2 \aleph^*_\ell(k)  (-1)^m i^{\ellvp+\ellvq}  \G^{\ell \ellvp \ellvq }_{-m  \mvp \mvq} \int \ud \bar \chi~\bar \chi^2  \W (\bar \chi)
\nonumber\\ 
&&\times \frac{1}{2} \bigg[ \WW_{\ellvp}^{i} \left(\bar \chi,  \eta, \frac{\p}{\p \bar \chi}, \frac{\p}{\p  \eta} \right)   \T^{i}(\bp, \eta) j_{\ellvp} (p \bar \chi) ~~ \WW_{\ellvq}^{j} \left(\bar \chi,  \eta, \frac{\p}{\p \bar \chi}, \frac{\p}{\p  \eta} \right)   \T^{j}(\bq, \eta) j_{\ellvq} (q \bar \chi) \nonumber\\
 &+& \WW_{\ellvq}^{i} \left(\bar \chi,  \eta, \frac{\p}{\p \bar \chi}, \frac{\p}{\p  \eta} \right)   \T^{i}(\bq, \eta) j_{\ellvq} (q \bar \chi) ~~ \WW_{\ellvp}^{j} \left(\bar \chi,  \eta, \frac{\p}{\p \bar \chi}, \frac{\p}{\p  \eta} \right)   \T^{j}(\bp, \eta) j_{\ellvp} (p \bar \chi) \bigg]  j_\ell (k  \bar \chi)\Bigg\} \nonumber\\
   &&\times ~Y_{\ellvp \mvp}^*(\hat \bp)   Y_{\ellvq \mvq}^*(\hat \bq) \Phi_{\rm p}(\bp) \Phi_{\rm p}(\bq)\;.
\eea
 Then, explicitly, we have
 \bea
 \M_{\ell m \ellvp \mvp \ellvq \mvq \barell \barm}^{ij(2)}(k; p, q) &=& \delta^K_{\barell 0} \delta^K_{\barm 0}(4\pi)^3 \aleph^*_\ell(k)  (-1)^m i^{\ellvp+\ellvq}  \G^{\ell \ellvp \ellvq }_{-m  \mvp \mvq} \int \ud \bar \chi~\bar \chi^2  \W (\bar \chi)
\nonumber\\ 
&&\times \frac{1}{2} \bigg[ \WW_{\ellvp}^{i} \left(\bar \chi,  \eta, \frac{\p}{\p \bar \chi}, \frac{\p}{\p  \eta} \right)   \T^{i}(\bp, \eta) j_{\ellvp} (p \bar \chi) ~~ \WW_{\ellvq}^{j} \left(\bar \chi,  \eta, \frac{\p}{\p \bar \chi}, \frac{\p}{\p  \eta} \right)   \T^{j}(\bq, \eta) j_{\ellvq} (q \bar \chi) \nonumber\\
 &+& \WW_{\ellvq}^{i} \left(\bar \chi,  \eta, \frac{\p}{\p \bar \chi}, \frac{\p}{\p  \eta} \right)   \T^{i}(\bq, \eta) j_{\ellvq} (q \bar \chi) ~~ \WW_{\ellvp}^{j} \left(\bar \chi,  \eta, \frac{\p}{\p \bar \chi}, \frac{\p}{\p  \eta} \right)   \T^{j}(\bp, \eta) j_{\ellvp} (p \bar \chi) \bigg]  j_\ell (k  \bar \chi)\;. \nonumber \\
 \eea
 Replacing the indices ${ij}$ by  the corresponding symbol for each additive contribution in Eq.(\ref{Deltag2-loc-2})--and thus making more explicit their physical meaning--below we list all  $ \M^{ij}_{\ell m \ellvp \mvp \ellvq \mvq \barell \barm} $:
\bea
&&\label{MdeltaPhi(2)}
\M_{\ell m \ellvp \mvp \ellvq \mvq \barell \barm}^{\delta_g \Phi(2)}(k; p, q)=\delta^K_{\barell 0} \delta^K_{\barm 0}(4\pi)^3 \aleph^*_\ell(k)  (-1)^m i^{\ellvp+\ellvq}  \G^{\ell \ellvp \ellvq }_{-m  \mvp \mvq} \int \ud \bar \chi~\bar \chi^2  \W (\bar \chi) \nonumber \\
&&\times {1\over2} \left[- b_e -1+ 4 \Q   +  \frac{\cH'}{\cH^2}  + \frac{2}{\bar \chi \cH} \left(1 - \Q\right)  \right]  \bigg[\T^{\delta_g}(\bp, \eta)  \T^{\Phi}(\bq, \eta) +\T^{\delta_g}(\bq, \eta)  \T^{\Phi}(\bp, \eta)  \bigg] j_{\ellvp} (p \bar \chi)j_{\ellvq} (q \bar \chi)  j_\ell (k \bar \chi)\;,\nonumber\\
  \eea
\bea
&&\M_{\ell m \ellvp \mvp \ellvq \mvq \barell \barm}^{\delta_g \p_\| v(2)}(k; p, q) =\delta^K_{\barell 0} \delta^K_{\barm 0}(4\pi)^3 \aleph^*_\ell(k)  (-1)^m i^{\ellvp+\ellvq}  \G^{\ell \ellvp \ellvq }_{-m  \mvp \mvq} \int \ud \bar \chi~\bar \chi^2  \W (\bar \chi) \left[ b_e - 2\Q   -  \frac{\cH'}{\cH^2}  - \frac{2}{\bar \chi \cH} \left(1 - \Q\right)  \right] \nonumber \\
&&\times {1\over2} \bigg\{\T^{\delta_g}(\bp, \eta)  \T^{v}(\bq, \eta) ~ j_{\ellvp} (p \bar \chi)\left[ \frac{\p}{\p \bar \chi}j_{\ellvq} (q \bar \chi)\right]  +\T^{\delta_g}(\bq, \eta)  \T^{v}(\bp, \eta)~ j_{\ellvq} (q \bar \chi) \left[\frac{\p}{\p \bar \chi} j_{\ellvp} (p \bar \chi)\right]   \bigg\}   j_\ell (k  \bar \chi)\;,
 \eea
\bea
&&\M_{\ell m \ellvp \mvp \ellvq \mvq \barell \barm}^{\delta_g \p_\|^2 v(2)}(k; p, q)= \delta^K_{\barell 0} \delta^K_{\barm 0}(4\pi)^3 \aleph^*_\ell(k)  (-1)^m i^{\ellvp+\ellvq}  \G^{\ell \ellvp \ellvq }_{-m  \mvp \mvq} \int \ud \bar \chi~\bar \chi^2  \W (\bar \chi) \left( - \frac{1}{\cH} \right) \nonumber \\
&&\times {1\over2} \bigg\{\T^{\delta_g}(\bp, \eta)  \T^{v}(\bq, \eta) ~ j_{\ellvp} (p \bar \chi)\left[ \frac{\p^2}{\p \bar \chi^2}j_{\ellvq} (q \bar \chi)\right]  +\T^{\delta_g}(\bq, \eta)  \T^{v}(\bp, \eta)~ j_{\ellvq} (q \bar \chi) \left[\frac{\p^2}{\p \bar \chi^2} j_{\ellvp} (p \bar \chi)\right]   \bigg\}   j_\ell (k \bar \chi)\;,
 \eea
\bea
&& \M_{\ell m \ellvp \mvp \ellvq \mvq \barell \barm}^{\delta_g \Phi'(2)}(k; p, q)=\delta^K_{\barell 0} \delta^K_{\barm 0}(4\pi)^3 \aleph^*_\ell(k)  (-1)^m i^{\ellvp+\ellvq}  \G^{\ell \ellvp \ellvq }_{-m  \mvp \mvq} \int \ud \bar \chi~\bar \chi^2  \W (\bar \chi) \left(  \frac{1}{\cH} \right) \nonumber \\
&&\times {1\over2}  \bigg\{\T^{\delta_g}(\bp, \eta) \left[\frac{\p}{\p \eta} \T^{\Phi}(\bq, \eta)\right] +\T^{\delta_g}(\bq, \eta)  \left[ \frac{\p}{\p \eta} \T^{\Phi}(\bp, \eta)\right]  \bigg\} j_{\ellvp} (p \bar \chi)j_{\ellvq} (q \bar \chi)  j_\ell (k\bar  \chi)\;,
 \eea
\bea
&&\M_{\ell m \ellvp \mvp \ellvq \mvq \barell \barm}^{\p_\|  \delta_g \Phi(2)}(k; p, q)= \delta^K_{\barell 0} \delta^K_{\barm 0}(4\pi)^3 \aleph^*_\ell(k)  (-1)^m i^{\ellvp+\ellvq}  \G^{\ell \ellvp \ellvq }_{-m  \mvp \mvq} \int \ud \bar \chi~\bar \chi^2  \W (\bar \chi) \left(  \frac{1}{\cH} \right) \nonumber \\
&&\times {1\over2} \bigg\{\T^{\delta_g}(\bp, \eta)  \T^{\Phi}(\bq, \eta) ~ \left[ \frac{\p}{\p \bar \chi}j_{\ellvp} (p \bar \chi)\right] j_{\ellvq} (q \bar \chi)  + \T^{\delta_g}(\bq, \eta) \T^{\Phi}(\bp, \eta)~\left[ \frac{\p}{\p \bar \chi}  j_{\ellvq} (q \bar \chi) \right]  j_{\ellvp} (p \bar \chi) \bigg\}   j_\ell (k  \bar \chi)\;,
 \eea
\bea
&& \M_{\ell m \ellvp \mvp \ellvq \mvq \barell \barm}^{\delta_g' \Phi(2)}(k; p, q)= \delta^K_{\barell 0} \delta^K_{\barm 0}(4\pi)^3 \aleph^*_\ell(k)  (-1)^m i^{\ellvp+\ellvq}  \G^{\ell \ellvp \ellvq }_{-m  \mvp \mvq} \int \ud \bar \chi~\bar \chi^2  \W (\bar \chi) \left( - \frac{1}{\cH} \right) \nonumber \\
&&\times {1\over2}  \bigg\{ \left[\frac{\p}{\p \eta} \T^{\delta_g}(\bp, \eta) \right]  \T^{\Phi}(\bq, \eta)+ \left[ \frac{\p}{\p \eta}T^{\delta_g}(\bq, \eta)  \right]  \T^{\Phi}(\bp, \eta) \bigg\} j_{\ellvp} (p \bar \chi)j_{\ellvq} (q \bar \chi)  j_\ell (k\bar  \chi)\;,
 \eea
\bea
&&\M_{\ell m \ellvp \mvp \ellvq \mvq \barell \barm}^{\p_\| \delta_g \p_\| v(2)}(k; p, q)= \delta^K_{\barell 0} \delta^K_{\barm 0}(4\pi)^3 \aleph^*_\ell(k)  (-1)^m i^{\ellvp+\ellvq}  \G^{\ell \ellvp \ellvq }_{-m  \mvp \mvq} \int \ud \bar \chi~\bar \chi^2  \W (\bar \chi) \left( - \frac{1}{\cH} \right) \nonumber \\
&&\times {1\over2} \bigg[\T^{\delta_g}(\bp, \eta)  \T^{v}(\bq, \eta)  +\T^{\delta_g}(\bq, \eta)  \T^{v}(\bp, \eta)   \bigg]   \left[\frac{\p}{\p \bar \chi} j_{\ellvp} (p \bar \chi)\right]  \left[ \frac{\p}{\p \bar \chi} j_{\ellvq} (q \bar \chi)\right]  j_\ell (k \bar \chi)\;,
 \eea 
\bea
&&\M_{\ell m \ellvp \mvp \ellvq \mvq \barell \barm}^{\delta_g' \p_\| v(2)}(k; p, q)=  \delta^K_{\barell 0} \delta^K_{\barm 0}(4\pi)^3 \aleph^*_\ell(k)  (-1)^m i^{\ellvp+\ellvq}  \G^{\ell \ellvp \ellvq }_{-m  \mvp \mvq} \int \ud \bar \chi~\bar \chi^2  \W (\bar \chi) \left(  \frac{1}{\cH} \right) \nonumber \\
&&\times {1\over2} \bigg\{ \left[\frac{\p}{\p \eta} \T^{\delta_g}(\bp, \eta) \right]   \T^{v}(\bq, \eta) ~ j_{\ellvp} (p \bar \chi)\left[ \frac{\p}{\p \bar \chi}j_{\ellvq} (q \bar \chi)\right]  +\left[ \frac{\p}{\p \eta}T^{\delta_g}(\bq, \eta)  \right]  \T^{v}(\bp, \eta)~ j_{\ellvq} (q \bar \chi) \left[\frac{\p}{\p \bar \chi} j_{\ellvp} (p \bar \chi)\right]   \bigg\}   j_\ell (k \bar \chi)\;,\nonumber\\
 \eea
\bea
&& \M_{\ell m \ellvp \mvp \ellvq \mvq \barell \barm}^{\Phi^2(2)}(k; p, q)=\delta^K_{\barell 0} \delta^K_{\barm 0}(4\pi)^3 \aleph^*_\ell(k)  (-1)^m i^{\ellvp+\ellvq}  \G^{\ell \ellvp \ellvq }_{-m  \mvp \mvq} \int \ud \bar \chi~\bar \chi^2  \W (\bar \chi) ~ {1\over2} \Bigg[ -5 + 4b_e    +b_e^2  + \frac{\p b_e}{\p \ln \bar a}   \nonumber \\
&& +8 \Q  +16 \Q^2    - 8 b_e \Q   - 16 \frac{\p \Q}{\p \ln \bar L}      -8 \frac{\p \Q}{\p \ln \bar a}  +2 \left( -2 -  b_e  + 4 \Q \right) \frac{\cH'}{\cH^2} -\frac{\cH'' }{\cH^3} +3\left( \frac{\cH' }{\cH^2} \right)^2 + \frac{6}{\bar \chi} \frac{\cH' }{\cH^3} \left(1 - \Q\right)   + \frac{4}{\bar \chi \cH} \bigg(    - b_e  \nonumber \\
&&   - 4\Q^2   + b_e \Q    + 4\frac{\p \Q}{\p \ln \bar L}   + \frac{\p \Q}{\p \ln \bar a} \bigg)     +  \frac{2}{\bar \chi^2 \cH^2} \bigg(1-\Q +2\Q^2  -2\frac{\p \Q}{\p \ln \bar L}\bigg)     \Bigg]   \T^{\Phi}(\bp, \eta) \T^{\Phi}(\bq, \eta) ~ j_{\ellvp} (p \bar \chi)j_{\ellvq} (q \bar \chi)  j_\ell (k \bar \chi)\;, \nonumber \\
 \eea
\bea
&&\M_{\ell m \ellvp \mvp \ellvq \mvq \barell \barm}^{\Phi \p_\| v (2)}(k; p, q)= \delta^K_{\barell 0} \delta^K_{\barm 0}(4\pi)^3 \aleph^*_\ell(k)  (-1)^m i^{\ellvp+\ellvq}  \G^{\ell \ellvp \ellvq }_{-m  \mvp \mvq} \int \ud \bar \chi~\bar \chi^2  \W (\bar \chi)  \Bigg[ - b_e  - 6\Q  - b_e^2  + 6 \Q b_e \nonumber \\
&& - 8 \Q^2   + 8 \frac{\p \Q}{\p \ln \bar L}   +6 \frac{\p \Q}{\p \ln \bar a}     - \frac{\p b_e}{\p \ln \bar a} + \left(1+ 2b_e -6 \Q \right) \frac{\cH' }{\cH^2}+\frac{\cH'' }{\cH^3} -3\left( \frac{\cH' }{\cH^2} \right)^2  - \frac{6}{\bar \chi} \frac{\cH' }{\cH^3} \left(1 - \Q\right)  \nonumber \\
&&+  \frac{2}{\bar \chi \cH}  \left(-1+ 2b_e  + \Q +6\Q^2    - 2b_e \Q    - 6\frac{\p \Q}{\p \ln \bar L}   - 2 \frac{\p \Q}{\p \ln \bar a} \right)  +  \frac{2}{\bar \chi^2 \cH^2} \bigg(- 1 +\Q - 2\Q^2  + 2\frac{\p \Q}{\p \ln \bar L}\bigg) \Bigg] \nonumber \\
&&\times {1\over2} \Bigg\{\T^{\Phi}(\bp, \eta)  \T^{v}(\bq, \eta) ~ j_{\ellvp} (p \bar \chi)\left[ \frac{\p}{\p \bar \chi}j_{\ellvq} (q \bar \chi)\right]  +\T^{\Phi}(\bq, \eta)  \T^{v}(\bp, \eta)~ j_{\ellvq} (q \bar \chi) \left[\frac{\p}{\p \bar \chi} j_{\ellvp} (p \bar \chi)\right]   \Bigg\}   j_\ell (k \bar \chi)\;,
 \eea
 \bea
&&\M_{\ell m \ellvp \mvp \ellvq \mvq \barell \barm}^{\Phi \p_\|^2 v (2)}(k; p, q)= \delta^K_{\barell 0} \delta^K_{\barm 0}(4\pi)^3 \aleph^*_\ell(k)  (-1)^m i^{\ellvp+\ellvq}  \G^{\ell \ellvp \ellvq }_{-m  \mvp \mvq} \int \ud \bar \chi~\bar \chi^2  \W (\bar \chi)  \frac{1}{\cH}\bigg[ -1- 5 \Q    +   2b_e -  3 \frac{\cH'}{\cH^2} \nonumber \\
&& -  \frac{4 \left(1 - \Q\right) }{\bar \chi \cH}   \bigg]  {1\over2} \Bigg\{\T^{\Phi}(\bp, \eta)  \T^{v}(\bq, \eta) ~ j_{\ellvp} (p \bar \chi)\left[ \frac{\p^2}{\p \bar \chi^2}j_{\ellvq} (q \bar \chi)\right]  +\T^{\Phi}(\bq, \eta)  \T^{v}(\bp, \eta)~ j_{\ellvq} (q \bar \chi) \left[\frac{\p^2}{\p \bar \chi^2} j_{\ellvp} (p \bar \chi)\right]   \Bigg\}   j_\ell (k \bar \chi)\;, \nonumber\\
 \eea
 \bea
&& \M_{\ell m \ellvp \mvp \ellvq \mvq \barell \barm}^{\Phi \p_\|^3 v (2)}(k; p, q)= \delta^K_{\barell 0} \delta^K_{\barm 0}(4\pi)^3 \aleph^*_\ell(k)  (-1)^m i^{\ellvp+\ellvq}  \G^{\ell \ellvp \ellvq }_{-m  \mvp \mvq} \int \ud \bar \chi~\bar \chi^2  \W (\bar \chi) \left(- \frac{1}{\cH^2}\right) \nonumber\\
&& \times {1\over2} \Bigg\{\T^{\Phi}(\bp, \eta)  \T^{v}(\bq, \eta) ~ j_{\ellvp} (p \bar \chi)\left[ \frac{\p^3}{\p \bar \chi^3}j_{\ellvq} (q \bar \chi)\right]  +\T^{\Phi}(\bq, \eta)  \T^{v}(\bp, \eta)~ j_{\ellvq} (q \bar \chi) \left[\frac{\p^3}{\p \bar \chi^3} j_{\ellvp} (p \bar \chi)\right]   \Bigg\}   j_\ell (k \bar \chi)\;, 
 \eea
\bea
&&\M_{\ell m \ellvp \mvp \ellvq \mvq \barell \barm}^{\Phi \Phi' (2)}(k; p, q)= \delta^K_{\barell 0} \delta^K_{\barm 0}(4\pi)^3 \aleph^*_\ell(k)  (-1)^m i^{\ellvp+\ellvq}  \G^{\ell \ellvp \ellvq }_{-m  \mvp \mvq} \int \ud \bar \chi~\bar \chi^2  \W (\bar \chi) ~\nonumber \\
&& \times \frac{1}{\cH}\bigg(2  - 2b_e +4 \Q   + 3\frac{\cH' }{\cH^2}    +   \frac{4 \left(1 - \Q\right) }{\bar \chi \cH}   \bigg)
\frac{\p}{\p \eta} \left[   \T^{\Phi}(\bp, \eta) \T^{\Phi}(\bq, \eta)\right] ~ j_{\ellvp} (p \bar \chi)j_{\ellvq} (q \bar \chi)  j_\ell (k \bar \chi)\;, 
 \eea
\bea
&&\M_{\ell m \ellvp \mvp \ellvq \mvq \barell \barm}^{\Phi \p_\| \Phi (2)}(k; p, q)= \delta^K_{\barell 0} \delta^K_{\barm 0}(4\pi)^3 \aleph^*_\ell(k)  (-1)^m i^{\ellvp+\ellvq}  \G^{\ell \ellvp \ellvq }_{-m  \mvp \mvq} \int \ud \bar \chi~\bar \chi^2  \W (\bar \chi) ~ {1\over \cH} \left(-1 +2 \Q \right) \nonumber \\
&&    \T^{\Phi}(\bp, \eta) \T^{\Phi}(\bq, \eta) ~\frac{\p}{\p \bar \chi} \left[ j_{\ellvp} (p \bar \chi)j_{\ellvq} (q \bar \chi)\right]  j_\ell (k \bar \chi)\;, 
 \eea
\bea
&&\M_{\ell m \ellvp \mvp \ellvq \mvq \barell \barm}^{\Phi \p_\|^2 \Phi (2)}(k; p, q)= \delta^K_{\barell 0} \delta^K_{\barm 0}(4\pi)^3 \aleph^*_\ell(k)  (-1)^m i^{\ellvp+\ellvq}  \G^{\ell \ellvp \ellvq }_{-m  \mvp \mvq} \int \ud \bar \chi~\bar \chi^2  \W (\bar \chi) \left( - \frac{1}{\cH^2}\right) \nonumber\\
&&\times {1\over 2}  \T^{\Phi}(\bp, \eta) \T^{\Phi}(\bq, \eta) ~ \Bigg\{ j_{\ellvp} (p \bar \chi)\left[ \frac{\p^2}{\p \bar \chi^2} j_{\ellvq} (q \bar \chi)\right]  + j_{\ellvq} (q \bar \chi) \left[\frac{\p^2}{\p \bar \chi^2} j_{\ellvp} (p \bar \chi)\right]   \Bigg\}   j_\ell (k \bar \chi)\;,
 \eea
\bea
&&\M_{\ell m \ellvp \mvp \ellvq \mvq \barell \barm}^{(\Phi')^2 (2)}(k; p, q)=  \delta^K_{\barell 0} \delta^K_{\barm 0}(4\pi)^3 \aleph^*_\ell(k)  (-1)^m i^{\ellvp+\ellvq}  \G^{\ell \ellvp \ellvq }_{-m  \mvp \mvq} \int \ud \bar \chi~\bar \chi^2  \W (\bar \chi) \left(  \frac{1}{\cH^2}\right)  \nonumber \\
&& \times  \left[  \frac{\p}{\p \eta} \T^{\Phi}(\bp, \eta)\right]  \left[  \frac{\p}{\p \eta} \T^{\Phi}(\bq, \eta)\right] ~ j_{\ellvp} (p \bar \chi)j_{\ellvq} (q \bar \chi)  j_\ell (k \bar \chi)\;, 
\eea
\bea
&&\M_{\ell m \ellvp \mvp \ellvq \mvq \barell \barm}^{\Phi \p_\| \Phi' (2)}(k; p, q)=   \delta^K_{\barell 0} \delta^K_{\barm 0}(4\pi)^3 \aleph^*_\ell(k)  (-1)^m i^{\ellvp+\ellvq}  \G^{\ell \ellvp \ellvq }_{-m  \mvp \mvq} \int \ud \bar \chi~\bar \chi^2  \W (\bar \chi) \left(  \frac{1}{\cH^2} \right) \nonumber \\
&&\times {1\over2} \bigg\{\T^{\Phi}(\bp, \eta)   \left[\frac{\p}{\p \eta} \T^{\Phi}(\bq, \eta) \right]  ~ j_{\ellvp} (p \bar \chi)\left[ \frac{\p}{\p \bar \chi}j_{\ellvq} (q \bar \chi)\right]  + T^{\Phi}(\bq, \eta)    \left[ \frac{\p}{\p \eta} \T^{\Phi}(\bp, \eta)\right]~ j_{\ellvq} (q \bar \chi) \left[\frac{\p}{\p \bar \chi} j_{\ellvp} (p \bar \chi)\right]   \bigg\}   j_\ell (k \bar \chi)\;,\nonumber\\
\eea
\bea
&&\M_{\ell m \ellvp \mvp \ellvq \mvq \barell \barm}^{\Phi \Phi''(2)}(k; p, q)= \delta^K_{\barell 0} \delta^K_{\barm 0}(4\pi)^3 \aleph^*_\ell(k)  (-1)^m i^{\ellvp+\ellvq}  \G^{\ell \ellvp \ellvq }_{-m  \mvp \mvq} \int \ud \bar \chi~\bar \chi^2  \W (\bar \chi) \left( - \frac{1}{\cH^2} \right) \nonumber \\
&&\times {1\over2}  \bigg\{\T^{\Phi}(\bp, \eta) \left[\frac{\p^2}{\p \eta^2} \T^{\Phi}(\bq, \eta)\right] +\T^{\Phi}(\bq, \eta)  \left[ \frac{\p^2}{\p \eta^2} \T^{\Phi}(\bp, \eta)\right]  \bigg\} j_{\ellvp} (p \bar \chi)j_{\ellvq} (q \bar \chi)  j_\ell (k\bar  \chi)\;,
\eea
\bea
&&\M_{\ell m \ellvp \mvp \ellvq \mvq \barell \barm}^{(\p_\| v)^2 (2)}(k; p, q)= \delta^K_{\barell 0} \delta^K_{\barm 0}(4\pi)^3 \aleph^*_\ell(k)  (-1)^m i^{\ellvp+\ellvq}  \G^{\ell \ellvp \ellvq }_{-m  \mvp \mvq} \int \ud \bar \chi~\bar \chi^2  \W (\bar \chi)\nonumber \\
&& \times {1\over2}  \Bigg[ 1  -2 b_e +b_e^2+ \frac{\p b_e}{\p \ln \bar a} + 8 \Q  -4 \Q b_e+4 \Q^2  -4\frac{\p \Q}{\p \ln \bar L}-4\frac{\p \Q}{\p \ln \bar a} + 2\left(1- b_e + 2\Q \right) \frac{\cH' }{\cH^2}  -\frac{\cH'' }{\cH^3}  +3\left( \frac{\cH' }{\cH^2} \right)^2 \nonumber \\
&&  + \frac{6}{\bar \chi} \frac{\cH' }{\cH^3} \left(1 - \Q\right) + \frac{2}{\bar \chi \cH} \left( 2 - 2b_e -2 \Q  + 2b_e \Q  - 4 \Q^2 +4\frac{\p \Q}{\p \ln \bar L} +2 \frac{\p \Q}{\p \ln \bar a} \right)  +  \frac{2}{\bar \chi^2 \cH^2} \bigg(1-\Q +2\Q^2  -2\frac{\p \Q}{\p \ln \bar L}\bigg)     \Bigg]   \nonumber \\
&&\times   \T^{v}(\bp, \eta)   \T^{v}(\bq, \eta)     \left[\frac{\p}{\p \bar \chi} j_{\ellvp} (p \bar \chi)\right]  \left[ \frac{\p}{\p \bar \chi} j_{\ellvq} (q \bar \chi)\right]  j_\ell (k \bar \chi)\;, 
\eea
\bea
&&\M_{\ell m \ellvp \mvp \ellvq \mvq \barell \barm}^{(\p_\|^2 v)^2 (2)}(k; p, q)=  \delta^K_{\barell 0} \delta^K_{\barm 0}(4\pi)^3 \aleph^*_\ell(k)  (-1)^m i^{\ellvp+\ellvq}  \G^{\ell \ellvp \ellvq }_{-m  \mvp \mvq} \int \ud \bar \chi~\bar \chi^2  \W (\bar \chi) \left(  \frac{1}{\cH^2} \right) \nonumber \\
&&\times   \T^{v}(\bp, \eta)   \T^{v}(\bq, \eta)     \left[\frac{\p^2}{\p \bar \chi^2} j_{\ellvp} (p \bar \chi)\right]  \left[ \frac{\p^2}{\p \bar \chi^2} j_{\ellvq} (q \bar \chi)\right]  j_\ell (k \bar \chi)\;, 
\eea
\bea
&&\M_{\ell m \ellvp \mvp \ellvq \mvq \barell \barm}^{\Phi' \p_\| v (2)}(k; p, q)= \delta^K_{\barell 0} \delta^K_{\barm 0}(4\pi)^3 \aleph^*_\ell(k)  (-1)^m i^{\ellvp+\ellvq}  \G^{\ell \ellvp \ellvq }_{-m  \mvp \mvq} \int \ud \bar \chi~\bar \chi^2  \W (\bar \chi)  \left( \frac{1}{ \cH} \right)\bigg[-1+  2b_e - 2 \Q   -  3 \frac{\cH'}{\cH^2}   \nonumber \\
&& -\frac{4  \left(1 - \Q\right) }{\bar \chi \cH}  \bigg]\times  {1\over2} \Bigg\{\left[ \frac{\p}{\p \eta} \T^{\Phi}(\bp, \eta)\right]  \T^{v}(\bq, \eta) ~ j_{\ellvp} (p \bar \chi)\left[ \frac{\p}{\p \bar \chi}j_{\ellvq} (q \bar \chi)\right]  + \left[ \frac{\p}{\p \eta} \T^{\Phi}(\bq, \eta)\right]  \T^{v}(\bp, \eta)~ j_{\ellvq} (q \bar \chi) \left[\frac{\p}{\p \bar \chi} j_{\ellvp} (p \bar \chi)\right]   \Bigg\}   \nonumber \\
&&  j_\ell (k \bar \chi)\;,
\eea
\bea
&&\M_{\ell m \ellvp \mvp \ellvq \mvq \barell \barm}^{\p_\| v \p_\|^2 v (2)}(k; p, q)=  \delta^K_{\barell 0} \delta^K_{\barm 0}(4\pi)^3 \aleph^*_\ell(k)  (-1)^m i^{\ellvp+\ellvq}  \G^{\ell \ellvp \ellvq }_{-m  \mvp \mvq} \int \ud \bar \chi~\bar \chi^2  \W (\bar \chi)\left( \frac{1}{\cH}\right) \bigg[ 1 - 2 b_e +4 \Q  +  3 \frac{\cH'}{\cH^2}  \nonumber\\
&&+ \frac{4  \left(1 - \Q\right)}{\bar \chi \cH} \bigg]
\times {1\over 2}  \T^{v}(\bp, \eta) \T^{v}(\bq, \eta) ~ \Bigg\{ \left[\frac{\p}{\p \bar \chi} j_{\ellvp} (p \bar \chi)\right] \left[ \frac{\p^2}{\p \bar \chi^2} j_{\ellvq} (q \bar \chi)\right]  + \left[\frac{\p}{\p \bar \chi} j_{\ellvq} (q \bar \chi) \right] \left[\frac{\p^2}{\p \bar \chi^2} j_{\ellvp} (p \bar \chi)\right]   \Bigg\}   j_\ell (k \bar \chi)\;,
\eea
\bea
&&\M_{\ell m \ellvp \mvp \ellvq \mvq \barell \barm}^{\p_\| \Phi \p_\| v (2)}(k; p, q)= 2\delta^K_{\barell 0} \delta^K_{\barm 0}(4\pi)^3 \aleph^*_\ell(k)  (-1)^m i^{\ellvp+\ellvq}  \G^{\ell \ellvp \ellvq }_{-m  \mvp \mvq} \int \ud \bar \chi~\bar \chi^2  \W (\bar \chi)   \frac{\left(1-\Q\right)}{ \cH }     \nonumber \\
&&\times {1 \over 2} \bigg[\T^{\Phi}(\bp, \eta)  \T^{v}(\bq, \eta)  +\T^{\Phi}(\bq, \eta)  \T^{v}(\bp, \eta)   \bigg]    \left[\frac{\p}{\p \bar \chi} j_{\ellvp} (p \bar \chi)\right]  \left[ \frac{\p}{\p \bar \chi} j_{\ellvq} (q \bar \chi)\right]  j_\ell (k \bar \chi)\;, 
\eea
\bea
&&\M_{\ell m \ellvp \mvp \ellvq \mvq \barell \barm}^{\p_\|^2 \Phi \p_\| v (2)}(k; p, q)= \delta^K_{\barell 0} \delta^K_{\barm 0}(4\pi)^3 \aleph^*_\ell(k)  (-1)^m i^{\ellvp+\ellvq}  \G^{\ell \ellvp \ellvq }_{-m  \mvp \mvq} \int \ud \bar \chi~\bar \chi^2  \W (\bar \chi) \left(\frac{1}{\cH^2}\right) \nonumber \\
&&\times {1\over2} \Bigg\{\T^{\Phi}(\bp, \eta)  \T^{v}(\bq, \eta) ~\left[ \frac{\p^2}{\p \bar \chi^2} j_{\ellvp} (p \bar \chi) \right]  \left[ \frac{\p}{\p \bar \chi} j_{\ellvq} (q \bar \chi)\right]  +\T^{\Phi}(\bq, \eta)  \T^{v}(\bp, \eta)~\left[ \frac{\p^2}{\p \bar \chi^2} j_{\ellvq} (q \bar \chi) \right]\left[\frac{\p}{\p \bar \chi} j_{\ellvp} (p \bar \chi)\right]   \Bigg\}   j_\ell (k \bar \chi)\;, \nonumber \\
\eea
\bea
&&\M_{\ell m \ellvp \mvp \ellvq \mvq \barell \barm}^{\p_\| \Phi' \p_\| v (2)}(k; p, q)=  \delta^K_{\barell 0} \delta^K_{\barm 0}(4\pi)^3 \aleph^*_\ell(k)  (-1)^m i^{\ellvp+\ellvq}  \G^{\ell \ellvp \ellvq }_{-m  \mvp \mvq} \int \ud \bar \chi~\bar \chi^2  \W (\bar \chi) \left(-{1\over \cH^2}  \right)\nonumber \\
&&\times {1 \over 2} \Bigg\{\left[ \frac{\p}{\p \eta} \T^{\Phi}(\bp, \eta)\right]  \T^{v}(\bq, \eta)  + \left[ \frac{\p}{\p \eta} \T^{\Phi}(\bq, \eta)\right]  \T^{v}(\bp, \eta)   \Bigg\}   \left[\frac{\p}{\p \bar \chi} j_{\ellvp} (p \bar \chi)\right]  \left[ \frac{\p}{\p \bar \chi} j_{\ellvq} (q \bar \chi)\right]  j_\ell (k \bar \chi)\;, 
\eea
\bea
&&\M_{\ell m \ellvp \mvp \ellvq \mvq \barell \barm}^{\Phi'' \p_\| v (2)}(k; p, q)=  \delta^K_{\barell 0} \delta^K_{\barm 0}(4\pi)^3 \aleph^*_\ell(k)  (-1)^m i^{\ellvp+\ellvq}  \G^{\ell \ellvp \ellvq }_{-m  \mvp \mvq} \int \ud \bar \chi~\bar \chi^2  \W (\bar \chi) \left({1\over \cH^2}  \right)   \nonumber \\
&& \times  {1\over2} \Bigg\{\left[ \frac{\p^2}{\p \eta^2} \T^{\Phi}(\bp, \eta)\right]  \T^{v}(\bq, \eta) ~ j_{\ellvp} (p \bar \chi)\left[ \frac{\p}{\p \bar \chi}j_{\ellvq} (q \bar \chi)\right]  + \left[ \frac{\p^2}{\p \eta^2} \T^{\Phi}(\bq, \eta)\right]  \T^{v}(\bp, \eta)~ j_{\ellvq} (q \bar \chi) \left[\frac{\p}{\p \bar \chi} j_{\ellvp} (p \bar \chi)\right]   \Bigg\}   \nonumber \\
&&\times   j_\ell (k \bar \chi)\;,
\eea
\bea
&&\M_{\ell m \ellvp \mvp \ellvq \mvq \barell \barm}^{\p_\| v \p_\|^3 v (2)}(k; p, q)=  \delta^K_{\barell 0} \delta^K_{\barm 0}(4\pi)^3 \aleph^*_\ell(k)  (-1)^m i^{\ellvp+\ellvq}  \G^{\ell \ellvp \ellvq }_{-m  \mvp \mvq} \int \ud \bar \chi~\bar \chi^2  \W (\bar \chi)\left( \frac{1}{\cH^2}\right)  \nonumber\\
&&\times {1\over 2}  \T^{v}(\bp, \eta) \T^{v}(\bq, \eta) ~ \Bigg\{ \left[\frac{\p}{\p \bar \chi} j_{\ellvp} (p \bar \chi)\right] \left[ \frac{\p^3}{\p \bar \chi^3} j_{\ellvq} (q \bar \chi)\right]  + \left[\frac{\p}{\p \bar \chi} j_{\ellvq} (q \bar \chi) \right] \left[\frac{\p^3}{\p \bar \chi^3} j_{\ellvp} (p \bar \chi)\right]   \Bigg\}   j_\ell (k \bar \chi)\;,
\eea
\bea
&&\M_{\ell m \ellvp \mvp \ellvq \mvq \barell \barm}^{ \Phi' \p_\|^2 v (2)}(k; p, q)=-2 \delta^K_{\barell 0} \delta^K_{\barm 0}(4\pi)^3 \aleph^*_\ell(k)  (-1)^m i^{\ellvp+\ellvq}  \G^{\ell \ellvp \ellvq }_{-m  \mvp \mvq} \int \ud \bar \chi~\bar \chi^2  \W (\bar \chi)  \left(  \frac{1}{ \cH^2} \right)\nonumber \\
&& \times  {1\over2} \Bigg\{\left[ \frac{\p}{\p \eta} \T^{\Phi}(\bp, \eta)\right]  \T^{v}(\bq, \eta) ~ j_{\ellvp} (p \bar \chi)\left[ \frac{\p^2}{\p \bar \chi^2}j_{\ellvq} (q \bar \chi)\right]  + \left[ \frac{\p}{\p \eta} \T^{\Phi}(\bq, \eta)\right]  \T^{v}(\bp, \eta)~ j_{\ellvq} (q \bar \chi) \left[\frac{\p^2}{\p \bar \chi^2} j_{\ellvp} (p \bar \chi)\right]   \Bigg\}   \nonumber \\
&& \times j_\ell (k \bar \chi)\;,
\eea
\bea\label{M-TQ}
&&\M_{\ell m \ellvp \mvp \ellvq \mvq \barell \barm}^{\Q^{(1)} \Phi (2)}(k; p, q)= 2\delta^K_{\barell 0} \delta^K_{\barm 0}(4\pi)^3 \aleph^*_\ell(k)  (-1)^m i^{\ellvp+\ellvq}  \G^{\ell \ellvp \ellvq }_{-m  \mvp \mvq} \int \ud \bar \chi~\bar \chi^2  \W (\bar \chi) ~ \left( 2 - \frac{1}{\bar \chi \cH} \right)  \nonumber \\
&&\times {1\over2}  \bigg[\T^{\Q^{(1)}}(\bp, \eta)  \T^{\Phi}(\bq, \eta) +\T^{\Q^{(1)}}(\bq, \eta)  \T^{\Phi}(\bp, \eta)  \bigg] j_{\ellvp} (p \bar \chi)j_{\ellvq} (q \bar \chi)  j_\ell (k \bar \chi)\;,\\
&&\M_{\ell m \ellvp \mvp \ellvq \mvq \barell \barm}^{\Q^{(1)} \p_\| v (2)} (k; p, q)=-2 \delta^K_{\barell 0} \delta^K_{\barm 0}(4\pi)^3 \aleph^*_\ell(k)  (-1)^m i^{\ellvp+\ellvq}  \G^{\ell \ellvp \ellvq }_{-m  \mvp \mvq} \int \ud \bar \chi~\bar \chi^2  \W (\bar \chi) ~ \left( 1 - \frac{1}{\bar \chi \cH} \right)   \nonumber \\
&&\times {1\over2} \bigg\{\T^{\Q^{(1)}}(\bp, \eta)  \T^{v}(\bq, \eta) ~ j_{\ellvp} (p \bar \chi)\left[ \frac{\p}{\p \bar \chi}j_{\ellvq} (q \bar \chi)\right]  +\T^{\Q^{(1)}}(\bq, \eta)  \T^{v}(\bp, \eta)~ j_{\ellvq} (q \bar \chi) \left[\frac{\p}{\p \bar \chi} j_{\ellvp} (p \bar \chi)\right]   \bigg\}   j_\ell (k  \bar \chi)\;. 
\eea

\subsection{Terms from $\Delta_{g\, {\rm loc}-3}^{(2)} $ [see Eq.\ (\ref{Deltag2-loc-3})]} \label{Deltag2-loc-2-par}

In  $\Delta_{g\, {\rm loc}-3}^{(2)} $ [see Eq.\ (\ref{Deltag2-loc-3})], we find all local projection terms with transverse partial derivatives and we can write these terms as 
\begin{equation}
{1 \over 2} \Delta^{rs(2)}(\bx, \eta)  = \left[\p_{\perp i}\WW^{r}\left(\bar \chi,  \eta, \frac{\p}{\p  \bar \chi}, \frac{\p}{\p  \eta}\right) \Delta^{r(1)}(\bx, \eta) \right]~\left[\p_{\perp}^i \WW^{s}\left(\bar \chi,  \eta, \frac{\p}{\p \bar \chi}, \frac{\p}{\p  \eta} \right)  \Delta^{s(1)}(\bx, \eta) \right]\;,
\end{equation}
where here the indices $r,s,$ and $rs$ have the same meaning of the superscript $i,j$ and $ij$ of previous subsection, $\Delta^{s(1)}$ was already defined in the previous subsection, and
\bea
\p_{\perp i} \WW^{j}\left(\bar \chi,  \eta, \frac{\p}{\p  \bar \chi}, \frac{\p}{\p  \eta} \right) \Delta^{(1)j}(\bx, \eta) = \int \frac{\ud^3\bk}{(2\pi)^3}   \left[\p_{\perp i} \WW^{j}\left(\bar \chi,  \eta, \frac{\p}{\p \bar \chi}, \frac{\p}{\p  \eta}\right)   \T^{j}(\bk, \eta) e^{i \bk \bx} \right]\Phi_{\rm p}(\bk) \\
= \sum_{\ell m} 4\pi i^\ell  \left[\p_{\perp i}Y_{\ell m}(\hat \bn) \right] \int \frac{\ud^3\bk}{(2\pi)^3}  \left[\WW^{j} \left(\bar \chi,  \eta, \frac{\p}{\p \bar \chi}, \frac{\p}{\p  \eta}, \right)   \T^{j}(\bk, \eta) j_\ell (k \bar \chi) \right] Y_{\ell m}^*(\hat \bk) \Phi_{\rm p}(\bk)\;.
\eea
Starting from (see Appendix \ref{spin})
\bea
\label{pert-ortYellm}
&&\p_{\perp i}Y_{\ell m}(\hat \bn)= {1 \over \bar \chi} {}^{(2)}\nabla_i Y_{\ell m}(\hat \bn) = {1 \over \bar \chi} \left(  m_{+ i}  m_{- }^j  {}^{(2)}\nabla_{j}  Y_{\ell m}(\hat \bn) +m_{-  i}  m_{+}^j  {}^{(2)}\nabla_{j}  Y_{\ell m}(\hat \bn) \right) \nonumber\\
&&=- {1 \over \bar \chi} \sqrt{\frac{1}{2}} \left(  m_{+ i}~  \eth  Y_{\ell m}(\hat \bn) +m_{-  i} ~  \bar \eth Y_{\ell m}(\hat \bn) \right)
={1 \over \bar \chi} \sqrt{\frac{\ell(\ell+1)}{2}}\left[- m_{+i}~ {}_{1}{Y_{\ell m}}(\hat \bn) + m_{-i} ~{}_{-1}{Y_{\ell m}}(\hat \bn) \right]\;,
\eea
where we used 
\[\bm_{\pm}\bm_{\mp}=1\;, \quad \quad  \bm_{\pm}\bm_{\pm}=0\;,  \quad \quad \Perp^j_i= m_{+ i}  m_{- }^j  +m_{-  i}  m_{+}^j\;, \]
\be
\label{spin1Ylm}
 ~\eth Y_{\ell m} = \sqrt{\ell(\ell+1)}~ {}_1Y_{\ell m}   \quad \quad {\rm and}  \quad \quad \bar \eth Y_{\ell m} =- \sqrt{\ell(\ell+1)} ~ {}_{-1}Y_{\ell m} \;,
\ee
 we find
\bea
&&{1 \over 2} \Delta^{rs(2)}(\bx, \eta)  =-\sum_{\ellvp \mvp \ellvq \mvq} (4 \pi)^2  i^{\ellvp+\ellvq} \frac{\sqrt{ \ellvp(\ellvp+1) \ellvq(\ellvq+1)}}{2 \bar \chi^2} \left[{}_{1}{Y_{\ellvp \mvp}}(\hat \bn) ~{}_{-1}{Y_{\ellvq \mvq}}(\hat \bn)+  {}_{-1}{Y_{\ellvp \mvp}}(\hat \bn) ~{}_{+1}{Y_{\ellvq \mvq}}(\hat \bn) \right] \nonumber\\
&&\times  \int \frac{\ud^3 \bp}{(2\pi)^3} \frac{\ud^3 \bq}{(2\pi)^3} \frac{1}{2}\Bigg\{ \bigg[ \WW^{r} \left(\bar \chi,  \eta, \frac{\p}{\p \bar \chi}, \frac{\p}{\p  \eta}, \right)   \T^{r}(\bp, \eta) j_{\ellvp} (p \bar \chi) \bigg]\bigg[ \WW^{s} \left(\bar \chi,  \eta, \frac{\p}{\p \bar \chi}, \frac{\p}{\p  \eta}, \right)   \T^{s}(\bq, \eta) j_{\ellvp} (q \bar \chi)\bigg] \nonumber\\
 &&+\bigg[ \WW^{r} \left(\bar \chi,  \eta, \frac{\p}{\p \bar \chi}, \frac{\p}{\p  \eta}, \right)   \T^{r}(\bq, \eta) j_{\ellvq} (q \bar \chi)\bigg]  \bigg[\WW^{s} \left(\bar \chi,  \eta, \frac{\p}{\p \bar \chi}, \frac{\p}{\p  \eta}, \right)   \T^{s}(\bp, \eta) j_{\ellvp} (p \bar \chi) \bigg]\Bigg\}  Y_{\ellvp \mvp}^*(\hat \bp)Y_{\ellvq \mvq}^*(\hat \bq) \nonumber\\
&&\times\Phi_{\rm p}(\bp) \Phi_{\rm p}(\bq)\;. 
\eea
Here $ {}^{(2)}\nabla_i$ is the covariant derivate on the unit sphere, $\eth$~($\bar \eth$) is the spin raising (lowering) operator and ${}_{s}{Y_{\ell m}}$ are spin-weighted spherical harmonics (for more details, see Appendix \ref{spin}).
Using
\be
\label{s_1s_2-0}
{}_{s_1}{Y_{\ell_1 m_1}}(\hat \bn) ~ {}_{s_2}{Y_{\ell_2 m_2}}(\hat \bn)= \sum_{\tell \tm \ts} (-1)^{\tm+\ts}~ {}_{-\ts}Y_{\tell -\tm}(\hat \bn) ~\I_{\ell_1 \ell_2 \tell}^{-s_1 -s_2 -\ts} 
\left(
\begin{array}{ccc}
\ell_1 & \ell_2 & \tell \\
m_1 & m_2 & \tm
\end{array}
\right)\;,
\ee
where we have defined
\be
\label{s_1s_2s_3}
 \I_{\ell_1 \ell_2 \ell_3}^{s_1 s_2 s_3} = \sqrt{\frac{(2\ell_1+1)(2\ell_2+1)(2\ell_3+1)}{4\pi}} \left(
\begin{array}{ccc}
\ell_1 & \ell_2 & \ell_3 \\
s_1 & s_2 & s_3 
\end{array}
\right)\;,
\ee
we have
\be
\left[{}_{1}{Y_{\ellvp \mvp}}(\hat \bn) ~{}_{-1}{Y_{\ellvq \mvq}}(\hat \bn)+  {}_{-1}{Y_{\ellvp \mvp}}(\hat \bn) ~{}_{+1}{Y_{\ellvq \mvq}}(\hat \bn) \right]=\sum_{\tell \tm } (-1)^{\tm}~ Y_{\tell -\tm}(\hat \bn) \left(\I_{\ellvp \ellvq \tell}^{-1 1 0}  + \I_{\ellvp \ellvq \tell}^{1-10}  \right)\left(
\begin{array}{ccc}
\ellvp & \ellvq & \tell \\
\mvp & \mvq & \tm
\end{array}
\right)
\ee
i.e. we can set $\ts=0$ (because $s_1+s_2+\ts=0$ where $\ts=s_3$). This implies that we can project $\Delta^{a(2)}(\bx)$ on $|k\ell m\rangle$ space and finally obtain
\bea
\label{spin1-Delta_klm}
\frac{1}{2}\Delta_{\ell m}^{rs(2)}(k)&=&\sum_{\ellvp \mvp \ellvq \mvq} \int \frac{\ud^3 \bp}{(2\pi)^3} \frac{\ud^3 \bq}{(2\pi)^3} \Bigg\{ (4\pi)^2 \aleph^*_\ell(k)  (-1)^{m+1}  i^{\ellvp+\ellvq} \frac{\sqrt{ \ellvp(\ellvp+1) \ellvq(\ellvq+1)}}{2} \left(
\begin{array}{ccc}
\ellvp & \ellvq & \ell \\
\mvp & \mvq & -m
\end{array}
\right)
\nonumber\\ 
&& \times   \left(\I_{\ellvp \ellvq \ell}^{-1 1 0}  + \I_{\ellvp \ellvq \ell}^{1-10}  \right)  \int \ud \bar \chi~  \W (\bar \chi)
 ~~\frac{1}{2}\Bigg[ \bigg( \WW^{r} \left(\bar \chi,  \eta, \frac{\p}{\p \bar \chi}, \frac{\p}{\p  \eta}, \right)   \T^{r}(\bp, \eta) j_{\ellvp} (p \bar \chi) \bigg)\nonumber\\
 && \times \bigg( \WW^{s} \left(\bar \chi,  \eta, \frac{\p}{\p \bar \chi}, \frac{\p}{\p  \eta}, \right)   \T^{s}(\bq, \eta) j_{\ellvp} (q \bar \chi)\bigg) +\bigg( \WW^{r} \left(\bar \chi,  \eta, \frac{\p}{\p \bar \chi}, \frac{\p}{\p  \eta}, \right)   \T^{r}(\bq, \eta) j_{\ellvq} (q \bar \chi)\bigg)  \nonumber\\
 && \times\bigg(\WW^{s} \left(\bar \chi,  \eta, \frac{\p}{\p \bar \chi}, \frac{\p}{\p  \eta}, \right)   \T^{s}(\bp, \eta) j_{\ellvp} (p \bar \chi) \bigg) \Bigg] j_\ell (k \bar \chi)\Bigg\} ~ Y_{\ellvp \mvp}^*(\hat \bp)  Y_{\ellvq \mvq}^*(\hat \bq)  \Phi_{\rm p}(\bp) \Phi_{\rm p}(\bq)\;. 
\eea

Here we may further simplify the notation in the following way.
We note that 
\[\left(
\begin{array}{ccc}
\ellvp & \ellvq & \tell \\
-1 & 1 & 0
\end{array}
\right) 
=
(-1)^{\ellvp + \ellvq + \tell}
\left(
\begin{array}{ccc}
\ellvp & \ellvq & \tell \\
1 & -1 & 0
\end{array}
\right) 
\]
then we find
\[\I_{\ellvp \ellvq \barell}^{-1 1 0}  + \I_{\ellvp \ellvq \tell}^{1-10} =\big[1+ (-1)^{\ellvp+ \ellvq +\tell} \big] \I_{\ellvp \ellvq \tell}^{1 -1 0}= \left\{ \begin{array}{ll}
2\,\I_{\ellvp \ellvq \tell}^{1 -1 0} ~~~~~~~\ellvp+ \ellvq +\tell ~~{\rm even} \\ \\
0 ~~~~~~~~~~~~~~~ \ellvp+ \ellvq +\tell ~~{\rm odd}\;.
\end{array} \right.  \]
Given that $\ellvp+ \ellvq +\barell$ is even, we can use the following identity \cite{Varshalovich:1988ye}
\[
\left(
\begin{array}{ccc}
\ellvp & \ellvq & \tell \\
1 & -1 & 0
\end{array}
\right) 
= \frac{[\tell(\tell+1)-\ellvp(\ellvp+1)-\ellvq(\ellvq+1)]}{2\sqrt{ \ellvp(\ellvp+1) \ellvq(\ellvq+1)}}\left(
\begin{array}{ccc}
\ellvp & \ellvq & \tell \\
0 & 0 & 0
\end{array}
\right) 
\]
and defining \cite{Hu:2000ee}
\be \label{I-tensor}
I_{\ell \ell_1 \ell_2}^{m m_1 m_2}=(-1)^m  \frac{[ \ell_1 ( \ell_1+1)+ \ell_2( \ell_2+1)- \ell(\ell+1)]}{2} ~ \G^{\ell \ell_1 \ell_2 }_{-m m_1 m_2}\;,
\ee
we can rewrite Eq.\ (\ref{spin1-Delta_klm}) as
\bea
\label{spin1-Delta_klm-2}
\frac{1}{2}\Delta_{\ell m}^{rs(2)}(k)&=&\sum_{\ellvp \mvp \ellvq \mvq} \int \frac{\ud^3 \bp}{(2\pi)^3} \frac{\ud^3 \bq}{(2\pi)^3} \Bigg\{ (4\pi)^2 \aleph^*_\ell(k)   i^{\ellvp+\ellvq} ~ I_{\ell \ellvp \ellvp}^{m \mvp \mvq}     \int \ud \bar \chi~  \W (\bar \chi) \nonumber\\ 
&& \times 
 ~~\frac{1}{2}\Bigg[ \bigg( \WW^{r} \left(\bar \chi,  \eta, \frac{\p}{\p \bar \chi}, \frac{\p}{\p  \eta}, \right)   \T^{r}(\bp, \eta) j_{\ellvp} (p \bar \chi) \bigg)  \bigg( \WW^{s} \left(\bar \chi,  \eta, \frac{\p}{\p \bar \chi}, \frac{\p}{\p  \eta}, \right)   \T^{s}(\bq, \eta) j_{\ellvp} (q \bar \chi)\bigg) \nonumber\\
 && +\bigg( \WW^{r} \left(\bar \chi,  \eta, \frac{\p}{\p \bar \chi}, \frac{\p}{\p  \eta}, \right)   \T^{r}(\bq, \eta) j_{\ellvq} (q \bar \chi)\bigg)  \bigg(\WW^{s} \left(\bar \chi,  \eta, \frac{\p}{\p \bar \chi}, \frac{\p}{\p  \eta}, \right)   \T^{s}(\bp, \eta) j_{\ellvp} (p \bar \chi) \bigg) \Bigg] j_\ell (k \bar \chi)\Bigg\} \nonumber \\
 &&\times ~ Y_{\ellvp \mvp}^*(\hat \bp)  Y_{\ellvq \mvq}^*(\hat \bq)  \Phi_{\rm p}(\bp) \Phi_{\rm p}(\bq)\;. 
\eea
and,  explicitly, we have
 \bea
 \label{M-Deltag2-loc-3}
 &&\M_{\ell m \ellvp \mvp \ellvq \mvq \barell \barm}^{rs(2)}(k; p, q) = \delta^K_{\barell 0} \delta^K_{\barm 0}(4\pi)^3\aleph^*_\ell(k)   i^{\ellvp+\ellvq} ~ I_{\ell \ellvp \ellvp}^{m \mvp \mvq}     \int \ud \bar \chi~  \W (\bar \chi) \nonumber\\ 
&& \times 
 ~~\frac{1}{2}\Bigg[ \bigg( \WW^{r} \left(\bar \chi,  \eta, \frac{\p}{\p \bar \chi}, \frac{\p}{\p  \eta}, \right)   \T^{r}(\bp, \eta) j_{\ellvp} (p \bar \chi) \bigg)  \bigg( \WW^{s} \left(\bar \chi,  \eta, \frac{\p}{\p \bar \chi}, \frac{\p}{\p  \eta}, \right)   \T^{s}(\bq, \eta) j_{\ellvp} (q \bar \chi)\bigg) \nonumber\\
 && +\bigg( \WW^{r} \left(\bar \chi,  \eta, \frac{\p}{\p \bar \chi}, \frac{\p}{\p  \eta}, \right)   \T^{r}(\bq, \eta) j_{\ellvq} (q \bar \chi)\bigg)  \bigg(\WW^{s} \left(\bar \chi,  \eta, \frac{\p}{\p \bar \chi}, \frac{\p}{\p  \eta}, \right)   \T^{s}(\bp, \eta) j_{\ellvp} (p \bar \chi) \bigg) \Bigg] j_\ell (k \bar \chi) \;. 
 \eea
 Applying Eq.(\ref{M-Deltag2-loc-3}) for each term in Eq.(\ref{Deltag2-loc-3}) we find
 \bea\label{vphi}
 \M_{\ell m \ellvp \mvp \ellvq \mvq \barell \barm}^{\p_\perp v \p_\perp \Phi (2)}(k; p, q) &=&  \delta^K_{\barell 0} \delta^K_{\barm 0}(4\pi)^3\aleph^*_\ell(k)   i^{\ellvp+\ellvq} ~ I_{\ell \ellvp \ellvp}^{m \mvp \mvq}     \int \ud \bar \chi~  \W (\bar \chi) \nonumber\\ 
& \times& ~~  \left({1 \over \cH}\right)  {1 \over 2} \bigg[\T^{\Phi}(\bp, \eta)  \T^{v}(\bq, \eta)  +\T^{\Phi}(\bq, \eta)  \T^{v}(\bp, \eta)   \bigg] j_{\ellvp} (p \bar \chi)j_{\ellvq} (q \bar \chi)  j_\ell (k \bar \chi)\;,
 \eea
 \bea
  \M_{\ell m \ellvp \mvp \ellvq \mvq \barell \barm}^{\p_\perp v \p_\perp \p_\| v (2)}(k; p, q) &=&-2 \delta^K_{\barell 0} \delta^K_{\barm 0}(4\pi)^3\aleph^*_\ell(k)   i^{\ellvp+\ellvq} ~ I_{\ell \ellvp \ellvp}^{m \mvp \mvq}     \int \ud \bar \chi~  \W (\bar \chi) \nonumber\\ 
& \times &
   \left({1 \over \cH}\right)   \T^{v}(\bp, \eta)  \T^{v}(\bq, \eta)  ~\frac{\p}{\p \bar \chi} \left[ j_{\ellvp} (p \bar \chi)j_{\ellvq} (q \bar \chi)\right]  j_\ell (k \bar \chi)\;,
 \eea
  \bea\label{vv}
 \M_{\ell m \ellvp \mvp \ellvq \mvq \barell \barm}^{\p_\perp v \p_\perp v (2)}(k; p, q) &=& \delta^K_{\barell 0} \delta^K_{\barm 0}(4\pi)^3\aleph^*_\ell(k)   i^{\ellvp+\ellvq} ~ I_{\ell \ellvp \ellvp}^{m \mvp \mvq}     \int \ud \bar \chi~  \W (\bar \chi) \nonumber\\ 
& \times &
 {1 \over 2}\left( -1 + b_e -   \frac{\cH'}{\cH^2} -2 \Q   \right) \T^{v}(\bp, \eta)  \T^{v}(\bq, \eta)  ~  j_{\ellvp} (p \bar \chi)j_{\ellvq} (q \bar \chi)  j_\ell (k \bar \chi)\;.
 \eea

\subsection{Terms from  $\Delta_{g\, {\rm int-1}}^{(2)}$ [see Eq.\ (\ref{Deltag2-int-1})]}

$\Delta_{g\, {\rm int-1}}^{(2)}$ contains convergence, ISW and STD terms. Then, in Eq.\ (\ref{Deltag2-int-1}), we have two possible contributions: \\
{\it (i)}
\begin{equation}
{1 \over 2} \Delta^{j(2)}(\bx, \eta)  =\left[\sum_{k_j} \WW^{k_j}\left(\bar \chi,  \eta, \frac{\p}{\p \bar \chi}, \frac{\p}{\p  \eta}\right)  \Delta^{(1)k_j}(\bx, \eta) \right] \times \int_0^{\bar \chi} \ud \tilde \chi ~ \WW^{j}\left(\bar \chi,\tilde \chi,  \eta, \tilde \eta, \frac{\p}{\p  \tilde \chi}, \frac{\p}{\p  \tilde \eta}, \triangle_{\hat  \bn}\right)  \Phi(\tilde \bx, \tilde \eta)
\end{equation}
or {\it (ii)}
\begin{equation}
{1 \over 2} \Delta^{ij(2)}(\bx, \eta) =\left[\int_0^{\bar \chi} \ud \tilde \chi ~ \WW^{i}\left(\bar \chi,\tilde \chi,  \eta, \tilde \eta, \frac{\p}{\p  \tilde \chi}, \frac{\p}{\p  \tilde \eta}, \triangle_{\hat  \bn}\right) \Phi(\tilde \bx, \tilde \eta)\right]
\times\left[ \int_0^{\bar \chi} \ud \tilde \chi ~ \WW^{j}\left(\bar \chi,\tilde \chi,  \eta, \tilde \eta, \frac{\p}{\p  \tilde \chi}, \frac{\p}{\p  \tilde \eta}, \triangle_{\hat  \bn}\right) \Phi(\tilde \bx, \tilde \eta)\right]\;.
\end{equation}
For the case {\it (i)} we obtain
\bea
&&{1 \over 2} \Delta^{j(2)}(\bx, \eta) =\int \frac{\ud^3 \bp}{(2\pi)^3} \frac{\ud^3 \bq}{(2\pi)^3}  \Bigg\{ \sum_{\ellvp \mvp \ellvq \mvq} (4\pi)^2 i^{\ellvp+\ellvq}  \int_0^{\bar \chi} \ud \tilde \chi ~  \sum_{k_j}  \frac{1}{2} \Bigg[ \bigg( \WW^{k_j}\left(\bar \chi,  \eta, \frac{\p}{\p \bar \chi}, \frac{\p}{\p  \eta}\right)  \T^{k_j}(\bp, \eta) j_{\ellvp} (p \bar \chi) \bigg) \nonumber\\
&&\bigg( \WW_{\ellvq}^{j}\left(\bar \chi,\tilde \chi,  \eta, \tilde \eta, \frac{\p}{\p  \tilde \chi}, \frac{\p}{\p  \tilde \eta} \right)  \T^{\Phi}(\bq, \tilde \eta) j_{\ellvq} (q \tilde \chi)\bigg) +  \bigg( \WW^{k_j}\left(\bar \chi,  \eta, \frac{\p}{\p \bar \chi}, \frac{\p}{\p  \eta}\right)  \T^{k_j}(\bq, \eta) j_{\ellvq} (q \bar \chi) \bigg) \nonumber\\
&&\bigg( \WW_{\ellvp}^{j}\left(\bar \chi,\tilde \chi,  \eta, \tilde \eta, \frac{\p}{\p  \tilde \chi}, \frac{\p}{\p  \tilde \eta} \right)  \T^{\Phi}(\bp, \tilde \eta) j_{\ellvp} (p \tilde \chi)\bigg)\Bigg]Y_{\ellvp \mvp}(\hat \bn) Y_{\ellvq \mvq}(\hat \bn) \Bigg\} ~ Y_{\ellvp \mvp}^*(\hat \bp) Y_{\ellvq \mvq}^*(\hat \bq) ~ \Phi_{\rm p}(\bp) \Phi_{\rm p}(\bq)\;,
\eea
and, using the Gaunt integral Eq.(\ref{gaunt}), $\Delta^{j(2)}(\bx)/2 $ projected in $|k\ell m\rangle$ space turns out
\bea
&&{1 \over 2} \Delta^{j(2)}_{\ell m}(\bk) =
\sum_{\ellvp \mvp \ellvq \mvq} \int \frac{\ud^3 \bp}{(2\pi)^3} \frac{\ud^3 \bq}{(2\pi)^3} \Bigg\{ (4\pi)^2 \aleph^*_\ell(k)  (-1)^m i^{\ellvp+\ellvq}  \G^{\ell \ellvp \ellvq }_{-m  \mvp \mvq} \int \ud \bar \chi~\bar \chi^2  \W (\bar \chi) \int_0^{\bar \chi} \ud \tilde \chi ~  \sum_{k_j}
\nonumber\\ 
&&\times   ~ \frac{1}{2} \Bigg[ \bigg( \WW^{k_j}\left(\bar \chi,  \eta, \frac{\p}{\p \bar \chi}, \frac{\p}{\p  \eta}\right)  \T^{k_j}(\bp, \eta) j_{\ellvp} (p \bar \chi) \bigg) \bigg( \WW_{\ellvq}^{j}\left(\bar \chi,\tilde \chi,  \eta, \tilde \eta, \frac{\p}{\p  \tilde \chi}, \frac{\p}{\p  \tilde \eta} \right)  \T^{\Phi}(\bq, \tilde \eta) j_{\ellvq} (q \tilde \chi)\bigg) \nonumber\\
&& +  \bigg( \WW^{k_j}\left(\bar \chi,  \eta, \frac{\p}{\p \bar \chi}, \frac{\p}{\p  \eta}\right)  \T^{k_j}(\bq, \eta) j_{\ellvq} (q \bar \chi) \bigg) \bigg( \WW_{\ellvp}^{j}\left(\bar \chi,\tilde \chi,  \eta, \tilde \eta, \frac{\p}{\p  \tilde \chi}, \frac{\p}{\p  \tilde \eta} \right)  \T^{\Phi}(\bp, \tilde \eta) j_{\ellvp} (p \tilde \chi)\bigg)\Bigg]  j_\ell (k \bar \chi) \Bigg\} \nonumber\\
&&\times ~ Y_{\ellvp \mvp}^*(\hat \bp) Y_{\ellvq \mvq}^*(\hat \bq) ~ \Phi_{\rm p}(\bp) \Phi_{\rm p}(\bq)\;,
\eea
and the transfer function can be written in the following way
 \bea
&& \M_{\ell m \ellvp \mvp \ellvq \mvq \barell \barm}^{j(2)}(k; p, q) = \delta^K_{\barell 0} \delta^K_{\barm 0}(4\pi)^3 \aleph^*_\ell(k)  (-1)^m i^{\ellvp+\ellvq}  \G^{\ell \ellvp \ellvq }_{-m  \mvp \mvq} \int \ud \bar \chi~\bar \chi^2  \W (\bar \chi)\int_0^{\bar \chi} \ud \tilde \chi ~  \sum_{k_j}
~~~~~~~~~~~~~~~~~~\nonumber\\ 
&&\times   ~ \frac{1}{2} \Bigg\{ \bigg[ \WW^{k_j}\left(\bar \chi,  \eta, \frac{\p}{\p \bar \chi}, \frac{\p}{\p  \eta}\right)  \T^{k_j}(\bp, \eta) j_{\ellvp} (p \bar \chi) \bigg] \bigg[ \WW_{\ellvq}^{j}\left(\bar \chi,\tilde \chi,  \eta, \tilde \eta, \frac{\p}{\p  \tilde \chi}, \frac{\p}{\p  \tilde \eta} \right)  \T^{\Phi}(\bq, \tilde \eta) j_{\ellvq} (q \tilde \chi)\bigg] \nonumber\\
&& + \bigg[ \WW^{k_j}\left(\bar \chi,  \eta, \frac{\p}{\p \bar \chi}, \frac{\p}{\p  \eta}\right)  \T^{k_j}(\bq, \eta) j_{\ellvq} (q \bar \chi) \bigg] \bigg[ \WW_{\ellvp}^{j}\left(\bar \chi,\tilde \chi,  \eta, \tilde \eta, \frac{\p}{\p  \tilde \chi}, \frac{\p}{\p  \tilde \eta} \right)  \T^{\Phi}(\bp, \tilde \eta) j_{\ellvp} (p \tilde \chi)\bigg]\Bigg\} j_\ell (k \bar \chi) \;.
 \eea
Likewise, for {\it (ii)}, we have
 \bea
&& \M_{\ell m \ellvp \mvp \ellvq \mvq \barell \barm}^{ij(2)}(k; p, q) = \delta^K_{\barell 0} \delta^K_{\barm 0}(4\pi)^3 \aleph^*_\ell(k)  (-1)^m i^{\ellvp+\ellvq}  \G^{\ell \ellvp \ellvq }_{-m  \mvp \mvq} \int \ud \bar \chi~\bar \chi^2  \W (\bar \chi)\int_0^{\bar \chi} \ud  \chi_{\bp}\int_0^{\bar \chi} \ud  \chi_{\bq}
~~~~~~~~~~~~~~~~~~\nonumber\\ 
&&\times   ~ \frac{1}{2} \Bigg\{ \bigg[ \WW_{\ellvp}^{i}\left(\bar \chi,\chi_{\bp},  \eta,  \eta_{\bp}, \frac{\p}{\p  \chi_{\bp}}, \frac{\p}{\p \eta_{\bp}} \right)  \T^{\Phi}(\bp, \eta_{\bp}) j_{\ellvp} (p  \chi_{\bp})\bigg]  \bigg[ \WW_{\ellvq}^{j}\left(\bar \chi, \chi_{\bq},  \eta,  \eta_{\bq}, \frac{\p}{\p   \chi_{\bq}}, \frac{\p}{\p   \eta_{\bq}} \right)  \T^{\Phi}(\bq, \eta_{\bq}) j_{\ellvq} (q  \chi_{\bq}) \bigg] \nonumber\\
&& + \bigg[\WW_{\ellvq}^{i}\left(\bar \chi, \chi_{\bq},  \eta,  \eta_{\bq}, \frac{\p}{\p   \chi_{\bq}}, \frac{\p}{\p   \eta_{\bq}} \right)  \T^{\Phi}(\bq, \eta_{\bq}) j_{\ellvq} (q  \chi_{\bq}) \bigg] \bigg[ \WW_{\ellvp}^{j}\left(\bar \chi,\chi_{\bp},  \eta,  \eta_{\bp}, \frac{\p}{\p  \chi_{\bp}}, \frac{\p}{\p \eta_{\bp}} \right)  \T^{\Phi}(\bp, \eta_{\bp}) j_{\ellvp} (p  \chi_{\bp})\bigg]\Bigg\} j_\ell (k \bar \chi)\;.\nonumber\\
 \eea

Explicitly, we list the terms computed from Eq.\ (\ref{Deltag2-int-1})
 \bea\label{Ixall}
&& \M_{\ell m \ellvp \mvp \ellvq \mvq \barell \barm}^{I(2)}(k; p, q) = \delta^K_{\barell 0} \delta^K_{\barm 0}(4\pi)^3 \aleph^*_\ell(k)  (-1)^m i^{\ellvp+\ellvq}  \G^{\ell \ellvp \ellvq }_{-m  \mvp \mvq} \int \ud \bar \chi~\bar \chi^2  \W (\bar \chi)\int_0^{\bar \chi} \ud \tilde \chi ~  
~~~~~~~~~~~~~~~~~~\nonumber\\ 
&&\times   ~  \Bigg\{\bigg[- 2b_e + 2 b_e \Q  - 4 \Q^2    + 4 \frac{\p \Q}{\p \ln \bar L} +2 \frac{\p \Q}{\p \ln \bar a} + 2  \frac{\cH'}{\cH^2} \left(1 - \Q\right)  + \frac{2}{\bar \chi \cH}\bigg(2\Q^2- 2\frac{\p \Q}{\p \ln \bar L} \bigg)   \bigg]   \nonumber\\
&& \times \bigg[ \T^{\Phi}(\bp, \eta)  \left(-\frac{\p}{\p  \tilde \eta} \T^{\Phi}(\bq, \tilde \eta) \right) j_{\ellvp} (p \bar \chi)  j_{\ellvq} (q \tilde \chi)  +  \T^{\Phi}(\bq, \eta) \left(-\frac{\p}{\p  \tilde \eta} \T^{\Phi}(\bp, \tilde \eta) \right)  j_{\ellvp} (p \tilde \chi)  j_{\ellvq} (q \bar \chi)  \bigg]\nonumber\\
&&+ \frac{1}{\cH} \left[  1 -2 b_e +4 \Q + 3 \frac{\cH' }{\cH^2}      +  \frac{4\left(1 - \Q\right)}{\bar \chi \cH}  \right] \bigg[ \T^{v}(\bp, \eta) \left(\frac{\p^2}{\p \bar \chi^2}  j_{\ellvp} (p \bar \chi)\right) \left(-\frac{\p}{\p  \tilde \eta} \T^{\Phi}(\bq, \tilde \eta) \right) j_{\ellvq} (q \tilde \chi) \nonumber\\
&& + \T^{v}(\bq, \eta) \left(\frac{\p^2}{\p \bar \chi^2} j_{\ellvq} (q \bar \chi) \right) \left(-\frac{\p}{\p  \tilde \eta} \T^{\Phi}(\bp, \tilde \eta) \right)  j_{\ellvp} (p \tilde \chi)\bigg]+ \frac{1}{\cH}\left[  -1  + 2 b_e  - 2 \Q - 3\frac{\cH' }{\cH^2}    -   \frac{4 \left(1 - \Q\right) }{\bar \chi \cH}  \right]   \nonumber\\
&& \times \bigg[\left( \frac{\p}{\p  \eta} \T^{\Phi}(\bp, \eta) \right) j_{\ellvp} (p \bar \chi)   \left(-\frac{\p}{\p  \tilde \eta}  \T^{\Phi}(\bq, \tilde \eta) \right) j_{\ellvq} (q \tilde \chi) +\left( \frac{\p}{\p  \eta} \T^{\Phi}(\bq, \eta)\right) j_{\ellvq} (q \bar \chi)  \left(-\frac{\p}{\p  \tilde \eta} \T^{\Phi}(\bp, \tilde \eta) \right)  j_{\ellvp} (p \tilde \chi)\bigg]\nonumber\\
&&+ \frac{ \left( 1 - 2 \Q  \right)}{\cH}  \bigg[  \T^{\Phi}(\bp, \eta) \left(\frac{\p}{\p \bar \chi} j_{\ellvp} (p \bar \chi) \right)  \left(-\frac{\p}{\p  \tilde \eta}  \T^{\Phi}(\bq, \tilde \eta) \right) j_{\ellvq} (q \tilde \chi) +  \T^{\Phi}(\bq, \eta)\left(\frac{\p}{\p \bar \chi}  j_{\ellvq} (q \bar \chi)\right)  \left(-\frac{\p}{\p  \tilde \eta} \T^{\Phi}(\bp, \tilde \eta) \right)  j_{\ellvp} (p \tilde \chi)\bigg]\nonumber\\
&&+ \frac{ 1}{\cH^2}  \bigg[  \T^{\Phi}(\bp, \eta) \left(\frac{\p^2}{\p \bar \chi^2} j_{\ellvp} (p \bar \chi) \right)  \left(-\frac{\p}{\p  \tilde \eta}  \T^{\Phi}(\bq, \tilde \eta) \right) j_{\ellvq} (q \tilde \chi) +  \T^{\Phi}(\bq, \eta)\left(\frac{\p^2}{\p \bar \chi^2}  j_{\ellvq} (q \bar \chi)\right)  \left(-\frac{\p}{\p  \tilde \eta} \T^{\Phi}(\bp, \tilde \eta) \right)  j_{\ellvp} (p \tilde \chi)\bigg]\nonumber\\
&&+ \frac{1}{\cH^2} \bigg[ \T^{v}(\bp, \eta) \left(\frac{\p^3}{\p \bar \chi^3}  j_{\ellvp} (p \bar \chi)\right) \left(-\frac{\p}{\p  \tilde \eta} \T^{\Phi}(\bq, \tilde \eta) \right) j_{\ellvq} (q \tilde \chi)  + \T^{v}(\bq, \eta) \left(\frac{\p^3}{\p \bar \chi^3} j_{\ellvq} (q \bar \chi) \right) \left(-\frac{\p}{\p  \tilde \eta} \T^{\Phi}(\bp, \tilde \eta) \right)  j_{\ellvp} (p \tilde \chi)\bigg] \nonumber\\
&&  - \frac{1}{\cH^2} \bigg[\left( \frac{\p}{\p  \eta} \T^{\Phi}(\bp, \eta) \right)\left(\frac{\p}{\p \bar \chi} j_{\ellvp} (p \bar \chi) \right)   \left(-\frac{\p}{\p  \tilde \eta}  \T^{\Phi}(\bq, \tilde \eta) \right) j_{\ellvq} (q \tilde \chi) \nonumber\\
&& +\left( \frac{\p}{\p  \eta} \T^{\Phi}(\bq, \eta)\right) \left(\frac{\p}{\p \bar \chi}  j_{\ellvq} (q \bar \chi)\right)   \left(-\frac{\p}{\p  \tilde \eta} \T^{\Phi}(\bp, \tilde \eta) \right)  j_{\ellvp} (p \tilde \chi)\bigg]+   \frac{1}{\cH^2}\bigg[\left( \frac{\p^2}{\p  \eta^2} \T^{\Phi}(\bp, \eta) \right) j_{\ellvp} (p \bar \chi)   \left(-\frac{\p}{\p  \tilde \eta}  \T^{\Phi}(\bq, \tilde \eta) \right) j_{\ellvq} (q \tilde \chi)  \nonumber\\
&& +\left( \frac{\p^2}{\p  \eta^2} \T^{\Phi}(\bq, \eta)\right) j_{\ellvq} (q \bar \chi)  \left(-\frac{\p}{\p  \tilde \eta} \T^{\Phi}(\bp, \tilde \eta) \right)  j_{\ellvp} (p \tilde \chi)\bigg]+ \left[ b_e - 2\Q   -  \frac{\cH'}{\cH^2}  - \frac{2}{\bar \chi \cH} \left(1 - \Q\right)  \right]\nonumber\\
&&\times   \bigg[   \T^{\delta_g}(\bp, \eta) j_{\ellvp} (p \bar \chi)   \left(-\frac{\p}{\p  \tilde \eta}  \T^{\Phi}(\bq, \tilde \eta) \right) j_{\ellvq} (q \tilde \chi)+  \T^{\delta_g}(\bq, \eta) j_{\ellvq} (q \bar \chi)  \left(-\frac{\p}{\p  \tilde \eta} \T^{\Phi}(\bp, \tilde \eta) \right)  j_{\ellvp} (p \tilde \chi)\bigg]\nonumber\\
&&- \frac{1}{\cH}  \bigg[  \T^{\delta_g}(\bp, \eta) \left(\frac{\p}{\p \bar \chi} j_{\ellvp} (p \bar \chi) \right)  \left(-\frac{\p}{\p  \tilde \eta}  \T^{\Phi}(\bq, \tilde \eta) \right) j_{\ellvq} (q \tilde \chi) +  \T^{\delta_g}(\bq, \eta)\left(\frac{\p}{\p \bar \chi}  j_{\ellvq} (q \bar \chi)\right)  \left(-\frac{\p}{\p  \tilde \eta} \T^{\Phi}(\bp, \tilde \eta) \right)  j_{\ellvp} (p \tilde \chi)\bigg]\nonumber\\
&& +\frac{1}{\cH} \bigg[\left( \frac{\p}{\p  \eta} \T^{\delta_g}(\bp, \eta) \right) j_{\ellvp} (p \bar \chi)   \left(-\frac{\p}{\p  \tilde \eta}  \T^{\Phi}(\bq, \tilde \eta) \right) j_{\ellvq} (q \tilde \chi) +\left( \frac{\p}{\p  \eta} \T^{\delta_g}(\bq, \eta)\right) j_{\ellvq} (q \bar \chi)  \left(-\frac{\p}{\p  \tilde \eta} \T^{\Phi}(\bp, \tilde \eta) \right)  j_{\ellvp} (p \tilde \chi)\bigg] \nonumber \\
 && -2 \left(1 - \frac{1}{\bar \chi \cH}\right) \bigg[ \T^{\Q}(\bp, \eta)  \left(-\frac{\p}{\p  \tilde \eta} \T^{\Phi}(\bq, \tilde \eta) \right) j_{\ellvp} (p \bar \chi)  j_{\ellvq} (q \tilde \chi)  +  \T^{\Q}(\bq, \eta) \left(-\frac{\p}{\p  \tilde \eta} \T^{\Phi}(\bp, \tilde \eta) \right)  j_{\ellvp} (p \tilde \chi)  j_{\ellvq} (q \bar \chi)  \bigg]
\Bigg\} j_\ell (k \bar \chi)\;,\nonumber\\
 \eea
  \bea
&& \M_{\ell m \ellvp \mvp \ellvq \mvq \barell \barm}^{I^2(2)}(k; p, q) = 2\delta^K_{\barell 0} \delta^K_{\barm 0}(4\pi)^3 \aleph^*_\ell(k)  (-1)^m i^{\ellvp+\ellvq}  \G^{\ell \ellvp \ellvq }_{-m  \mvp \mvq} \int \ud \bar \chi~\bar \chi^2  \W (\bar \chi)\int_0^{\bar \chi} \ud  \chi_{\bp}\int_0^{\bar \chi} \ud  \chi_{\bq}
~~~~~~~~~~~~~~~~~~\nonumber\\ 
&&\times ~\Bigg\{b_e+b_e^2 + \frac{\p b_e}{\p \ln \bar a}  +2\Q-4 \Q b_e+4 \Q^2  -4\frac{\p \Q}{\p \ln \bar L}-4\frac{\p \Q}{\p \ln \bar a}  
+ \left(-1-2 b_e + 4\Q \right) \frac{\cH' }{\cH^2} -\frac{\cH'' }{\cH^3} +3\left( \frac{\cH' }{\cH^2} \right)^2   \nonumber \\
&&  + \frac{6}{\bar \chi} \frac{\cH' }{\cH^3} \left(1 - \Q\right)  + \frac{2}{\bar \chi \cH} \left[-1  - 2b_e + \Q - 4 \Q^2   + 2b_e \Q  +4\frac{\p \Q}{\p \ln \bar L} +2 \frac{\p \Q}{\p \ln \bar a} \right]  +  \frac{2}{\bar \chi^2 \cH^2} \bigg(1-\Q +2\Q^2  -2\frac{\p \Q}{\p \ln \bar L}\bigg)    \Bigg\} \nonumber\\
&&\times  \left(  \frac{\p}{\p \eta_{\bp}}  \T^{\Phi}(\bp, \eta_{\bp}) \right)  \left(\frac{\p}{\p   \eta_{\bq}}  \T^{\Phi}(\bq, \eta_{\bq}) \right)   j_{\ellvp} (p  \chi_{\bp}) j_{\ellvq} (q  \chi_{\bq}) j_\ell (k \bar \chi) \;,
 \eea
 \bea
&& \M_{\ell m \ellvp \mvp \ellvq \mvq \barell \barm}^{IT(2)}(k; p, q) = \delta^K_{\barell 0} \delta^K_{\barm 0}(4\pi)^3 \aleph^*_\ell(k)  (-1)^m i^{\ellvp+\ellvq}  \G^{\ell \ellvp \ellvq }_{-m  \mvp \mvq} \int \ud \bar \chi~\bar \chi^2  \W (\bar \chi)\int_0^{\bar \chi} \ud  \chi_{\bp}\int_0^{\bar \chi} \ud  \chi_{\bq}
~~~~~~~~~~~~~~~~~~\nonumber\\ 
&& \times \Bigg\{- \frac{2}{\bar \chi}\bigg[  - \left(b_e-b_e\Q+2 \Q^2 -2\frac{\p \Q}{\p \ln \bar L}-\frac{\p \Q}{\p \ln \bar a} \right) +  \frac{\cH'}{\cH^2} \left(1 - \Q\right)   + \frac{1}{\bar \chi \cH}  \left(1-\Q +2\Q^2  -2\frac{\p \Q}{\p \ln \bar L}\right) \bigg] \nonumber\\ 
&&\times   ~\bigg[\left(-  \frac{\p}{\p \eta_{\bp}}  \T^{\Phi}(\bp, \eta_{\bp}) \right)  \T^{\Phi}(\bq, \eta_{\bq})   + \left(- \frac{\p}{\p   \eta_{\bq}}  \T^{\Phi}(\bq, \eta_{\bq}) \right)   \T^{\Phi}(\bp, \eta_{\bp}) \bigg] j_{\ellvp} (p  \chi_{\bp})   j_{\ellvq} (q  \chi_{\bq}) j_\ell (k \bar \chi) \Bigg\}\;,
 \eea
 \bea
&& \M_{\ell m \ellvp \mvp \ellvq \mvq \barell \barm}^{T(2)}(k; p, q) = \delta^K_{\barell 0} \delta^K_{\barm 0}(4\pi)^3 \aleph^*_\ell(k)  (-1)^m i^{\ellvp+\ellvq}  \G^{\ell \ellvp \ellvq }_{-m  \mvp \mvq} \int \ud \bar \chi~\bar \chi^2  \W (\bar \chi)\int_0^{\bar \chi} \ud \tilde \chi ~  
~~~~~~~~~~~~~~~~\nonumber\\ 
&&\times   ~ \Bigg\{- \frac{2}{\bar \chi} \left( -1 + \Q+2 \Q^2- 2\frac{\p \Q}{\p \ln \bar L} \right)  \bigg[ \T^{\Phi}(\bp, \eta)  \T^{\Phi}(\bq, \tilde \eta)  j_{\ellvp} (p \bar \chi)  j_{\ellvq} (q \tilde \chi)  +  \T^{\Phi}(\bq, \eta)  \T^{\Phi}(\bp, \tilde \eta)   j_{\ellvp} (p \tilde \chi)  j_{\ellvq} (q \bar \chi)  \bigg]\nonumber\\
&&- \left[ - b_e + 2 \Q  +   \frac{\cH'}{\cH^2} +  \frac{4 \left(1 - \Q\right)}{\bar \chi \cH} \right] \bigg[ \T^{v}(\bp, \eta) \left(\frac{\p^2}{\p \bar \chi^2}  j_{\ellvp} (p \bar \chi)\right) \T^{\Phi}(\bq, \tilde \eta)  j_{\ellvq} (q \tilde \chi) \nonumber\\
&&  + \T^{v}(\bq, \eta) \left(\frac{\p^2}{\p \bar \chi^2} j_{\ellvq} (q \bar \chi) \right) \T^{\Phi}(\bp, \tilde \eta)   j_{\ellvp} (p \tilde \chi)\bigg]   -2\left[ b_e-2 \Q  -   \frac{\cH'}{\cH^2}  -\frac{3\left(1  - \Q \right)}{\bar \chi \cH}  \right]  \nonumber\\
&&  \times  \bigg[\left( \frac{\p}{\p  \eta} \T^{\Phi}(\bp, \eta) \right) j_{\ellvp} (p \bar \chi)  \T^{\Phi}(\bq, \tilde \eta)    j_{\ellvq} (q \tilde \chi)+\left( \frac{\p}{\p  \eta} \T^{\Phi}(\bq, \eta)\right) j_{\ellvq} (q \bar \chi) \T^{\Phi}(\bp, \tilde \eta)   j_{\ellvp} (p \tilde \chi)\bigg]\nonumber\\
&&- \left[1+b_e - 4 \Q - \frac{\cH'}{\cH^2} - \left(1 - \Q\right) \frac{2}{\bar \chi \cH} \right]  \bigg[  \T^{\Phi}(\bp, \eta) \left(\frac{\p}{\p \bar \chi} j_{\ellvp} (p \bar \chi) \right) \T^{\Phi}(\bq, \tilde \eta)    j_{\ellvq} (q \tilde \chi) \nonumber\\
&&+  \T^{\Phi}(\bq, \eta)\left(\frac{\p}{\p \bar \chi}  j_{\ellvq} (q \bar \chi)\right)   \T^{\Phi}(\bp, \tilde \eta)    j_{\ellvp} (p \tilde \chi)\bigg] - \frac{1}{\cH} \bigg[ \T^{v}(\bp, \eta) \left(\frac{\p^3}{\p \bar \chi^3}  j_{\ellvp} (p \bar \chi)\right)\T^{\Phi}(\bq, \tilde \eta)  j_{\ellvq} (q \tilde \chi)   \nonumber\\
&&+ \T^{v}(\bq, \eta) \left(\frac{\p^3}{\p \bar \chi^3} j_{\ellvq} (q \bar \chi) \right) \T^{\Phi}(\bp, \tilde \eta)  j_{\ellvp} (p \tilde \chi)\bigg] + \frac{1}{\cH} \bigg[\left( \frac{\p}{\p  \eta} \T^{\Phi}(\bp, \eta) \right)\left(\frac{\p}{\p \bar \chi} j_{\ellvp} (p \bar \chi) \right)  \T^{\Phi}(\bq, \tilde \eta)    j_{\ellvq} (q \tilde \chi) \nonumber\\
&&  +\left( \frac{\p}{\p  \eta} \T^{\Phi}(\bq, \eta)\right) \left(\frac{\p}{\p \bar \chi}  j_{\ellvq} (q \bar \chi)\right)   \T^{\Phi}(\bp, \tilde \eta)  j_{\ellvp} (p \tilde \chi)\bigg]\nonumber\\
&& + \frac{2}{\bar \chi}  \left(1- \Q \right)   \bigg[   \T^{\delta_g}(\bp, \eta) j_{\ellvp} (p \bar \chi)   \T^{\Phi}(\bq, \tilde \eta)    j_{\ellvq} (q \tilde \chi)+  \T^{\delta_g}(\bq, \eta) j_{\ellvq} (q \bar \chi)  \T^{\Phi}(\bp, \tilde \eta)    j_{\ellvp} (p \tilde \chi)\bigg]\nonumber\\
&&+  \bigg[  \T^{\delta_g}(\bp, \eta) \left(\frac{\p}{\p \bar \chi} j_{\ellvp} (p \bar \chi) \right)   \T^{\Phi}(\bq, \tilde \eta)   j_{\ellvq} (q \tilde \chi) +  \T^{\delta_g}(\bq, \eta)\left(\frac{\p}{\p \bar \chi}  j_{\ellvq} (q \bar \chi)\right)   \T^{\Phi}(\bp, \tilde \eta)    j_{\ellvp} (p \tilde \chi)\bigg] \nonumber\\
&& - \frac{2}{\bar \chi}  \bigg[ \T^{\Q}(\bp, \eta)  \T^{\Phi}(\bq, \tilde \eta)  j_{\ellvp} (p \bar \chi)  j_{\ellvq} (q \tilde \chi)  +  \T^{\Q}(\bq, \eta)  \T^{\Phi}(\bp, \tilde \eta)   j_{\ellvp} (p \tilde \chi)  j_{\ellvq} (q \bar \chi)  \bigg]\Bigg\} j_\ell (k \bar \chi) \;,
\eea
  \bea
&& \M_{\ell m \ellvp \mvp \ellvq \mvq \barell \barm}^{T^2(2)}(k; p, q) =4 \delta^K_{\barell 0} \delta^K_{\barm 0}(4\pi)^3 \aleph^*_\ell(k)  (-1)^m i^{\ellvp+\ellvq}  \G^{\ell \ellvp \ellvq }_{-m  \mvp \mvq} \int \ud \bar \chi~  \W (\bar \chi) \bigg(1-\Q +2\Q^2    -2\frac{\p \Q}{\p \ln \bar L}\bigg)
~~~~~~~~~~\nonumber\\ 
&&\times ~ \int_0^{\bar \chi} \ud  \chi_{\bp}\int_0^{\bar \chi} \ud  \chi_{\bq} ~  \T^{\Phi}(\bp, \eta_{\bp})  \T^{\Phi}(\bq, \eta_{\bq})    j_{\ellvp} (p  \chi_{\bp}) j_{\ellvq} (q  \chi_{\bq})  j_\ell (k \bar \chi)\;,
 \eea
 \bea
 \M_{\ell m \ellvp \mvp \ellvq \mvq \barell \barm}^{\kappa(2)}(k; p, q) &=& \delta^K_{\barell 0} \delta^K_{\barm 0}(4\pi)^3 \aleph^*_\ell(k)  (-1)^m i^{\ellvp+\ellvq}  \G^{\ell \ellvp \ellvq }_{-m  \mvp \mvq} \int \ud \bar \chi~\bar \chi^2  \W (\bar \chi)\int_0^{\bar \chi} \ud \tilde \chi \nonumber \\
&&\times ~\Bigg\{  \bigg[ -1 - 3 \Q  + 2\Q^2 - 2\frac{\p \Q}{\p \ln \bar L}  +2b_e  -   2\frac{\cH'}{\cH^2} -  \frac{5\left(1 - \Q\right)}{\bar \chi \cH}    \bigg]  \nonumber \\
&& \times  \bigg[ \T^{\Phi} (\bp, \eta) j_{\ellvp} (p \bar \chi)  \left(- \ellvq (\ellvq + 1)\frac{\bar \chi - \tilde \chi}{\bar \chi\tilde \chi }  \right)  \T^{\Phi}(\bq, \tilde \eta) j_{\ellvq} (q \tilde \chi) \nonumber\\
&&+ \T^{\Phi} (\bq, \eta) j_{\ellvq} (q \bar \chi)   \left(- \ellvp (\ellvp + 1)\frac{\bar \chi - \tilde \chi}{\bar \chi\tilde \chi }  \right) \T^{\Phi}(\bp, \tilde \eta) j_{\ellvp} (p \tilde \chi)\bigg] \nonumber \\
&&+  \frac{\left(1- \Q \right)}{\bar \chi\cH} \bigg[  \T^{v}(\bp, \eta)\left(\frac{\p}{\p \bar \chi} j_{\ellvp} (p \bar \chi) \right)  \left(- \ellvq (\ellvq + 1)\frac{\bar \chi - \tilde \chi}{\bar \chi\tilde \chi }  \right)  \T^{\Phi}(\bq, \tilde \eta) j_{\ellvq} (q \tilde \chi) \nonumber\\
&&+   \T^{v}(\bq, \eta) \left(\frac{\p}{\p \bar \chi}  j_{\ellvq} (q \bar \chi)\right)    \left(- \ellvp (\ellvp + 1)\frac{\bar \chi - \tilde \chi}{\bar \chi\tilde \chi }  \right) \T^{\Phi}(\bp, \tilde \eta) j_{\ellvp} (p \tilde \chi)\bigg] \nonumber \\
&&+  \frac{\left(1- \Q \right)}{\cH} \bigg[  \T^{v}(\bp, \eta)\left(\frac{\p^2}{\p \bar \chi^2} j_{\ellvp} (p \bar \chi) \right)  \left(- \ellvq (\ellvq + 1)\frac{\bar \chi - \tilde \chi}{\bar \chi\tilde \chi }  \right)  \T^{\Phi}(\bq, \tilde \eta) j_{\ellvq} (q \tilde \chi) \nonumber\\
&&+   \T^{v}(\bq, \eta) \left(\frac{\p^2}{\p \bar \chi^2}  j_{\ellvq} (q \bar \chi)\right)    \left(- \ellvp (\ellvp + 1)\frac{\bar \chi - \tilde \chi}{\bar \chi\tilde \chi }  \right) \T^{\Phi}(\bp, \tilde \eta) j_{\ellvp} (p \tilde \chi)\bigg] \nonumber \\
&& -\frac{\left(1- \Q \right)}{\cH}    \bigg[ \left( \frac{\p}{\p  \eta} \T^{\Phi}(\bp, \eta) \right) j_{\ellvp} (p \bar \chi)  \left(- \ellvq (\ellvq + 1)\frac{\bar \chi - \tilde \chi}{\bar \chi\tilde \chi }  \right)  \T^{\Phi}(\bq, \tilde \eta) j_{\ellvq} (q \tilde \chi) \nonumber\\
&& + \left( \frac{\p}{\p  \eta} \T^{\Phi}(\bq, \eta)\right) j_{\ellvq} (q \bar \chi)  \left(- \ellvp (\ellvp + 1)\frac{\bar \chi - \tilde \chi}{\bar \chi\tilde \chi }  \right) \T^{\Phi}(\bp, \tilde \eta) j_{\ellvp} (p \tilde \chi)\bigg] \nonumber \\
&&-  \left(1- \Q \right)  \bigg[ \T^{\delta_g}(\bp, \eta) j_{\ellvp} (p \bar \chi)  \left(- \ellvq (\ellvq + 1)\frac{\bar \chi - \tilde \chi}{\bar \chi\tilde \chi }  \right)  \T^{\Phi}(\bq, \tilde \eta) j_{\ellvq} (q \tilde \chi) \nonumber\\
&&+  \T^{\delta_g}(\bq, \eta) j_{\ellvq} (q \bar \chi)   \left(- \ellvp (\ellvp + 1)\frac{\bar \chi - \tilde \chi}{\bar \chi\tilde \chi }  \right) \T^{\Phi}(\bp, \tilde \eta) j_{\ellvp} (p \tilde \chi)\bigg] \nonumber \\
&&+  \bigg[ \T^{\Q} (\bp, \eta) j_{\ellvp} (p \bar \chi)  \left(- \ellvq (\ellvq + 1)\frac{\bar \chi - \tilde \chi}{\bar \chi\tilde \chi }  \right)  \T^{\Phi}(\bq, \tilde \eta) j_{\ellvq} (q \tilde \chi) \nonumber\\
&&+\T^{\Q} (\bq, \eta) j_{\ellvq} (q \bar \chi)   \left(- \ellvp (\ellvp + 1)\frac{\bar \chi - \tilde \chi}{\bar \chi\tilde \chi }  \right) \T^{\Phi}(\bp, \tilde \eta) j_{\ellvp} (p \tilde \chi)\bigg]
 \Bigg\} j_\ell (k \bar \chi)\;,
 \eea
 \bea
 \M_{\ell m \ellvp \mvp \ellvq \mvq \barell \barm}^{\int_{\tilde \chi} \tilde \chi \Phi' \kappa (2)}(k; p, q) &=&2 \delta^K_{\barell 0} \delta^K_{\barm 0}(4\pi)^3 \aleph^*_\ell(k)  (-1)^m i^{\ellvp+\ellvq}  \G^{\ell \ellvp \ellvq }_{-m  \mvp \mvq} \int \ud \bar \chi~\bar \chi  \W (\bar \chi)  (1-\Q)\int_0^{\bar \chi} \ud  \chi_{\bp}\int_0^{\bar \chi} \ud  \chi_{\bq}
~~~~~~~~\nonumber\\ 
&\times&   ~  \Bigg[  \chi_{\bp}  \left( \frac{\p}{\p  \eta_{\bp} } \T^{\Phi}(\bp, \eta_{\bp} ) \right) j_{\ellvp} (p  \chi_{\bp})  \left(- \ellvq (\ellvq + 1)\frac{\bar \chi - \chi_{\bq}}{\bar \chi \chi_{\bq}}  \right) \T^{\Phi}(\bq, \eta_{\bq}) j_{\ellvq} (q  \chi_{\bq})  \nonumber\\
&+&\chi_{\bq}  \left( \frac{\p}{\p  \eta_{\bq} }  \T^{\Phi}(\bq, \eta_{\bq}) \right) j_{\ellvq} (q  \chi_{\bq})  \left(- \ellvp (\ellvp + 1)\frac{\bar \chi - \chi_{\bp}}{\bar \chi \chi_{\bp}}  \right)   \T^{\Phi}(\bp, \eta_{\bp}) j_{\ellvp} (p  \chi_{\bp})\Bigg] j_\ell (k \bar \chi)\;, \nonumber\\
 \eea
  \bea
 \M_{\ell m \ellvp \mvp \ellvq \mvq \barell \barm}^{T \kappa (2)}(k; p, q) &=& 4 \delta^K_{\barell 0} \delta^K_{\barm 0}(4\pi)^3 \aleph^*_\ell(k)  (-1)^m i^{\ellvp+\ellvq}  \G^{\ell \ellvp \ellvq }_{-m  \mvp \mvq} \int \ud \bar \chi~\bar \chi  \W (\bar \chi)  \int_0^{\bar \chi} \ud  \chi_{\bp}\int_0^{\bar \chi} \ud  \chi_{\bq}
~~~~~~~~\nonumber\\ 
&\times& \bigg(1-\Q +\Q^2    -\frac{\p \Q}{\p \ln \bar L}\bigg)~  \Bigg[  \ellvq (\ellvq + 1)\frac{\bar \chi - \chi_{\bq}}{\bar \chi \chi_{\bq}}      \T^{\Phi}(\bp, \eta_{\bp} )  j_{\ellvp} (p  \chi_{\bp})     \T^{\Phi}(\bq, \eta_{\bq}) j_{\ellvq} (q  \chi_{\bq})  \nonumber\\
&+& \ellvp (\ellvp + 1)\frac{\bar \chi - \chi_{\bp}}{\bar \chi \chi_{\bp}}   \T^{\Phi}(\bq, \eta_{\bq})  j_{\ellvq} (q  \chi_{\bq})   \T^{\Phi}(\bp, \eta_{\bp}) j_{\ellvp} (p  \chi_{\bp})\Bigg] j_\ell (k \bar \chi)\;,
 \eea
  \bea
&& \M_{\ell m \ellvp \mvp \ellvq \mvq \barell \barm}^{\kappa^2(2)}(k; p, q) = \delta^K_{\barell 0} \delta^K_{\barm 0}(4\pi)^3 \aleph^*_\ell(k)  (-1)^m i^{\ellvp+\ellvq}  \G^{\ell \ellvp \ellvq }_{-m  \mvp \mvq} \int \ud \bar \chi~\bar \chi^2  \W (\bar \chi)\left(1-\Q +2\Q^2- 2\frac{\p \Q}{\p \ln \bar L}\right) 
~~~~~~\nonumber\\ 
&&\times \int_0^{\bar \chi} \ud  \chi_{\bp}\int_0^{\bar \chi} \ud  \chi_{\bq}  ~ \bigg[  \ellvq (\ellvq + 1)  \ellvp (\ellvp + 1) \frac{\bar \chi - \chi_{\bq}}{\bar \chi \chi_{\bq}}\frac{\bar \chi - \chi_{\bp}}{\bar \chi \chi_{\bp}}  ~   \T^{\Phi}(\bp, \eta_{\bp}) \T^{\Phi}(\bq, \eta_{\bq})  j_{\ellvp} (p  \chi_{\bp})  j_{\ellvq} (q  \chi_{\bq}) j_\ell (k \bar \chi) \bigg]\;.
 \eea
   \bea
&& \M_{\ell m \ellvp \mvp \ellvq \mvq \barell \barm}^{I\kappa(2)}(k; p, q) = \delta^K_{\barell 0} \delta^K_{\barm 0}(4\pi)^3 \aleph^*_\ell(k)  (-1)^m i^{\ellvp+\ellvq}  \G^{\ell \ellvp \ellvq }_{-m  \mvp \mvq} \int \ud \bar \chi~\bar \chi^2  \W (\bar \chi)
~~~~~~~~~~~~~~~~~~\nonumber\\ 
&&\times  \bigg[ 2 - b_e + b_e\Q - 2\Q   - 2 \Q^2 + 2\frac{\p \Q}{\p \ln \bar L} + \frac{\p \Q}{\p \ln \bar a} +  \frac{\cH'}{\cH^2} \left(1 - \Q\right) 
  + \frac{1}{\bar \chi \cH}  \left(3-3\Q +2\Q^2  -2\frac{\p \Q}{\p \ln \bar L}\right) \bigg]\nonumber\\
&&\times \int_0^{\bar \chi} \ud  \chi_{\bp}\int_0^{\bar \chi} \ud  \chi_{\bq}  ~  \Bigg[  \ellvq (\ellvq + 1)\frac{\bar \chi - \chi_{\bq}}{\bar \chi \chi_{\bq}}  \left( \frac{\p}{\p \eta_{\bp}}   \T^{\Phi}(\bp, \eta_{\bp})\right) j_{\ellvp} (p  \chi_{\bp}) \T^{\Phi}(\bq, \eta_{\bq}) j_{\ellvq} (q  \chi_{\bq})  \nonumber\\
&& +  \ellvp (\ellvp + 1)\frac{\bar \chi - \chi_{\bp}}{\bar \chi \chi_{\bp}}   \left( \frac{\p}{\p \eta_{\bq}}   \T^{\Phi}(\bq, \eta_{\bq}) \right) j_{\ellvq} (q  \chi_{\bq})    \T^{\Phi}(\bp, \eta_{\bp}) j_{\ellvp} (p  \chi_{\bp})\Bigg] j_\ell (k \bar \chi)\;,
 \eea 
  \bea
&& \M_{\ell m \ellvp \mvp \ellvq \mvq \barell \barm}^{I\nabla^2_\perp T (2)}(k; p, q) = \delta^K_{\barell 0} \delta^K_{\barm 0}(4\pi)^3 \aleph^*_\ell(k)  (-1)^m i^{\ellvp+\ellvq}  \G^{\ell \ellvp \ellvq }_{-m  \mvp \mvq} \int \ud \bar \chi~\bar \chi  \W (\bar \chi) \bigg[-2 (1-\Q) \left(1 + \frac{1}{\bar \chi \cH}\right) \bigg]~~~~~~\nonumber\\ 
&&\times   ~\int_0^{\bar \chi} \ud  \chi_{\bp}\int_0^{\bar \chi} \ud  \chi_{\bq} \Bigg[ \left(-\frac{\p}{\p \eta_{\bp}}   \T^{\Phi}(\bp, \eta_{\bp})\right) \ellvq(\ellvq+1)  \T^{\Phi}(\bq, \eta_{\bq})  +\left(-\frac{\p}{\p \eta_{\bq}} \T^{\Phi}(\bq, \eta_{\bq})\right) \ellvp (\ellvp+1) \T^{\Phi}(\bp, \eta_{\bp}) \Bigg] \nonumber\\ 
&&\times  j_{\ellvp} (p  \chi_{\bp})  j_{\ellvq} (q  \chi_{\bq})  j_\ell (k \bar \chi)  \;,
 \eea
   \bea
&& \M_{\ell m \ellvp \mvp \ellvq \mvq \barell \barm}^{ \int_{\tilde \chi} \tilde \chi \Phi'  \nabla^2_\perp T (2)}(k; p, q) = \delta^K_{\barell 0} \delta^K_{\barm 0}(4\pi)^3 \aleph^*_\ell(k)  (-1)^m i^{\ellvp+\ellvq}  \G^{\ell \ellvp \ellvq }_{-m  \mvp \mvq} \int \ud \bar \chi~  \W (\bar \chi) \bigg[-2 (1-\Q) \bigg] \int_0^{\bar \chi} \ud  \chi_{\bp}\int_0^{\bar \chi} \ud  \chi_{\bq}
\nonumber\\ 
&&\times   ~  \Bigg[  \chi_{\bp} \left(\frac{\p}{\p \eta_{\bp}}   \T^{\Phi}(\bp, \eta_{\bp})\right) \ellvq(\ellvq+1)  \T^{\Phi}(\bq, \eta_{\bq})  + \chi_{\bq} \left(\frac{\p}{\p \eta_{\bq}} \T^{\Phi}(\bq, \eta_{\bq})\right)  \ellvp (\ellvp+1)  \T^{\Phi}(\bp, \eta_{\bp})\Bigg]  j_{\ellvp} (p  \chi_{\bp}) j_{\ellvq} (q  \chi_{\bq}) j_\ell (k \bar \chi) \;, \nonumber \\
 \eea
   \bea
&& \M_{\ell m \ellvp \mvp \ellvq \mvq \barell \barm}^{T\nabla^2_\perp T (2)}(k; p, q) = \delta^K_{\barell 0} \delta^K_{\barm 0}(4\pi)^3 \aleph^*_\ell(k)  (-1)^m i^{\ellvp+\ellvq}  \G^{\ell \ellvp \ellvq }_{-m  \mvp \mvq} \int \ud \bar \chi~ \W (\bar \chi)  [2(1-\Q)]  \int_0^{\bar \chi} \ud  \chi_{\bp}\int_0^{\bar \chi} \ud  \chi_{\bq}~~~~~~\nonumber\\ 
&&\times   ~ \left[ \ellvq(\ellvq+1)   + \ellvp (\ellvp+1)  \right]  \T^{\Phi}(\bp, \eta_{\bp})  \T^{\Phi}(\bq, \eta_{\bq})  j_{\ellvp} (p  \chi_{\bp})  j_{\ellvq} (q  \chi_{\bq})  j_\ell (k \bar \chi)  \;,
 \eea
  \bea\label{nablaperpTxall}
&& \M_{\ell m \ellvp \mvp \ellvq \mvq \barell \barm}^{\nabla^2_\perp T(2)}(k; p, q) = \delta^K_{\barell 0} \delta^K_{\barm 0}(4\pi)^3 \aleph^*_\ell(k)  (-1)^m i^{\ellvp+\ellvq}  \G^{\ell \ellvp \ellvq }_{-m  \mvp \mvq} \int \ud \bar \chi~  \W (\bar \chi)  \frac{\left(1- \Q \right)}{\cH} \int_0^{\bar \chi} \ud \tilde \chi ~~~~~~ \nonumber \\
 &\times& ~     \Bigg\{ \bigg[ \T^{\Phi} (\bp, \eta)  j_{\ellvp} (p \bar \chi)\ellvq (\ellvq+1)\T^{\Phi}(\bq, \tilde \eta) j_{\ellvq} (q \tilde \chi) + \T^{\Phi} (\bq, \eta)  j_{\ellvq} (q \bar \chi) \ellvp (\ellvp+1) \T^{\Phi}(\bp, \tilde \eta) j_{\ellvp} (p \tilde \chi)  \bigg] \nonumber \\
 &-& \bigg[ \T^{v} (\bp, \eta) \left(\frac{\p}{\p \bar \chi} j_{\ellvp} (p \bar \chi) \right) \ellvq (\ellvq+1) \T^{\Phi}(\bq, \tilde \eta) j_{\ellvq} (q \tilde \chi) 
 + \T^{v} (\bq, \eta) \left(\frac{\p}{\p \bar \chi} j_{\ellvq} (q \bar \chi) \right) \ellvp (\ellvp+1) \T^{\Phi}(\bp, \tilde \eta) j_{\ellvp} (p \tilde \chi)  \bigg]  \Bigg\} j_\ell (k \bar \chi)\;. \nonumber\\
 \eea
 

 \subsection{Terms from  $\Delta_{g\, {\rm int-2}}^{(2)}$ [see Eq.\ (\ref{Deltag2-int-2})]}

Here we have terms like $\p_{\perp i} A^{(1)} \p_{\perp}^{i}B^{(1)}$, where $A^{(1)}$ or $B^{(1)}$ can be a local or an integrated term at first order.
From $\Delta_{g\, {\rm int-2}}^{(2)}$ we can write these terms as
{\it (i)}
\begin{equation}
{1 \over 2} \Delta^{j(2)}(\bx, \eta) =\left[\sum_{k_j} \p_{\perp i} \WW^{k_j}\left(\bar \chi,  \eta, \frac{\p}{\p \bar \chi}, \frac{\p}{\p  \eta}\right)  \Delta^{(1)k_j}(\bx, \eta)\right] \times \int_0^{\bar \chi} \ud \tilde \chi ~\tilde\p_{\perp}^i  \WW^{j}\left(\bar \chi,\tilde \chi,  \eta, \tilde \eta, \frac{\p}{\p  \tilde \chi}, \frac{\p}{\p  \tilde \eta}, \right) \Phi(\tilde \bx, \tilde \eta)
\end{equation}
 or {\it (ii)}
\begin{equation}
{1 \over 2} \Delta^{rs(2)}(\bx, \eta) =\left[\int_0^{\bar \chi} \ud \tilde \chi ~\tilde  \p_{\perp i} \WW^{r}\left(\bar \chi,\tilde \chi,  \eta, \tilde \eta, \frac{\p}{\p  \tilde \chi}, \frac{\p}{\p  \tilde \eta}\right) \Phi(\tilde \bx, \tilde \eta)\right]
\times \left[ \int_0^{\bar \chi} \ud \tilde \chi ~\tilde\p_{\perp}^i \WW^{s}\left(\bar \chi,\tilde \chi,  \eta, \tilde \eta, \frac{\p}{\p  \tilde \chi}, \frac{\p}{\p  \tilde \eta} \right) \Phi(\tilde \bx, \tilde \eta)\right]\;,
\end{equation}
where here the index $a=rs$. Following the prescription in Sec.\ \ref{Deltag2-loc-2-par} we obtain 
\bea
&&\int_0^{\bar \chi} \ud \tilde \chi ~\tilde\p_{\perp}^i  \WW^{j}\left(\bar \chi,\tilde \chi,  \eta, \tilde \eta, \frac{\p}{\p  \tilde \chi}, \frac{\p}{\p  \tilde \eta}\right) \Phi(\tilde \bx, \tilde \eta)= \int \frac{\ud^3\bk}{(2\pi)^3}  \int_0^{\bar \chi} \ud \tilde \chi ~\left[\tilde\p_{\perp}^i  \WW^{j}\left(\bar \chi,\tilde \chi,  \eta, \tilde \eta, \frac{\p}{\p  \tilde \chi}, \frac{\p}{\p  \tilde \eta}\right)    \T^{j}(\bk, \tilde \eta) e^{i \bk \tilde \bx} \right]\Phi_{\rm p}(\bk) \nonumber \\
&&= \sum_{\ell m}  \int \frac{\ud^3\bk}{(2\pi)^3}  \left( 4\pi i^\ell \right)    \int_0^{\bar \chi} \ud \tilde \chi ~ \left[\tilde \p^i_{\perp }Y_{\ell m}(\hat \bn) \right] \left[ \WW^{j}\left(\bar \chi,\tilde \chi,  \eta, \tilde \eta, \frac{\p}{\p  \tilde \chi}, \frac{\p}{\p  \tilde \eta}\right)    \T^{j}(\bk,\tilde \eta)   j_\ell (k \tilde \chi) \right] Y_{\ell m}^*(\hat \bk) \Phi_{\rm p}(\bk) \nonumber\\
&&= \sum_{\ell m}  \int \frac{\ud^3\bk}{(2\pi)^3}    \int_0^{\bar \chi} \ud \tilde \chi {4\pi i^\ell  \over \tilde \chi} \sqrt{\frac{\ell(\ell+1)}{2}}\left(- m^i_{+}~ {}_{1}{Y_{\ell m}}(\hat \bn) + m^i_{-} \,{}_{-1}{Y_{\ell m}}(\hat \bn)\right)  \left[ \WW^{j}\left(\bar \chi,\tilde \chi,  \eta, \tilde \eta, \frac{\p}{\p  \tilde \chi}, \frac{\p}{\p  \tilde \eta}\right)    \T^{j}(\bk,\tilde \eta)   j_\ell (k \tilde \chi) \right] \nonumber\\
&&\times  ~Y_{\ell m}^*(\hat \bk) \Phi_{\rm p}(\bk)\;.
\eea

Then, for {\it (i)} we find
\bea
&&{1 \over 2} \Delta^{j(2)}(\bx, \eta) =\int \frac{\ud^3 \bp}{(2\pi)^3} \frac{\ud^3 \bq}{(2\pi)^3} \Bigg\{   \int_0^{\bar \chi} \ud \tilde \chi   \sum_{\ellvp \mvp \ellvq \mvq} (4\pi)^2(-1) i^{\ellvp+\ellvq}  \frac{\sqrt{ \ellvp(\ellvp+1) \ellvq(\ellvq+1)}}{2 \bar \chi \tilde \chi}  \nonumber\\
&&\times  \sum_{k_j}   \frac{1}{2} \Bigg[ \bigg( \WW^{k_j}\left(\bar \chi,  \eta, \frac{\p}{\p \bar \chi}, \frac{\p}{\p  \eta}\right)  \T^{k_j}(\bp, \eta) j_{\ellvp} (p \bar \chi) \bigg)\bigg( \WW_{\ellvq}^{j}\left(\bar \chi,\tilde \chi,  \eta, \tilde \eta, \frac{\p}{\p  \tilde \chi}, \frac{\p}{\p  \tilde \eta} \right)  \T^{\Phi}(\bq, \tilde \eta) j_{\ellvq} (q \tilde \chi)\bigg) \nonumber\\
&& +  \bigg( \WW^{k_j}\left(\bar \chi,  \eta, \frac{\p}{\p \bar \chi}, \frac{\p}{\p  \eta}\right)  \T^{k_j}(\bq, \eta) j_{\ellvq} (q \bar \chi) \bigg) \bigg( \WW_{\ellvp}^{j}\left(\bar \chi,\tilde \chi,  \eta, \tilde \eta, \frac{\p}{\p  \tilde \chi}, \frac{\p}{\p  \tilde \eta} \right)  \T^{\Phi}(\bp, \tilde \eta) j_{\ellvp} (p \tilde \chi)\bigg)\Bigg] \nonumber\\
&& \times \left[{}_{1}{Y_{\ellvp \mvp}}(\hat \bn) ~{}_{-1}{Y_{\ellvq \mvq}}(\hat \bn)+  {}_{-1}{Y_{\ellvp \mvp}}(\hat \bn) ~{}_{+1}{Y_{\ellvq \mvq}}(\hat \bn) \right] \Bigg\} ~ Y_{\ellvp \mvp}^*(\hat \bp) Y_{\ellvq \mvq}^*(\hat \bq) ~ \Phi_{\rm p}(\bp) \Phi_{\rm p}(\bq)\nonumber\\
&& =\int \frac{\ud^3 \bp}{(2\pi)^3} \frac{\ud^3 \bq}{(2\pi)^3} \Bigg\{  \int_0^{\bar \chi} \frac{ \ud \tilde \chi }{\bar \chi \tilde \chi}  \sum_{\ellvp \mvp \ellvq \mvq \tell \tm} (4\pi)^2 i^{\ellvp+\ellvq}    ~ I^{- \tm \mvp \mvq}_{\tell \ellvp \ellvq}   \nonumber\\
&&\times  \sum_{k_j}  \frac{1}{2} \Bigg[ \bigg( \WW^{k_j}\left(\bar \chi,  \eta, \frac{\p}{\p \bar \chi}, \frac{\p}{\p  \eta}\right)  \T^{k_j}(\bp, \eta) j_{\ellvp} (p \bar \chi) \bigg)\bigg( \WW_{\ellvq}^{j}\left(\bar \chi,\tilde \chi,  \eta, \tilde \eta, \frac{\p}{\p  \tilde \chi}, \frac{\p}{\p  \tilde \eta} \right)  \T^{\Phi}(\bq, \tilde \eta) j_{\ellvq} (q \tilde \chi)\bigg) \nonumber\\
&& +  \bigg( \WW^{k_j}\left(\bar \chi,  \eta, \frac{\p}{\p \bar \chi}, \frac{\p}{\p  \eta}\right)  \T^{k_j}(\bq, \eta) j_{\ellvq} (q \bar \chi) \bigg) \bigg( \WW_{\ellvp}^{j}\left(\bar \chi,\tilde \chi,  \eta, \tilde \eta, \frac{\p}{\p  \tilde \chi}, \frac{\p}{\p  \tilde \eta} \right)  \T^{\Phi}(\bp, \tilde \eta) j_{\ellvp} (p \tilde \chi)\bigg)\Bigg]  Y_{\barell -\barm} (\hat \bn) \Bigg\}\nonumber\\
&& \times ~ Y_{\ellvp \mvp}^*(\hat \bp) Y_{\ellvq \mvq}^*(\hat \bq) ~ \Phi_{\rm p}(\bp) \Phi_{\rm p}(\bq) \nonumber\\
\eea
and, for {\it (ii)},
\bea
&&{1 \over 2} \Delta^{rs(2)}(\bx) = \int \frac{\ud^3 \bp}{(2\pi)^3} \frac{\ud^3 \bq}{(2\pi)^3}  \Bigg\{   \int_0^{\bar \chi} \ud  \chi_\bp  \int_0^{\bar \chi} \ud  \chi_\bq      \sum_{\ellvp \mvp \ellvq \mvq} (4 \pi)^2 (-1)  i^{\ellvp+\ellvq} \frac{\sqrt{ \ellvp(\ellvp+1) \ellvq(\ellvq+1)}}{4 \chi_\bp \chi_\bq    }\nonumber\\
&&\times \Bigg[ \bigg( \WW_{\ellvp}^{r}\left(\bar \chi,\chi_{\bp},  \eta,  \eta_{\bp}, \frac{\p}{\p  \chi_{\bp}}, \frac{\p}{\p \eta_{\bp}} \right)  \T^{\Phi}(\bp, \eta_{\bp}) j_{\ellvp} (p  \chi_{\bp})\bigg)  \bigg( \WW_{\ellvq}^{s}\left(\bar \chi, \chi_{\bq},  \eta,  \eta_{\bq}, \frac{\p}{\p   \chi_{\bq}}, \frac{\p}{\p   \eta_{\bq}} \right)  \T^{\Phi}(\bq, \eta_{\bq}) j_{\ellvq} (q  \chi_{\bq}) \bigg) \nonumber\\
&& + \bigg(\WW_{\ellvq}^{r}\left(\bar \chi, \chi_{\bq},  \eta,  \eta_{\bq}, \frac{\p}{\p   \chi_{\bq}}, \frac{\p}{\p   \eta_{\bq}} \right)  \T^{\Phi}(\bq, \eta_{\bq}) j_{\ellvq} (q  \chi_{\bq}) \bigg) \bigg( \WW_{\ellvp}^{s}\left(\bar \chi,\chi_{\bp},  \eta,  \eta_{\bp}, \frac{\p}{\p  \chi_{\bp}}, \frac{\p}{\p \eta_{\bp}} \right)  \T^{\Phi}(\bp, \eta_{\bp}) j_{\ellvp} (p  \chi_{\bp})\bigg)\Bigg] \nonumber\\
&&\times \left[{}_{1}{Y_{\ellvp \mvp}}(\hat \bn) ~{}_{-1}{Y_{\ellvq \mvq}}(\hat \bn)+  {}_{-1}{Y_{\ellvp \mvp}}(\hat \bn) ~{}_{+1}{Y_{\ellvq \mvq}}(\hat \bn) \right] \Bigg\}  Y_{\ellvp \mvp}^*(\hat \bp)Y_{\ellvq \mvq}^*(\hat \bq) \Phi_{\rm p}(\bp) \Phi_{\rm p}(\bq) \nonumber\\
&&= \int \frac{\ud^3 \bp}{(2\pi)^3} \frac{\ud^3 \bq}{(2\pi)^3}  \Bigg\{  \int_0^{\bar \chi} \frac{\ud  \chi_\bp}{ \chi_\bp}  \int_0^{\bar \chi} \frac{\ud  \chi_\bq}{ \chi_\bq}     \sum_{\ellvp \mvp \ellvq \mvq \tell \tm} (4\pi)^2~ i^{\ellvp+\ellvq}    I^{- \tm \mvp \mvq}_{\tell \ellvp \ellvq}  \nonumber\\
&&\times \frac{1}{2   }  \Bigg[ \bigg( \WW_{\ellvp}^{r}\left(\bar \chi,\chi_{\bp},  \eta,  \eta_{\bp}, \frac{\p}{\p  \chi_{\bp}}, \frac{\p}{\p \eta_{\bp}} \right)  \T^{\Phi}(\bp, \eta_{\bp}) j_{\ellvp} (p  \chi_{\bp})\bigg) \bigg( \WW_{\ellvq}^{s}\left(\bar \chi, \chi_{\bq},  \eta,  \eta_{\bq}, \frac{\p}{\p   \chi_{\bq}}, \frac{\p}{\p   \eta_{\bq}} \right)  \T^{\Phi}(\bq, \eta_{\bq}) j_{\ellvq} (q  \chi_{\bq}) \bigg)   \nonumber\\
&& + \bigg(\WW_{\ellvq}^{r}\left(\bar \chi, \chi_{\bq},  \eta,  \eta_{\bq}, \frac{\p}{\p   \chi_{\bq}}, \frac{\p}{\p   \eta_{\bq}} \right)  \T^{\Phi}(\bq, \eta_{\bq}) j_{\ellvq} (q  \chi_{\bq}) \bigg)  \bigg( \WW_{\ellvp}^{s}\left(\bar \chi,\chi_{\bp},  \eta,  \eta_{\bp}, \frac{\p}{\p  \chi_{\bp}}, \frac{\p}{\p \eta_{\bp}} \right)  \T^{\Phi}(\bp, \eta_{\bp}) j_{\ellvp} (p  \chi_{\bp})\bigg)\Bigg]  \nonumber \\
&&\times  Y_{\barell -\barm} (\hat \bn)  \Bigg\}  Y_{\ellvp \mvp}^*(\hat \bp)Y_{\ellvq \mvq}^*(\hat \bq) \Phi_{\rm p}(\bp) \Phi_{\rm p}(\bq)\;,
\eea
where we have directly used the following identity:
\be
\label{I}
 \left[{}_{1}{Y_{\ellvp \mvp}}(\hat \bn) ~{}_{-1}{Y_{\ellvq \mvq}}(\hat \bn)+  {}_{-1}{Y_{\ellvp \mvp}}(\hat \bn) ~{}_{+1}{Y_{\ellvq \mvq}}(\hat \bn) \right]=\sum_{\tell \tm}  \frac{-2}{\sqrt{ \ellvp(\ellvp+1) \ellvq(\ellvq+1)}} I^{- \tm \mvp \mvq}_{\tell \ellvp \ellvq}  Y_{\tell -\tm} (\hat \bn)\;, 
\ee
where $ I^{\tm \mvp \mvq}_{\tell \ellvp \ellvq} $ has already been defined in Eq.\ (\ref{I-tensor}).
Projecting these contributions in $|k\ell m\rangle$ space, we find, for {\it i)},
\bea
&&{1 \over 2} \Delta_{\ell m}^{j(2)}(k) =  \sum_{\ellvp \mvp \ellvq \mvq} \int \frac{\ud^3 \bp}{(2\pi)^3} \frac{\ud^3 \bq}{(2\pi)^3} \Bigg\{    (4\pi)^2 \aleph^*_\ell(k)  i^{\ellvp+\ellvq} ~ I^{m \mvp \mvq}_{\ell \ellvp \ellvq}   \int \ud \bar \chi~ \bar \chi \W (\bar \chi)  \int_0^{\bar \chi} \frac{\ud \tilde \chi}{\tilde \chi} \nonumber\\
&& \times
  \sum_{k_j}  \frac{1}{2} \Bigg[ \bigg( \WW^{k_j}\left(\bar \chi,  \eta, \frac{\p}{\p \bar \chi}, \frac{\p}{\p  \eta}\right)  \T^{k_j}(\bp, \eta) j_{\ellvp} (p \bar \chi) \bigg)   \bigg( \WW_{\ellvq}^{j}\left(\bar \chi,\tilde \chi,  \eta, \tilde \eta, \frac{\p}{\p  \tilde \chi}, \frac{\p}{\p  \tilde \eta} \right)  \T^{\Phi}(\bq, \tilde \eta) j_{\ellvq} (q \tilde \chi)\bigg) \nonumber\\
&& +  \bigg( \WW^{k_j}\left(\bar \chi,  \eta, \frac{\p}{\p \bar \chi}, \frac{\p}{\p  \eta}\right)  \T^{k_j}(\bq, \eta) j_{\ellvq} (q \bar \chi) \bigg)  \bigg( \WW_{\ellvp}^{j}\left(\bar \chi,\tilde \chi,  \eta, \tilde \eta, \frac{\p}{\p  \tilde \chi}, \frac{\p}{\p  \tilde \eta} \right)  \T^{\Phi}(\bp, \tilde \eta) j_{\ellvp} (p \tilde \chi)\bigg)\Bigg]    j_\ell(k \bar \chi)  \Bigg\} \nonumber\\
&& \times ~ Y_{\ellvp \mvp}^*(\hat \bp) Y_{\ellvq \mvq}^*(\hat \bq) ~ \Phi_{\rm p}(\bp) \Phi_{\rm p}(\bq) 
\eea
and, consequently,
\bea
&&  \M_{\ell m \ellvp \mvp \ellvq \mvq \barell \barm}^{j(2)}(k; p, q) = \delta^K_{\barell 0} \delta^K_{\barm 0} (4\pi)^3 \aleph^*_\ell(k) i^{\ellvp+\ellvq} ~  I^{m \mvp \mvq}_{\ell \ellvp \ellvq}   \int \ud \bar \chi~ \bar \chi \W (\bar \chi)  \int_0^{\bar \chi} \frac{\ud \tilde \chi}{\tilde \chi} \nonumber\\
&& \times    \sum_{k_j}  \frac{1}{2} \Bigg[ \bigg( \WW^{k_j}\left(\bar \chi,  \eta, \frac{\p}{\p \bar \chi}, \frac{\p}{\p  \eta}\right)  \T^{k_j}(\bp, \eta) j_{\ellvp} (p \bar \chi) \bigg) \bigg( \WW_{\ellvq}^{j}\left(\bar \chi,\tilde \chi,  \eta, \tilde \eta, \frac{\p}{\p  \tilde \chi}, \frac{\p}{\p  \tilde \eta} \right)  \T^{\Phi}(\bq, \tilde \eta) j_{\ellvq} (q \tilde \chi)\bigg)  \nonumber\\
&&+  \bigg( \WW^{k_j}\left(\bar \chi,  \eta, \frac{\p}{\p \bar \chi}, \frac{\p}{\p  \eta}\right)  \T^{k_j}(\bq, \eta) j_{\ellvq} (q \bar \chi) \bigg)  \times \bigg( \WW_{\ellvp}^{j}\left(\bar \chi,\tilde \chi,  \eta, \tilde \eta, \frac{\p}{\p  \tilde \chi}, \frac{\p}{\p  \tilde \eta} \right)  \T^{\Phi}(\bp, \tilde \eta) j_{\ellvp} (p \tilde \chi)\bigg)\Bigg]   j_\ell(k \bar \chi)\;. \nonumber\\
\eea
Finally, for {\it (ii)}, we have
\bea
&&\frac{1}{2}\Delta_{\ell m}^{rs(2)}(k)=\sum_{\ellvp \mvp \ellvq \mvq} \int \frac{\ud^3 \bp}{(2\pi)^3} \frac{\ud^3 \bq}{(2\pi)^3} \Bigg\{ (4\pi)^2 \aleph^*_\ell(k)  ~  i^{\ellvp+\ellvq} ~  I^{m \mvp \mvq}_{\ell \ellvp \ellvq}  \int \ud \bar \chi~ \bar \chi^2 \W (\bar \chi)
 \int_0^{\bar \chi} \frac{\ud  \chi_\bp}{ \chi_\bp}  \int_0^{\bar \chi} \frac{\ud  \chi_\bq}{ \chi_\bq}   \nonumber\\ 
&& \times  
 ~\frac{1}{2}\Bigg[ \bigg( \WW_{\ellvp}^{r}\left(\bar \chi,\chi_{\bp},  \eta,  \eta_{\bp}, \frac{\p}{\p  \chi_{\bp}}, \frac{\p}{\p \eta_{\bp}} \right)  \T^{\Phi}(\bp, \eta_{\bp}) j_{\ellvp} (p  \chi_{\bp})\bigg)  \bigg( \WW_{\ellvq}^{s}\left(\bar \chi, \chi_{\bq},  \eta,  \eta_{\bq}, \frac{\p}{\p   \chi_{\bq}}, \frac{\p}{\p   \eta_{\bq}} \right)  \T^{\Phi}(\bq, \eta_{\bq}) j_{\ellvq} (q  \chi_{\bq}) \bigg)   \nonumber\\
&& + \bigg(\WW_{\ellvq}^{r}\left(\bar \chi, \chi_{\bq},  \eta,  \eta_{\bq}, \frac{\p}{\p   \chi_{\bq}}, \frac{\p}{\p   \eta_{\bq}} \right)  \T^{\Phi}(\bq, \eta_{\bq}) j_{\ellvq} (q  \chi_{\bq}) \bigg)  \bigg( \WW_{\ellvp}^{s}\left(\bar \chi,\chi_{\bp},  \eta,  \eta_{\bp}, \frac{\p}{\p  \chi_{\bp}}, \frac{\p}{\p \eta_{\bp}} \right)  \T^{\Phi}(\bp, \eta_{\bp}) j_{\ellvp} (p  \chi_{\bp})\bigg)\Bigg]\nonumber \\
&&\times ~ j_\ell (k \bar \chi)\Bigg\}  Y_{\ellvp \mvp}^*(\hat \bp)  Y_{\ellvq \mvq}^*(\hat \bq)  \Phi_{\rm p}(\bp) \Phi_{\rm p}(\bq) 
\eea
and
\bea
&&  \M_{\ell m \ellvp \mvp \ellvq \mvq \barell \barm}^{rs(2)}(k; p, q) =\delta^K_{\barell 0} \delta^K_{\barm 0} (4\pi)^3\aleph^*_\ell(k)   i^{\ellvp+\ellvq} ~ I_{\ell \ellvp \ellvp}^{m \mvp \mvq} \nonumber  \int \ud \bar \chi~ \bar \chi^2 \W (\bar \chi)
 \int_0^{\bar \chi} \frac{\ud  \chi_\bp}{ \chi_\bp}  \int_0^{\bar \chi} \frac{\ud  \chi_\bq}{ \chi_\bq}  \\ 
&& \times  
 ~\frac{1}{2}\Bigg[ \bigg( \WW_{\ellvp}^{r}\left(\bar \chi,\chi_{\bp},  \eta,  \eta_{\bp}, \frac{\p}{\p  \chi_{\bp}}, \frac{\p}{\p \eta_{\bp}} \right)  \T^{\Phi}(\bp, \eta_{\bp}) j_{\ellvp} (p  \chi_{\bp})\bigg)  \bigg( \WW_{\ellvq}^{s}\left(\bar \chi, \chi_{\bq},  \eta,  \eta_{\bq}, \frac{\p}{\p   \chi_{\bq}}, \frac{\p}{\p   \eta_{\bq}} \right)  \T^{\Phi}(\bq, \eta_{\bq}) j_{\ellvq} (q  \chi_{\bq}) \bigg)   \nonumber\\
&& + \bigg(\WW_{\ellvq}^{r}\left(\bar \chi, \chi_{\bq},  \eta,  \eta_{\bq}, \frac{\p}{\p   \chi_{\bq}}, \frac{\p}{\p   \eta_{\bq}} \right)  \T^{\Phi}(\bq, \eta_{\bq}) j_{\ellvq} (q  \chi_{\bq}) \bigg)  \bigg( \WW_{\ellvp}^{s}\left(\bar \chi,\chi_{\bp},  \eta,  \eta_{\bp}, \frac{\p}{\p  \chi_{\bp}}, \frac{\p}{\p \eta_{\bp}} \right)  \T^{\Phi}(\bp, \eta_{\bp}) j_{\ellvp} (p  \chi_{\bp})\bigg)\Bigg] j_\ell (k \bar \chi).\nonumber \\
\eea

Here below we list all transfer spherical multipole functions   for cases {\it (i)} and {\it (ii)} :
\bea\label{SperpxAll}
&&  \M_{\ell m \ellvp \mvp \ellvq \mvq \barell \barm}^{ S_{\perp}(2)}(k; p, q) =- \delta^K_{\barell 0} \delta^K_{\barm 0} (4\pi)^3\aleph^*_\ell(k)   i^{\ellvp+\ellvq} ~ I_{\ell \ellvp \ellvp}^{m \mvp \mvq}\int \ud \bar \chi~ \bar \chi^2 \W (\bar \chi)  \int_0^{\bar \chi} \frac{\ud \tilde \chi}{\tilde \chi} \nonumber\\
&& \times  
    \Bigg\{
 \Bigg[  \T^{\delta_g}(\bp, \eta) j_{\ellvp} (p \bar \chi)   \T^{\Phi}(\bq, \tilde \eta) j_{\ellvq} (q \tilde \chi)+  \T^{\delta_g} (\bq, \eta) j_{\ellvq} (q \bar \chi)   \T^{\Phi}(\bp, \tilde \eta) j_{\ellvp} (p \tilde \chi)\Bigg] \nonumber\\
&&   -  \left[ 1 + b_e-4 \Q  -   \frac{\cH'}{\cH^2} -2 \frac{ \left(1 - \Q\right)}{\bar \chi \cH} \right]  \Bigg[ \T^{\Phi}(\bp, \eta) j_{\ellvp} (p \bar \chi)   \T^{\Phi}(\bq, \tilde \eta) j_{\ellvq} (q \tilde \chi) +  \T^{\Phi}(\bq, \eta) j_{\ellvq} (q \bar \chi) \T^{\Phi}(\bp, \tilde \eta) j_{\ellvp} (p \tilde \chi)\Bigg] \nonumber\\
&&  +  \left[ b_e-2 \Q  -   \frac{\cH'}{\cH^2} - 2\frac{ \left(1 - \Q\right)}{\bar \chi \cH} \right] \Bigg[  \T^{v}(\bp, \eta) \bigg( \frac{\p}{\p \bar \chi} j_{\ellvp} (p \bar \chi) \bigg)   \T^{\Phi}(\bq, \tilde \eta) j_{\ellvq} (q \tilde \chi) +    \T^{v} (\bq, \eta) \bigg( \frac{\p}{\p \bar \chi} j_{\ellvq} (q \bar \chi) \bigg)   \T^{\Phi}(\bp, \tilde \eta) j_{\ellvp} (p \tilde \chi)\bigg)\Bigg] \nonumber\\
&&  + \frac{1}{\cH}\Bigg[ \bigg( \frac{\p}{\p  \eta} \T^{\Phi}(\bp, \eta)\bigg)  j_{\ellvp} (p \bar \chi)  \T^{\Phi}(\bq, \tilde \eta) j_{\ellvq} (q \tilde \chi) + \bigg( \frac{\p}{\p  \eta} \T^{\Phi} (\bq, \eta)\bigg) j_{\ellvq} (q \bar \chi)  \T^{\Phi}(\bp, \tilde \eta) j_{\ellvp} (p \tilde \chi)\Bigg] \nonumber\\
&&  -  \frac{1}{\cH}\Bigg[  \T^{v}(\bp, \eta) \bigg( \frac{\p^2}{\p \bar \chi^2} j_{\ellvp} (p \bar \chi) \bigg)   \T^{\Phi}(\bq, \tilde \eta) j_{\ellvq} (q \tilde \chi) +    \T^{v} (\bq, \eta) \bigg( \frac{\p^2}{\p \bar \chi^2} j_{\ellvq} (q \bar \chi) \bigg)   \T^{\Phi}(\bp, \tilde \eta) j_{\ellvp} (p \tilde \chi)\bigg)\Bigg] \Bigg\} j_\ell(k \bar \chi)\;,
\eea 
\bea
&&  \M_{\ell m \ellvp \mvp \ellvq \mvq \barell \barm}^{ \p_{\perp} T(2)}(k; p, q) =-\delta^K_{\barell 0} \delta^K_{\barm 0} (4\pi)^3\aleph^*_\ell(k)   i^{\ellvp+\ellvq} ~ I_{\ell \ellvp \ellvp}^{m \mvp \mvq}  \int \ud \bar \chi~\bar \chi  \W (\bar \chi)  \int_0^{\bar \chi} \ud \tilde \chi \nonumber\\
&& \times  \Bigg\{-\Bigg[  \T^{\delta_g}(\bp, \eta) j_{\ellvp} (p \bar \chi)   \T^{\Phi}(\bq, \tilde \eta) j_{\ellvq} (q \tilde \chi)+  \T^{\delta_g} (\bq, \eta) j_{\ellvq} (q \bar \chi)   \T^{\Phi}(\bp, \tilde \eta) j_{\ellvp} (p \tilde \chi)\Bigg] \nonumber\\
  && - \frac{1}{\cH}\Bigg[ \bigg( \frac{\p}{\p  \eta} \T^{\Phi}(\bp, \eta)\bigg)  j_{\ellvp} (p \bar \chi)  \T^{\Phi}(\bq, \tilde \eta) j_{\ellvq} (q \tilde \chi) + \bigg( \frac{\p}{\p  \eta} \T^{\Phi} (\bq, \eta)\bigg) j_{\ellvq} (q \bar \chi)  \T^{\Phi}(\bp, \tilde \eta) j_{\ellvp} (p \tilde \chi)\Bigg] \nonumber\\
  && +   \left[ 2 + b_e-4 \Q  -   \frac{\cH'}{\cH^2} -2 \frac{ \left(1 - \Q\right)}{\bar \chi \cH} \right]  \Bigg[ \T^{\Phi}(\bp, \eta) j_{\ellvp} (p \bar \chi)   \T^{\Phi}(\bq, \tilde \eta) j_{\ellvq} (q \tilde \chi) +  \T^{\Phi}(\bq, \eta) j_{\ellvq} (q \bar \chi) \T^{\Phi}(\bp, \tilde \eta) j_{\ellvp} (p \tilde \chi)\Bigg] \nonumber\\
  &&  -  \left[ b_e-2 \Q  -   \frac{\cH'}{\cH^2} + \frac{2 \Q}{\bar \chi \cH} \right] \Bigg[  \T^{v}(\bp, \eta) \bigg( \frac{\p}{\p \bar \chi} j_{\ellvp} (p \bar \chi) \bigg)   \T^{\Phi}(\bq, \tilde \eta) j_{\ellvq} (q \tilde \chi) +    \T^{v} (\bq, \eta) \bigg( \frac{\p}{\p \bar \chi} j_{\ellvq} (q \bar \chi) \bigg)   \T^{\Phi}(\bp, \tilde \eta) j_{\ellvp} (p \tilde \chi)\bigg)\Bigg] \nonumber\\
    &&  + \frac{1}{\bar \chi} \left[ b_e-2 \Q  -   \frac{\cH'}{\cH^2} + \frac{2 \Q}{\bar \chi \cH} \right] \Bigg[  \T^{v}(\bp, \eta)  j_{\ellvp} (p \bar \chi)    \T^{\Phi}(\bq, \tilde \eta) j_{\ellvq} (q \tilde \chi) +    \T^{v} (\bq, \eta)  j_{\ellvq} (q \bar \chi)   \T^{\Phi}(\bp, \tilde \eta) j_{\ellvp} (p \tilde \chi)\bigg)\Bigg] \nonumber\\
&&  +  \frac{1}{\cH}\Bigg[  \T^{v}(\bp, \eta) \bigg( \frac{\p^2}{\p \bar \chi^2} j_{\ellvp} (p \bar \chi) \bigg)   \T^{\Phi}(\bq, \tilde \eta) j_{\ellvq} (q \tilde \chi) +    \T^{v} (\bq, \eta) \bigg( \frac{\p^2}{\p \bar \chi^2} j_{\ellvq} (q \bar \chi) \bigg)   \T^{\Phi}(\bp, \tilde \eta) j_{\ellvp} (p \tilde \chi)\bigg)\Bigg] \Bigg\} j_\ell(k \bar \chi)\;,
\eea 
\bea
&&  \M_{\ell m \ellvp \mvp \ellvq \mvq \barell \barm}^{S_{\perp} \p_{\perp} T(2)}(k; p, q) =-2 \delta^K_{\barell 0} \delta^K_{\barm 0} (4\pi)^3\aleph^*_\ell(k)   i^{\ellvp+\ellvq} ~ I_{\ell \ellvp \ellvp}^{m \mvp \mvq}   \int \ud \bar \chi~ \bar \chi \W (\bar \chi)(1-\Q)
 \int_0^{\bar \chi} \frac{\ud  \chi_\bp}{ \chi_\bp}  \int_0^{\bar \chi} \frac{\ud  \chi_\bq}{ \chi_\bq}  ~~~~~~~~~~~~~
 \nonumber\\ 
&& \times   ~\bigg[   \chi_{\bq}  \T^{\Phi}(\bp, \eta_{\bp}) j_{\ellvp} (p  \chi_{\bp})  \T^{\Phi}(\bq, \eta_{\bq}) j_{\ellvq} (q  \chi_{\bq}) +  \chi_{\bp}  \T^{\Phi}(\bq, \eta_{\bq}) j_{\ellvq} (q  \chi_{\bq})  \T^{\Phi}(\bp, \eta_{\bp}) j_{\ellvp} (p  \chi_{\bp})\bigg]  j_\ell (k \bar \chi)\;,
\eea
\bea
&&  \M_{\ell m \ellvp \mvp \ellvq \mvq \barell \barm}^{S_{\perp} \p_{\perp}  I (2)}(k; p, q) =- 2 \delta^K_{\barell 0} \delta^K_{\barm 0} (4\pi)^3\aleph^*_\ell(k)   i^{\ellvp+\ellvq} ~ I_{\ell \ellvp \ellvp}^{m \mvp \mvq}   \int \ud \bar \chi~ \bar \chi^2 \W (\bar \chi)
 (1-\Q)
 \int_0^{\bar \chi} \frac{\ud  \chi_\bp}{ \chi_\bp}  \int_0^{\bar \chi} \frac{\ud  \chi_\bq}{ \chi_\bq}  ~~~~~~~~~~~~ \nonumber\\ 
&& \times  
 ~\bigg[   \chi_{\bq}  \T^{\Phi}(\bp, \eta_{\bp}) j_{\ellvp} (p  \chi_{\bp})  \bigg( \frac{\p}{\p  \eta_{\bq}} \T^{\Phi} (\bq, \eta_{\bq})\bigg) j_{\ellvq} (q  \chi_{\bq})+  \chi_{\bp}  \T^{\Phi}(\bq, \eta_{\bq}) j_{\ellvq} (q  \chi_{\bq})  \bigg( \frac{\p}{\p  \eta_{\bp}} \T^{\Phi}(\bp, \eta_{\bp})\bigg) j_{\ellvp} (p  \chi_{\bp})\bigg]  j_\ell (k \bar \chi)\;, \nonumber\\
\eea
\bea
&&  \M_{\ell m \ellvp \mvp \ellvq \mvq \barell \barm}^{S_{\perp} \p_{\perp}  \int \tilde \chi \Phi' (2)}(k; p, q) =2 \delta^K_{\barell 0} \delta^K_{\barm 0} (4\pi)^3\aleph^*_\ell(k)   i^{\ellvp+\ellvq} ~ I_{\ell \ellvp \ellvp}^{m \mvp \mvq}  \int \ud \bar \chi~ \bar \chi \W (\bar \chi)
 (1-\Q)
 \int_0^{\bar \chi} \frac{\ud  \chi_\bp}{ \chi_\bp}  \int_0^{\bar \chi} \frac{\ud  \chi_\bq}{ \chi_\bq}   ~~~~~~~~~~~~~~\nonumber\\ 
&& \times 
 ~\bigg[   \chi_{\bq}^2  \T^{\Phi}(\bp, \eta_{\bp}) j_{\ellvp} (p  \chi_{\bp})  \bigg( \frac{\p}{\p  \eta_{\bq}} \T^{\Phi} (\bq, \eta_{\bq})\bigg) j_{\ellvq} (q  \chi_{\bq}) +  \chi_{\bp}^2  \T^{\Phi}(\bq, \eta_{\bq}) j_{\ellvq} (q  \chi_{\bq})  \bigg( \frac{\p}{\p  \eta_{\bp}} \T^{\Phi}(\bp, \eta_{\bp})\bigg) j_{\ellvp} (p  \chi_{\bp})\bigg]  j_\ell (k \bar \chi) \;, \nonumber \\
\eea
\bea
\label{M-theta^2}
&&  \M_{\ell m \ellvp \mvp \ellvq \mvq \barell \barm}^{ \vartheta_{ij}\vartheta^{ij}(2)}(k; p, q) =\delta^K_{\barell 0} \delta^K_{\barm 0} (4\pi)^3\aleph^*_\ell(k)   i^{\ellvp+\ellvq} ~ I_{\ell \ellvp \ellvp}^{m \mvp \mvq}  \int \ud \bar \chi~ \W (\bar \chi) (1-\Q)
 \int_0^{\bar \chi} \frac{\ud  \chi_\bp}{ \chi_\bp}  \int_0^{\bar \chi} \frac{\ud  \chi_\bq}{ \chi_\bq} ~~~~~~~~~~~~~~~~~~
\nonumber\\ 
&& \times     \left(\bar \chi-\chi_{\bp}\right) \left(\bar \chi-\chi_{\bq}\right)
  \T^{\Phi}(\bp, \eta_{\bp}) j_{\ellvp} (p  \chi_{\bp})    \T^{\Phi}(\bq, \eta_{\bq}) j_{\ellvq} (q  \chi_{\bq})    j_\ell (k \bar \chi)\;,
\eea
\bea\label{SperpSperp}
&&  \M_{\ell m \ellvp \mvp \ellvq \mvq \barell \barm}^{ S_{\perp} S_{\perp} (2)}(k; p, q) =-2 \delta^K_{\barell 0} \delta^K_{\barm 0} (4\pi)^3\aleph^*_\ell(k)   i^{\ellvp+\ellvq} ~ I_{\ell \ellvp \ellvp}^{m \mvp \mvq}   \int \ud \bar \chi~\bar \chi^2 \W (\bar \chi) (1-\Q)  \int_0^{\bar \chi} \frac{\ud  \chi_\bp}{ \chi_\bp}  \int_0^{\bar \chi} \frac{\ud  \chi_\bq}{ \chi_\bq}  ~~~~~~~~~~~~~~~~
 \nonumber\\ 
&& \times    \T^{\Phi}(\bp, \eta_{\bp}) j_{\ellvp} (p  \chi_{\bp})    \T^{\Phi}(\bq, \eta_{\bq}) j_{\ellvq} (q  \chi_{\bq}) 
  j_\ell (k \bar \chi) \;.
\eea


\subsection{Terms from $ \Delta_{g\, {\rm int-3}}^{(2)}$ [see Eq.\ (\ref{Deltag2-int-3}) (only for scalar terms)]}

In $ \Delta_{g\, {\rm int-3}}^{(2)}$ we find integrated terms as, for example, ISW and STD at second order.
In particular, for the first two additive terms of Eq.\ (\ref{Deltag2-int-3}), we can use the following expression
\bea\label{Deltag2-int-3-klm}
 \frac{1}{2}\Delta^{a(2)}_{\ell m} =  \int \ud^3 \bx  ~   \W (\bar \chi) \aleph^*_\ell(k)  j_\ell(k\chi)  Y^*_{\ell m} ({\hat \bn})    ~
 \int_0^{\bar \chi}\ud\tilde\chi ~ \WW^{a}\left(\bar \chi,\tilde \chi,  \eta, \tilde \eta, \frac{\p}{\p  \tilde \chi}, \frac{\p}{\p  \tilde \eta} \right)\frac{1}{2}\left( \Phi^{ (2)}(\tilde \bx, \tilde \eta) + \Psi^{ (2)}(\tilde \bx, \tilde \eta) \right)\;.
 \eea
In this case, applying the same prescription used in Sec.\ \ref{Sec:Deltag2-loc-1}, we find quickly 
 \bea\label{T^(2)}
 \M_{\ell m \ellvp \mvp \ellvq \mvq \barell \barm}^{T^{(2)}(2)}(k; p, q) &=& 2 (4\pi)^3 \aleph^*_\ell(k)  (-1)^{m} i^{\ellvp+\ellvq} (2\bar \ell +1)^{-1} \G^{\ell \ellvp \ellvq }_{-m  \mvp \mvq} \int \ud \bar \chi~\bar \chi  \W (\bar \chi)~(1-\Q)
 \int_0^{\bar \chi}\ud\tilde\chi 
\nonumber\\ 
&\times & \left[\frac{1}{2}  F_{\bar \ell}^{\Phi (2)}(p, q;\tilde \eta)+ \frac{1}{2}  F_{\bar \ell}^{\Psi (2)}(p, q; \tilde \eta)\right]  \T^{\Phi}(\bp, \tilde \eta)\T^{\Phi}(\bq, \tilde \eta )  j_{\ellvp} (p \tilde \chi) j_{\ellvq} (q \tilde \chi) j_\ell (k  \bar \chi)  \nonumber\\
\eea
and 
 \bea\label{I^(2)}
&& \M_{\ell m \ellvp \mvp \ellvq \mvq \barell \barm}^{I^{(2)}(2)}(k; p, q) =  (4\pi)^3 \aleph^*_\ell(k)  (-1)^{m+1} i^{\ellvp+\ellvq} (2\bar \ell +1)^{-1} \G^{\ell \ellvp \ellvq }_{-m  \mvp \mvq} \int \ud \bar \chi~\bar \chi^2  \W (\bar \chi)~\bigg[ b_e-2 \Q  -   \frac{\cH'}{\cH^2}
\nonumber\\ 
&& - 2 \frac{\left(1 - \Q\right)}{\bar \chi \cH} \bigg] \times  \int_0^{\bar \chi}\ud\tilde\chi  ~\frac{\p}{\p  \tilde \eta} \left\{ \left[\frac{1}{2}  F_{\bar \ell}^{\Phi (2)}(p, q;\tilde \eta)+ \frac{1}{2}  F_{\bar \ell}^{\Psi (2)}(p, q; \tilde \eta)\right]  \T^{\Phi}(\bp, \tilde \eta)\T^{\Phi}(\bq, \tilde \eta )\right\}  j_{\ellvp} (p \tilde \chi) j_{\ellvq} (q \tilde \chi) j_\ell (k  \bar \chi) \;. \nonumber\\
\eea
Using the same approach, the last two additive terms of  Eq.\ (\ref{Deltag2-int-3}) can be written together in the following way
\bea \label{Delta_kappa-2}
&&\frac{1}{2} \Delta^{[2\p_\perp S^{(2)} -\nabla_\perp^2 T^{(2)}] (2)}(\bx, \eta) = - (1-\Q) \int_0^{\bar \chi}\ud\tilde\chi (\bar \chi - \tilde \chi) \frac{\tilde \chi}{\bar \chi} \,   \tilde \nabla^2_\perp \left[\frac{1}{2}\Phi^{(2)}(\tilde \bx, \tilde \eta)+\frac{1}{2}\Psi^{(2)}(\tilde \bx, \tilde \eta)\right] =-(1-\Q)\int_0^{\bar \chi}\ud\tilde\chi  \frac{(\bar \chi - \tilde \chi)}{\bar \chi \tilde \chi} \nonumber\\
&&\times \sum_{\ellvp \mvp \ellvq \mvq \bar \ell \bar m}   (4\pi)^3  i^{\ellvp+\ellvq} (2\bar \ell +1)^{-1}  ~{}^{(2)}\nabla^2 \left[  Y_{\ellvp \mvp} (\hat \bn) Y_{\ellvq \mvq} (\hat \bn)  \right]    \int \frac{\ud^3 \bp}{(2\pi)^3} \frac{\ud^3 \bq}{(2\pi)^3}   \Bigg\{
\left[\frac{1}{2}  F_{\bar \ell}^{\Phi (2)}(p, q;\tilde \eta)+ \frac{1}{2}  F_{\bar \ell}^{\Psi (2)}(p, q; \tilde \eta)\right]  \nonumber\\
&&\times \T^{\Phi}(\bp, \tilde \eta)\T^{\Phi}(\bq, \tilde \eta )  j_{\ellvp} (p \tilde \chi) j_{\ellvq} (q \tilde \chi)  \Bigg\}   Y_{\ellvp \mvp}^*(\hat \bp)  Y_{\barell \barm} (\hat \bp) Y_{\ellvq \mvq}^*(\hat \bq) Y_{\barell \barm}^*(\hat \bq) \Phi_{\rm p}(\bp) \Phi_{\rm p}(\bq)\;.
\eea
From Eq. (\ref{pert-ortYellm}), we note that
\bea
{}^{(2)}\nabla^2 \left[  Y_{\ellvp \mvp} (\hat \bn) Y_{\ellvq \mvq} (\hat \bn)  \right]  &=& - \ellvp (\ellvp +1) Y_{\ellvp \mvp} (\hat \bn) Y_{\ellvq \mvq} (\hat \bn) - \ellvq (\ellvq +1) Y_{\ellvp \mvp} (\hat \bn) Y_{\ellvq \mvq} (\hat \bn) \nonumber\\
 &&  - \sqrt{\ellvp (\ellvp +1)}\sqrt{\ellvq (\ellvq +1) } \left[{}_{1}{Y_{\ellvp \mvp}}(\hat \bn) ~{}_{-1}{Y_{\ellvq \mvq}}(\hat \bn)+  {}_{-1}{Y_{\ellvp \mvp}}(\hat \bn) ~{}_{+1}{Y_{\ellvq \mvq}}(\hat \bn) \right] \;, \nonumber\\
 \eea
 and using Eqs.\  (\ref{gaunt}) and (\ref{I}) we find
 \bea \label{kappa^(2)}
&&  \M_{\ell m \ellvp \mvp \ellvq \mvq \barell \barm}^{[2\p_\perp S^{(2)} -\nabla_\perp^2 T^{(2)}](2)}(k; p, q) = (4\pi)^3 \aleph^*_\ell(k)   i^{\ellvp+\ellvq}(2\bar \ell +1)^{-1} \Bigg\{-2I_{\ell \ell_1 \ell_2}^{m m_1 m_2} + \bigg[ \ellvp (\ellvp +1) +\ellvq (\ellvq +1) \bigg] (-1)^m  \G^{\ell \ellvp \ellvq }_{-m  \mvp \mvq}\Bigg\} \nonumber\\
&&\times   \int \ud \bar \chi~\bar \chi  \W (\bar \chi)~(1-\Q)\int_0^{\bar \chi}\ud\tilde\chi  \frac{(\bar \chi - \tilde \chi)}{ \tilde \chi}\left[\frac{1}{2}  F_{\bar \ell}^{\Phi (2)}(p, q;\tilde \eta)+ \frac{1}{2}  F_{\bar \ell}^{\Psi (2)}(p, q; \tilde \eta)\right]   \T^{\Phi}(\bp, \tilde \eta)\T^{\Phi}(\bq, \tilde \eta )  j_{\ellvp} (p \tilde \chi) j_{\ellvq} (q \tilde \chi) j_\ell (k  \bar \chi)\;. \nonumber\\
 \eea
 
Let us conclude by noting that, in Appendix \ref{NewcalM2}, we compute also these terms in a different way.
 
\subsection{Terms from $ \Delta_{g\, {\rm int-4}}^{(2)}$ [see Eq.\ (\ref{Deltag2-int-4})]}

In $ \Delta_{g\, {\rm int-4}}^{(2)}$, each contribution is an integral along the line of sight of the product between a local and an integrated term (or two local terms).
Specifically, we have the following possible terms
{\it (i)} 
\be
{1 \over 2} \Delta^{ij(2)}(\bx, \eta) =\int_0^{\bar \chi} \ud \tilde \chi ~\Bigg\{ \left[\WW^{i}\left(\bar \chi,\tilde \chi,  \eta, \tilde \eta, \frac{\p}{\p  \tilde \chi}, \frac{\p}{\p  \tilde \eta}, \triangle_{\hat  \bn}\right) \Phi(\tilde \bx, \tilde \eta) \right]
\times \left[\WW^{j}\left(\bar \chi,\tilde \chi,  \eta, \tilde \eta, \frac{\p}{\p  \tilde \chi}, \frac{\p}{\p  \tilde \eta}, \triangle_{\hat  \bn}\right) \Phi(\tilde \bx, \tilde \eta)\right] \Bigg\}\;,
\ee
{\it (ii)}
\begin{equation}
{1 \over 2} \Delta^{uv(2)}(\bx, \eta) =\int_0^{\bar \chi} \ud \tilde \chi ~\Bigg\{ \left[~\tilde  \p_{\perp i} \WW^{u}\left(\bar \chi,\tilde \chi,  \eta, \tilde \eta, \frac{\p}{\p  \tilde \chi}, \frac{\p}{\p  \tilde \eta}, \triangle_{\hat  \bn}\right) \Phi(\tilde \bx, \tilde \eta) \right]
\times \left[\tilde\p_{\perp}^i\WW^{v}\left(\bar \chi,\tilde \chi,  \eta, \tilde \eta, \frac{\p}{\p  \tilde \chi}, \frac{\p}{\p  \tilde \eta}, \triangle_{\hat  \bn}\right) \Phi(\tilde \bx, \tilde \eta)\right] \Bigg\}\;,
\ee
{\it (iii)}
\begin{equation}
{1 \over 2} \Delta^{ij(2)}(\bx, \eta) =\int_0^{\bar \chi} \ud \tilde \chi ~\Bigg\{ \left[\WW^{i}\left(\bar \chi,\tilde \chi,  \eta, \tilde \eta, \frac{\p}{\p  \tilde \chi}, \frac{\p}{\p  \tilde \eta}, \triangle_{\hat  \bn}\right) \Phi(\tilde \bx, \tilde \eta) \right]
\times \left[ \int_0^{\tilde \chi} \ud \tilde{\tilde \chi} ~ \WW^{j}\left(\tilde \chi, \tilde{\tilde \chi} ,  \tilde \eta, \tilde{ \tilde \eta}, \frac{\p}{\p   \tilde{\tilde \chi}}, \frac{\p}{\p  \tilde{\tilde \eta}}, \triangle_{\hat  \bn}\right) \Phi\left(\tilde{\tilde \bx}, \tilde{\tilde \eta}\right) \right]\Bigg\}\;,
\end{equation}
and {\it (iv)}
\bea
{1 \over 2} \Delta^{uv(2)}(\bx, \eta) =\int_0^{\bar \chi}  \ud \tilde \chi && \Bigg\{ \left[~\tilde  \p_{\perp i} \WW^{u}\left(\bar \chi,\tilde \chi,  \eta, \tilde \eta, \frac{\p}{\p  \tilde \chi}, \frac{\p}{\p  \tilde \eta}, \triangle_{\hat  \bn}\right) \Phi(\tilde \bx, \tilde \eta) \right] \nonumber \\
&&\times \left[ \int_0^{\tilde \chi} \ud \tilde{\tilde \chi} ~ \tilde{\tilde \p}_{\perp}^i\WW^{v}\left(\tilde \chi, \tilde{\tilde \chi} ,  \tilde \eta, \tilde{ \tilde \eta}, \frac{\p}{\p   \tilde{\tilde \chi}}, \frac{\p}{\p  \tilde{\tilde \eta}}, \triangle_{\hat  \bn}\right) \Phi\left(\tilde{\tilde \bx}, \tilde{\tilde \eta}\right) \right]\Bigg\}\;.
\eea

Following the approach used in the previous subsection we find, for {\it (i)},
\bea
\frac{1}{2}\Delta_{\ell m}^{ij(2)}(k)&=&\sum_{\ellvp \mvp \ellvq \mvq} \int \frac{\ud^3 \bp}{(2\pi)^3} \frac{\ud^3 \bq}{(2\pi)^3} \Bigg\{ (4\pi)^2 \aleph^*_\ell(k)  (-1)^m i^{\ellvp+\ellvq}  \G^{\ell \ellvp \ellvq }_{-m  \mvp \mvq} \int \ud \bar \chi~\bar \chi^2  \W (\bar \chi) \int_0^{\bar \chi} \ud \tilde \chi
\nonumber\\ 
&&\times \frac{1}{2} \bigg[ \WW_{\ellvp}^{i} \left(\bar \chi,  \eta, \tilde \chi, \tilde \eta , \frac{\p}{\p \tilde \chi}, \frac{\p}{\p \tilde \eta} \right)    \T^{\Phi}(\bp, \tilde \eta) j_{\ellvp} (p \tilde \chi) ~~ \WW_{\ellvq}^{j} \left(\bar \chi,  \eta, \tilde \chi, \tilde \eta , \frac{\p}{\p \tilde \chi}, \frac{\p}{\p \tilde \eta} \right)    \T^{\Phi}(\bq, \tilde \eta) j_{\ellvq} (q \tilde \chi) \nonumber\\
 &&+ \WW_{\ellvq}^{i}\left(\bar \chi,  \eta, \tilde \chi, \tilde \eta , \frac{\p}{\p \tilde \chi}, \frac{\p}{\p \tilde \eta} \right)   \T^{\Phi}(\bq, \tilde \eta) j_{\ellvq} (q \tilde \chi) ~~ \WW_{\ellvp}^{j}\left(\bar \chi,  \eta, \tilde \chi, \tilde \eta , \frac{\p}{\p \tilde \chi}, \frac{\p}{\p \tilde \eta} \right)   \T^{\Phi}   (\bp, \tilde \eta)\bigg] \nonumber\\
 &&\times Y_{\ellvp \mvp}^*(\hat \bp)   Y_{\ellvq \mvq}^*(\hat \bq) \Phi_{\rm p}(\bp) \Phi_{\rm p}(\bq)
\eea
 and, therefore,
 \bea
 &&\M_{\ell m \ellvp \mvp \ellvq \mvq \barell \barm}^{ij(2)}(k; p, q) = \delta^K_{\barell 0} \delta^K_{\barm 0}(4\pi)^3 \aleph^*_\ell(k)  (-1)^m i^{\ellvp+\ellvq}  \G^{\ell \ellvp \ellvq }_{-m  \mvp \mvq} \int \ud \bar \chi~\bar \chi^2  \W (\bar \chi)\int_0^{\bar \chi} \ud \tilde \chi ~~~~~~~~~~~~~~~~~~~~~~~~~~~
\nonumber\\ 
&&\times ~ \frac{1}{2} \bigg[ \WW_{\ellvp}^{i} \left(\bar \chi,  \eta, \tilde \chi, \tilde \eta , \frac{\p}{\p \tilde \chi}, \frac{\p}{\p \tilde \eta} \right)    \T^{\Phi}(\bp, \tilde \eta) j_{\ellvp} (p \tilde \chi) ~~ \WW_{\ellvq}^{j} \left(\bar \chi,  \eta, \tilde \chi, \tilde \eta , \frac{\p}{\p \tilde \chi}, \frac{\p}{\p \tilde \eta} \right)  \T^{\Phi}(\bq, \tilde \eta) j_{\ellvq} (q \tilde \chi) \nonumber\\
 &&+~ \WW_{\ellvq}^{i}\left(\bar \chi,  \eta, \tilde \chi, \tilde \eta , \frac{\p}{\p \tilde \chi}, \frac{\p}{\p \tilde \eta} \right)   \T^{\Phi}(\bq, \tilde \eta) j_{\ellvq} (q \tilde \chi) ~~ \WW_{\ellvp}^{j}\left(\bar \chi,  \eta, \tilde \chi, \tilde \eta , \frac{\p}{\p \tilde \chi}, \frac{\p}{\p \tilde \eta} \right)   \T^{\Phi}(\bp, \tilde \eta) j_{\ellvp} (p \tilde \chi) \bigg]  j_\ell (k  \bar \chi)\;.
 \eea
 For  {\it (ii)}, we obtain
 \bea
\frac{1}{2}\Delta_{\ell m}^{uv(2)}(k)&=&\sum_{\ellvp \mvp \ellvq \mvq} \int \frac{\ud^3 \bp}{(2\pi)^3} \frac{\ud^3 \bq}{(2\pi)^3} \Bigg\{ (4\pi)^2 \aleph^*_\ell(k)   i^{\ellvp+\ellvq} ~ I_{\ell \ellvp \ellvp}^{m \mvp \mvq}     \int \ud \bar \chi~\bar \chi^2  \W (\bar \chi)\int_0^{\bar \chi} \frac{\ud \tilde \chi}{\tilde \chi^2}
\nonumber\\ 
&&\times ~\frac{1}{2} \bigg[ \WW_{\ellvp}^{u} \left(\bar \chi,  \eta, \tilde \chi, \tilde \eta , \frac{\p}{\p \tilde \chi}, \frac{\p}{\p \tilde \eta} \right)    \T^{\Phi}(\bp, \tilde \eta) j_{\ellvp} (p \tilde \chi) ~~ \WW_{\ellvq}^{v} \left(\bar \chi,  \eta, \tilde \chi, \tilde \eta , \frac{\p}{\p \tilde \chi}, \frac{\p}{\p \tilde \eta} \right) \T^{\Phi}(\bq, \tilde \eta) j_{\ellvq} (q \tilde \chi) \nonumber\\
 &&+ ~ \WW_{\ellvq}^{u}\left(\bar \chi,  \eta, \tilde \chi, \tilde \eta , \frac{\p}{\p \tilde \chi}, \frac{\p}{\p \tilde \eta} \right)   \T^{\Phi}(\bq, \tilde \eta) j_{\ellvq} (q \tilde \chi) ~~ \WW_{\ellvp}^{v}\left(\bar \chi,  \eta, \tilde \chi, \tilde \eta , \frac{\p}{\p \tilde \chi}, \frac{\p}{\p \tilde \eta} \right)  \T^{\Phi}(\bp, \tilde \eta) j_{\ellvp} (p \tilde \chi) \bigg]  j_\ell (k  \bar \chi)\Bigg\} \nonumber \\
 &&\times ~ Y_{\ellvp \mvp}^*(\hat \bp)  Y_{\ellvq \mvq}^*(\hat \bq)  \Phi_{\rm p}(\bp) \Phi_{\rm p}(\bq) 
\eea
and, consequently, 
 \bea
 &&\M_{\ell m \ellvp \mvp \ellvq \mvq \barell \barm}^{uv(2)}(k; p, q) = \delta^K_{\barell 0} \delta^K_{\barm 0}(4\pi)^3\aleph^*_\ell(k)   i^{\ellvp+\ellvq} ~ I_{\ell \ellvp \ellvp}^{m \mvp \mvq}     \int \ud \bar \chi~\bar \chi^2  \W (\bar \chi)\int_0^{\bar \chi} \frac{\ud \tilde \chi}{\tilde \chi^2}
\nonumber\\ 
&&\times ~\frac{1}{2} \bigg[ \WW_{\ellvp}^{u} \left(\bar \chi,  \eta, \tilde \chi, \tilde \eta , \frac{\p}{\p \tilde \chi}, \frac{\p}{\p \tilde \eta} \right)    \T^{\Phi}(\bp, \tilde \eta) j_{\ellvp} (p \tilde \chi) ~~ \WW_{\ellvq}^{v} \left(\bar \chi,  \eta, \tilde \chi, \tilde \eta , \frac{\p}{\p \tilde \chi}, \frac{\p}{\p \tilde \eta} \right)    \T^{\Phi}(\bq, \tilde \eta) j_{\ellvq} (q \tilde \chi) \nonumber\\
 &&+ ~ \WW_{\ellvq}^{u}\left(\bar \chi,  \eta, \tilde \chi, \tilde \eta , \frac{\p}{\p \tilde \chi}, \frac{\p}{\p \tilde \eta} \right)   \T^{\Phi}(\bq, \tilde \eta) j_{\ellvq} (q \bar \chi) ~~ \WW_{\ellvp}^{v}\left(\bar \chi,  \eta, \tilde \chi, \tilde \eta , \frac{\p}{\p \tilde \chi}, \frac{\p}{\p \tilde \eta} \right)   \T^{\Phi}(\bp, \tilde \eta) j_{\ellvp} (p \tilde \chi) \bigg]  j_\ell (k  \bar \chi) \;. 
 \eea
Instead for {\it (iii)} we have
\bea
&&\frac{1}{2}\Delta_{\ell m}^{ij(2)}(k)=\sum_{\ellvp \mvp \ellvq \mvq} \int \frac{\ud^3 \bp}{(2\pi)^3} \frac{\ud^3 \bq}{(2\pi)^3} \Bigg\{ (4\pi)^2 \aleph^*_\ell(k)  (-1)^m i^{\ellvp+\ellvq}  \G^{\ell \ellvp \ellvq }_{-m  \mvp \mvq} \int \ud \bar \chi~\bar \chi^2  \W (\bar \chi) \int_0^{\bar \chi} \ud \tilde \chi \int_0^{\tilde \chi} \ud \tilde{\tilde \chi}
\nonumber\\ 
&&\times~ \frac{1}{2} \bigg[ \WW_{\ellvp}^{i} \left(\bar \chi,  \eta, \tilde \chi, \tilde \eta , \frac{\p}{\p \tilde \chi}, \frac{\p}{\p \tilde \eta} \right)    \T^{\Phi}(\bp, \tilde \eta) j_{\ellvp} (p \tilde \chi) ~~ \WW_{\ellvq}^{j} \left(\tilde \chi, \tilde{\tilde \chi} ,  \tilde \eta, \tilde{ \tilde \eta}, \frac{\p}{\p   \tilde{\tilde \chi}}, \frac{\p}{\p  \tilde{\tilde \eta}}\right)   \T^{\Phi}\left(\bq, \tilde{\tilde \eta}\right) j_{\ellvq} \left(q \tilde{\tilde \chi}\right) \nonumber\\
 &&+~ \WW_{\ellvq}^{i}\left(\bar \chi,  \eta, \tilde \chi, \tilde \eta , \frac{\p}{\p \tilde \chi}, \frac{\p}{\p \tilde \eta} \right)   \T^{\Phi}(\bq, \tilde \eta) j_{\ellvq} (q \bar \chi) ~~ \WW_{\ellvp}^{j} \left(\tilde \chi, \tilde{\tilde \chi} ,  \tilde \eta, \tilde{ \tilde \eta}, \frac{\p}{\p   \tilde{\tilde \chi}}, \frac{\p}{\p  \tilde{\tilde \eta}}\right)   \T^{\Phi} \left(\bp, \tilde {\tilde \eta}\right) j_{\ellvp} \left(p \tilde{ \tilde \chi}\right) \bigg]  j_\ell (k  \bar \chi)\Bigg\} \nonumber\\
   &&\times~ Y_{\ellvp \mvp}^*(\hat \bp)   Y_{\ellvq \mvq}^*(\hat \bq) \Phi_{\rm p}(\bp) \Phi_{\rm p}(\bq)
\eea
and so
 \bea
 &&\M_{\ell m \ellvp \mvp \ellvq \mvq \barell \barm}^{ij(2)}(k; p, q) = \delta^K_{\barell 0} \delta^K_{\barm 0}(4\pi)^3 \aleph^*_\ell(k)  (-1)^m i^{\ellvp+\ellvq}  \G^{\ell \ellvp \ellvq }_{-m  \mvp \mvq}  \int \ud \bar \chi~\bar \chi^2  \W (\bar \chi) \int_0^{\bar \chi} \ud \tilde \chi \int_0^{\tilde \chi} \ud \tilde{\tilde \chi}
\nonumber\\ 
&&\times~ \frac{1}{2} \bigg[ \WW_{\ellvp}^{i} \left(\bar \chi,  \eta, \tilde \chi, \tilde \eta , \frac{\p}{\p \tilde \chi}, \frac{\p}{\p \tilde \eta} \right)    \T^{\Phi}(\bp, \tilde \eta) j_{\ellvp} (p \tilde \chi) ~~ \WW_{\ellvq}^{j} \left(\tilde \chi, \tilde{\tilde \chi} ,  \tilde \eta, \tilde{ \tilde \eta}, \frac{\p}{\p   \tilde{\tilde \chi}}, \frac{\p}{\p  \tilde{\tilde \eta}}\right)    \T^{\Phi}\left(\bq, \tilde{\tilde \eta}\right) j_{\ellvq} \left(q \tilde{\tilde \chi}\right) \nonumber\\
 &&+~ \WW_{\ellvq}^{i}\left(\bar \chi,  \eta, \tilde \chi, \tilde \eta , \frac{\p}{\p \tilde \chi}, \frac{\p}{\p \tilde \eta} \right)  \T^{\Phi}(\bq, \tilde \eta) j_{\ellvq} (q \tilde \chi) ~~ \WW_{\ellvp}^{j} \left(\tilde \chi, \tilde{\tilde \chi} ,  \tilde \eta, \tilde{ \tilde \eta}, \frac{\p}{\p   \tilde{\tilde \chi}}, \frac{\p}{\p  \tilde{\tilde \eta}}\right)  \T^{\Phi} \left(\bp, \tilde {\tilde \eta}\right) j_{\ellvp} \left(p \tilde{ \tilde \chi}\right) \bigg]  j_\ell (k  \bar \chi) \;. \nonumber \\
 \eea
  Finally,  for {\it (iv)}, it turns out
   \bea
\frac{1}{2}\Delta_{\ell m}^{uv(2)}(k)&=&\sum_{\ellvp \mvp \ellvq \mvq} \int \frac{\ud^3 \bp}{(2\pi)^3} \frac{\ud^3 \bq}{(2\pi)^3} \Bigg\{ (4\pi)^2 \aleph^*_\ell(k)   i^{\ellvp+\ellvq} ~ I_{\ell \ellvp \ellvp}^{m \mvp \mvq}     \int \ud \bar \chi~\bar \chi^2  \W (\bar \chi) \int_0^{\bar \chi} \frac{\ud \tilde \chi}{\tilde \chi}  \int_0^{\tilde \chi} \frac{\ud \tilde{\tilde \chi}}{ \tilde{\tilde \chi}}
\nonumber\\ 
&&\times ~\frac{1}{2} \bigg[ \WW_{\ellvp}^{u} \left(\bar \chi,  \eta, \tilde \chi, \tilde \eta , \frac{\p}{\p \tilde \chi}, \frac{\p}{\p \tilde \eta} \right)    \T^{\Phi}(\bp, \tilde \eta) j_{\ellvp} (p \tilde \chi) ~~ \WW_{\ellvq}^{v} \left(\tilde \chi, \tilde{\tilde \chi} ,  \tilde \eta, \tilde{ \tilde \eta}, \frac{\p}{\p   \tilde{\tilde \chi}}, \frac{\p}{\p  \tilde{\tilde \eta}}\right)    \T^{\Phi}\left(\bq, \tilde{\tilde \eta}\right) j_{\ellvq} \left(q \tilde{\tilde \chi}\right)  \nonumber\\
 &&+ ~ \WW_{\ellvq}^{u}\left(\bar \chi,  \eta, \tilde \chi, \tilde \eta , \frac{\p}{\p \tilde \chi}, \frac{\p}{\p \tilde \eta} \right)   \T^{\Phi}(\bq, \tilde \eta) j_{\ellvq} (q \bar \chi) ~~ \WW_{\ellvp}^{v} \left(\tilde \chi, \tilde{\tilde \chi} ,  \tilde \eta, \tilde{ \tilde \eta}, \frac{\p}{\p   \tilde{\tilde \chi}}, \frac{\p}{\p  \tilde{\tilde \eta}}\right)     \T^{\Phi} \left(\bp, \tilde {\tilde \eta}\right) j_{\ellvp} \left(p \tilde{ \tilde \chi}\right)  \bigg]  j_\ell (k  \bar \chi)\Bigg\} \nonumber \\
 &&\times ~ Y_{\ellvp \mvp}^*(\hat \bp)  Y_{\ellvq \mvq}^*(\hat \bq)  \Phi_{\rm p}(\bp) \Phi_{\rm p}(\bq)
\eea
and  hence
\bea
 &&\M_{\ell m \ellvp \mvp \ellvq \mvq \barell \barm}^{uv(2)}(k; p, q) = \delta^K_{\barell 0} \delta^K_{\barm 0}(4\pi)^3\aleph^*_\ell(k)   i^{\ellvp+\ellvq} ~ I_{\ell \ellvp \ellvp}^{m \mvp \mvq}      \int \ud \bar \chi~\bar \chi^2  \W (\bar \chi) \int_0^{\bar \chi} \frac{\ud \tilde \chi}{\tilde \chi}  \int_0^{\tilde \chi} \frac{\ud \tilde{\tilde \chi}}{ \tilde{\tilde \chi}}
\nonumber\\ 
&&\times ~\frac{1}{2} \bigg[ \WW_{\ellvp}^{u} \left(\bar \chi,  \eta, \tilde \chi, \tilde \eta , \frac{\p}{\p \tilde \chi}, \frac{\p}{\p \tilde \eta} \right)    \T^{\Phi}(\bp, \tilde \eta) j_{\ellvp} (p \tilde \chi) ~~ \WW_{\ellvq}^{v} \left(\tilde \chi, \tilde{\tilde \chi} ,  \tilde \eta, \tilde{ \tilde \eta}, \frac{\p}{\p   \tilde{\tilde \chi}}, \frac{\p}{\p  \tilde{\tilde \eta}}\right)      \T^{\Phi}\left(\bq, \tilde{\tilde \eta}\right) j_{\ellvq} \left(q \tilde{\tilde \chi}\right)  \nonumber\\
 &&+ ~ \WW_{\ellvq}^{u}\left(\bar \chi,  \eta, \tilde \chi, \tilde \eta , \frac{\p}{\p \tilde \chi}, \frac{\p}{\p \tilde \eta} \right) \T^{\Phi}(\bq, \tilde \eta) j_{\ellvq} (q \bar \chi) ~~ \WW_{\ellvp}^{v} \left(\tilde \chi, \tilde{\tilde \chi} ,  \tilde \eta, \tilde{ \tilde \eta}, \frac{\p}{\p   \tilde{\tilde \chi}}, \frac{\p}{\p  \tilde{\tilde \eta}}\right)    \T^{\Phi}\left(\bp, \tilde {\tilde \eta}\right) j_{\ellvp} \left(p \tilde{ \tilde \chi}\right)  \bigg]  j_\ell (k  \bar \chi) \;. 
 \eea
 
 Here below we write explicitly  all terms contained in  Eq.\ (\ref{Deltag2-int-4}):
  \bea\label{phiperpS}
 &&\M_{\ell m \ellvp \mvp \ellvq \mvq \barell \barm}^{\int [ \Phi \p_{\perp}S_{\perp}](2)}(k; p, q) =2 \delta^K_{\barell 0} \delta^K_{\barm 0}(4\pi)^3 \aleph^*_\ell(k)  (-1)^m i^{\ellvp+\ellvq}  \G^{\ell \ellvp \ellvq }_{-m  \mvp \mvq}  \int \ud \bar \chi~\bar \chi^2  \W (\bar \chi) \int_0^{\bar \chi} \frac{ \ud \tilde \chi }{\tilde \chi } \int_0^{\tilde \chi}\frac{ \ud \tilde{\tilde \chi}}{\tilde{\tilde \chi}} ~~~~~~~~~~~~~~~~~~\nonumber \\
 &&\times \left[ b_e-3 + \Q  -   \frac{\cH'}{\cH^2} - 2\frac{\left(1 - \Q\right) }{\bar \chi \cH} +3\frac{\left(1 - \Q\right)\tilde \chi}{\bar \chi} \right] 
 \bigg[  \ellvq(\ellvq+1) \T^{\Phi}(\bp, \tilde \eta) j_{\ellvp} (p \tilde \chi)  \T^{\Phi}\left(\bq, \tilde{\tilde \eta}\right) j_{\ellvq} \left(q \tilde{\tilde \chi}\right)  \nonumber\\ 
&&+~ \ellvp(\ellvp+1)  \T^{\Phi}(\bq, \tilde \eta) j_{\ellvq} (q \tilde \chi)  \T^{\Phi} \left(\bp, \tilde {\tilde \eta}\right) j_{\ellvp} \left(p \tilde{ \tilde \chi}\right) \bigg]  j_\ell (k  \bar \chi) \;,
 \eea
  \bea
 &&\M_{\ell m \ellvp \mvp \ellvq \mvq \barell \barm}^{\int[ \p_\| \Phi \kappa](2)}(k; p, q) =-2 \delta^K_{\barell 0} \delta^K_{\barm 0}(4\pi)^3 \aleph^*_\ell(k)  (-1)^m i^{\ellvp+\ellvq}  \G^{\ell \ellvp \ellvq }_{-m  \mvp \mvq}  \int \ud \bar \chi~\bar \chi^2  \W (\bar \chi) \int_0^{\bar \chi} \frac{ \ud \tilde \chi }{\tilde \chi } \int_0^{\tilde \chi} \ud \tilde{\tilde \chi}%
 \frac{ (\tilde \chi -  \tilde{\tilde \chi } ) }{ \tilde{\tilde \chi} } ~~
 \nonumber \\
 &&\times \left[ 2-b_e +   \frac{\cH'}{\cH^2} + 2\frac{\left(1 - \Q\right) }{\bar \chi \cH} -\frac{\left(1 - \Q\right)\tilde \chi}{\bar \chi} \right] 
 \Bigg\{  \ellvq(\ellvq+1) \T^{\Phi}(\bp, \tilde \eta)\left[\frac{\p}{\p \tilde \chi} j_{\ellvp} (p \tilde \chi)\right]  \T^{\Phi}\left(\bq, \tilde{\tilde \eta}\right) j_{\ellvq} \left(q \tilde{\tilde \chi}\right)  \nonumber\\ 
&&+~ \ellvp(\ellvp+1)  \T^{\Phi}(\bq, \tilde \eta) \left[\frac{\p}{\p \tilde \chi} j_{\ellvq} (q \tilde \chi)\right]   \T^{\Phi} \left(\bp, \tilde {\tilde \eta}\right) j_{\ellvp} \left(p \tilde{ \tilde \chi}\right) \Bigg\}  j_\ell (k  \bar \chi) \;,
 \eea
   \bea
 &&\M_{\ell m \ellvp \mvp \ellvq \mvq \barell \barm}^{\int[  \Phi' \kappa](2)}(k; p, q) =2 \delta^K_{\barell 0} \delta^K_{\barm 0}(4\pi)^3 \aleph^*_\ell(k)  (-1)^m i^{\ellvp+\ellvq}  \G^{\ell \ellvp \ellvq }_{-m  \mvp \mvq}  \int \ud \bar \chi~\bar \chi^2  \W (\bar \chi) \int_0^{\bar \chi} \frac{ \ud \tilde \chi }{\tilde \chi } \int_0^{\tilde \chi} \ud \tilde{\tilde \chi}%
 \frac{ (\tilde \chi -  \tilde{\tilde \chi } ) }{ \tilde{\tilde \chi} } ~~~~~~~~~
 \nonumber \\
 &&\times \left[ 2-b_e +   \frac{\cH'}{\cH^2} + 2\frac{\left(1 - \Q\right) }{\bar \chi \cH} -\frac{\left(1 - \Q\right)\tilde \chi}{\bar \chi} \right] 
 \Bigg\{  \ellvq(\ellvq+1) \left[\frac{\p}{\p \tilde \eta}\T^{\Phi}(\bp, \tilde \eta) \right] j_{\ellvp} (p \tilde \chi) \T^{\Phi}\left(\bq, \tilde{\tilde \eta}\right) j_{\ellvq} \left(q \tilde{\tilde \chi}\right)  \nonumber\\ 
&&+~ \ellvp(\ellvp+1) \left[\frac{\p}{\p \tilde \eta} \T^{\Phi}(\bq, \tilde \eta) \right]    j_{\ellvq} (q \tilde \chi) \T^{\Phi} \left(\bp, \tilde {\tilde \eta}\right) j_{\ellvp} \left(p \tilde{ \tilde \chi}\right) \Bigg\}  j_\ell (k  \bar \chi) \;,
 \eea
   \bea
 &&\M_{\ell m \ellvp \mvp \ellvq \mvq \barell \barm}^{\int [ \Phi \kappa](2)}(k; p, q) =-2 \delta^K_{\barell 0} \delta^K_{\barm 0}(4\pi)^3 \aleph^*_\ell(k)  (-1)^m i^{\ellvp+\ellvq}  \G^{\ell \ellvp \ellvq }_{-m  \mvp \mvq}  \int \ud \bar \chi~\bar \chi^2  \W (\bar \chi) \left[ b_e-2 -   \frac{\cH'}{\cH^2} - 2\frac{\left(1 - \Q\right) }{\bar \chi \cH}  \right] \nonumber \\
 &&\times ~\int_0^{\bar \chi} \frac{ \ud \tilde \chi }{\tilde \chi^2 } \int_0^{\tilde \chi} \ud \tilde{\tilde \chi} \frac{(\tilde \chi -  \tilde{\tilde \chi } ) }{\tilde{\tilde \chi}}
 \bigg[  \ellvq(\ellvq+1) \T^{\Phi}(\bp, \tilde \eta) j_{\ellvp} (p \tilde \chi)  \T^{\Phi}\left(\bq, \tilde{\tilde \eta}\right) j_{\ellvq} \left(q \tilde{\tilde \chi}\right)   \nonumber\\ 
&&+~ \ellvp(\ellvp+1)  \T^{\Phi}(\bq, \tilde \eta) j_{\ellvq} (q \tilde \chi)  \T^{\Phi} \left(\bp, \tilde {\tilde \eta}\right) j_{\ellvp} \left(p \tilde{ \tilde \chi}\right) \bigg] j_\ell (k  \bar \chi) \;,
 \eea
 \bea
 &&\M_{\ell m \ellvp \mvp \ellvq \mvq \barell \barm}^{\int \Phi^2 (2)}(k; p, q) =-4 \delta^K_{\barell 0} \delta^K_{\barm 0}(4\pi)^3 \aleph^*_\ell(k)  (-1)^m i^{\ellvp+\ellvq}  \G^{\ell \ellvp \ellvq }_{-m  \mvp \mvq} \int \ud \bar \chi~\bar \chi  \W (\bar \chi)  (1-\Q) \int_0^{\bar \chi} \ud \tilde \chi ~~~~~~~~~~~~~~~~~~~~~
\nonumber\\ 
&&\times ~   \T^{\Phi}(\bp, \tilde \eta) \T^{\Phi}(\bq, \tilde \eta)  j_{\ellvp} (p \tilde \chi) j_{\ellvq} (q \tilde \chi)  j_\ell (k  \bar \chi)\;,
 \eea
 \bea
 &&\M_{\ell m \ellvp \mvp \ellvq \mvq \barell \barm}^{\int[ \Phi'T](2)}(k; p, q) = 4\delta^K_{\barell 0} \delta^K_{\barm 0}(4\pi)^3 \aleph^*_\ell(k)  (-1)^m i^{\ellvp+\ellvq}  \G^{\ell \ellvp \ellvq }_{-m  \mvp \mvq}  \int \ud \bar \chi~\bar \chi  \W (\bar \chi) (1-\Q) \int_0^{\bar \chi} \ud \tilde \chi \int_0^{\tilde \chi} \ud \tilde{\tilde \chi}~~~~~~~~~~
\nonumber\\ 
&&\times~ \left\{    \left[\frac{\p}{\p \tilde \eta}\T^{\Phi}(\bp, \tilde \eta) \right]   j_{\ellvp} (p \tilde \chi)  \T^{\Phi}\left(\bq, \tilde{\tilde \eta}\right) j_{\ellvq} \left(q \tilde{\tilde \chi}\right) + \left[\frac{\p}{\p \tilde \eta} \T^{\Phi}(\bq, \tilde \eta) \right]  j_{\ellvq} (q \tilde \chi) \T^{\Phi} \left(\bp, \tilde {\tilde \eta}\right) j_{\ellvp} \left(p \tilde{ \tilde \chi}\right) \right\}  j_\ell (k  \bar \chi) \;,
 \eea
 \bea
 &&\M_{\ell m \ellvp \mvp \ellvq \mvq \barell \barm}^{\int[\p_\perp \Phi \p_\perp T](2)}(k; p, q) =- \delta^K_{\barell 0} \delta^K_{\barm 0}(4\pi)^3\aleph^*_\ell(k)   i^{\ellvp+\ellvq} ~ I_{\ell \ellvp \ellvp}^{m \mvp \mvq}      \int \ud \bar \chi~\bar \chi^2  \W (\bar \chi)(1-\Q) \int_0^{\bar \chi} \frac{\ud \tilde \chi}{\tilde \chi^2} \left(3-5\frac{\tilde \chi}{\bar \chi}\right)  \int_0^{\tilde \chi}   \ud \tilde{\tilde \chi}
\nonumber\\ 
&&\times ~ \bigg[  \T^{\Phi}(\bp, \tilde \eta) j_{\ellvp} (p \tilde \chi)      \T^{\Phi}\left(\bq, \tilde{\tilde \eta}\right) j_{\ellvq} \left(q \tilde{\tilde \chi}\right)  + ~ \T^{\Phi}(\bq, \tilde \eta) j_{\ellvq} (q \tilde \chi)    \T^{\Phi}\left(\bp, \tilde {\tilde \eta}\right) j_{\ellvp} \left(p \tilde{ \tilde \chi}\right)  \bigg]  j_\ell (k  \bar \chi) \;, 
 \eea
  \bea
 &&\M_{\ell m \ellvp \mvp \ellvq \mvq \barell \barm}^{\int[ \Phi''T](2)}(k; p, q) = -2\delta^K_{\barell 0} \delta^K_{\barm 0}(4\pi)^3 \aleph^*_\ell(k)  (-1)^m i^{\ellvp+\ellvq}  \G^{\ell \ellvp \ellvq }_{-m  \mvp \mvq}  \int \ud \bar \chi~\bar \chi^2  \W (\bar \chi)  \left[ b_e-2 \Q  -   \frac{\cH'}{\cH^2} - 2 \frac{\left(1 - \Q\right)}{\bar \chi \cH} \right] 
\nonumber\\ 
&&\times~ \int_0^{\bar \chi} \ud \tilde \chi \int_0^{\tilde \chi} \ud \tilde{\tilde \chi} \Bigg\{    \bigg[\frac{\p^2}{\p \tilde \eta^2}\T^{\Phi}(\bp, \tilde \eta) \bigg]   j_{\ellvp} (p \tilde \chi)  \T^{\Phi}\left(\bq, \tilde{\tilde \eta}\right) j_{\ellvq} \left(q \tilde{\tilde \chi}\right) + \left[\frac{\p^2}{\p \tilde \eta^2} \T^{\Phi}(\bq, \tilde \eta) \right]  j_{\ellvq} (q \tilde \chi) \T^{\Phi} \left(\bp, \tilde {\tilde \eta}\right) j_{\ellvp} \left(p \tilde{ \tilde \chi}\right) \Bigg\}  j_\ell (k  \bar \chi) \;, \nonumber\\
 \eea
  \bea
 &&\M_{\ell m \ellvp \mvp \ellvq \mvq \barell \barm}^{\int [\Phi' \Phi](2)}(k; p, q) = 2\delta^K_{\barell 0} \delta^K_{\barm 0}(4\pi)^3 \aleph^*_\ell(k)  (-1)^m i^{\ellvp+\ellvq}  \G^{\ell \ellvp \ellvq }_{-m  \mvp \mvq} \int \ud \bar \chi~\bar \chi^2  \W (\bar \chi)  \left[ b_e-2 \Q  -   \frac{\cH'}{\cH^2} - 2 \frac{\left(1 - \Q\right)}{\bar \chi \cH} \right]  \nonumber\\ 
&&\times ~ \int_0^{\bar \chi} \ud \tilde \chi   \Bigg\{ \left[\frac{\p}{\p \tilde \eta}\T^{\Phi}(\bp, \tilde \eta) \right]  ~ \T^{\Phi}(\bq, \tilde \eta) + \left[\frac{\p}{\p \tilde \eta} \T^{\Phi}(\bq, \tilde \eta) \right]   ~   \T^{\Phi}(\bp, \tilde \eta)  \Bigg\} j_{\ellvp} (p \tilde \chi) j_{\ellvq} (q \tilde \chi)  j_\ell (k  \bar \chi) \;,
 \eea
  \bea
 &&\M_{\ell m \ellvp \mvp \ellvq \mvq \barell \barm}^{\int[ \Phi'I](2)}(k; p, q) = -2\delta^K_{\barell 0} \delta^K_{\barm 0}(4\pi)^3 \aleph^*_\ell(k)  (-1)^m i^{\ellvp+\ellvq}  \G^{\ell \ellvp \ellvq }_{-m  \mvp \mvq}  \int \ud \bar \chi~\bar \chi^2  \W (\bar \chi) \left[ b_e-2 \Q  -   \frac{\cH'}{\cH^2} - 2 \frac{\left(1 - \Q\right)}{\bar \chi \cH} \right]  
\nonumber\\ 
&&\times \int_0^{\bar \chi} \ud \tilde \chi \int_0^{\tilde \chi} \ud \tilde{\tilde \chi} \Bigg\{    \left[\frac{\p}{\p \tilde \eta}\T^{\Phi}(\bp, \tilde \eta) \right]   j_{\ellvp} (p \tilde \chi) \left[\frac{\p}{\p \tilde {\tilde \eta}} \T^{\Phi}\left(\bq, \tilde{\tilde \eta}\right)  \right] j_{\ellvq} \left(q \tilde{\tilde \chi}\right) + \left[\frac{\p}{\p \tilde \eta} \T^{\Phi}(\bq, \tilde \eta) \right]  j_{\ellvq} (q \tilde \chi) \left[\frac{\p}{\p \tilde{ \tilde \eta}}\T^{\Phi} \left(\bp, \tilde {\tilde \eta}\right) \right]  j_{\ellvp} \left(p \tilde{ \tilde \chi}\right) \Bigg\} \nonumber \\
&& \times~  j_\ell (k  \bar \chi) \;,
 \eea
 \bea
 &&\M_{\ell m \ellvp \mvp \ellvq \mvq \barell \barm}^{\int[\p_\perp \Phi' S_\perp] (2)}(k; p, q) = 2\delta^K_{\barell 0} \delta^K_{\barm 0}(4\pi)^3\aleph^*_\ell(k)   i^{\ellvp+\ellvq} ~ I_{\ell \ellvp \ellvp}^{m \mvp \mvq}      \int \ud \bar \chi~\bar \chi^2  \W (\bar \chi) \int_0^{\bar \chi} \ud \tilde \chi    \int_0^{\tilde \chi} \frac{\ud \tilde{\tilde \chi}}{ \tilde{\tilde \chi}}
~~~~~~~~~~~~~~~~~~~~~~~~~~~~~\nonumber\\ 
&&\times ~\left[ 1+ b_e-3 \Q  -   \frac{\cH'}{\cH^2} - 2\frac{\left(1 - \Q\right) }{\bar \chi \cH}-\left(1 - \Q\right) \frac{\tilde \chi}{\bar \chi} \right]   \Bigg\{   \left[\frac{\p}{\p \tilde \eta}\T^{\Phi}(\bp, \tilde \eta) \right]   j_{\ellvp} (p \tilde \chi)     \T^{\Phi}\left(\bq, \tilde{\tilde \eta}\right) j_{\ellvq} \left(q \tilde{\tilde \chi}\right)  \nonumber\\
 &&+  \left[\frac{\p}{\p \tilde \eta}\T^{\Phi}(\bq, \tilde \eta)\right]    j_{\ellvq} (q \tilde \chi)     \T^{\Phi}\left(\bp, \tilde {\tilde \eta}\right) j_{\ellvp} \left(p \tilde{ \tilde \chi}\right)  \Bigg\}  j_\ell (k  \bar \chi) \;, 
 \eea
  \bea
 &&\M_{\ell m \ellvp \mvp \ellvq \mvq \barell \barm}^{\int[\p_\perp \Phi' \p_\perp T] (2)}(k; p, q) = -2\delta^K_{\barell 0} \delta^K_{\barm 0}(4\pi)^3\aleph^*_\ell(k)   i^{\ellvp+\ellvq} ~ I_{\ell \ellvp \ellvp}^{m \mvp \mvq}      \int \ud \bar \chi~\bar \chi^2  \W (\bar \chi) \int_0^{\bar \chi}  \frac{\ud \tilde \chi}{\tilde \chi}    \int_0^{\tilde \chi}\ud \tilde{\tilde \chi}
~~~~~~~~~~~~~~~~~~~~~~~~~~~~~\nonumber\\ 
&&\times ~\left[ 2+ b_e-4 \Q  -   \frac{\cH'}{\cH^2} - 2\frac{\left(1 - \Q\right) }{\bar \chi \cH}-2\left(1 - \Q\right) \frac{\tilde \chi}{\bar \chi} \right]   \Bigg\{   \left[\frac{\p}{\p \tilde \eta}\T^{\Phi}(\bp, \tilde \eta) \right]   j_{\ellvp} (p \tilde \chi)     \T^{\Phi}\left(\bq, \tilde{\tilde \eta}\right) j_{\ellvq} \left(q \tilde{\tilde \chi}\right)  \nonumber\\
 &&+  \left[\frac{\p}{\p \tilde \eta}\T^{\Phi}(\bq, \tilde \eta)\right]    j_{\ellvq} (q \tilde \chi)     \T^{\Phi}\left(\bp, \tilde {\tilde \eta}\right) j_{\ellvp} \left(p \tilde{ \tilde \chi}\right)  \Bigg\}  j_\ell (k  \bar \chi) \;, 
 \eea
  \bea
 &&\M_{\ell m \ellvp \mvp \ellvq \mvq \barell \barm}^{\int [\triangle_\Omega \Phi I](2)}(k; p, q) =-2 \delta^K_{\barell 0} \delta^K_{\barm 0}(4\pi)^3 \aleph^*_\ell(k)  (-1)^m i^{\ellvp+\ellvq}  \G^{\ell \ellvp \ellvq }_{-m  \mvp \mvq}  \int \ud \bar \chi~\bar \chi^2  \W (\bar \chi)(1-\Q) \int_0^{\bar \chi} \frac{\ud \tilde \chi}{ \tilde \chi}\int_0^{\tilde \chi} \ud \tilde{\tilde \chi}
\nonumber\\ 
&&\times   \Bigg\{ \ellvp(\ellvp+1)  \T^{\Phi}(\bp, \tilde \eta) j_{\ellvp} (p \tilde \chi)
 \left[\frac{\p}{\p \tilde {\tilde \eta}} \T^{\Phi}\left(\bq, \tilde{\tilde \eta}\right)  \right]   j_{\ellvq} \left(q \tilde{\tilde \chi}\right) + \ellvq(\ellvq+1)   \T^{\Phi}(\bq, \tilde \eta) j_{\ellvq} (q \tilde \chi)  \left[\frac{\p}{\p \tilde {\tilde \eta}} \T^{\Phi} \left(\bp, \tilde{\tilde \eta}\right)  \right] j_{\ellvp} \left(p \tilde{ \tilde \chi}\right) \Bigg\}\nonumber\\
 &&\times~  j_\ell (k  \bar \chi) \;,
 \eea
   \bea
 &&\M_{\ell m \ellvp \mvp \ellvq \mvq \barell \barm}^{\int [\Delta_\Omega \Phi  \int  \tilde{\tilde \chi}  \Phi {'}  ](2)}(k; p, q) =2 \delta^K_{\barell 0} \delta^K_{\barm 0}(4\pi)^3 \aleph^*_\ell(k)  (-1)^m i^{\ellvp+\ellvq}  \G^{\ell \ellvp \ellvq }_{-m  \mvp \mvq}  \int \ud \bar \chi~\bar \chi  \W (\bar \chi)(1-\Q) \int_0^{\bar \chi} \frac{\ud \tilde \chi}{ \tilde \chi}\int_0^{\tilde \chi} \tilde{\tilde \chi} \ud \tilde{\tilde \chi}
\nonumber\\ 
&&\times   \Bigg\{ \ellvp(\ellvp+1)  \T^{\Phi}(\bp, \tilde \eta) j_{\ellvp} (p \tilde \chi)
 \left[\frac{\p}{\p \tilde {\tilde \eta}} \T^{\Phi}\left(\bq, \tilde{\tilde \eta}\right)  \right]   j_{\ellvq} \left(q \tilde{\tilde \chi}\right) + \ellvq(\ellvq+1)   \T^{\Phi}(\bq, \tilde \eta) j_{\ellvq} (q \tilde \chi)  \left[\frac{\p}{\p \tilde {\tilde \eta}} \T^{\Phi} \left(\bp, \tilde{\tilde \eta}\right)  \right] j_{\ellvp} \left(p \tilde{ \tilde \chi}\right) \Bigg\}\nonumber\\
 &&\times~  j_\ell (k  \bar \chi) \;,
 \eea
 \bea
 &&\M_{\ell m \ellvp \mvp \ellvq \mvq \barell \barm}^{\int [ \p_\perp\Phi \p_\perp I ](2)}(k; p, q) =2 \delta^K_{\barell 0} \delta^K_{\barm 0}(4\pi)^3\aleph^*_\ell(k)   i^{\ellvp+\ellvq} ~ I_{\ell \ellvp \ellvp}^{m \mvp \mvq}      \int \ud \bar \chi~\bar \chi^2  \W (\bar \chi) (1-\Q) \int_0^{\bar \chi} \frac{\ud \tilde \chi}{\tilde \chi}  \int_0^{\tilde \chi} \ud \tilde{\tilde \chi}~~~~~~~~~~~~~~~~~~~~\nonumber\\ 
&&\times ~  \Bigg\{ \T^{\Phi}(\bp, \tilde \eta) j_{\ellvp} (p \tilde \chi)
 \left[\frac{\p}{\p \tilde {\tilde \eta}} \T^{\Phi}\left(\bq, \tilde{\tilde \eta}\right)  \right]   j_{\ellvq} \left(q \tilde{\tilde \chi}\right) +  \T^{\Phi}(\bq, \tilde \eta) j_{\ellvq} (q \tilde \chi)  \left[\frac{\p}{\p \tilde {\tilde \eta}} \T^{\Phi} \left(\bp, \tilde{\tilde \eta}\right)  \right] j_{\ellvp} \left(p \tilde{ \tilde \chi}\right) \Bigg\}  j_\ell (k  \bar \chi) \;,
 \eea
 \bea
 &&\M_{\ell m \ellvp \mvp \ellvq \mvq \barell \barm}^{\int [ \p_\perp\Phi   \tilde \p_\perp \int \tilde{\tilde \chi} \Phi {'} ](2)}(k; p, q) = - 2 \delta^K_{\barell 0} \delta^K_{\barm 0}(4\pi)^3\aleph^*_\ell(k)   i^{\ellvp+\ellvq} ~ I_{\ell \ellvp \ellvp}^{m \mvp \mvq}      \int \ud \bar \chi~\bar \chi  \W (\bar \chi) (1-\Q) \int_0^{\bar \chi} \frac{\ud \tilde \chi}{\tilde \chi}  \int_0^{\tilde \chi} \tilde{\tilde \chi} \ud \tilde{\tilde \chi}~~~~~~~~~~~\nonumber\\ 
&&\times ~  \Bigg\{ \T^{\Phi}(\bp, \tilde \eta) j_{\ellvp} (p \tilde \chi)
 \left[\frac{\p}{\p \tilde {\tilde \eta}} \T^{\Phi}\left(\bq, \tilde{\tilde \eta}\right)  \right]   j_{\ellvq} \left(q \tilde{\tilde \chi}\right) +  \T^{\Phi}(\bq, \tilde \eta) j_{\ellvq} (q \tilde \chi)  \left[\frac{\p}{\p \tilde {\tilde \eta}} \T^{\Phi} \left(\bp, \tilde{\tilde \eta}\right)  \right] j_{\ellvp} \left(p \tilde{ \tilde \chi}\right) \Bigg\}  j_\ell (k  \bar \chi) \;,
   \eea
  \bea
 &&\M_{\ell m \ellvp \mvp \ellvq \mvq \barell \barm}^{ \int [\p_\perp \Phi \p_\perp \Phi ](2)}(k; p, q) = 4\delta^K_{\barell 0} \delta^K_{\barm 0}(4\pi)^3\aleph^*_\ell(k)   i^{\ellvp+\ellvq} ~ I_{\ell \ellvp \ellvp}^{m \mvp \mvq}     \int \ud \bar \chi~\bar \chi  \W (\bar \chi)  (1-\Q)\int_0^{\bar \chi} \frac{\ud \tilde \chi}{\tilde \chi }\left( \bar \chi - \tilde \chi\right)~~~~~~~~~~~~~~~~~~~~
\nonumber\\ 
&&\times ~   \T^{\Phi}(\bp, \tilde \eta) \T^{\Phi}(\bq, \tilde \eta)  j_{\ellvp} (p \tilde \chi) j_{\ellvq} (q \tilde \chi)  j_\ell (k  \bar \chi)\;, 
 \eea
  \bea
 &&\M_{\ell m \ellvp \mvp \ellvq \mvq \barell \barm}^{  \int[ \Phi \Delta_\Omega \Phi](2)}(k; p, q) =-2 \delta^K_{\barell 0} \delta^K_{\barm 0}(4\pi)^3 \aleph^*_\ell(k)  (-1)^m i^{\ellvp+\ellvq} \left[ \ellvq(\ellvq+1) +\ellvp(\ellvp+1) \right] \G^{\ell \ellvp \ellvq }_{-m  \mvp \mvq}
~~~~~~~~~~~~~~~~
\nonumber\\ 
&&\times ~  \int \ud \bar \chi~\bar \chi  \W (\bar \chi) ~ (1-\Q) \int_0^{\bar \chi} \frac{\ud \tilde \chi}{\tilde \chi }\left( \bar \chi - \tilde \chi\right)  \T^{\Phi}(\bp, \tilde \eta) \T^{\Phi}(\bq, \tilde \eta)  j_{\ellvp} (p \tilde \chi) j_{\ellvq} (q \tilde \chi)  j_\ell (k  \bar \chi) \;,
 \eea
   \bea
 &&\M_{\ell m \ellvp \mvp \ellvq \mvq \barell \barm}^{\int [\triangle_\Omega \Phi T](2)}(k; p, q) =2 \delta^K_{\barell 0} \delta^K_{\barm 0}(4\pi)^3 \aleph^*_\ell(k)  (-1)^m i^{\ellvp+\ellvq}  \G^{\ell \ellvp \ellvq }_{-m  \mvp \mvq}  \int \ud \bar \chi~\bar \chi  \W (\bar \chi)(1-\Q) \int_0^{\bar \chi} \frac{\ud \tilde \chi}{ \tilde \chi}\left( \bar \chi - \tilde \chi\right)\int_0^{\tilde \chi} \ud \tilde{\tilde \chi}
\nonumber\\ 
&&\times   \left[ \ellvp(\ellvp+1)  \T^{\Phi}(\bp, \tilde \eta) j_{\ellvp} (p \tilde \chi)
 \T^{\Phi}\left(\bq, \tilde{\tilde \eta}\right)    j_{\ellvq} \left(q \tilde{\tilde \chi}\right) + \ellvq(\ellvq+1)   \T^{\Phi}(\bq, \tilde \eta) j_{\ellvq} (q \tilde \chi)  \T^{\Phi} \left(\bp, \tilde{\tilde \eta}\right) j_{\ellvp} \left(p \tilde{ \tilde \chi}\right) \right]  j_\ell (k  \bar \chi) \;, 
 \eea
  \bea
 &&\M_{\ell m \ellvp \mvp \ellvq \mvq \barell \barm}^{\int[\p_\perp \Phi S_\perp] (2)}(k; p, q) = \delta^K_{\barell 0} \delta^K_{\barm 0}(4\pi)^3\aleph^*_\ell(k)   i^{\ellvp+\ellvq} ~ I_{\ell \ellvp \ellvp}^{m \mvp \mvq}      \int \ud \bar \chi~\bar \chi  \W (\bar \chi)  \left(1 - \Q\right)   \ \int_0^{\bar \chi} \frac{\ud \tilde \chi}{ \tilde \chi} \left( \bar \chi - \tilde \chi\right)  \int_0^{\tilde \chi} \frac{\ud \tilde{\tilde \chi}}{ \tilde{\tilde \chi}}~~~~~~~~~~\nonumber\\ 
&&\times \left[ \T^{\Phi}(\bp, \tilde \eta)  j_{\ellvp} (p \tilde \chi)     \T^{\Phi}\left(\bq, \tilde{\tilde \eta}\right) j_{\ellvq} \left(q \tilde{\tilde \chi}\right)+  \T^{\Phi}(\bq, \tilde \eta)   j_{\ellvq} (q \tilde \chi)     \T^{\Phi}\left(\bp, \tilde {\tilde \eta}\right) j_{\ellvp} \left(p \tilde{ \tilde \chi}\right)  \right]   j_\ell (k  \bar \chi) \;,
 \eea
   \bea
 &&\M_{\ell m \ellvp \mvp \ellvq \mvq \barell \barm}^{\int[ \p_{\perp} \tilde \nabla^2_{\perp} \Phi  S_\perp ] (2)}(k; p, q) = -2\delta^K_{\barell 0} \delta^K_{\barm 0}(4\pi)^3\aleph^*_\ell(k)   i^{\ellvp+\ellvq} ~ I_{\ell \ellvp \ellvp}^{m \mvp \mvq}      \int \ud \bar \chi~\bar \chi  \W (\bar \chi)  \left(1 - \Q\right)   \ \int_0^{\bar \chi} \frac{\ud \tilde \chi}{ \tilde \chi} \left( \bar \chi - \tilde \chi\right)  \int_0^{\tilde \chi} \frac{\ud \tilde{\tilde \chi}}{ \tilde{\tilde \chi}}
~~~~~~~~~\nonumber\\ 
&&\times   \left[ \ellvp(\ellvp+1)  \T^{\Phi}(\bp, \tilde \eta) j_{\ellvp} (p \tilde \chi)
 \T^{\Phi}\left(\bq, \tilde{\tilde \eta}\right)    j_{\ellvq} \left(q \tilde{\tilde \chi}\right) + \ellvq(\ellvq+1)   \T^{\Phi}(\bq, \tilde \eta) j_{\ellvq} (q \tilde \chi)  \T^{\Phi} \left(\bp, \tilde{\tilde \eta}\right) j_{\ellvp} \left(p \tilde{ \tilde \chi}\right) \right]  j_\ell (k  \bar \chi) \;,
 \eea
   \bea
 &&\M_{\ell m \ellvp \mvp \ellvq \mvq \barell \barm}^{\int[ \p_{\perp} \tilde \nabla^2_{\perp} \Phi  \p_\perp T ] (2)}(k; p, q) = 2\delta^K_{\barell 0} \delta^K_{\barm 0}(4\pi)^3\aleph^*_\ell(k)   i^{\ellvp+\ellvq} ~ I_{\ell \ellvp \ellvp}^{m \mvp \mvq}      \int \ud \bar \chi~\bar \chi  \W (\bar \chi)  \left(1 - \Q\right)   \ \int_0^{\bar \chi} \frac{\ud \tilde \chi}{ \tilde \chi^2} \left( \bar \chi - \tilde \chi\right)  \int_0^{\tilde \chi}\ud \tilde{\tilde \chi}
~~~~~~~~~\nonumber\\ 
&&\times   \left[ \ellvp(\ellvp+1)  \T^{\Phi}(\bp, \tilde \eta) j_{\ellvp} (p \tilde \chi)
 \T^{\Phi}\left(\bq, \tilde{\tilde \eta}\right)    j_{\ellvq} \left(q \tilde{\tilde \chi}\right) + \ellvq(\ellvq+1)   \T^{\Phi}(\bq, \tilde \eta) j_{\ellvq} (q \tilde \chi)  \T^{\Phi} \left(\bp, \tilde{\tilde \eta}\right) j_{\ellvp} \left(p \tilde{ \tilde \chi}\right) \right]  j_\ell (k  \bar \chi) \;,
 \eea 
    \bea
 &&\M_{\ell m \ellvp \mvp \ellvq \mvq \barell \barm}^{\int [ \triangle_\Omega \Phi ~ \kappa](2)}(k; p, q) =-2 \delta^K_{\barell 0} \delta^K_{\barm 0}(4\pi)^3 \aleph^*_\ell(k)  (-1)^m i^{\ellvp+\ellvq}  \ellvp(\ellvp+1) \ellvq(\ellvq+1)  \G^{\ell \ellvp \ellvq }_{-m  \mvp \mvq}  \int \ud \bar \chi~\bar \chi  \W (\bar \chi) \left(1 - \Q\right) \nonumber \\
 &&\times ~\int_0^{\bar \chi} \frac{ \ud \tilde \chi }{\tilde \chi^2 } \left( \bar \chi - \tilde \chi\right) \int_0^{\tilde \chi}\frac{ \ud \tilde{\tilde \chi} }{\tilde{\tilde \chi}}(\tilde \chi -  \tilde{\tilde \chi } ) 
 \bigg[  \T^{\Phi}(\bp, \tilde \eta) j_{\ellvp} (p \tilde \chi)  \T^{\Phi}\left(\bq, \tilde{\tilde \eta}\right) j_{\ellvq} \left(q \tilde{\tilde \chi}\right)   +  \T^{\Phi}(\bq, \tilde \eta) j_{\ellvq} (q \tilde \chi)  \T^{\Phi} \left(\bp, \tilde {\tilde \eta}\right) j_{\ellvp} \left(p \tilde{ \tilde \chi}\right) \bigg] j_\ell (k  \bar \chi) \;, \nonumber\\
 \eea
    \bea
 &&\M_{\ell m \ellvp \mvp \ellvq \mvq \barell \barm}^{\int[  \Phi' \p_\perp S_\perp](2)}(k; p, q) = - 2 \delta^K_{\barell 0} \delta^K_{\barm 0}(4\pi)^3 \aleph^*_\ell(k)  (-1)^m i^{\ellvp+\ellvq}  \G^{\ell \ellvp \ellvq }_{-m  \mvp \mvq}  \int \ud \bar \chi~\bar \chi  \W (\bar \chi) \left(1 - \Q\right) \int_0^{\bar \chi}  \ud \tilde \chi \left( \bar \chi - \tilde \chi\right)  \int_0^{\tilde \chi}
 \frac{  \ud \tilde{\tilde \chi} }{ \tilde{\tilde \chi} } 
 \nonumber \\
 &&\times  
 \Bigg\{  \ellvq(\ellvq+1) \left[\frac{\p}{\p \tilde \eta}\T^{\Phi}(\bp, \tilde \eta) \right] j_{\ellvp} (p \tilde \chi) \T^{\Phi}\left(\bq, \tilde{\tilde \eta}\right) j_{\ellvq} \left(q \tilde{\tilde \chi}\right) + \ellvp(\ellvp+1) \left[\frac{\p}{\p \tilde \eta} \T^{\Phi}(\bq, \tilde \eta) \right]    j_{\ellvq} (q \tilde \chi) \T^{\Phi} \left(\bp, \tilde {\tilde \eta}\right) j_{\ellvp} \left(p \tilde{ \tilde \chi}\right) \Bigg\}  j_\ell (k  \bar \chi) , \nonumber\\
 \eea
     \bea
 &&\M_{\ell m \ellvp \mvp \ellvq \mvq \barell \barm}^{\int[  \Phi'  \triangle_\Omega T](2)}(k; p, q) =  2 \delta^K_{\barell 0} \delta^K_{\barm 0}(4\pi)^3 \aleph^*_\ell(k)  (-1)^m i^{\ellvp+\ellvq}  \G^{\ell \ellvp \ellvq }_{-m  \mvp \mvq}  \int \ud \bar \chi~\bar \chi  \W (\bar \chi) \left(1 - \Q\right)  \int_0^{\bar \chi}  \frac{ \ud \tilde \chi }{\tilde \chi } \left( \bar \chi - \tilde \chi\right)  \int_0^{\tilde \chi} \ud \tilde{\tilde \chi}
 \nonumber \\
 &&\times  
 \Bigg\{  \ellvq(\ellvq+1) \left[\frac{\p}{\p \tilde \eta}\T^{\Phi}(\bp, \tilde \eta) \right] j_{\ellvp} (p \tilde \chi) \T^{\Phi}\left(\bq, \tilde{\tilde \eta}\right) j_{\ellvq} \left(q \tilde{\tilde \chi}\right) + \ellvp(\ellvp+1) \left[\frac{\p}{\p \tilde \eta} \T^{\Phi}(\bq, \tilde \eta) \right]    j_{\ellvq} (q \tilde \chi) \T^{\Phi} \left(\bp, \tilde {\tilde \eta}\right) j_{\ellvp} \left(p \tilde{ \tilde \chi}\right) \Bigg\}  j_\ell (k  \bar \chi) , \nonumber\\
 \eea
      \bea\label{phinablaT}
 &&\M_{\ell m \ellvp \mvp \ellvq \mvq \barell \barm}^{\int[  \Phi  \triangle_\Omega T](2)}(k; p, q) =  2 \delta^K_{\barell 0} \delta^K_{\barm 0}(4\pi)^3 \aleph^*_\ell(k)  (-1)^m i^{\ellvp+\ellvq}  \G^{\ell \ellvp \ellvq }_{-m  \mvp \mvq}  \int \ud \bar \chi~\bar \chi  \W (\bar \chi) \left(1 - \Q\right)  \int_0^{\bar \chi}  \frac{ \ud \tilde \chi }{\tilde \chi^2 } \left( \bar \chi - \tilde \chi\right)  \int_0^{\tilde \chi} \ud \tilde{\tilde \chi}
 \nonumber \\
 &&\times  
\bigg[  \ellvq(\ellvq+1) \T^{\Phi}(\bp, \tilde \eta) j_{\ellvp} (p \tilde \chi)  \T^{\Phi}\left(\bq, \tilde{\tilde \eta}\right) j_{\ellvq} \left(q \tilde{\tilde \chi}\right)   + \ellvp(\ellvp+1)  \T^{\Phi}(\bq, \tilde \eta) j_{\ellvq} (q \tilde \chi)  \T^{\Phi} \left(\bp, \tilde {\tilde \eta}\right) j_{\ellvp} \left(p \tilde{ \tilde \chi}\right) \bigg] j_\ell (k  \bar \chi) \;.
 \eea 

\subsection{Terms from $ \Delta_{g\, {\rm int-5}}^{(2)}$  [see Eq.\ (\ref{Deltag2-int-5})]}

We immediately see that $ \Delta_{g\, {\rm int-5}}^{(2)}$ contains all symmetric trace-free terms with orthogonal partial derivatives. In order to compute correctly all these relations, it is useful to rewrite $ \bar \p^{i}_{\perp} \bar \p^{j}_{\perp}$ with covariant derivates ${}^{(2)}\nabla_i$ and, then, the usual spin-raising and spin-lowering operators (for more details, see Appendix \ref{spin}). In particular, we have
\be
 \bar \p_{\perp (i} \bar \p_{\perp j)} \Phi =\frac{1}{\bar \chi^2} \left(- \hat n_{(i} {}^{(2)}\nabla_{j)} \Phi +{}^{(2)}\nabla_{(i} {}^{(2)}\nabla_{j)} \Phi \right) \;,
\ee
where
\be
m_{\pm}^i m_{\pm}^j \bar \p_{\perp (i} \bar \p_{\perp j)} \Phi = \frac{1}{\bar \chi^2} m_{\pm}^i m_{\pm}^j  {}^{(2)} \nabla_{(i} {}^{(2)}\nabla_{j)}\Phi
\ee
and using $\Perp_{ij} = 2 m_{+ (i} m_{-j)}$, we find
\be
\label{shear-local}
{}^{(2)}\nabla_{(i} {}^{(2)}\nabla_{j)}  \Phi - \frac{1}{2} \Perp_{ij} {}^{(2)}\nabla^2 \Phi = {}_2\Phi~ m_{+ i} m_{+ j}  +  {}_{-2} \Phi~ m_{- i} m_{- j} \;,
\ee
where
\be
{}_2\Phi =m_{- }^i m_{- }^j {}^{(2)}\nabla_{i} {}^{(2)}\nabla_{j}  \Phi =\frac{1}{2} ~\eth^2\Phi  \quad \quad {\rm and}  \quad \quad {}_{-2} \Phi = m_{+ }^i m_{+ }^j {}^{(2)}\nabla_{i} {}^{(2)}\nabla_{j}  \Phi =\frac{1}{2}~  \bar \eth^2\Phi \;.
\ee
Using
\be
{}_1\Phi =m_{- }^i  {}^{(2)}\nabla_{i} \Phi =- \sqrt{\frac{1}{2}} ~\eth \Phi   \quad \quad {\rm and}  \quad \quad {}_{-1} \Phi=m_{+}^i  {}^{(2)}\nabla_{i} \Phi =- \sqrt{\frac{1}{2}}~  \bar \eth \Phi \;,
\ee
then 
\be
\label{local-shear-3D}
\bar \p_{\perp (i} \bar \p_{\perp j)} \Phi - \frac{1}{2} \Perp_{ij} \nabla_{\perp}^2\Phi = \frac{1}{\bar \chi^2} \left[ \sqrt{\frac{1}{2}} ~
\hat n_{(i} \left(m_{+j)}  \eth \Phi +m_{-j)}  \bar \eth\Phi  \right) + \frac{1}{2} \left( \eth^2\Phi ~ m_{+ i} m_{+ j}  +  \bar \eth^2\Phi~ m_{- i} m_{- j} \right)\right]\;.
\ee

Now taking into account Eq.\ (\ref{spin1Ylm}) and
\be
\label{spin2Ylm}
\eth^2  Y_{\ell m}(\hat \bn)=  \sqrt{\frac{(\ell +2)!}{(\ell - 2)!}} ~{}_2Y_{\ell m}(\hat \bn)  \quad \quad {\rm and}  \quad \quad \bar \eth^2  Y_{\ell m}(\hat \bn)=  \sqrt{\frac{(\ell +2)!}{(\ell - 2)!}}~ {}_{-2} Y_{\ell m}(\hat \bn)\;,
\ee
Eq.\ (\ref{shear-local}) turns out
\be
 {}_2\Phi~ m_{+ i} m_{+ j}  +  {}_{-2} \Phi~ m_{- i} m_{- j} 
  = \sum_{\ell m} (2\pi) i^\ell \sqrt{\frac{(\ell +2)!}{(\ell - 2)!}} \bigg( {}_2Y_{\ell m}(\hat \bn)~ m_{+ i} m_{+ j}  +   {}_{-2} Y_{\ell m}(\hat \bn)~ m_{- i} m_{- j} \bigg) \int \frac{\ud^3\bk}{(2\pi)^3}   j_\ell (k \bar \chi)  Y_{\ell m}^*(\hat \bk) \Phi(\bk)\;.
\ee
Using these results we can focus on the following terms: {\it (i)}
\be
 \Delta^{\gamma^2(2)}=-2  \left(1- \Q \right)\big|\gamma^{(1)}\big|^2=-  \left(1- \Q \right) \gamma_{ij}^{(1)} \gamma^{ij(1)}
\ee
and, {\it (ii)}  a shear term  related to  post-Born term perturbations
\be
 \Delta^{\gamma{\rm Post-Born} (2)}= 8\left(1- \Q \right)  \int_0^{\bar \chi} \ud \tilde \chi (\bar \chi- \tilde \chi) \frac{\tilde \chi}{\bar \chi} \left( \tilde \p_{\perp (j} \tilde \p_{\perp m)} \Phi - \frac{1}{2} \Perp_{jm}  \tilde \nabla^2_{\perp} \Phi\right) \times \Bigg[\left( \tilde \p^{(j}_{\perp} \tilde \p^{m)}_{\perp} - \frac{1}{2} \Perp^{jm}  \tilde \nabla^2_{\perp} \right)\int_0^{\tilde \chi} \ud \tilde{\tilde \chi}  
  ( \tilde \chi -  \tilde{\tilde \chi}) \frac{\tilde \chi}{ \tilde{\tilde \chi}}\Phi\Bigg] \;.
\ee

For {\it (i)} and using Eqs.\ (\ref{gamp}) and (\ref{local-shear-3D}) we find
\bea
\gamma_{ij}^{(1)} &=&   2\int_0^{\bar \chi} \ud \tilde \chi  \frac{(\bar \chi- \tilde \chi)}{\bar \chi \tilde \chi}  \left[ \sqrt{\frac{1}{2}} ~
\hat n_{(i} \left(m_{+j)}  \eth \Phi +m_{-j)}  \bar \eth\Phi  \right) + \frac{1}{2} \left( \eth^2\Phi ~ m_{+ i} m_{+ j}  +  \bar \eth^2\Phi~ m_{- i} m_{- j} \right)\right] \nonumber \\
&=&   -2 n_{(i} \left[ \int_0^{\bar \chi} \ud \tilde \chi  \frac{(\bar \chi- \tilde \chi)}{\bar \chi \tilde \chi} 
 \left(m_{+j)}~ {}_1\Phi +m_{-j)}~ {}_{-1}\Phi  \right) \right]
+  \left({}_2\gamma^{(1)} ~ m_{+ i} m_{+ j}  + {}_{-2}\gamma^{(1)}~ m_{- i} m_{- j} \right) \nonumber \\
&=&   2 n_{(i}  \left(m_{+j)}~ {}_1\gamma^{(1)} +m_{-j)}~ {}_{-1}\gamma^{(1)}  \right) 
+  \left({}_2\gamma^{(1)} ~ m_{+ i} m_{+ j}  + {}_{-2}\gamma^{(1)}~ m_{- i} m_{- j} \right)\;,
\eea
where
\bea
{}_1\gamma^{(1)} &=&-\int_0^{\bar \chi} \ud \tilde \chi \left[ \frac{\left(\bar \chi-\tilde \chi\right)}{ \bar \chi \tilde \chi} ~ {}_1\Phi \right] =  \sqrt{\frac{1}{2}} \int_0^{\bar \chi} \ud \tilde \chi \left[ \frac{\left(\bar \chi-\tilde \chi\right)}{ \bar \chi \tilde \chi} \eth \Phi \right] \nonumber\\
&=&  \sum_{\ell m} (4\pi) i^\ell \sqrt{\frac{\ell(\ell +1)}{2}} ~ {}_1Y_{\ell m}(\hat \bn)    \int \frac{\ud^3\bk}{(2\pi)^3}     Y_{\ell m}^*(\hat \bk) \Phi_{\rm p}   (\bk)   \int_0^{\bar \chi} \ud \tilde \chi \frac{\left(\bar \chi-\tilde \chi\right)}{ \bar \chi \tilde \chi}   j_\ell (k \tilde \chi) \T^{\Phi}(\bk, \tilde \eta)\;, \\
{}_{-1}\gamma^{(1)} &=&-\int_0^{\bar \chi} \ud \tilde \chi \left[ \frac{\left(\bar \chi-\tilde \chi\right)}{ \bar \chi \tilde \chi} ~ {}_{-1}\Phi \right] =  \sqrt{\frac{1}{2}} \int_0^{\bar \chi} \ud \tilde \chi \left[ \frac{\left(\bar \chi-\tilde \chi\right)}{ \bar \chi \tilde \chi} \bar \eth \Phi \right] \nonumber\\
&=& -\sum_{\ell m} (4\pi) i^\ell \sqrt{\frac{\ell(\ell +1)}{2}} ~ {}_{-1}Y_{\ell m}(\hat \bn)    \int \frac{\ud^3\bk}{(2\pi)^3}     Y_{\ell m}^*(\hat \bk) \Phi_{\rm p}   (\bk)   \int_0^{\bar \chi} \ud \tilde \chi \frac{\left(\bar \chi-\tilde \chi\right)}{ \bar \chi \tilde \chi}   j_\ell (k \tilde \chi) \T^{\Phi}(\bk, \tilde \eta)\;, \\
{}_2\gamma^{(1)} &=& m_{- }^i m_{- }^j \gamma_{ij}^{(1)} = 2 \int_0^{\bar \chi} \ud \tilde \chi \left[ \frac{\left(\bar \chi-\tilde \chi\right)}{ \bar \chi \tilde \chi}  {}_2\Phi \right]= \int_0^{\bar \chi} \ud \tilde \chi \left[ \frac{\left(\bar \chi-\tilde \chi\right)}{ \bar \chi \tilde \chi}  \eth^2\Phi \right] 
 \nonumber\\
&=& \sum_{\ell m} (4\pi) i^\ell \sqrt{\frac{(\ell +2)!}{(\ell - 2)!}} ~ {}_2Y_{\ell m}(\hat \bn)    \int \frac{\ud^3\bk}{(2\pi)^3}     Y_{\ell m}^*(\hat \bk) \Phi_{\rm p}   (\bk)   \int_0^{\bar \chi} \ud \tilde \chi \frac{\left(\bar \chi-\tilde \chi\right)}{ \bar \chi \tilde \chi}   j_\ell (k \tilde \chi) \T^{\Phi}(\bk, \tilde \eta)\;, \\
{}_{-2}\gamma^{(1)} &=& m_{+ }^i m_{+ }^j \gamma_{ij}^{(1)}= 2 \int_0^{\bar \chi} \ud \tilde \chi \left[ \frac{\left(\bar \chi-\tilde \chi\right)}{ \bar \chi \tilde \chi}  {}_2\Phi \right]= \int_0^{\bar \chi} \ud \tilde \chi \left[ \frac{\left(\bar \chi-\tilde \chi\right)}{ \bar \chi \tilde \chi} \bar \eth^2\Phi \right]
 \nonumber\\
&=& \sum_{\ell m} (4\pi) i^\ell \sqrt{\frac{(\ell +2)!}{(\ell - 2)!}} ~  {}_{-2} Y_{\ell m}(\hat \bn)  \int \frac{\ud^3\bk}{(2\pi)^3}     Y_{\ell m}^*(\hat \bk) \Phi_{\rm p}   (\bk)  \int_0^{\bar \chi} \ud \tilde \chi \frac{\left(\bar \chi-\tilde \chi\right)}{ \bar \chi \tilde \chi}   j_\ell (k \tilde \chi) \T^{\Phi}(\bk, \tilde \eta) \;,
\eea
and
\be
\label{Deltagamma2}
\frac{1}{2} \Delta^{\gamma^2(2)}=- \left(1- \Q \right)\left[{}_1\gamma^{(1)} {}_{-1}\gamma^{(1)} + {}_{-1}\gamma^{(1)} {}_1 \gamma^{(1)} \right]
 - \frac{\left(1- \Q \right)}{2} \left[{}_2\gamma^{(1)} {}_{-2}\gamma^{(1)} + {}_{-2}\gamma^{(1)} {}_2 \gamma^{(1)} \right]\;.
\ee
Therefore Eq. (\ref{Deltagamma2}) reads
\bea
&&{1 \over 2} \Delta^{\gamma^2(2)} (\bx)  =
 \int \frac{\ud^3 \bp}{(2\pi)^3} \frac{\ud^3 \bq}{(2\pi)^3}  \Bigg\{    \left(1- \Q \right)   \sum_{\ellvp \mvp \ellvq \mvq} (4 \pi)^2 ~ i^{\ellvp+\ellvq} 
 \frac{\sqrt{ \ellvp(\ellvp+1) \ellvq(\ellvq+1)}}{2 }\nonumber\\
&&\times  \left[{}_{1}{Y_{\ellvp \mvp}}(\hat \bn) ~{}_{-1}{Y_{\ellvq \mvq}}(\hat \bn)+  {}_{-1}{Y_{\ellvp \mvp}}(\hat \bn) ~{}_{+1}{Y_{\ellvq \mvq}}(\hat \bn) \right] 
 \int_0^{\bar \chi} \ud  \chi_\bp  \int_0^{\bar \chi} \ud  \chi_\bq   ~ \frac{\left(\bar \chi-\chi_\bp \right)}{ \bar \chi \chi_\bp} \frac{\left(\bar \chi-\chi_\bq \right)}{ \bar \chi\chi_\bq}  ~ \T^{\Phi}(\bp, \eta_{\bp})  \T^{\Phi}(\bq, \eta_{\bq})   \nonumber\\
&& \times ~  j_{\ellvp} (p  \chi_{\bp}) \, j_{\ellvq} (q  \chi_{\bq}) \Bigg\} ~ Y_{\ellvp \mvp}^*(\hat \bp)Y_{\ellvq \mvq}^*(\hat \bq) \Phi_{\rm p}(\bp) \Phi_{\rm p}(\bq) \nonumber\\
&&+  \int \frac{\ud^3 \bp}{(2\pi)^3} \frac{\ud^3 \bq}{(2\pi)^3}  \Bigg\{  - \frac{\left(1- \Q \right)}{2}   \sum_{\ellvp \mvp \ellvq \mvq} (4 \pi)^2 ~ i^{\ellvp+\ellvq}  \sqrt{\frac{(\ellvp +2)!}{(\ellvp - 2)!}}  \sqrt{\frac{(\ellvq +2)!}{(\ellvq - 2)!}}      \nonumber\\
&&\times  \left[{}_{2}{Y_{\ellvp \mvp}}(\hat \bn) ~{}_{-2}{Y_{\ellvq \mvq}}(\hat \bn)+  {}_{-2}{Y_{\ellvp \mvp}}(\hat \bn) ~{}_{+2}{Y_{\ellvq \mvq}}(\hat \bn) \right] \int_0^{\bar \chi} \ud  \chi_\bp  \int_0^{\bar \chi} \ud  \chi_\bq   ~ \frac{\left(\bar \chi-\chi_\bp \right)}{ \bar \chi \chi_\bp} \frac{\left(\bar \chi-\chi_\bq \right)}{ \bar \chi\chi_\bq}  ~ \T^{\Phi}(\bp, \eta_{\bp})  \T^{\Phi}(\bq, \eta_{\bq})   \nonumber\\
&& \times ~  j_{\ellvp} (p  \chi_{\bp}) \, j_{\ellvq} (q  \chi_{\bq}) \Bigg\} ~ Y_{\ellvp \mvp}^*(\hat \bp)Y_{\ellvq \mvq}^*(\hat \bq) \Phi_{\rm p}(\bp) \Phi_{\rm p}(\bq)\;.
\eea

Using Eqs.\  (\ref{s_1s_2-0}) and (\ref{s_1s_2s_3}) we have
\be
\left[{}_{2}{Y_{\ellvp \mvp}}(\hat \bn) ~{}_{-2}{Y_{\ellvq \mvq}}(\hat \bn)+  {}_{-2}{Y_{\ellvp \mvp}}(\hat \bn) ~{}_{+2}{Y_{\ellvq \mvq}}(\hat \bn) \right]=\sum_{\tell \tm } (-1)^{\tm}~ Y_{\tell -\tm}(\hat \bn) \left(\I_{\ellvp \ellvq \tell}^{-2 2 0}  + \I_{\ellvp \ellvq \tell}^{2-20}  \right)\left(
\begin{array}{ccc}
\ellvp & \ellvq & \tell \\
\mvp & \mvq & \tm
\end{array}
\right)\;.
\ee
Noting that
\[\I_{\ellvp \ellvq \barell}^{-2 2 0}  + \I_{\ellvp \ellvq \tell}^{2-20} =\big[1+ (-1)^{\ellvp+ \ellvq +\tell} \big] \I_{\ellvp \ellvq \tell}^{2 -2 0}= \left\{ \begin{array}{ll}
2\,\I_{\ellvp \ellvq \tell}^{2 -2 0} ~~~~~~~\ellvp+ \ellvq +\tell ~~{\rm even} \\ \\
0 ~~~~~~~~~~~~~~~ \ellvp+ \ellvq +\tell ~~{\rm odd}\;,
\end{array} \right.  \]
and taking into account Eq.\ (\ref{I}), we can easily project $\Delta^{\gamma^2(2)}(\bx)$ on $|k\ell m\rangle$ space. Indeed
\bea
\frac{1}{2}\Delta_{\ell m}^{\gamma^2(2)}(k)&=&\sum_{\ellvp \mvp \ellvq \mvq} \int \frac{\ud^3 \bp}{(2\pi)^3} \frac{\ud^3 \bq}{(2\pi)^3} \Bigg\{  (4\pi)^2 \aleph^*_\ell(k) ~  i^{\ellvp+\ellvq} ~\Bigg[  I^{m \mvp \mvq}_{\ell \ellvp \ellvq}  + (-1)^m  \G^{\ell \ellvp \ellvq }_{-m  \mvp \mvq} ~ \left(
\begin{array}{ccc}
\ellvp & \ellvq & \ell \\
2 & -2 & 0
\end{array}
\right) \nonumber\\ 
&\times&
\left(
\begin{array}{ccc}
\ellvp & \ellvq & \ell \\
0 & 0 & 0
\end{array}
\right)^{-1}     \sqrt{\frac{(\ellvp +2)!}{(\ellvp - 2)!}}  \sqrt{\frac{(\ellvq +2)!}{(\ellvq - 2)!}}   \Bigg]
    \int \ud \bar \chi~  \W (\bar \chi) \left[ -  \left(1- \Q \right)   \right] \int_0^{\bar \chi}\frac{ \ud  \chi_\bp}{  \chi_\bp}  \int_0^{\bar \chi} \frac{\ud  \chi_\bq}{\chi_\bq}   ~ \left(\bar \chi-\chi_\bp \right)\left(\bar \chi-\chi_\bq \right)   \nonumber\\
& \times & ~ \T^{\Phi}(\bp, \eta_{\bp})  \T^{\Phi}(\bq, \eta_{\bq}) ~ j_{\ellvp} (p  \chi_{\bp}) \, j_{\ellvq} (q  \chi_{\bq}) \, j_\ell (k \bar \chi)\Bigg\} ~ Y_{\ellvp \mvp}^*(\hat \bp)Y_{\ellvq \mvq}^*(\hat \bq) \Phi_{\rm p}(\bp) \Phi_{\rm p}(\bq)\;.
\eea
and, finally, we find
 \bea
 \label{M-Deltag2-gamma-sq}
 \M_{\ell m \ellvp \mvp \ellvq \mvq \barell \barm}^{\gamma^2(2)}(k; p, q) &=&  \delta^K_{\barell 0} \delta^K_{\barm 0}(4\pi)^3  \aleph^*_\ell(k) ~  i^{\ellvp+\ellvq} ~\Bigg[  I^{m \mvp \mvq}_{\ell \ellvp \ellvq}  + (-1)^m  \G^{\ell \ellvp \ellvq }_{-m  \mvp \mvq} ~ \left(
\begin{array}{ccc}
\ellvp & \ellvq & \ell \\
2 & -2 & 0
\end{array}
\right) 
\left(
\begin{array}{ccc}
\ellvp & \ellvq & \ell \\
0 & 0 & 0
\end{array}
\right)^{-1}
\nonumber\\ 
&\times& \left.  \sqrt{\frac{(\ellvp +2)!}{(\ellvp - 2)!}}  \sqrt{\frac{(\ellvq +2)!}{(\ellvq - 2)!}}     \right]  \int \ud \bar \chi~  \W (\bar \chi) \left[ -  \left(1- \Q \right)   \right] \int_0^{\bar \chi}\frac{ \ud  \chi_\bp}{  \chi_\bp}  \int_0^{\bar \chi} \frac{\ud  \chi_\bq}{\chi_\bq}   ~ \left(\bar \chi-\chi_\bp \right)\left(\bar \chi-\chi_\bq \right)   \nonumber\\
& \times & ~ \T^{\Phi}(\bp, \eta_{\bp})  \T^{\Phi}(\bq, \eta_{\bq}) ~ j_{\ellvp} (p  \chi_{\bp}) \, j_{\ellvq} (q  \chi_{\bq}) \, j_\ell (k \bar \chi) \;.
\eea
\
Here we can immediately see that the first additive term within square brackets cancels 
Eq.(\ref{M-theta^2}).

For case {\it (ii)}, using the relations written above we can quickly obtain 
\bea
 &&\Delta^{\gamma{\rm Post-Born} (2)}= 8\left(1- \Q \right)  \int_0^{\bar \chi} \ud \tilde \chi \frac{(\bar \chi- \tilde \chi)} {\bar \chi \tilde \chi}\left[ \sqrt{\frac{1}{2}} ~\hat n_{(i} \left(m_{+j)}  \eth \Phi +m_{-j)}  \bar \eth\Phi  \right) + \frac{1}{2} \left( \eth^2\Phi ~ m_{+ i} m_{+ j}  +  \bar \eth^2\Phi~ m_{- i} m_{- j} \right)\right]\nonumber\\
&&  \times  \int_0^{\tilde \chi} \ud \tilde{\tilde \chi}    \frac{ ( \tilde \chi -  \tilde{\tilde \chi})} {\tilde \chi \tilde{\tilde \chi}} \left[ \sqrt{\frac{1}{2}} ~\hat n^{(i} \left(m^{j)}_{+}~  \eth \Phi + m^{j)}_{-}  ~\bar \eth\Phi  \right) + \frac{1}{2} \left( \eth^2\Phi ~ m^i_+ m^j_+  +  \bar \eth^2\Phi~ m^i_- m^j_- \right)\right] \;.
\eea
Projecting this relation on $|k\ell m\rangle$ space, it turns out
   \bea
\frac{1}{2}\Delta_{\ell m}^{\gamma{\rm Post-Born}} (k)&=&\sum_{\ellvp \mvp \ellvq \mvq} \int \frac{\ud^3 \bp}{(2\pi)^3} \frac{\ud^3 \bq}{(2\pi)^3} \Bigg\{ (4\pi)^2 \aleph^*_\ell(k)   i^{\ellvp+\ellvq} ~\Bigg[  I^{m \mvp \mvq}_{\ell \ellvp \ellvq}  + (-1)^m  \G^{\ell \ellvp \ellvq }_{-m  \mvp \mvq} ~ \left(
\begin{array}{ccc}
\ellvp & \ellvq & \ell \\
2 & -2 & 0
\end{array}
\right) \nonumber\\ 
&\times& \left.
\left(
\begin{array}{ccc}
\ellvp & \ellvq & \ell \\
0 & 0 & 0
\end{array}
\right)^{-1}  \sqrt{\frac{(\ellvp +2)!}{(\ellvp - 2)!}}  \sqrt{\frac{(\ellvq +2)!}{(\ellvq - 2)!}}     \right]   \int \ud \bar \chi~\bar \chi  \W (\bar \chi) (1-\Q) \int_0^{\bar \chi} \frac{\ud \tilde \chi}{\tilde \chi^2} (\bar \chi - \tilde \chi) \int_0^{\tilde \chi} \frac{\ud \tilde{\tilde \chi}}{ \tilde{\tilde \chi}} (\tilde \chi - \tilde{\tilde \chi})
\nonumber\\ 
 &&\times  \left[ \T^{\Phi}(\bp, \tilde \eta)  j_{\ellvp} (p \tilde \chi)     \T^{\Phi}\left(\bq, \tilde{\tilde \eta}\right) j_{\ellvq} \left(q \tilde{\tilde \chi}\right)+  \T^{\Phi}(\bq, \tilde \eta)   j_{\ellvq} (q \bar \chi)     \T^{\Phi}\left(\bp, \tilde {\tilde \eta}\right) j_{\ellvp} \left(p \tilde{ \tilde \chi}\right)  \right]   j_\ell (k  \bar \chi) \Bigg\}\nonumber\\
 &&\times ~ Y_{\ellvp \mvp}^*(\hat \bp)  Y_{\ellvq \mvq}^*(\hat \bq)  \Phi_{\rm p}(\bp) \Phi_{\rm p}(\bq)
\eea
and  hence
\bea\label{M-gamma-Post-Born}
 &&\M_{\ell m \ellvp \mvp \ellvq \mvq \barell \barm}^{\gamma{\rm Post-Born}(2)} (k; p, q) = \delta^K_{\barell 0} \delta^K_{\barm 0}(4\pi)^3 \aleph^*_\ell(k)   i^{\ellvp+\ellvq} ~\Bigg[  I^{m \mvp \mvq}_{\ell \ellvp \ellvq}  + (-1)^m  \G^{\ell \ellvp \ellvq }_{-m  \mvp \mvq} ~ \left(
\begin{array}{ccc}
\ellvp & \ellvq & \ell \\
2 & -2 & 0
\end{array}
\right) \nonumber\\ 
&\times& \left.
\left(
\begin{array}{ccc}
\ellvp & \ellvq & \ell \\
0 & 0 & 0
\end{array}
\right)^{-1}  \sqrt{\frac{(\ellvp +2)!}{(\ellvp - 2)!}}  \sqrt{\frac{(\ellvq +2)!}{(\ellvq - 2)!}}     \right]   \int \ud \bar \chi~\bar \chi  \W (\bar \chi) (1-\Q) \int_0^{\bar \chi} \frac{\ud \tilde \chi}{\tilde \chi^2} (\bar \chi - \tilde \chi) \int_0^{\tilde \chi} \frac{\ud \tilde{\tilde \chi}}{ \tilde{\tilde \chi}} (\tilde \chi - \tilde{\tilde \chi})
\nonumber\\ 
 &&\times  \left[ \T^{\Phi}(\bp, \tilde \eta)  j_{\ellvp} (p \tilde \chi)     \T^{\Phi}\left(\bq, \tilde{\tilde \eta}\right) j_{\ellvq} \left(q \tilde{\tilde \chi}\right)+  \T^{\Phi}(\bq, \tilde \eta)   j_{\ellvq} (q \bar \chi)     \T^{\Phi}\left(\bp, \tilde {\tilde \eta}\right) j_{\ellvp} \left(p \tilde{ \tilde \chi}\right)  \right]   j_\ell (k  \bar \chi)  \;. 
 \eea


\section{Bias}
\label{sec:bias}
To correctly incorporate galaxy bias in the expression for the overdensity (which appears e.g., in Eqs.~ (\ref{eq:deltagfirstorder}), (\ref{Deltag1}), (\ref{Deltag-2}), (\ref{Deltag1-esplicito}) etc.,  weÊ treat
the galaxy bias up to second order using the comoving-synchronous (CS) gauge.
We argue that the CS gauge is entirely appropriate to describe the matter overdensity at second order (for details, see \cite{Bertacca:2015mca}).
The bias should be defined in the rest frame of CDM, which is assumed to coincide with the rest frame of galaxies on large scales and can be computed using the peak-background split approach \cite{Press:1973iz}. In CS gauge, the spherical collapse model has an exact GR interpretation \cite{Matarrese:1997ay, Bartolo:2003gh, Wands:2009ex, Bartolo:2010rw}. Indeed, only in this frame we can resort to the peak-background split \cite{Sheth:1999mn, Schmidt:2010gw}  approach, in which, in the  Press-Schecter-inspired prescription \cite{Press:1973iz}, halos of a given mass, M, collapse when the linearly growing local density contrast (smoothed on the corresponding mass scale) reaches a critical value.
This is important for a self-consistent calculation of the so-called non-Gaussian halo bias  and for precision parameter estimates, introduced by nonlinear projection effects.

 In $\Lambda$CDM, the CDM rest frame is defined up to second order in the CS gauge--in which the galaxy and matter overdensities are gauge invariant.
The CS gauge is defined by $g^{\rm CS}_{00}=-1$,    $g^{\rm CS}_{0i}=0$ and $v_{\rm CS}^{i}=0$. At second order,
 \begin{eqnarray} 
 \label{Comoving-Synchronous_metric}
 \ud s^2 = a(\eta)^2\left[-\ud\eta^2+\left(\delta_{ij} +{h^{\rm CS}_{ij}}^{(1)}+\frac{1}{2}{h^{\rm CS}_{ij}}^{(2)}\right)\ud x^i\ud x^j\right],
\end{eqnarray}
where ${h^{\rm CS}_{ij}}^{(n)} =- 2 \psi_{\rm CS}^{(n)} \delta_{ij} + (\p_i\p_j- \delta_{ij}\nabla^2/3)\chi_{\rm CS}^{(n)}+\p_i {\chi^{\rm CS}_j}^{(n)}+ \p_j {\chi^{\rm CS}_i}+{\chi^{\rm CS}_{ij}}^{(n)}$, $\p_i  \chi_{\rm CS}^{i(n)}=\p_i \chi_{\rm CS}^{ij(n)}=0$ (where, for simplicity, we have removed the purely transverse space-dependent constant  $C^{\perp}_i$ in  $g_{0i}$; see  \cite{Rampf:2016wom}). 

Before proceeding into the main part of this section, some comments are in order.

 {\it (i)} 
The primordial non-Gaussianity (NG) has to be considered in all
second-order contributions. NG is set at primordial inflationary epochs on large scales. At later times
cosmological perturbations reenter the Hubble radius during the radiation or during the matter epoch. 
 Beyond linear order, integrated (or projection) effects couple  large and small scales. Thus the NG  large-scale information leaks into smaller scales. Therefore
 a complete general relativistic computation is required  to evaluate all the  observable imprint of PNG in the LSS.

{\it (ii)} The long-mode curvature perturbation modulates the matter overdensity, with an effective $f_{\rm NL} ^{\rm GR}=-5/3$ \cite{Bartolo:2015qva}, but this effect cancels out in the halo overdensity, when perturbations are evaluated at a fixed local scale $\tilde \bx$, rather than fixed global scale $\bx$.
More in detail, the small-scale density at a fixed {\it local  physical} scale is independent of the long-wavelength perturbation. Thus the long-wavelength mode has no effect on the small-scale variance of the density field smoothed on a fixed mass scale.
In other words, the long-wavelength perturbations are not observable locally if the distribution of the primordial metric perturbation\footnote{Here the comoving curvature perturbation $\zeta$ has the same definition and notation as in e.g. Refs.\cite{Baldauf:2011bh, dePutter:2015vga, Bruni:2014xma}. Finally $\zeta$ here is $-\zeta$ in \cite{Mukhanov:2005sc}, and $\zeta =-\R_c$ defined in  \cite{Bruni:2013qta}. } $\zeta$ is a Gaussian random field and we conclude that, within $\delta_{m  {\rm CS}}^{(2)} $, $f_{\rm NL} ^{\rm GR}=-5/3$ is, in the strictly squeezed limit, reabsorbed via a local coordinate transformation (see \cite{Dai:2015jaa, dePutter:2015vga, Bartolo:2015qva} and Appendix \ref{LS}). 
 
Taking into account the above discussion, we define the scale-independent bias in a completely general way at first and second order (considering scales down to the mildly nonlinear regime) as 
\cite{McDonald:2009dh, Baldauf:2012hs, Desjacques:2012eb, Biagetti:2014pha, Tellarini:2015faa, Desjacques:2016bnm}
\bea
\label{bias}
&&\delta_{g {\rm CS}}^{(1)}+\frac{1}{2} \delta_{g  {\rm CS}}^{(2)} =  b_{10}\delta_{m  {\rm CS}}^{(1)} +  b_{\nabla^2\delta} \nabla^2 \delta_{m  {\rm CS}}^{(1)}+ b_{01} \zeta  + \frac{1}{2} b_{10}  \delta_{m  {\rm CS}}^{(2)} + \frac{1}{2} b_{\nabla^2\delta} \nabla^2 \delta_{m  {\rm CS}}^{(2)}+\frac{1}{2} b_{20} \left( \delta_{m  {\rm CS}}^{(1)} \right)^2 + b_{11} \zeta \delta_{m  {\rm CS}}^{(1)}  + \frac{1}{2} b_{02} \zeta^2 \nonumber \\
&& +\frac{1}{2} b_{s^2} s^2  +  \frac{1}{2}  b_{(\nabla^2\delta)^2}  \left(\nabla^2\delta_{m  {\rm CS}}^{(1)}\right)^2  +b_{\delta\nabla^2\delta} \delta_{m  {\rm CS}}^{(1)} \nabla^2\delta_{m  {\rm CS}}^{(1)} +  b_{(\p\delta)^2} \delta^K_{ij} \left(\p_i \delta_{m  {\rm CS}}^{(1)}\right) \left(\p_j \delta_{m  {\rm CS}}^{(1)}\right) \;,
 \eea
 where\footnote{Let us note that if the single-field consistency conditions hold, then for example we have $b_{01} = 0$.} $b_{10},  b_{\nabla^2\delta}, b_{01}, b_{11}, b_{20},  b_{20}, b_{s^2},  b_{(\nabla^2\delta)^2}, b_{\delta\nabla^2\delta}, b_{(\p\delta)^2}$ depend on the conformal time\footnote{
Rigorously $b_{10},  b_{\nabla^2\delta}, b_{01}, b_{11}, b_{20},  b_{20}, b_{s^2},  b_{(\nabla^2\delta)^2}, b_{\delta\nabla^2\delta}, b_{(\p\delta)^2}$ should be also functions of space and time.
Then, in Fourier space,  $\delta_{g  {\rm CS}}(\bk)$ should be written in the following way: 
\bea
\delta_{g  {\rm CS}}^{(1)}(k\; \eta) &=& b_{10}(k\; \eta) \delta_{m  {\rm CS}}^{(1)}(k\; \eta) - b_{\nabla^2\delta}(k\; \eta)  k^2 \delta_{m  {\rm CS}}^{(1)}(k\; \eta) + b_{01}(k\; \eta) \zeta(k)\;,  \\
\delta_{g  {\rm CS}}^{(2)}(\bk\; \eta) &=&   \int \frac{\ud^3 \bp}{(2\pi)^3} \frac{\ud^3 \bq}{(2\pi)^3} ~(2\pi)^3\delta^D \left(\bp+\bq- \bk \right) \bigg[
 b_{10} (\bp, \bq, k ; \eta)  \delta_{m  {\rm CS}}^{(2)} (\bp, \bq, k ; \eta) -  b_{\nabla^2\delta}(\bp, \bq, k ; \eta) k^2  \delta_{m  {\rm CS}}^{(2)} (\bp, \bq, k ; \eta)  \nonumber \\
 &&+ b_{20} (\bp, \bq, k ; \eta) \delta_{m  {\rm CS}}^{(1)}(p ; \eta)    \delta_{m  {\rm CS}}^{(1)} (q ; \eta) +  b_{11} (\bp, \bq, k ; \eta)\left( \zeta( p) \delta_{m  {\rm CS}}^{(1)} (q ; \eta) + \zeta (q) \delta_{m  {\rm CS}}^{(1)} ( p ; \eta)  \right) +  b_{02} (\bp, \bq, k ; \eta) \zeta( p) \zeta( q) \nonumber \\
 && + b_{s^2} (\bp, \bq, k ; \eta) s( \bp; \eta)  s( \bq ; \eta)  + b_{(\nabla^2\delta)^2} (\bp, \bq, k ; \eta) p^2q^2\delta_{m  {\rm CS}}^{(1)}(p ; \eta)    \delta_{m  {\rm CS}}^{(1)} (q ; \eta) - b_{\delta\nabla^2\delta}(\bp, \bq, k ; \eta)   (p^2+q^2)\delta_{m  {\rm CS}}^{(1)}(p ; \eta)    \delta_{m  {\rm CS}}^{(1)} (q ; \eta)\nonumber \\
 &&  - 2 b_{(\p\delta)^2}(\bp, \bq, k ; \eta)  (\bp\cdot\bq) \delta_{m  {\rm CS}}^{(1)}(p ; \eta)    \delta_{m  {\rm CS}}^{(1)} (q ; \eta) \bigg]\;.
%
%
%
%
%
%
%
\eea}, $ \zeta $ is the comoving curvature perturbation and
 \be
 s^2 =  s_{ij} s^{ij}\;, \quad \quad {\rm with} \quad \quad   s_{ij}= \left(\frac{\p_i\p_j}{\nabla^2} - \frac{\delta^K_{ij}}{ 3} \right) \delta_{m  {\rm CS}}^{(1)}\;.
 \ee
 Let us point out that  $\delta_{g  {\rm CS}}^{(2)}$ depends also on primordial non-Gaussianity (see Appendix \ref{LS} for an analysis at $k \le k_{eq}$ where the primordial $f_{\rm NL}$ is written explicitly).
 
 Now we need to connect $\delta_{g {\rm CS}}$ in Eq. (\ref{bias}) with $\delta_g$ defined above in the Poisson gauge.
  From \cite{Bertacca:2014dra, Bertacca:2014wga, Bertacca:2014hwa} we know that
 \begin{eqnarray} 
 \label{dg1}
\delta_{g }&=&\delta_{g {\rm CS}}- b_e \cH v+ 3 \cH v ,\\
 \label{dg2}
\delta_{g }^{(2)} &=&  \delta_{g  {\rm CS}}^{(2)}- b_e \cH v^{(2)}+ 3 \cH v^{(2)} + \left( b_e \cH'-3 \cH' + \cH  b_e'  + b_e^2 \cH^2  -6  b_e  \cH^2 + 9 \cH^2 \right) v^2 + \cH b_e  v  {v}' - 3 \cH   v  {v}' \nonumber\\
&&  -2\cH b_e  v\delta_{g  {\rm CS}}   + 6 \cH  v\delta_{g  {\rm CS}} - 2  v {\delta_{g  {\rm CS}}}' - \frac{1}{2} \p^i \chi \left(- b_e \cH \p_i v+ 3 \cH \p_i v + 2 \p_i \delta_{g  {\rm CS}}\right) - \left(b_e-3\right) \cH  \nabla^{-2}\bigg( v \nabla^2 {v}' - {v}' \nabla^2 v  \nonumber \\
&&- 6 \p_i \Phi \p^i v - 6 \Phi \nabla^2 v  + \frac{1}{2} \p_i \chi^{(1)} \p^i \nabla^2 v + \frac{1}{2} \p_i v \p^i \nabla^2 \chi^{(1)} + \p_i \p_j \chi^{(1)} \p^i \p^j v\bigg).
\end{eqnarray} 
Note the useful relation\footnote{For simplicity we have defined $\chi^{(1)}=\chi_{\rm CS}^{(1)}$.} $v={\chi^{(1)}}'/2$. 
In particular, Eq.(\ref{dg2}) can be expressed as follows
 \begin{eqnarray} 
 \label{dg2-2}
\delta_{g }^{(2)} &=&  \delta_{g  {\rm CS}}^{(2)} - \p^i \chi^{(1)}  \p_i \delta_{g  {\rm CS}}  - (b_e-3) \cH v^{(2)} + \left[ (b_e -3) \cH'  + \cH  b_e'  + (b_e -3)^2 \cH^2 \right] v^2 +  (b_e-3) \cH v  {v}'  - 2  v {\delta_{g  {\rm CS}}}'  \nonumber\\
&&  -2\cH( b_e-3)  v\delta_{g  {\rm CS}}  - \left(b_e-3\right) \cH  \nabla^{-2}\bigg( v \nabla^2 {v}' - {v}' \nabla^2 v - 6 \p_i \Phi \p^i v - 6 \Phi \nabla^2 v \bigg)\;,
\end{eqnarray} 
where we used the relation $\nabla^2 ( \p_i \chi^{(1)} \p^i v)= \p_i \chi^{(1)} \p^i \nabla^2 v+ \p_i v \p^i \nabla^2 \chi^{(1)} +2  \p_i \p_j \chi^{(1)} \p^i \p^j v$. 
 Bearing in mind Eq.(\ref{bias}), Eqs.(\ref{dg1}) and (\ref{dg2-2}) become
\begin{eqnarray} 
  \label{dg1-2}
\delta_{g }&=&  b_{10}\delta_{m  {\rm CS}}^{(1)}+ b_{\nabla^2\delta} \nabla^2 \delta_{m  {\rm CS}}^{(1)} + b_{01} \zeta   - (b_e -3)\cH v ,\\  \label{dg2-3}
\delta_{g }^{(2)} &=&  b_{10}  \delta_{m  {\rm CS}}^{(2)} + b_{\nabla^2\delta} \nabla^2 \delta_{m  {\rm CS}}^{(2)} + b_{20} \left( \delta_{m  {\rm CS}}^{(1)} \right)^2 -  b_{10}  \p^i \chi^{(1)}  \p_i \delta_{m  {\rm CS}}^{(1)}-  b_{\nabla^2\delta}  \p^i \chi^{(1)}  \nabla^2 \p_i \delta_{m  {\rm CS}}^{(1)}+ b_{s^2} s^2  +   b_{(\nabla^2\delta)^2}  \left(\nabla^2\delta_{m  {\rm CS}}^{(1)}\right)^2  \nonumber\\
&&  + 2 b_{\delta\nabla^2\delta} \delta_{m  {\rm CS}}^{(1)} \nabla^2\delta_{m  {\rm CS}}^{(1)} +  2 b_{(\p\delta)^2} \p_i \delta_{m  {\rm CS}}^{(1)}~\p^i \delta_{m  {\rm CS}}^{(1)}   - (b_e-3) \cH v^{(2)}    -2 \left[b_{10}' +\cH( b_e-3)   b_{10}  \right] v \delta_{m  {\rm CS}}^{(1)}      \nonumber\\
&& -2 \left[b_{\nabla^2\delta}' +\cH( b_e-3)   b_{\nabla^2\delta} \right] v \nabla^2\delta_{m  {\rm CS}}^{(1)}     + 2b_{11} \zeta \delta_{m  {\rm CS}}^{(1)} -  b_{01}  \p^i \chi^{(1)}  \p_i  \zeta \   -2 \left[b_{01}' +\cH( b_e-3)   b_{01}  \right] v \zeta   \nonumber\\
&&
 - 2   b_{10} v   \,  {\delta_{m  {\rm CS}}^{(1)}}'   - 2   b_{\nabla^2\delta} v  \, \nabla^2 {\delta_{m  {\rm CS}}^{(1)}}'    + b_{02} \zeta^2 +  (b_e-3) \cH v  {v}'  + \left[ (b_e -3) \cH' + \cH  b_e' + (b_e -3)^2 \cH^2 \right] v^2   \nonumber\\
&&  
  - \left(b_e-3\right) \cH  \nabla^{-2}\bigg( v \nabla^2 {v}' - {v}' \nabla^2 v - 6 \p_i \Phi \p^i v - 6 \Phi \nabla^2 v \bigg)\;.\nonumber\\
\end{eqnarray} 
Now, using this definition for the bias in Eqs.\ (\ref{bias}) and (\ref{dg1}), the transfer function $ \T^{\delta_g}$, defined in Eq.\ (\ref{T}),  can be explicitly written as
\be
\label{Tdeltag}
 \T^{\delta_g} (\bk , \eta) = b_{10} (\eta,\bar L)  \T^{\delta_{\rm SC}}(\bk, \eta) - b_{\nabla^2\delta} (\eta,\bar L) k^2  \T^{\delta_{\rm SC}}(\bk, \eta)  + b_{01} ( \eta,\bar L)  \T^{\zeta}(\bk, \eta) - \left(b_e(\eta,\bar L)-3\right)\cH \T^{v }(\bk, \eta) \;.
 \ee
Finally  we can also rewrite the magnification bias, which first appeared in Sec.\ \ref{sec:Second-ordernumbercounts}, as
\begin{equation}\label{Q-first-order}
 \Q^{(1)}(\bar x^\alpha, \bar L)=-\frac{\p  \delta_g}{\p \ln \bar L} =-\frac{\p   \delta_{g {\rm CS}}}{\p \ln \bar L} + \cH \frac{\p b_e}{\p  \ln \bar L} v  =-\frac{\p   b_{10} (\bar x^0,\bar L)}{\p \ln \bar L} \,\delta_{m {\rm CS}}^{(1)} -\frac{\p  b_{\nabla^2\delta}  (\bar x^0,\bar L)}{\p \ln \bar L} \nabla^2\delta_{m {\rm CS}}^{(1)}  -\frac{\p   b_{01}(\bar x^0,\bar L)}{\p \ln \bar L} \,\zeta   + \cH \frac{\p b_e}{\p  \ln \bar L} v\;,
\end{equation}
 and, consequently, $\T^{ \Q^{(1)}}$, in Eq.\ (\ref{M-TQ}), turns out to be 
\begin{equation}
\T^{ \Q^{(1)}} (\bk, \eta, \bar L)=-\frac{\p   b_{10} (\eta,\bar L)}{\p \ln \bar L} \, \T^{\delta_{\rm SC}}(\bk,\eta)+\frac{\p  b_{\nabla^2\delta}  (\eta,\bar L)}{\p \ln \bar L} \,k^2 \T^{\delta_{\rm SC}}(\bk,\eta) -\frac{\p   b_{01}(\eta,\bar L)}{\p \ln \bar L} \, \T^{\zeta}(\bk, \eta)   + \cH \frac{\p b_e(\eta,\bar L)}{\p  \ln \bar L} \T^{v }(\bk, \eta)
\;.
\end{equation}

Appendix \ref {Transfer_functions} is devoted to write explicitly all the transfer functions $\T^b(\bk, \eta)$ and the kernels $F_{\bar \ell}^{b (2)}(p, q; \eta)$ for the (spatially flat) $\Lambda$CDM model (all these results can be quickly generalized for CDM+dynamical dark energy models).


\section{Conclusions}
\label{Sec:Conclusions}

In this paper we presented, for the first time, the full expression of the galaxy three-point function  (the bispectrum)   including all relativistic local and integrated terms at second order in the perturbations, in the wide-angle geometry (we never rely on the Limber and flat-sky approximations) and including the non-Gaussianity parameter $f_{\rm NL}$. We  also include the galaxy bias up to second order, including a tidal term and add all magnification corrections.
We believe that the calculations and expressions presented here, despite being cumbersome and appearing tedious and complicated, are a valuable  and timely contribution. Future and forthcoming galaxy surveys will  map the sky on ultralarge scales with unprecedented precision bringing LSS in regime where the simple Newtonian approximation is not longer sufficient. The next-to-leading-order correlation (such as the bispectrum)  despite being  a lower signal-to-noise quantity than the power spectrum, encloses key complementary physical and cosmological information, which should not be neglected if we are maximizing the scientific return from this observational effort. Given that the bispectrum on large scales will be measured,  the correct interpretation of the measurement relies on accurate and precise  theoretical modeling of the signal. It is for this reason that we believe it is  of fundamental importance at this point to have the full expression of the galaxy bispectrum,  and we embarked in this challenging  and time consuming task, despite its apparent complexity.

Working in spherical Bessel coordinates, we derived a compact expression for the bispectrum that encompasses all the physical contributions at first and second order, in curved sky and including integrated terms for radial configurations. We found that we can write the full GR bispectrum as
\bea
\label{eq:bispectrum_fourier_summary}
\langle \Delta_g (\bk_1) \Delta_g (\bk_2) \Delta_g(\bk_3)\rangle
 =\sum_{\ell_i  m_i=1}^{3}
\mathcal{U}_{\ell_1\ell_2 \ell_3 }(k_1,k_2,k_3) \times \mathcal{T_{SH}}^{\ell_i, m_i} \times
B_{\ell_1\ell_2 \ell_3 }(k_1,k_2,k_3) \, , \nonumber
\eea
after resumming over $\{\ell_1  m_1, \ell_2 m_2, \ell_3 m_3\}$, $\mathcal{U}$ is a scale-dependent angular function and $\mathcal{T_{SH}}^{\ell_i, m_i}$ are tripolar spherical harmonics; see Eq. (\ref{Bispectrum-2}).

In this case, the bispectrum in Fourier angular space is written as
\bea
B_{\ell_1\ell_2 \ell_3 }(k_1,k_2,k_3) = \nsumm_{abc}
\mathcal{C}_\ell^j
\int \frac{q_2^2\ud q_2}{(2\pi)^3} \frac{q_3^2\ud q_3}{(2\pi)^3}\left[ \K_{\ell_1 {\ellvp}_1  {\ellvq}_1 {\barell}_1 }^{a(2)}(k_1; q_2, q_3) \M_{\ell_2}^{b(1)}(k_2, q_2)  \M_{\ell_3}^{c(1)}(k_3, q_3)\right] P_\Phi(q_2) P_\Phi(q_3) + {\rm cyc}\, , \nonumber
\eea
where $\mathcal{C}_\ell^j$ 
is a combination of multipoles and 3j- and 6j-Wigner symbols, $P_{\Phi}$ denotes the primordial  (linear, Gaussian) power spectrum of the Bardeen gravitational potential and
$\K_{\ell_1 {\ellvp}_1  {\ellvq}_1 {\barell}_1 }^{a(2)}(k_1; q_2, q_3) \propto \M_{\ell_1 m_1 {\ellvp}_1 {\mvp}_1 {\ellvq}_1 {\mvq}_1 {\barell}_1 {\barm}_1}^{a(2)}(k_1; q_2, q_3) $  and $\M_{\ell_\iota}^{\iota(1)}(k_\iota, q_\iota)$  [see Eq.(\ref{M-1})] are spherical multipole functions  at second and first-order, respectively, containing all the physical effects.
For the explicit expressions of all the terms, a concise summary can be found in Sec.\ \ref{sec:summary} and the details in the rest of the text. In particular, $\M_{\ell_1 m_1 {\ellvp}_1 {\mvp}_1 {\ellvq}_1 {\mvq}_1 {\barell}_1 {\barm}_1}^{a(2)}(k_1; q_2, q_3)$ are in  Eqs. (\ref{delta_g^(2)}-\ref{v^(2)}), (\ref{MdeltaPhi(2)}-\ref{M-TQ}), (\ref{vphi}-\ref{vv}), (\ref{Ixall}-\ref{nablaperpTxall}),  (\ref{SperpxAll}-\ref{SperpSperp}),  (\ref{T^(2)}-\ref{I^(2)}),  (\ref{kappa^(2)}), (\ref{phiperpS}-\ref{phinablaT}), (\ref{M-Deltag2-gamma-sq}), (\ref{M-gamma-Post-Born}),  (\ref{delta_g^(2)-2}-\ref{v^(2)-2}), (\ref{M-T^(2)-2}-\ref{M-I^(2)-2}) and (\ref{M-kappa^(2)-2}).

It is evident that  the expressions provided here will likely  be impractical for most realistic applications or data analyses. Hence approximations will have to be made. The results presented  here provide the starting point  to devise suitable approximations and  a reference and benchmark to assess the validity, accuracy, performance, advantages and disadvantages  of any such approximation. For example for numerical and computational reasons, real data analyses should involve an optimal choice of which terms to include. 
 We envision that the results presented here will be a reference for assessing which physical effects are most important and which one are ultimately  negligible. The choice will depend on specific survey characteristics, on the physics to test, models considered or question at hand. Significant work remains to be done in evaluating the relative importance of the ({\it many}) different terms and physical contributions  to the overall signal in different regimes.  While preliminary investigation suggest that this is a viable program, we leave this effort to future work.

\section*{ACKNOWLEDGMENTS:}

We would like to thank Roland de Putter, Donghui Jeong, Eiichiro Komatsu, Roy Maartens, Juli\'an Mu\~noz, Guido Pettinari, Cristiano Porciani, Fabian Schmidt, Jean-Philippe Uzan and David Wands for discussions.
During the preparation of this work D.B. was supported by the Deutsche Forschungsgemeinschaft through the Transregio
33, The Dark Universe.
A.R. has received funding from the People Programme (Marie Curie Actions) of the European Union H2020 Programme
under REA Grant Agreement No. 706896 (COSMOFLAGS). Funding for this work was partially provided by the
Spanish MINECO under MDM-2014-0369 of ICCUB (Unidad de Excelencia ``Maria de Maeztu") and the Templeton
Foundation.
L.V. acknowledges support  by Spanish Mineco via AYA2014-58747-P AEI/FEDER UE and  MDM-2014-0369 of ICCUB (Unidad de Excelencia Maria de Maeztu). 
N.B., M.L. and S.M. thank ASI/INAF Agreement I/072/09/0 for the Planck LFI Activity of Phase E2 for partial financial support.

\appendix
\section{Useful relations in the Poisson gauge}\label{appendix:A}

Here we report  the expressions for the metric and the four-velocities in Poisson gauge \cite{Matarrese:1997ay, Malik:2008im} used in sec.~\ref{sec:Second-ordernumbercounts}.

From Eq.\ (\ref{Poiss-metric}) the perturbation of  the FRW metric metric $g_{\mu \nu}$ and $g^{\mu \nu}$ is
\begin{eqnarray}
\begin{array} {lll}
g_{00}=- a^2 \left(1+ 2 \Phi + \Phi^{(2)}\right),  & \quad& g^{00}=- a^{-2} \left[1-2  \Phi  - \Phi^{(2)} +4  {\Phi}^2\right], \\ \\
g_{0i}= a^2 \omega^{(2)}_i,  & \quad&  g^{0i}=a^{-2}\omega^{i (2)}, \\ \\
g_{ij}= a^2 \left(\delta_{ij} -2 \delta_{ij} \Phi  -\delta_{ij} \Psi^{(2)}+\hat h^{(2)}_{ij}/2 \right),  & \quad&   g^{ij}=a^{-2}  \left[ \delta^{ij}+2 \delta^{ij} \Phi  + \delta^{ij} \Psi^{(2)}-\hat h^{ij(2)}/2 +4\delta^{ij} (\Phi )^2\right], \\
  \end{array}  \nonumber \\
\end{eqnarray}

For the four-velocity $u^\mu$, we have
\begin{eqnarray}
\label{Poiss-u0i}
u_0&=&-a\left[1+\Phi +\frac{1}{2}\Phi^{(2)}-\frac{1}{2} {\Phi}^2+\frac{1}{2}v_k v^{k }\right] ,\\ 
u_i&=&a\left[v_i +\frac{1}{2}\left(v_i^{(2)}+2\omega_i^{(2)}\right)- 2 \Phi v_i \right], \\
u^0&=&\frac{1}{a}\left[1-\Phi -\frac{1}{2}\Phi^{(2)}+\frac{3}{2} {\Phi}^2+\frac{1}{2}v_k v^{k }\right] ,\\
u^i&=&\frac{1}{a}\left(v^{i }+\frac{1}{2}v^{i(2)}\right)\;.
\end{eqnarray}

Given $T_m^{\mu \nu}=\rho_m u^\mu u^\nu$, i.e. the cold dark matter stress-energy tensor,  for  first- and second-order perturbations we obtain
\begin{eqnarray}
\label{eqdelta-1}
&& \delta_m {'} + \p_i v^{i }- 3 \Phi  {'}=0\;, \nonumber \\  
\label{eqv-1}
&&v^{i }{'}+ \cH  v^{i }+ \p^i  \Phi  =0\;, \nonumber \\ 
\label{eqdelta-2}
&& \frac{1}{2} \delta_m^{(2)}{'} + \frac{1}{2}\p_i v^{i(2)}-\frac{3}{2} \Psi^{(2)}{'}+\frac{1}{4} \hat h^{i(2)}_i {'}-  \cH  v^{i } v _i + \left(\Phi  + \delta_m  \right)\p_i v^{i } +v^{i }  \p_i \delta_m    -3 \delta_m   \Phi {'} -3 v^{j } \p_j  \Phi - 6 \Phi  \Phi {'} =0\;, \nonumber \\
\label{eqv-2}
&& \left( \frac{1}{2} v^{i(2)} + \omega^{i(2)}\right){'}+ \cH  \left( \frac{1}{2} v^{i(2)}+ \omega^{i(2)} \right) + \frac{1}{2}  \p^i  \Phi^{(2)} +  v^{j } \p_j  v^{i }  -2 v^{i  } \Phi {'}  + \Phi \p^i  \Phi      =0 \;.
\label{ConsEq}
\end{eqnarray}

\section{Spherical Fourier Decomposition: $|k \ell m\rangle$ frame} \label{klm-basis}

Let us start with the plane-wave representation
\be
\langle \bx | \bk \rangle = e^{i \bx \bk}= \sum_{\ell m} (4 \pi) i^\ell  j_\ell(k\chi) Y^*_{\ell m} ({\hat \bk}) Y_{\ell m} ({\hat \bn})
\ee
where $ \bx = \bar \chi \hat \bn$ and $\bk= k \hat \bk$

and defining the complete radial and angular basis in a spherical Bessel Fourier
space as $ | k \ell m \rangle$,  its representation in configuration space is
\be
\langle \bx | k \ell m \rangle = \aleph_\ell(k) j_\ell(k\chi) Y_{\ell m} ({\hat \bn})\;.
\ee

This basis is orthonormal, i.e.
\be
\langle k\ell m | k' \ell 'm' \rangle = \mbox{\footnotesize $\mbox{\footnotesize $\beth$}$}_\ell(k) \delta^D (k-k') \delta_{\ell \ell'} \delta_{m m'}
\ee
and
\be
\langle\bk' | k \ell m \rangle= \gimel_\ell(k)  Y_{\ell m} ({\hat \bk}) \delta^D (k-k')\;.
\ee
At the moment, $\aleph_\ell(k), \mbox{\footnotesize $\beth$}_\ell(k)$ and $ \gimel_\ell(k) $ are  generic functions.  Obviously, as we will see below, these functions are closely related to each other.

The completeness conditions on all possible bases are defined in the following way:
\bea
1= \int \ud^3 \bx ~  |\bx \rangle \langle \bx| =   \int \frac{\ud^3 \bk}{(2\pi)^3}  |\bk \rangle \langle \bk | = \sum_{\ell m}  \int \frac{k^2 \ud k} {(2\pi)^3} \daleth_\ell(k) |k \ell m\rangle \langle k\ell m |
\eea
Here, by construction, $\mbox{\footnotesize $\beth$}_\ell(k)$ and $\daleth_\ell(k)$ are real, i.e.
\be
\mbox{\footnotesize $\beth$}_\ell(k) = \mbox{\footnotesize $\beth$}_\ell^*(k) \quad \quad {\rm and}  \quad \quad \daleth_\ell(k)= \daleth_\ell^*(k)\;.
\ee

Now, projecting the field $\phi=| \phi \rangle$, respectively on $| \bx \rangle$, $| \bk \rangle$ and $|  k\ell m   \rangle$, i.e.
\be
\langle \bx | \phi \rangle = \phi(\bx) \quad \quad \langle \bk | \phi \rangle = \phi(\bk) \quad \quad  \langle k\ell m | \phi \rangle = \phi_{\ell m}(k)\;,
\ee
and using the completeness conditions, we find
\bea
\phi(\bx) = \int \frac{\ud^3 \bk}{(2\pi)^3}~  \langle \bx |\bk \rangle \langle \bk | \phi \rangle = \int \frac{\ud^3 \bk}{(2\pi)^3}  e^{i \bx \bk} \phi(\bk)\;, \quad \quad
\phi(\bk) = \int \ud^3 \bx  ~ \langle \bk |\bx  \rangle \langle \bx|  \phi \rangle = \int \ud^3 \bx  ~   e^{- i \bx \bk} \phi(\bx)\;,
\eea
For $\phi_{\ell m}(k)$:
 \bea
 \label{phiklm}
 \phi_{\ell m}(k) &=&  \int \ud^3 \bx  ~ \langle k \ell m|\bx  \rangle \langle \bx|  \phi \rangle = \int \ud^3 \bx  ~  \aleph^*_\ell(k)  j_\ell(k\chi)  Y^*_{\ell m} ({\hat \bn}) \phi(\bx)  \\
 {\rm or} \quad \quad\quad \quad \quad \quad && \nonumber \\\label{phi-veck_to_phi-klm}
 &=&   \int \frac{\ud^3 \tilde \bk}{(2\pi)^3}  \langle k \ell m |\tilde \bk \rangle \langle \tilde \bk |  \phi \rangle = \left[\frac{k^2 \gimel_\ell^*(k)}{(2 \pi)^3}\right]   \int \ud^2 \hat \bk ~ Y_{\ell m}^*(\hat \bk) \phi(\bk) \;.
 \eea
 
Relations between $ \aleph_\ell(k), \mbox{\footnotesize $\beth$}_\ell(k), \gimel_\ell(k) $ and $\daleth_\ell(k)$:
{\it i)} From $\langle \bx | \bk \rangle$
\bea
\langle \bx | \bk \rangle= \sum_{\ell m}  \int \frac{\tilde k^2 \ud \tilde k} {(2\pi)^3} \daleth_\ell(\tilde k)  \langle \bx  | \tilde k \ell m\rangle \langle \tilde k\ell m |  \bk \rangle = \sum_{\ell m} \left[\frac{k^2 \aleph_\ell(k)  \gimel_\ell^*(k)  \daleth_\ell(k)}{(2 \pi)^3}\right]  j_\ell(k\chi) Y^*_{\ell m} ({\hat \bk}) Y_{\ell m} ({\hat \bn})
\eea
 we find
\be
\frac{k^2 \aleph_\ell(k)  \gimel_l^*(k)  \daleth_\ell(k)}{(2 \pi)^3}= (4 \pi) i^\ell \;. 
\ee
From the above relation, immediately, we obtain the following property
\be
 \aleph_\ell(k)  \gimel_l^*(k) = (-1)^\ell  \aleph_\ell^*(k)  \gimel_l (k) \;.
 \ee

{\it ii)} From $\langle k\ell m | k' \ell 'm' \rangle$
\bea
\langle k\ell m | k' \ell 'm' \rangle &=& \int \ud^3 \bx  ~ \langle k \ell m|\bx  \rangle \langle \bx| k' \ell 'm'   \rangle = \delta^K_{\ell \ell'} \delta^K_{m m'} \frac{\pi}{2 k^2}  \left|\aleph_\ell(k)\right|^2\delta^D (k-k') \\
 {\rm or} \quad \quad\quad \quad \quad \quad && \nonumber \\
 &=&   \int \frac{\ud^3 \tilde \bk}{(2\pi)^3}  \langle k \ell m |\tilde \bk \rangle \langle \tilde \bk |  k' \ell 'm' \rangle = \delta^K_{\ell \ell'} \delta^K_{m m'} \frac{k^2}{(2 \pi)^3}  \left|\gimel_\ell(k)\right|^2\delta^D (k-k')\;,
 \eea
where we have used
\[
\int \ud \bar \chi~\bar \chi^2  j_\ell (k \bar \chi)  j_\ell (k'  \bar \chi) =  \frac{\pi}{2 k^2} \delta^D (k-k')\;.
\]

Then we find
\be
\mbox{\footnotesize $\beth$}_\ell(k)= \frac{\pi}{2 k^2}  \left|\aleph_\ell(k)\right|^2=\frac{k^2}{(2 \pi)^3}  \left|\gimel_\ell(k)\right|^2\;.
\ee

In the literature, these coefficients are usually fixed in the following way: 
\bea \label{Function_explicitly}
\begin{array}{ccccc}
{\rm (1)} \quad\quad &  \aleph_\ell(k)=k\sqrt{\frac{2}{\pi}}\;, \quad \quad & \mbox{\footnotesize $\beth$}_\ell(k)=1 \;,\quad \quad &\gimel_\ell(k) = \frac{(2\pi)^{3/2}(-i)^\ell}{k} \;, \quad \quad & \daleth_\ell(k) = \frac{(2\pi)^3}{k^2}\;;  
 \\
{\rm (2)}  \quad\quad & \aleph_\ell(k)=4\pi i^\ell\;, \quad \quad & \mbox{\footnotesize $\beth$}_\ell(k)=\frac{(2\pi)^3}{k^2} \;, \quad \quad & \gimel_\ell(k)= \frac{(2\pi)^3}{k^2} \;,\quad \quad  & \daleth_\ell(k) =1 \;,
\end{array}
\eea
where in (1) see Refs.\ \cite{Verde:2000xj, Yoo:2013tc} and in (2) see Refs.\ \cite{Hamilton:1997zq, Dai:2012bc, Dai:2012ma}.

In the paper it is also useful to write the inverse of Eq.\ (\ref{phi-veck_to_phi-klm}). Using the results obtained in this section, we immediately find  
 \be\label{phi-klm_to_phi-veck}
  \phi(\bk)=\sum_{\ell m}  \int \frac{\tilde k^2 \ud \tilde k} {(2\pi)^3} \daleth_\ell(\tilde k) \langle \bk | \tilde k \ell m\rangle \phi_{\ell m}(\tilde k)= \sum_{\ell m} \frac{4\pi (-i)^\ell}{ \aleph^*_\ell(k)} Y_{\ell m} ({\hat \bk})  \phi_{\ell m}(k)\;.
 \ee

\section{Covariant derivative on unit sphere and the spin-weighted spherical harmonic decomposition}\label{spin}

In this appendix, we outline the decomposition of three dimensional quantities on a unity sphere in the observer tangent space and show how one can derive the relation between the covariant 2D derivative and 
spin-weighted spherical harmonics function. This discussion is based on Refs. \cite{Zaldarriaga:1996xe, Hu:1997hp, Hu:2000ee, Okamoto:2003zw, Castro:2005bg, Schmidt:2012ne, Lewis:2001hp, Pitrou:2015iya}.

 Starting from the orthonormal polar basis vectors $\{\bn, \e_\theta, \e_\varphi\}$,
 it is possible to define the following helicity basis $$\bm_{\pm}=\frac{\e_\theta \mp i \e_\varphi }{\sqrt{2}}$$  which are also called spin$\pm 1$ unit basis vector on the unity sphere. (In the literature, e.g. Refs.\ \cite{Okamoto:2003zw, Castro:2005bg}, $\bm_{\pm}$ is also denoted as $\bm_+ = \bar \bm $ and $\bm_-=\bm$.) Under a right-handed rotation of the coordinate system $\{\e_\theta, \e_\varphi\}$ around $\bn$ by an angle $\psi$, so that 
\bea
\e_\theta &\to \e_\theta' &=  \cos \psi \,  \e_\theta +\sin\psi \, \e_\varphi  \nonumber \\
\e_\varphi &\to \e_\varphi' &= -\sin \psi \, \e_\theta + \cos \psi \, \e_\varphi\;,
\nonumber
\eea 
$\bm_{\pm} $ transform as  $$\bm_{\pm} \to \bm_{\pm}' =\exp{(\pm i \psi)} \, \bm_{\pm}\;.$$
 In particular,  $\bm_{\pm}$ satisfy the following properties $$m_{\pm}^i= \Perp^i_{j} m_{\pm}^j, \quad  \quad \quad m_{\pm}^i m_{\pm i}=0, \quad  \quad \quad m_{\pm}^i m_{\mp i}=1, \quad  \quad \quad n^i m_{\pm i}=0\;, \quad  \quad \quad \Perp^j_i= m_{+ i}  m_{- }^j  +m_{-  i}  m_{+}^j \;.$$
Given a spin-weight $s$ complex function $_s f(\bn)$, it transforms as $_s f(\bn)\to e^{-is\psi} _s f(\bn)$. Then, defining the projected tensor
as
 $$T_{i_1...i_s}=  \Perp^{j_1}_{i_1}... \Perp^{j_s}_{i_s}~  T_{j_1...j_s} , $$
 $ T_{i_1...i_s} m_-^{i_1}...m_-^{i_s}$ transforms as a spin-weight $s$ object. Specifically, if $ T_{j_1...j_s} $ is  tensor symmetric and trace-free component of a rank $s$ tensor,  we can associate it with a $s$-spin weight function in the following way 
 \bea
{_s}f(\bn) &=&T_{i_1...i_s} m_-^{i_1}...m_-^{i_s}  \nonumber \\
_{-s}f(\bn) &=&T_{i_1...i_s} m_+^{i_1}...m_+^{i_s}
 \eea
or conversely
\be
T_{j_1...j_s}={_s}f(\bn) \, m_{+i_1}...m_{+i_s}+ {_{-s}}f(\bn)\, m_{-i_1}...m_{-i_s}\;.
\ee
 
  At this point it is important to define the covariant derivative on the unit sphere\footnote{In Ref.\ \cite{Pitrou:2015iya} ${}^{(2)}\nabla_i= D_i$.}
  \be
  \frac{1}{\bar \chi} {}^{(2)}\nabla_i T_{j_1...j_s} = \Perp_i^p  \Perp^{q_1}_{j_1} ... \Perp^{q_s}_{j_s}\p_{p} T_{q_1...q_s}\;.
  \ee
  From the above relation we note immediately that, $${\rm for}  ~ s \neq 0, \quad  \quad \quad \p_{\perp i} \neq  \frac{1}{\bar \chi} {}^{(2)}\nabla_i$$ and,   for $s=0$ we have 
 \be
  \p_{\perp i} f =  \frac{1}{\bar \chi} {}^{(2)}\nabla_i f \quad \quad {\rm and } \quad \quad \p_i f= \p_\| f +  \frac{1}{\bar \chi} {}^{(2)}\nabla_i f\;.
 \ee
  (Here $_0 f= f$.) Instead, with two derivates we find
  \be
    \p_{\perp i}   \p_{\perp j} f = - \frac{1}{\bar \chi^2} n_j  {}^{(2)} \nabla_i  f + \frac{1}{\bar \chi^2}  {}^{(2)} \nabla_i   {}^{(2)} \nabla_j f\;.
  \ee
  
  Now it is useful to relate $ {}^{(2)}\nabla_i$ to the usual  spin-raising and lowering operators, i.e. $\eth$ and $\bar \eth$, defined and analyzed for the first time in Refs.\ \cite{Penrose:1959vz, Newman:1961qr, Sachs:1962zza, Newman:1966ub, Goldberg:1966uu, Campbell:1971rm} (see also e.g. \cite{Dray:1984gy, Gomez:1996ge, Newman:2005hy, Boyle:2016tjj}). First of all, the spin operators are defined in the following way
  \bea
 \eth\, {_s}f  &=&  - \sin^s \theta \left(\p_\theta + \frac{i}{\sin \theta} \p_\varphi \right) \frac{1}{\sin^s \theta}\; {_s}f  \nonumber \\
 \bar \eth \, {_s}f  &=&  - \frac{1}{\sin^s \theta} \left(\p_\theta -  \frac{i}{\sin \theta} \p_\varphi \right) \sin^s \theta \; {_s}f \;, \nonumber
 \eea 
  and have the ability to transform under the angle $\psi$ rotation as
  \bea
 \eth\,  _s f(\bn) &\to& e^{-i(s+1)\psi}  \eth\,   _s f(\bn)\;, \nonumber \\
  \bar \eth \, _s f(\bn) &\to& e^{-i(s-1)\psi}   \bar \eth \,  _s f(\bn)   \;. \nonumber
  \eea
  Then the covariant derivate of $T_{j_1...j_s}$ can be written as follows:
    \bea
 {}^{(2)}\nabla_i T_{j_1...j_s}= [\mathscr{D}_i~ {_s}f(\bn) ]\, m_{+j_1}...m_{+j_s}+ [ \mathscr{D}_i~ {_{-s}}f (\bn)] \, m_{-j_1}...m_{-j_s}\;,
    \eea
    where
    \be\label{D}
  - \sqrt{2}  [\mathscr{D}_i ~ {_s}f(\bn)]= \eth\, {_s}f ~ m_{+ i}  + \bar  \eth\, {_s}f  ~ m_{- i}\;.
    \ee
Conversely, 
      \bea
 \eth\, {_s}f  &=&  -\sqrt{2}  m_-^j m_-^{i_1}...m_-^{i_s} ~ {}^{(2)}\nabla_j T_{i_1...i_s}\;,  \nonumber \\
 \bar \eth \, {_s}f  &=& -\sqrt{2}  m_+^j m_-^{i_1}...m_-^{i_s} ~ {}^{(2)}\nabla_j T_{i_1...i_s}  \;, \nonumber
 \eea 
 for $s\ge0$, and we have to replace $m_-^{i_s}$ with $m_+^{i_s}$ for $s<0$.
 
 In order to describe correctly objects of spin-weight $s$ we need to generalize the spherical harmonic basis $Y_{\ell m}(\theta, \varphi)$. This new basis is called spin-weight spherical harmonics $_s Y_{\ell m}(\theta, \varphi)$ and it can be related with the usual $ Y_{\ell m}(\theta, \varphi)={_0}Y_{\ell m}(\theta, \varphi)$ in the following way:
 \bea
{_s}Y_{\ell m}(\theta, \varphi) = \sqrt{\frac{(\ell - |s|)!}{(\ell + |s|)!}}
\left\{ \begin{array}{ll}
 \eth^s \,  Y_{\ell m}(\theta, \varphi) & {\rm for} ~ 0\le s\le \ell \\
 \\
(-1)^s ~ \bar  \eth^{|s|} \,  Y_{\ell m}(\theta, \varphi)  & {\rm  for} ~ -\ell \le s \le 0 
\end{array} \right.
\eea
 and satisfies the properties
 \bea
 {_s}Y_{\ell m}^*(\theta, \varphi) &=& (-1)^{m+s}  {_{-s}}Y_{\ell -m}(\theta, \varphi) \nonumber\\
 \eth   {_s}Y_{\ell m}(\theta, \varphi)&=& \sqrt{(\ell-s)(\ell+s+1)}  {_{s+1}}Y_{\ell m}(\theta, \varphi) \nonumber\\
\bar \eth   {_s}Y_{\ell m}(\theta, \varphi)&=& - \sqrt{(\ell+ s)(\ell-s+1)}  {_{s-1}}Y_{\ell m}(\theta, \varphi)\nonumber\\
   \bar \eth   {_s}  \eth   {_s} Y_{\ell m}(\theta, \varphi)  &=&-(\ell-s)(\ell+s+1)  {_s} Y_{\ell m}(\theta, \varphi)\;.
 \eea
 Then we find
 \be
   \bar \eth^s  \eth^s  Y_{\ell m}(\theta, \varphi) =     \eth^s  \bar \eth^s Y_{\ell m}(\theta, \varphi)  = (-1)^s \frac{(\ell + |s|)!}{(\ell - |s|)!} Y_{\ell m}(\theta, \varphi)\;,
 \ee
 and we can rewrite Eq.\ (\ref{D}) by using the spin-$s$ spherical harmonics, i.e.
 \be
   - \sqrt{2}  [\mathscr{D}_i ~ {_s}Y_{\ell m}(\theta, \varphi)]= \sqrt{(\ell-s)(\ell+s+1)} {_{s+1}}Y_{\ell m}(\theta, \varphi) ~ m_{+ i} - \sqrt{(\ell+ s)(\ell-s+1)}  {_{s-1}}Y_{\ell m}(\theta, \varphi)~ m_{- i}\;.
 \ee
Now we have all the required tools  to connect quickly the covariant derivative on the unit sphere with the spin-$s$ spherical harmonic basis.
 For example,
 \bea
 m^i_-   {}^{(2)}\nabla_i Y_{\ell m} &=& - \sqrt{\frac{(\ell-s)(\ell+s+1)}{2}}  {_{s+1}}Y_{\ell m}\;, \\
 m^i_+ {}^{(2)}\nabla_i Y_{\ell m}&=& \sqrt{\frac{(\ell+s)(\ell-s+1)}{2}} {_{s-1}}Y_{\ell m}\;,
 \eea
 and
 \be
 {}^{(2)}\nabla_i Y_{\ell m}(\theta, \varphi) = - \sqrt{\frac{\ell(\ell+1)}{2}}\left[_1Y_{\ell m}(\theta, \varphi) ~ m_{+ i} - _{-1}Y_{\ell m}(\theta, \varphi) ~ m_{- i}\right]\;.
 \ee
 
  Let us conclude this section with the orthogonality and completeness expression
 \bea
 \int \ud^2 \hat \bn ~ {_s}Y_{\ell_1 m_1} (\hat \bn)  {_s}Y^*_{\ell_2 m_2} (\hat \bn) =  \delta^K_{\ell_1 \ell_2} \delta^K_{m_1 m_2} \;, \\
 \sum_{\ell m}{_s}Y_{\ell_ m} (\theta_1, \varphi_1)   {_s}Y^*_{\ell m} (\theta_2, \varphi_2) = \delta^D(\varphi_1-\varphi_2) \delta^D (\theta_1-\theta_2) \;,
 \eea
 and the generalization of the Gaunt integral
 \bea
  \int \ud^2 \hat \bn ~ {_{s_1}}Y_{\ell_1 m_1} (\hat \bn)  {_{s_2}}Y_{\ell_2 m_2} (\hat \bn) {_{s_3}}Y_{\ell_3 m_3} (\hat \bn)= \I^{-s_1-s_2-s_3}_{\ell_1\ell_2\ell2} \left(
\begin{array}{ccc}
\ell_1 & \ell_2 & \ell_3 \\
m_1 & m_2 & m_3
\end{array}
\right) \;,
 \eea
 where $\I^{s_1 s_2 s_3}_{\ell_1\ell_2\ell2}$ has  already been defined in Eq.\ (\ref{s_1s_2s_3}).


\section{Properties of $\Upsilon^{[\#]abc}$}\label{Sec:Upsilon}

This section is devoted to compute and prove some subtle cancellations of the bispectrum building blocks $\Upsilon$ which are already mentioned and briefly discussed in Sec.\ \ref{sec:Bispectr-ScalarCase}.

\subsection{Proof that $\Upsilon^{[1]abc} \propto  \delta^K_{\ell_1 0} $}

Let us start with $\Upsilon^{[1]abc}$ defined in Eq.\ (\ref{Upsilon[1]}). For $i=2,3$ of Eq.\ (\ref{Deltag2-loc}) and $i=1,2,4,5$ of Eq.\ (\ref{Deltag2-int}), we have 
\[\M_{\ell_1 m_1 {\ellvp}_1 {\mvp}_1 {\ellvq}_1 {\mvq}_1 {\barell}_1 {\barm}_1}^{a(2)}(k_1; q_1, q_1)  \propto \delta^K_{{\barell}_1 0} \delta^K_{{\barm}_1 0}(4\pi)^3 \aleph^*_{\ell_1}(k)  (-1)^{m_1} i^{{\ellvp}_1+{\ellvq}_1}  \G^{\ell_1 {\ellvp}_1 {\ellvq}_1 }_{-m_1  {\mvp}_1 {\mvq}_1} \times ...\,.\]
Using $\delta^K_{{\barell}_1 0} \delta^K_{{\barm}_1 0}$, it is possible to rewrite the Gaunt integrals in  $\Upsilon^{[1]abc}$ as
\[ \delta^K_{{\barell}_1 0} \delta^K_{{\barm}_1 0} ~\G^{{\ellvp}_1 {\barell}_1 \ell_{1}' }_{-{\mvp}_1 {\barm}_1 m_{1}' }\G^{{\ellvq}_1 {\barell}_1 \ell_{1}' }_{{\mvq}_1 {\barm}_1 m_{1}' }  = \G^{{\ellvp}_1 0 \ell_{1}' }_{-{\mvp}_1 0 m_{1}' }\G^{{\ellvq}_1 0 \ell_{1}' }_{{\mvq}_1 0 m_{1}' }  \propto \delta^K_{ \ell_{1}' {\ellvp}_1} \delta^K_{ \ell_{1}' {\ellvq}_1} \delta^K_{ m_{1}'  {\mvp}_1} \delta^K_{ m_{1}' - {\mvq}_1}=\delta^K_{{\ellvp}_1 {\ellvq}_1} \delta^K_{{\mvp}_1- {\mvq}_1}\;.\]
 Then we find
 \bea
 \Upsilon^{[1]abc} &\propto& \delta^K_{\ell_2 \ell_3} \delta^K_{m_2 -m_3} (-1)^{\ell_2 + m_2} \sum_{{\ellvp}_1} (-1)^{{\ellvp}_1} \sqrt{2 {\ellvp}_1 +1} ~ C_{{\ellvp}_1 0 {\ellvp}_1 0}^{\ell_1 0} \sum_{{\mvp}_1  = -{\ellvp}_1}^{{\ellvp}_1}  C_{{\ellvp}_1 -{\mvp}_1 {\ellvp}_1 0}^{\ell_1 -{\mvp}_1} \times ...  \nonumber \\
 &\propto&  ~ \delta^K_{\ell_2 \ell_3} \delta^K_{m_2 -m_3} \delta^K_{\ell_1 0}\;,
 \eea
 because \cite{Varshalovich:1988ye}
 \[\sum_{{\mvp}_1=-{\ellvp}_1}^{{\ellvp}_1}  C_{{\ellvp}_1 -{\mvp}_1 {\ellvp}_1 0}^{\ell_1 -{\mvp}_1} = (2 {\ellvp}_1 +1)  ~ \delta^K_{\ell_1 0}\;,\]
 where $C_{a \alpha b \beta}^{c \gamma}$ are the ClebschÐGordan coefficients which are related to Wigner 3-j symbols in the following way
 \be
 \sqrt{2c+1} 
 \left(
\begin{array}{ccc}
a & b & c \\
\alpha & \beta & \gamma
\end{array}
\right) = (-1)^{c+\gamma+2a} ~ C_{a -\alpha b -\beta}^{c \gamma}\;.
 \ee
 Here we consider only $\ell_1 >0$, then we can discard  $\Upsilon^{[1]abc}$ for all the terms considered.
 
 Instead, for $i=1$ of Eq.\ (\ref{Deltag2-loc}) and $i=3$ of Eq.\ (\ref{Deltag2-int}), $ \M_{\ell m \ellvp \mvp \ellvq \mvq \barell \barm}^{ a(2)}(k; p, q)$ is independent with $\bar m_1$ and we  find  directly that  \cite{Varshalovich:1988ye}
 \be
 \sum_{\bar m_1 = -\bar \ell_1}^{\bar \ell_1}\sum_{m_{1}' = -\ell_{1}'}^{\ell_{1}'} \G^{{\ellvp}_1 {\barell}_1 \ell_{1}' }_{-{\mvp}_1 {\barm}_1 m_{1}' }\G^{{\ellvq}_1 {\barell}_1 \ell_{1}' }_{{\mvq}_1 {\barm}_1 m_{1}' }  \propto C_{\bar \ell_1 0 {\ell_1}' 0}^{{\ellvp}_1 0} C_{\bar \ell_1 0 {\ell_1}' 0}^{{\ellvq}_1 0} \sum_{\bar m_1 = -\bar \ell_1}^{\bar \ell_1}\sum_{m_{1}' = -\ell_{1}'}^{\ell_{1}'}  C_{\bar \ell_1 -\bar m_1 {\ell_1}' -{m_1}'}^{{\ellvp}_1 -{\mvp}_1} C_{\bar \ell_1 -\bar m_1 {\ell_1}' -{m_1}'}^{{\ellvq}_1 {\mvq}_1} \propto  \delta^K_{{\ellvp}_1 {\ellvq}_1} \delta^K_{{\mvp}_1- {\mvq}_1}\;.
 \ee
 Taking into account the above relation and $ \M_{\ell m \ellvp \mvp \ellvq \mvq \barell \barm}^{ a(2)}(k_1; p_1, q_1) \propto (-1)^{-{\mvq}_1}  \G^{\ell_1 {\ellvp}_1 {\ellvq}_1 }_{-m_1  {\mvp}_1 {\mvq}_1} $ we have
 \be
  \delta^K_{{\ellvp}_1 {\ellvq}_1} \delta^K_{{\mvp}_1- {\mvq}_1}  (-1)^{-{\mvq}_1} \G^{\ell_1 {\ellvp}_1 {\ellvq}_1 }_{-m_1  {\mvp}_1 {\mvq}_1} \propto C_{{\ellvq}_1 -{\mvq}_1 \ell_1 0}^{{\ellvq}_1 -{\mvq}_1}
 \ee
and, finally, we obtain
 \be
 \Upsilon^{[1]abc} \propto ..... \times \sum_{\bar \ell_1 \ell_1' {\ellvq}_1}   .... \times \sum_{{\mvq}_1= -{\ellvq}_1}^{{\ellvq}_1}   C_{{\ellvq}_1 -{\mvq}_1 \ell_1 0}^{{\ellvq}_1 -{\mvq}_1} \propto \delta^K_{\ell_1 0}\;.
 \ee 
 
 \subsection{Proof that $\Upsilon^{[2]abc}=\Upsilon^{[3]abc}$}
 For simplicity, let us start to write again Eq.(\ref{Gammas}) 
\be
 \bigg\langle \frac{1}{2}\Delta^{a(2)}_{\ell_1  m_1}(k_1)  \Delta^b_{\ell_2 m_2} (k_2)   \Delta^c_{\ell_3 m_3}(k_3) \bigg\rangle =
 ~ \Upsilon^{[1]abc}_{\ell_1 m_1 \ell_2 m_2 \ell_3 m_3}(k_1,k_2,k_3) + \Upsilon^{[2]abc}_{\ell_1 m_1 \ell_2 m_2 \ell_3 m_3}(k_1,k_2,k_3) + \Upsilon^{[3]abc}_{\ell_1 m_1 \ell_2 m_2 \ell_3 m_3}(k_1,k_2,k_3)
 \ee
 where
 $ \Upsilon^{[j]abc}_{\ell_1 m_1 \ell_2 m_2 \ell_3 m_3}$, for $j=1,2,3$, are the bispectrum building blocks and contain all of the information on the bispectrum in redshift space. Explicitly, we have 
\bea
\label{Upsilon[1]-2}
\Upsilon^{[1]abc}_{\ell_1 m_1 \ell_2 m_2 \ell_3 m_3}(k_1,k_2,k_3) = \sum_{{\ellvp}_1 {\mvp}_1 {\ellvq}_1 {\mvq}_1 {\barell}_1 {\barm}_1 \ell_{1}' m_{1}'} \delta^K_{m_2-m_3}  \delta^K_{\ell_2 \ell_3}  (-1)^{\ell_{1}' +\ell_2+m_2+{\mvp}_1} \G^{{\ellvp}_1 {\barell}_1 \ell_{1}' }_{-{\mvp}_1 {\barm}_1 m_{1}' }\G^{{\ellvq}_1 {\barell}_1 \ell_{1}' }_{{\mvq}_1 {\barm}_1 m_{1}' } \nonumber \\
  \int \frac{q_1^2\ud q_1}{(2\pi)^3}   \frac{q_2^2\ud q_2}{(2\pi)^3}\left[\M_{\ell_1 m_1 {\ellvp}_1 {\mvp}_1 {\ellvq}_1 {\mvq}_1 {\barell}_1 {\barm}_1}^{a(2)}(k_1; q_1, q_1) ~ \M_{\ell_2}^{b(1)}(k_2, q_2)  ~\M_{\ell_3}^{c(1)}(k_3, q_2)\right] P_\Phi(q_1) P_\Phi(q_2)\;,
\eea
\bea
\label{Upsilon[2]-2}
\Upsilon^{[2]abc}_{\ell_1 m_1 \ell_2 m_2 \ell_3 m_3}(k_1,k_2,k_3) = \sum_{{\ellvp}_1 {\mvp}_1 {\ellvq}_1 {\mvq}_1 {\barell}_1 {\barm}_1}  (-1)^{\ell_2 +\ell_3-m_2-m_3+{\mvp}_1+{\mvq}_1+{\barm}_1} \G^{{\ellvp}_1 {\barell}_1 \ell_{2} }_{-{\mvp}_1 {\barm}_1 -m_2 }\G^{{\ellvq}_1 {\barell}_1 \ell_{3} }_{{\mvq}_1 {\barm}_1 m_{3}}~~~~~~~ \nonumber \\
  \int \frac{q_2^2\ud q_2}{(2\pi)^3}   \frac{q_3^2\ud q_3}{(2\pi)^3}\left[\M_{\ell_1 m_1 {\ellvp}_1 {\mvp}_1 {\ellvq}_1 {\mvq}_1 {\barell}_1 {\barm}_1}^{a(2)}(k_1; q_2, q_3) ~ \M_{\ell_2}^{b(1)}(k_2, q_2)  ~\M_{\ell_3}^{c(1)}(k_3, q_3)\right] P_\Phi(q_2) P_\Phi(q_3)\;,
\eea
and
\bea
\label{Upsilon[3]}
\Upsilon^{[3]abc}_{\ell_1 m_1 \ell_2 m_2 \ell_3 m_3}(k_1,k_2,k_3) = \sum_{{\ellvp}_1 {\mvp}_1 {\ellvq}_1 {\mvq}_1 {\barell}_1 {\barm}_1}   (-1)^{\ell_2 +\ell_3-m_2-m_3+{\mvp}_1+{\mvq}_1+{\barm}_1}
 \G^{{\ellvp}_1 {\barell}_1 \ell_{3} }_{-{\mvp}_1 {\barm}_1 -m_3 }\G^{{\ellvq}_1 {\barell}_1 \ell_{2} }_{{\mvq}_1 {\barm}_1 m_{2}} ~~~~~~ \nonumber \\
  \int \frac{q_2^2\ud q_2}{(2\pi)^3}   \frac{q_3^2\ud q_3}{(2\pi)^3}\left[\M_{\ell_1 m_1 {\ellvp}_1 {\mvp}_1 {\ellvq}_1 {\mvq}_1 {\barell}_1 {\barm}_1}^{a(2)}(k_1; q_3, q_2) ~ \M_{\ell_2}^{b(1)}(k_2, q_2)  ~\M_{\ell_3}^{c(1)}(k_3, q_3)\right] P_\Phi(q_2) P_\Phi(q_3)\;,
\eea

 Here below we prove that $\Upsilon^{[2]abc}_{\ell_1 m_1 \ell_2 m_2 \ell_3 m_3}(k_1,k_2,k_3)=\Upsilon^{[3]abc}_{\ell_1 m_1 \ell_2 m_2 \ell_3 m_3}(k_1,k_2,k_3)$. First of all, by construction,  we note that
 \be
  \M_{\ell m \ellvp \mvp \ellvq \mvq \barell \barm}^{ a(2)}(k; p, q) = \M_{\ell m \ellvq \mvq \ellvp \mvp \barell \barm}^{ a(2)}(k;q, p) \;.
 \ee
 Then, for example, using Eq. (\ref{Upsilon[3]})
 \bea
\Upsilon^{[3]abc}_{\ell_1 m_1 \ell_2 m_2 \ell_3 m_3}(k_1,k_2,k_3) = \sum_{{\ellvp}_1 {\mvp}_1 {\ellvq}_1 {\mvq}_1 {\barell}_1 {\barm}_1}   (-1)^{\ell_2 +\ell_3-m_2-m_3+{\mvp}_1+{\mvq}_1+{\barm}_1}
 \G^{{\ellvp}_1 {\barell}_1 \ell_{3} }_{-{\mvp}_1 {\barm}_1 -m_3 }\G^{{\ellvq}_1 {\barell}_1 \ell_{2} }_{{\mvq}_1 {\barm}_1 m_{2}} ~~~~~~ \nonumber \\
  \int \frac{q_2^2\ud q_2}{(2\pi)^3}   \frac{q_3^2\ud q_3}{(2\pi)^3}\left[\M_{\ell_1 m_1 {\ellvp}_1 {\mvp}_1 {\ellvq}_1 {\mvq}_1 {\barell}_1 {\barm}_1}^{a(2)}(k_1; q_3, q_2) ~ \M_{\ell_2}^{b(1)}(k_2, q_2)  ~\M_{\ell_3}^{c(1)}(k_3, q_3)\right] P_\Phi(q_2) P_\Phi(q_3) \nonumber \\
=\sum_{{\ellvp}_1 {\mvp}_1 {\ellvq}_1 {\mvq}_1 {\barell}_1 {\barm}_1}   (-1)^{\ell_2 +\ell_3-m_2-m_3+{\mvp}_1+{\mvq}_1+{\barm}_1}
 \G^{{\ellvp}_1 {\barell}_1 \ell_{3} }_{-{\mvp}_1 {\barm}_1 -m_3 }\G^{{\ellvq}_1 {\barell}_1 \ell_{2} }_{{\mvq}_1 {\barm}_1 m_{2}} ~~~~~~ \nonumber \\
  \int \frac{q_2^2\ud q_2}{(2\pi)^3}   \frac{q_3^2\ud q_3}{(2\pi)^3}\left[\M_{\ell_1 m_1 {\ellvq}_1 {\mvq}_1   {\ellvp}_1 {\mvp}_1 {\barell}_1 {\barm}_1}^{a(2)}(k_1; q_2, q_3) ~ \M_{\ell_2}^{b(1)}(k_2, q_2)  ~\M_{\ell_3}^{c(1)}(k_3, q_3)\right] P_\Phi(q_2) P_\Phi(q_3) \nonumber \\
  = \sum_{{\ellvq}_1 {\mvq}_1 {\ellvp}_1 {\mvp}_1 {\barell}_1 {\barm}_1}   (-1)^{\ell_2 +\ell_3-m_2-m_3+{\mvq}_1+{\mvp}_1+{\barm}_1}
 \G^{{\ellvq}_1 {\barell}_1 \ell_{3} }_{-{\mvq}_1 {\barm}_1 -m_3 }\G^{{\ellvp}_1 {\barell}_1 \ell_{2} }_{{\mvp}_1 {\barm}_1 m_{2}} ~~~~~~ \nonumber \\
  \int \frac{q_2^2\ud q_2}{(2\pi)^3}   \frac{q_3^2\ud q_3}{(2\pi)^3}\left[\M_{\ell_1 m_1 {\ellvp}_1 {\mvp}_1   {\ellvq}_1 {\mvq}_1 {\barell}_1 {\barm}_1}^{a(2)}(k_1; q_2, q_3) ~ \M_{\ell_2}^{b(1)}(k_2, q_2)  ~\M_{\ell_3}^{c(1)}(k_3, q_3)\right] P_\Phi(q_2) P_\Phi(q_3) \;,\nonumber 
  \eea
where, in the last step, we renamed ${\ellvp}_1 \to {\ellvq}_1$, ${\ellvq}_1 \to {\ellvp}_1$, $ {\mvp}_1 \to  {\mvq}_1$, and $ {\mvq}_1 \to  {\mvp}_1$. Now, it is easy to see that
\be
 \G^{{\ellvq}_1 {\barell}_1 \ell_{3} }_{-{\mvq}_1 {\barm}_1 -m_3 }\G^{{\ellvp}_1 {\barell}_1 \ell_{2} }_{{\mvp}_1 {\barm}_1 m_{2}} = \G^{{\ellvq}_1 {\barell}_1 \ell_{3} }_{{\mvq}_1 -{\barm}_1 m_3 }\G^{{\ellvp}_1 {\barell}_1 \ell_{2} }_{-{\mvp}_1 -{\barm}_1 -m_{2}} \;.\nonumber
\ee 
For the terms considered in this work,  $ \M_{\ell m \ellvp \mvp \ellvq \mvq \barell \barm}^{ a(2)}$ does not depend on the sign of $\bar m_1$ and we can redefine $\bar m_1$ as $- \bar m_1$. This concludes the proof.

\subsection{Proof of isotropy of Eq.(\ref{Bispectrum-general})}

Let us start with  $\Upsilon^{[2]abc}$. From the results obtained in Sec. \ref{calM-second_order}, it is possible to redefine the kernel at second order in the following way
\[\M_{\ell_1 m_1 {\ellvp}_1 {\mvp}_1 {\ellvq}_1 {\mvq}_1 {\barell}_1 {\barm}_1}^{a(2)}(k_1; q_2, q_3)= (-1)^{m_1}
 \G^{\ell_1 {\ellvp}_1 {\ellvq}_1 }_{-m_1  {\mvp}_1 {\mvq}_1}  \K_{\ell_1 {\ellvp}_1  {\ellvq}_1 {\barell}_1 }^{a(2)}(k_1; q_2, q_3)\]
and we find\footnote{Here $(-1)^{-m_i}=(-1)^{m_i}$ because all $m_i$ are integers.}
\bea
&&\sum_{ {\mvp}_1 {\mvq}_1  {\barm}_1}   (-1)^{-m_2-m_3 +m_1+{\mvp}_1+{\mvq}_1+{\barm}_1}  \G^{\ell_1 {\ellvp}_1 {\ellvq}_1 }_{-m_1  {\mvp}_1 {\mvq}_1}  \G^{{\ellvp}_1 {\barell}_1 \ell_{2} }_{-{\mvp}_1 {\barm}_1 -m_2 }\G^{{\ellvq}_1 {\barell}_1 \ell_{3} }_{{\mvq}_1 {\barm}_1 m_{3}}  \nonumber\\
&&= \left(
\begin{array}{ccc}
\ell_1 &\ell_2& \ell_3 \\
m_1 & m_2 &m_3
\end{array}
\right)
\sqrt{\frac{(2\ell_1+1)(2\ell_2+1)(2\ell_3+1)}{4\pi}} 
\times \Bigg[ 
(-1)^{-({\barell}_1 +  {\ellvq}_1 +{\ellvp}_1)} \frac{(2{\barell}_1+1)(2{\ellvq}_1+1)(2{\ellvp}_1+1)}{4\pi}
 \nonumber\\
&&\times
\left(
\begin{array}{ccc}
{\ellvq}_1 & \ell_1 & {\ellvp}_1 \\
0 & 0 & 0
\end{array}
\right)
\left(
\begin{array}{ccc}
{\ellvp}_1 & \ell_2 & {\barell}_1 \\
0 & 0 & 0
\end{array}
\right)
\left(
\begin{array}{ccc}
 {\barell}_1 & \ell_3 & {\ellvq}_1 \\
0 & 0 & 0
\end{array}
\right)
  \left\{
\begin{array}{ccc}
\ell_1 &\ell_2& \ell_3 \\
{\barell}_1 &  {\ellvq}_1 &{\ellvp}_1 
\end{array}
\right\} 
\Bigg]\;,
\eea
where we have used the identity
\bea
 \sum_{ m_4 m_5  m_6} (-1)^{\sum_{i=4}^6 (l_i-m_i)} \left(
\begin{array}{ccc}
\ell_5 & \ell_1 & \ell_6 \\
m_5 & -m_1 & -m_6
\end{array}
\right)
\left(
\begin{array}{ccc}
\ell_6 & \ell_2 & \ell_4 \\
m_6 & -m_2 & -m_4
\end{array}
\right)
\left(
\begin{array}{ccc}
 \ell_4 & \ell_3 & \ell_5 \\
m_4 & -m_3 & -m_5
\end{array}
\right) 
\nonumber \\
=
\left(
\begin{array}{ccc}
\ell_1 &\ell_2& \ell_3 \\
m_1 & m_2 &m_3
\end{array}
\right)
  \left\{
\begin{array}{ccc}
\ell_1 &\ell_2& \ell_3 \\
\ell_4 &  \ell_5 & \ell_6 
\end{array}
\right\} \;. ~~~~~~~~~~~~~~~~~~~~~~~~~~~~~
 \eea
Here 
 $ \left\{
\begin{array}{ccc}
j_1 & j_2& j_{12} \\
j_3 &  j & j_{23} 
\end{array}
\right\} $
is the definition of the Wigner $6j$ symbols. Due to the properties of the Gaunt integrals we note that $\ell_1 +\ell_2+\ell_3$ is even.
Then we can  recast Eq.(\ref{Upsilon[2]}) as
\bea
\label{Upsilon[2]-2}
\Upsilon^{[2]abc}_{\ell_1 m_1 \ell_2 m_2 \ell_3 m_3}(k_1,k_2,k_3) =   \left(
\begin{array}{ccc}
\ell_1 &\ell_2& \ell_3 \\
m_1 & m_2 &m_3
\end{array}
\right)
\sqrt{\frac{(2\ell_1+1)(2\ell_2+1)(2\ell_3+1)}{4\pi}} 
 \sum_{{\ellvp}_1  {\ellvq}_1 {\barell}_1} (-1)^{-({\barell}_1 +  {\ellvq}_1 +{\ellvp}_1)} \nonumber \\
\times  \frac{(2{\barell}_1+1)(2{\ellvq}_1+1)(2{\ellvp}_1+1)}{4\pi}  \left(
\begin{array}{ccc}
{\ellvq}_1 & \ell_1 & {\ellvp}_1 \\
0 & 0 & 0
\end{array}
\right)
\left(
\begin{array}{ccc}
{\ellvp}_1 & \ell_2 & {\barell}_1 \\
0 & 0 & 0
\end{array}
\right)
\left(
\begin{array}{ccc}
 {\barell}_1 & \ell_3 & {\ellvq}_1 \\
0 & 0 & 0
\end{array}
\right)
  \left\{
\begin{array}{ccc}
\ell_1 &\ell_2& \ell_3 \\
{\barell}_1 &  {\ellvq}_1 &{\ellvp}_1 
\end{array}
\right\}  ~~~~~~~~~~~~\nonumber\\
\times  (-1)^{\ell_2 +\ell_3} \int \frac{q_2^2\ud q_2}{(2\pi)^3}   \frac{q_3^2\ud q_3}{(2\pi)^3}\left[ \K_{\ell_1 {\ellvp}_1  {\ellvq}_1 {\barell}_1 }^{a(2)}(k_1; q_2, q_3)  ~ \M_{\ell_2}^{b(1)}(k_2, q_2)  ~\M_{\ell_3}^{c(1)}(k_3, q_3)\right] P_\Phi(q_2) P_\Phi(q_3)~~~~~~~~~~~~~~~~
\eea
Taking into account that $\Upsilon^{[2]abc}=\Upsilon^{[3]abc}$ and using the same prescription for all additive terms in Eq.(\ref{Bispectrum-general}) we obtain, at second order, the final result
\be
\langle \Delta^g_{\ell_1  m_1} (k_1) \Delta^g_{\ell_2 m_2}(k_2) \Delta^g_{\ell_3 m_3}(k_2)\rangle = \left(
\begin{array}{ccc}
\ell_1 &\ell_2& \ell_3 \\
m_1 & m_2 &m_3
\end{array}
\right)
B_{\ell_1\ell_2 \ell_3 }(k_1,k_2,k_3)\;.
\ee


\section{Transfer functions at first and second order}\label{Transfer_functions}

\subsection{Transfer functions at first-order}
At background level,
 \bea
 \cH^2&=&a^2 \cH_0^2 \left(\Omo a^{-3}+\OLo\right), \\
 \frac{\cH'}{\cH^2}&=&\left(1-\frac{3}{2} \Omo a^{-3}\right),\\
 \bar \rho_m ' &=& -3 \cH \bar \rho_m \;,
\eea
where $\Omo=8\pi \bar \rho_{\rm m 0}/3 \cH_0^2$, $\OLo=1-\Omo$ and we set $a_0=1$. A useful relation is the following
$$\left(\frac{\cH}{a}\right)'+\frac{3}{2} \frac{\Omega_m}{a} \cH^2 =0\;,$$
where $\Omega_{\rm m} =8\pi \bar \rho_{\rm m 0}/3 a^2\cH^2= a^{-3} \Omo$ or, equivalently, $a \cH^2 \Omega_{\rm m} = \cH_0^2 \Omo$.
and, at the first perturbative order, for $k<k_{\rm eq}$, where $k_{\rm eq}=0.073{\rm Mpc}^{-1}\Omo h^2$, for our aims the relevant equations are
\bea
\Phi''+3\cH\Phi'+ (2\cH' + \cH^2)\Phi &=&0, \label{Phi-first}  \\
\Phi' + \cH \Phi&=&- 4\pi G \bar \rho_m a^2 v = -\frac{3}{2a}\cH_0^2 \Omo v\;, \label{v-first}\\
 v'+\cH v+\Phi &=& 0, \label{v-first2}\\
\nabla^2 \Phi &=& 4\pi G \bar \rho_m a^2 \delta_{m  {\rm CS}}^{(1)}=\frac{3}{2a}\cH_0^2 \Omo \delta_{m  {\rm CS}}^{(1)} ,\label{eq:poisson}
\eea
Defining $\Phi(\bk,\eta) = D(\eta) \Phi_0(\bk)/a$ where $D(\eta_0)=1$, and $\Phi_0(\bk)=\Phi(\bk,\eta_0)$, and combining Eqs. (\ref{Phi-first}) and (\ref{eq:poisson}) we find the well-known differential equation
\be \label{D-eq}
{\delta^{(1)}_{m {\rm CS}}}'' + \cH {\delta^{(1)}_{m {\rm CS}}}'- \frac{3}{2}\cH^2 \Omega_{\rm m}\delta^{(1)}_{m {\rm CS}} =0 \quad \quad {\rm or,~equivalently,} \quad \quad   D'' + \cH D'-\frac{3}{2a}\cH_0^2 \Omo  D =0 \;.
\ee 
Here selecting only $D(\eta)$ the growth mode we find $\delta_{m  {\rm CS}}^{(1)}=D\delta_{m  {\rm CS}0}^{(1)}$, and as defining the growth factor $f=\ud \ln D / \ud \ln a$ it is easy to see that
\be
\left(f+ \frac{3}{2} \Omega_{\rm m}\right) \cH^2 D= {\rm const}.
\ee 
and we can write
\be
D=\frac{\left(f_0+ \frac{3}{2} \Omo\right)}{\left(f+ \frac{3}{2} \Omega_{\rm m}\right)}\frac{\cH_0^2}{\cH^2}=\frac{5D_{\rm in}}{2\left(f+ \frac{3}{2} \Omega_{\rm m}\right)}\frac{\cH_{\rm in}^2}{\cH^2}\;,
\ee
where we use the subscripts ${\rm in}$ as generic time during the Einstein de Sitter (EdS) period (for example during the recombination epoch), i.e. when $\Omega_{\rm m\, in}=1,~D_{\rm in}\propto a_{\rm in},~ f_{\rm in}=1$.  
Solving directly Eq. (\ref{D-eq}), we immediately find 
\be
D(\eta)=\frac{\cH(\eta) \left[1 +\frac{5 \cH_{\rm in}^3}{2 a_{\rm in}} \int_{a_{\rm in}}^{a(\eta)} \frac{\ud \tilde a}{\cH^3(\tilde a)}\right]}{a(\eta) \cH_0 \left[1 +\frac{5 \cH_{\rm in}^3}{2 a_{\rm in} } \int_{a_{\rm in}}^1 \frac{\ud \tilde a}{\cH^3(\tilde a)}\right]}\;.
\ee
From Eq.(\ref{v-first}), for the velocity potential we have
\be
\label{v-first-2}
v=-\frac{2}{3\cH \Omega_{\rm m}}f  \Phi
\ee
and from Eq.(\ref{v-first2}), for the derivate of the velocity potential
\be
v'=\frac{2}{3 \Omega_{\rm m}} \left(f- \frac{3}{2} \Omega_{\rm m}\right) \Phi\;.
\ee

In order to compute  all $\T^{i}$, we also need the expression for the density contrast in the Poisson gauge, i.e.
\be
\frac{3}{2}\frac{\cH_0^2\Omo}{a}\delta_{m}^{(1)} = \left(\nabla^2-3f\cH^2\right)\Phi
\ee
and the comoving curvature perturbation at first-order:
\be\label{zeta}
-\zeta (\bx)=\Phi+\frac{2}{3}\frac{1}{1+w}\left(\Phi + \cH^{-1} \Phi' \right)\;,
\ee
 where $w$ is the the total equation state which for $\eta \gtrsim \eta_{\rm rec}$ can be written as $w=\Omega_{\rm m}-1$ in $\Lambda$CDM model and we can write
 $$\zeta (\bx)=-\frac{2}{3} \frac{ \left(f+ \frac{3}{2} \Omega_{\rm m}\right) }{ \Omega_{\rm m}} \Phi\;.$$
 For completeness let us also write explicitly linear metric perturbations quantities in SC gauge with $\zeta$. From \cite{Matarrese:1997ay, Bartolo:2010rw, Bruni:2013qta} 
 \be
 -\zeta (\bx)= \psi^{(1)}  + \frac{1}{6} \nabla^2 \chi^{(1)} \quad {\rm and}   \quad \chi^{(1)}  = \frac{2}{\left(f+ \frac{3}{2} \Omega_{\rm m}\right) \cH^2}  \zeta = - \frac{4}{3 \Omega_{\rm m} \cH^2} \Phi \quad {\rm or} \quad  -\psi^{(1)} =\frac{1}{3}\frac{1}{\left(f+ \frac{3}{2} \Omega_{\rm m}\right) \cH^2} \nabla^2 \zeta + \zeta\;.
 \ee
 
Finally, for scales $k>k_{\rm eq}$, we have to consider the evolution of the perturbation modes that enter the horizon before and around the epoch of matter-radiation equality, i.e. the {\it Meszaros effect}.\footnote{Note that for $\eta<\eta_{\rm rec}$ and $k \gtrsim  k_{\rm eq}$ is not valid Eq.\ (\ref{zeta}); for example see \cite{Mukhanov:2005sc}.}
At linear order and for  $\eta \gtrsim \eta_{\rm rec}$, it is possible to implement these scales analytically by using the transfer function $T_m(k)$ defined by 
Eisenstein \& Hu \cite{Eisenstein:1997jh} (or BBKS by \cite{Bardeen:1985tr}), i.e.
\be\label{Phi_k-1}
\Phi(\bk,\eta) = \frac{D(\eta)}{a} \Phi_0(\bk)  = \frac{9}{10} \frac{D(\eta)}{D_{\rm in}}\frac{a_{\rm in}}{a} T_m(k) \Phi_{\rm p}(\bk) =  -\frac{3}{5} \frac{D(\eta)}{D_{\rm in}}\frac{a_{\rm in}}{a} T_m(k) \zeta_{\rm p}(\bk) \;, 
\ee
where $ \Phi_{\rm p}(\bk)=-2 \zeta(\bk)/3 $\;. Here  $\zeta(\bk)$ is the primordial curvature perturbation set at the inflation epoch.

In conclusion,  we find
\bea
\T^{\Phi}(\bk,\eta) &=& \frac{9}{10} \frac{D(\eta)}{D_{\rm in}}\frac{a_{\rm in}}{a} T_m(k)\;, \\
 \T^{\delta_{\rm SC}}(\bk,\eta) &=&  -\frac{3}{5\cH_0^2 \Omo} \frac{a_{\rm in}}{D_{\rm in}} D(\eta)  k^2 T_m(k)\;,\\
  \T^{v}(\bk,\eta) &=&  -\frac{3}{5\cH_0^2 \Omo} \frac{a_{\rm in}}{D_{\rm in}}  \cH f D(\eta) T_m(k)\;, \\
  T^{\zeta}(\bk,\eta) &=&  -\frac{3}{5 \Omega_{\rm m}} \left(f+ \frac{3}{2} \Omega_{\rm m}\right) \frac{D(\eta)}{D_{\rm in}}\frac{a_{\rm in}}{a} T_m(k) \;, \\
   T^{\chi}(\bk,\eta) &=& - \frac{6}{ 5\Omega_{\rm m} \cH^2}  \frac{D(\eta)}{D_{\rm in}}\frac{a_{\rm in}}{a} T_m(k)\;, \\
   \T^{\delta_{\rm m}}(\bk,\eta) &=& -\frac{3}{5\cH_0^2 \Omo} \frac{a_{\rm in}}{D_{\rm in}} D(\eta)  \left( k^2 + 3 f \cH^2\right) T_m(k) \;,\\
   \T^{\delta_g} (\bk , \eta)  &=& -\frac{3}{5} \frac{D(\eta)}{D_{\rm in}}\frac{a_{\rm in}}{a}\left\{a \frac{\left[ b_{10} k^2 - b_{\nabla^2\delta} k^4+ (3-b_e)f \cH^2\right]}{\cH_0^2 \Omo}  + \frac{ \left(f+ \frac{3}{2} \Omega_{\rm m}\right) }{ \Omega_{\rm m}}b_{01}\right\}T_m(k)\;, \\
   \T^{ \Q^{(1)}} (\bk, \eta, \bar L)  & =&   \frac{3}{5} \frac{D(\eta)}{D_{\rm in}}\frac{a_{\rm in}}{a}\left\{a \frac{ \left[  \frac{\p   b_{10} (\bk, \eta,\bar L)}{\p \ln \bar L}  k^2 -\frac{\p   b_{\nabla^2\delta} (\bk, \eta,\bar L)}{\p \ln \bar L}  k^4     - \frac{\p b_e( \eta,\bar L)}{\p  \ln \bar L} f \cH^2\right]  }{\cH_0^2 \Omo}  + \frac{ \left(f+ \frac{3}{2} \Omega_{\rm m}\right) }{ \Omega_{\rm m}}\frac{\p   b_{01}(\bk, \eta,\bar L)}{\p \ln \bar L}\right\}T_m(k)   \;. \nonumber\\
\eea

\subsection{Transfer functions at second order}\label{F_2}

In this subsection we give some examples in which we provide a prescription to compute kernels at second order $F_{\bar \ell}^{b (2)}(p, q; \eta)$, starting from numerical or analytical results found in literature.

\subsubsection{Large scales, i.e. for $p,q<k_{\rm eq}$} \label{LS}

Here below, as an illustrative example, we will write explicitly all kernels  $F_{\bar \ell}^{b (2)}(p, q; \eta)$.
In particular, we will use the analytical results obtained in Refs. \cite{Bartolo:2010rw,  Bruni:2013qta, Villa:2015ppa, Tram:2016cpy}.

Starting from \cite{Villa:2015ppa} (see also \cite{Bartolo:2005kv})
\bea
\label{Phi2}
\Phi^{(2)}(\bx , \eta)&=&\left[3\left(\frac{D}{a}\right)^2 - 2\frac{D}{a}\frac{D_{\rm in}}{a_{\rm in}} \left(f_{\rm NL} + \frac{5}{6}\right)+ \frac{2}{3} \frac{D^2 f^2}{a^2 \Omega_{\rm m}} \right] \left[\Phi_0(\bx)\right]^2 + 12 \left[2 \left(\frac{D}{a}\right)^2 - \frac{5}{3}\frac{D}{a}\frac{D_{\rm in}}{a_{\rm in}}+ \frac{2}{3} \frac{D^2 f^2}{a^2 \Omega_{\rm m}} \right]\Theta_0 (\bx) \nonumber\\
&&+\frac{2}{3} \frac{D^2}{a^2 \Omega_{\rm m}} \p_i \Phi_0(\bx) \p^i\Phi_0(\bx) - \frac{4}{3}\frac{D^2(1+\F/D^2 )}{a^2 \Omega_{\rm m}}  \Xi_0 (\bx) \;, \\
\label{Psi2}
\Psi^{(2)}(\bx , \eta)&=&\left[-\left(\frac{D}{a}\right)^2 - 2\frac{D}{a}\frac{D_{\rm in}}{a_{\rm in}} \left(f_{\rm NL} + \frac{5}{6}\right)+ \frac{2}{3} \frac{D^2 f^2}{a^2 \Omega_{\rm m}} \right] \left[\Phi_0(\bx)\right]^2 + 12 \left[ \left(\frac{D}{a}\right)^2 - \frac{5}{3}\frac{D}{a}\frac{D_{\rm in}}{a_{\rm in}}\right]\Theta_0 (\bx) \nonumber\\
&&+\frac{2}{3} \frac{D^2}{a^2 \Omega_{\rm m}} \p_i \Phi_0(\bx) \p^i\Phi_0(\bx) - \frac{4}{3}\frac{D^2(1+\F/D^2 )}{a^2 \Omega_{\rm m}}  
\Xi_0 (\bx) \;, \\
\label{v2}
v^{(2)}(\bx , \eta)&=&\frac{2}{3}\frac{D f}{a \cH \Omega_{\rm m}}  \left[\frac{D}{a} + 2\frac{D_{\rm in}}{a_{\rm in}} \left(f_{\rm NL} - \frac{5}{3}\right) \right] \left[\Phi_0(\bx)\right]^2 -8 \frac{D^2 f}{a^2 \cH \Omega_{\rm m}} \Theta_0 (\bx) -\frac{4}{9} \frac{D^2f}{a^2\cH^3 \Omega^2_{\rm m}} \p_i \Phi_0(\bx) \p^i\Phi_0(\bx)\nonumber\\
&& + \frac{8}{9}\frac{\F' }{a^2 \cH^4 \Omega_{\rm m}^2}  \Xi_0 (\bx) \;,
\eea
where $$\Theta_0 (\bx)=\frac{1}{6}\nabla^{-2}   \left[ \p_i \Phi_0(\bx) \p^i\Phi_0(\bx) - 3\nabla^{-2} \p_i \p_j \left( \p^i\Phi_0(\bx)   \p^j\Phi_0(\bx) \right)\right] =  \nabla^{-2}   \left[ \Xi_0 (\bx) -\frac{1}{3}\p_i \Phi_0(\bx) \p^i\Phi_0(\bx)\right]$$ 
and\footnote{In \cite{Bartolo:2005kv, Villa:2015ppa,  Tram:2016cpy} $\Psi_0$ is equal to our definition $\Xi_0 $.}
$$  \Xi_0 (\bx)=- \frac{1}{2}\nabla^{-2} \left[ \left(\nabla^2 \Phi_0(\bx) \right)^2 - \p_i \p_j\Phi_0(\bx) \p^i \p^j\Phi_0(\bx)  \right]   \;.$$
Here $\F$ is the solution of the following differential equation
\be
\F''+\cH\F'-\frac{3}{2}\cH^2 \Omega_{\rm m} \F=\frac{3}{2}\cH^2 \Omega_{\rm m} D^2\;.
\ee
Instead, for the density contrast at second order in the CS gauge one finds \cite{Bartolo:2010rw,  Bruni:2013qta, Villa:2015ppa}
 \bea
 \label{deltamSC2-total}
  \delta_{m  {\rm CS}}^{(2)}&=&-{8\over 3} \frac{D}{a \cH^2 \Omega_{\rm m}}\frac{D_{\rm in}}{a_{\rm in}} \left[ \left(f_{\rm NL} + \frac{5}{12}\right)\p_i \Phi_0(\bx) \p^i\Phi_0(\bx) +  \left(f_{\rm NL}-{5\over 3}\right)  \Phi_0(\bx) \nabla^2 \Phi_0(\bx)\right] \nonumber\\
  &&+{4 \over 9}  \frac{D^2}{a^2 \cH^4 \Omega^2_{\rm m}}\left[\left(1+{\F \over D^2} \right) \left(\nabla^2 \Phi_0(\bx) \right)^2+ \left(1-{\F \over D^2} \right) \p_i \p_j\Phi_0(\bx) \p^i \p^j\Phi_0(\bx) \right]\;.
  \eea
  To the aim of computing the halo density contrast to second order the CS gauge second-order density contrast will have to be multiplied by the linear bias term. 
 As widely discussed in the recent literature, however (see \cite{Dai:2015jaa, dePutter:2015vga, Bartolo:2015qva}) this procedure makes unavoidable the use of some sort of ``peak background" or ``short-long" splitting, where local coordinates are defined within a patch whose size is much larger than the typical halo Lagrangian radius but much smaller than the distance over which correlations are
computed; as a consequence of this short-long splitting, inherent in the halo bias approach, the $-5/3$ additive term in the first line of the above expression, which would act as a local $f_{\rm NL}$-like contribution,  
can be removed (in the strict squeezed limit) by a local (i.e. within the patch) coordinate transformation, 
yielding the effective expression 
   \bea
 \label{deltamSC2}
  \delta_{m  {\rm CS}}^{(2)}&\simeq&-{8\over 3} \frac{D}{a \cH^2 \Omega_{\rm m}}\frac{D_{\rm in}}{a_{\rm in}} \left[ \left(f_{\rm NL} + \frac{5}{12}\right)\p_i \Phi_0(\bx) \p^i\Phi_0(\bx) +  f_{\rm NL}  \Phi_0(\bx) \nabla^2 \Phi_0(\bx)\right] \nonumber\\
  &&+{4 \over 9}  \frac{D^2}{a^2 \cH^4 \Omega^2_{\rm m}}\left[ \left(1+{\F \over D^2} \right) \left(\nabla^2 \Phi_0(\bx) \right)^2+ \left(1- {\F \over D^2} \right) \p_i \p_j\Phi_0(\bx) \p^i \p^j\Phi_0(\bx) \right]
  \eea
which can be used in connection with halo bias calculations, up to negligible corrections of order $\left(\lambda_{\rm S}/\lambda_{\rm L}\right)^2$, with $\lambda_{\rm S}$ and $\lambda_{\rm L}$ 
typical scales much smaller and much larger than the patch size, respectively. 
  
Transforming to Fourier space Eqs.(\ref{Phi2}), (\ref{Psi2}), (\ref{v2}), (\ref{deltamSC2}), and applying Eqs.(\ref{F2-b}) and (\ref{Phi_k-1}) we find the following kernels
 \bea
 F^{\Phi (2)}(\bp, \bq, \tilde k; \eta) &=& 3 - 2\frac{a}{a_{\rm in}}\frac{D_{\rm in}}{D } \left(f_{\rm NL} + \frac{5}{6}\right)+ \frac{2}{3} \frac{f^2}{ \Omega_{\rm m}}  + 12 \left(2 - \frac{5}{3}\frac{a}{a_{\rm in}}\frac{D_{\rm in}}{D }+ \frac{2}{3} \frac{f^2}{ \Omega_{\rm m}} \right) F^\Theta (\bp, \bq, \tilde k) \nonumber\\
&&+\frac{2}{3} \frac{1}{\Omega_{\rm m}} p\,q \, (\hat \bp \cdot \hat \bq) - \frac{4}{3}\frac{(1+\F/D^2 )}{ \Omega_{\rm m}}  F^\Xi (\bp, \bq, \tilde k) \;, \\
 &&\nonumber\\
  F^{\Psi (2)}(\bp, \bq, \tilde k; \eta) &=& -1 - 2\frac{a}{a_{\rm in}}\frac{D_{\rm in}}{D } \left(f_{\rm NL} + \frac{5}{6}\right)+ \frac{2}{3} \frac{f^2}{ \Omega_{\rm m}}  + 12 \left(1 - \frac{5}{3}\frac{a}{a_{\rm in}}\frac{D_{\rm in}}{D } \right) F^\Theta (\bp, \bq, \tilde k) \nonumber\\
&&+\frac{2}{3} \frac{1}{\Omega_{\rm m}} p\,q \, (\hat \bp \cdot \hat \bq) - \frac{4}{3}\frac{(1+\F/D^2 )}{ \Omega_{\rm m}}  F^\Xi (\bp, \bq, \tilde k) \;, \\
 &&\nonumber\\
  F^{v (2)}(\bp, \bq, \tilde k; \eta) &=&   \frac{2}{3}\frac{ f}{ \cH \Omega_{\rm m}}  \left[1+ 2\frac{a}{a_{\rm in}}\frac{D_{\rm in}}{D } \left(f_{\rm NL} - \frac{5}{3}\right) \right] 
  - 8 \frac{ f}{ \cH \Omega_{\rm m}} F^\Theta (\bp, \bq, \tilde k)  -\frac{4}{9} \frac{f}{\cH^3 \Omega^2_{\rm m}}  p\,q \, (\hat \bp \cdot \hat \bq)\nonumber\\
&& + \frac{8}{9}\frac{\F' }{D^2 \cH^4 \Omega_{\rm m}^2}  F^\Xi (\bp, \bq, \tilde k)\;,\\
 &&\nonumber\\
  F^{\delta_{m  {\rm CS}} (2)}(\bp, \bq, \tilde k; \eta) &=& {4\over 3} \frac{1}{ \cH^2 \Omega_{\rm m}}\frac{a}{a_{\rm in}}\frac{D_{\rm in}}{D }\left[2 \left(f_{\rm NL} + \frac{5}{12}\right) p\,q \, (\hat \bp \cdot \hat \bq)+  f_{\rm NL}\left(p^2 + q^2\right)\right] \nonumber\\
  &&+{4 \over 9}  \frac{1}{\cH^4 \Omega^2_{\rm m}}\left[\left(1+{\F \over D^2} \right)\, p^2\,q^2+\left(1-{\F \over D^2} \right)\,p^2\,q^2\, (\hat \bp \cdot \hat \bq)^2  \right]\;,
  \eea
where (see also \cite{Tram:2016cpy}),
 \bea \label{FTheta}
F^\Theta (\bp, \bq, \tilde k) &=&\frac{1}{2}\left\{\frac{1}{3} \frac{p\,q}{ \tilde k^2}(\hat \bp \cdot \hat \bq)- \frac{p^2\,q^2}{ \tilde k^4}\left[ 1+ \left(\frac{p}{q}+\frac{q}{p}\right)(\hat \bp \cdot \hat \bq)+(\hat \bp \cdot \hat \bq)^2\right]\right\}=-\frac{p\,q}{\tilde k^2} \left[ \frac{F^\Xi (\bp, \bq, \tilde k)}{p\,q}  +\frac{1}{3}(\hat \bp \cdot \hat \bq) \right]\;, \nonumber \\ \\ 
\label{FXi}
 F^\Xi (\bp, \bq, \tilde k) &=&\frac{1}{2}\frac{p^2\,q^2}{ \tilde k^2}\bigg[ 1 - (\hat \bp \cdot \hat \bq)^2\bigg] \;.
 \eea
 Finally, from Eq. (\ref{dg2-3}) we have
  \bea
  \label{Fdg2}
&&F^{\delta_{g }(2)}(\bp, \bq, \tilde k; \eta)  =  b_{10}   F^{\delta_{m  {\rm CS}} (2)}(\bp, \bq, \tilde k; \eta)   - (b_e-3) \cH   F^{v (2)}(\bp, \bq, \tilde k; \eta)    -\frac{2f}{ \Omega_{\rm m}}(b_e -3)\frac{(\bp + \bq)^2}{\tilde k^2}   \nonumber\\
 && + \frac{4b_{02}}{9 \Omega^2_{\rm m} }\left(f+ \frac{3}{2} \Omega_{\rm m}\right)^2  +   \frac{4 f^2}{9 \Omega^2_{\rm m} } \left[(b_e -3)^2  + \frac{\p b_e}{\p \ln a} \right] +\frac{2f}{3 \Omega_{\rm m}}(b_e -3)(1-f) - \frac{8 f}{9 \cH\Omega^2_{\rm m} }\left(f+ \frac{3}{2} \Omega_{\rm m}\right)\left[b_{01}' +\cH( b_e-3)   b_{01}  \right] \nonumber\\
&& +\left[  - \frac{4f^2 b_{10}}{9 \cH^2 \Omega^2_{\rm m} }  -\frac{4f}{9 \cH^3 \Omega_{\rm m}^2 }  \left[b_{10}' +\cH( b_e-3)   b_{10}  \right]   + \frac{4b_{11}}{9 \cH^2 \Omega^2_{\rm m} }\left(f+ \frac{3}{2} \Omega_{\rm m}\right)  \right]p\,q \left(\frac{p}{q}+\frac{q}{p}\right)   + \frac{4b_{20}}{9 \cH^4 \Omega^2_{\rm m} } p^2\,q^2  \nonumber\\
&& + \frac{4b_{10}}{9 \cH^4 \Omega_{\rm m}^2 } p^2\,q^2 \left(\frac{p}{q}+\frac{q}{p}\right)(\hat \bp \cdot \hat \bq) + \frac{8b_{01}}{9 \cH^2 \Omega^2_{\rm m} }\left(f+ \frac{3}{2} \Omega_{\rm m}\right)  p\,q (\hat \bp \cdot \hat \bq)  -\frac{b_{s^2}}{3} +b_{s^2}(\hat \bp \cdot \hat \bq)^2 +   b_{\nabla^2\delta} F^{\nabla^2 \delta_{m  {\rm CS}} (2)}(\bp, \bq, \tilde k; \eta)  \nonumber\\
&& - \frac{4b_{\nabla^2\delta}}{9 \cH^4 \Omega_{\rm m}^2 } p^3\,q^3 \left(\frac{p^2}{q^2}+\frac{q^2}{p^2}\right) (\hat \bp \cdot \hat \bq) +  \frac{4b_{(\nabla^2\delta)^2} }{9 \cH^4 \Omega^2_{\rm m} } p^4\,q^4 - \frac{4b_{\delta\nabla^2\delta}}{9 \cH^4 \Omega_{\rm m}^2 } p^3\,q^3 \left(\frac{p}{q}+\frac{q}{p}\right) - \frac{8b_{(\p\delta)^2}}{9 \cH^4 \Omega_{\rm m}^2 } p^3\,q^3 (\hat \bp \cdot \hat \bq) \nonumber\\
&&-\frac{4f}{9 \cH^3 \Omega_{\rm m}^2 } \left[b_{\nabla^2\delta}' +\cH( b_e-3)   b_{\nabla^2\delta} \right] p\,q \left(\frac{p}{q}+\frac{q}{p}\right) 
+ \frac{4f^2 b_{\nabla^2\delta}}{9 \cH^3 \Omega_{\rm m}^2 } p^2\,q^2 \left(\frac{p^2}{q^2}+\frac{q^2}{p^2}\right)\;,
 \eea
where
\be\label{Fnabla^2delta_mCS}
F^{\nabla^2 \delta_{m  {\rm CS}} (2)}(\bp, \bq, \tilde k; \eta) = - \tilde k^2 F^{\delta_{m  {\rm CS}} (2)}(\bp, \bq, \tilde k; \eta) \;.
\ee

Now, using Eq.(\ref{projLegendre}) we can rewrite the above relations in the following way:
 \bea
 F_{\ell}^{\Phi (2)}(p, q ; \eta) &=& \left[3 - 2\frac{a}{a_{\rm in}}\frac{D_{\rm in}}{D } \left(f_{\rm NL} + \frac{5}{6}\right)+ \frac{2}{3} \frac{f^2}{ \Omega_{\rm m}} \right] \delta^K_{\ell0} +\frac{2}{3} \frac{1}{\Omega_{\rm m}} p\,q   \delta^K_{\ell1} + 12 \left(2 - \frac{5}{3}\frac{a}{a_{\rm in}}\frac{D_{\rm in}}{D }+ \frac{2}{3} \frac{f^2}{ \Omega_{\rm m}} \right) F_{\ell}^\Theta (p,q) \nonumber\\
&& - \frac{4}{3}\frac{(1+\F/D^2 )}{ \Omega_{\rm m}}  F_{\ell}^\Xi (p, q) \;, \\
  F_{\ell}^{\Psi (2)}( p, q ; \eta) &=& \left[ -1 - 2\frac{a}{a_{\rm in}}\frac{D_{\rm in}}{D } \left(f_{\rm NL} + \frac{5}{6}\right)+ \frac{2}{3} \frac{f^2}{ \Omega_{\rm m}}  \right] \delta_{\ell0}^K +\frac{2}{3} \frac{1}{\Omega_{\rm m}} p\,q \delta^K_{\ell1} + 12 \left(1 - \frac{5}{3}\frac{a}{a_{\rm in}}\frac{D_{\rm in}}{D } \right)  F_{\ell}^\Theta (p,q) \nonumber\\
&&     - \frac{4}{3}\frac{(1+\F/D^2 )}{ \Omega_{\rm m}}  F_{\ell}^\Xi (p, q) \;, \\
  F_{\ell}^{v (2)}(p, q;\eta) &=&   \frac{2}{3}\frac{ f}{ \cH \Omega_{\rm m}}  \left[1+ 2\frac{a}{a_{\rm in}}\frac{D_{\rm in}}{D } \left(f_{\rm NL} - \frac{5}{3}\right) \right] \delta_{\ell0}^K  -\frac{4}{9} \frac{f}{\cH^3 \Omega^2_{\rm m}}  p\,q \, \delta^K_{\ell1}   - 8 \frac{ f}{ \cH \Omega_{\rm m}} F_{\ell}^\Theta (p,q)  + \frac{8}{9}\frac{\F' }{D^2 \cH^4 \Omega_{\rm m}^2}  F_{\ell}^\Xi (p, q)    \;,\nonumber\\ \\
 &&\nonumber\\
  F_{\ell}^{\delta_{m  {\rm CS}} (2)}(p, q;\eta)  &=& {4\over 3} \frac{1}{ \cH^2 \Omega_{\rm m}}\frac{a}{a_{\rm in}}\frac{D_{\rm in}}{D }\left[  f_{\rm NL}\left(p^2 + q^2\right)  \delta_{\ell0}^K  +2 \left(f_{\rm NL} + \frac{5}{12}\right) p\,q \,  \delta^K_{\ell1} \right] \nonumber\\
  &&+{8 \over 27}  \frac{1}{\cH^4 \Omega^2_{\rm m}} p^2\,q^2 \left[ \left(2+{\F \over D^2} \right) \delta_{\ell0}^K + \left(1-{\F \over D^2} \right) \delta^K_{\ell2} \right]\;,\\
  F_{\ell}^{\delta_{g }(2)}(p, q;\eta)  &=&  b_{10}    F_{\ell}^{\delta_{m  {\rm CS}} (2)}(p, q,\eta)   - (b_e-3) \cH    F_{\ell}^{v (2)}(p, q,\eta)   
  +   b_{\nabla^2\delta} F_{\ell}^{\nabla^2 \delta_{m  {\rm CS}} (2)}(p, q;\eta) \nonumber\\
&&  +\Bigg\{ \frac{4b_{02}}{9 \Omega^2_{\rm m} }\left(f+ \frac{3}{2} \Omega_{\rm m}\right)^2   +   \frac{4 f^2}{9 \Omega^2_{\rm m} } \left[(b_e -3)^2  + \frac{\p b_e}{\p \ln a} \right]  -\frac{2f}{3 \Omega_{\rm m}}(b_e -3)(f+2) + \frac{4b_{20}}{9 \cH^4 \Omega^2_{\rm m} } p^2\,q^2 \nonumber\\
 && - \frac{8 f}{9 \cH\Omega^2_{\rm m} }\left(f+ \frac{3}{2} \Omega_{\rm m}\right)\left[b_{01}' +\cH( b_e-3)   b_{01}  \right] +\bigg[  - \frac{4f^2 b_{10}}{9 \cH^2 \Omega^2_{\rm m} }  -\frac{4f}{9 \cH^3 \Omega_{\rm m}^2 }  \left[b_{10}' +\cH( b_e-3)   b_{10}  \right] \nonumber\\
&&   + \frac{4b_{11}}{9 \cH^2 \Omega^2_{\rm m} }\left(f+ \frac{3}{2} \Omega_{\rm m}\right) - \frac{4b_{\delta\nabla^2\delta}}{9 \cH^4 \Omega_{\rm m}^2 } p^2\,q^2 -\frac{4f}{9 \cH^3 \Omega_{\rm m}^2 } \left[b_{\nabla^2\delta}' +\cH( b_e-3)   b_{\nabla^2\delta} \right] \bigg]p\,q \left(\frac{p}{q}+\frac{q}{p}\right)   \nonumber\\
&&+  \frac{4b_{(\nabla^2\delta)^2} }{9 \cH^4 \Omega^2_{\rm m} } p^4\,q^4  + \frac{4f^2 b_{\nabla^2\delta}}{9 \cH^3 \Omega_{\rm m}^2 } p^2\,q^2 \left(\frac{p^2}{q^2}+\frac{q^2}{p^2}\right) \Bigg\} \delta_{\ell 0}+\bigg[ \frac{4b_{10}}{9 \cH^4 \Omega_{\rm m}^2 } p\,q \left(\frac{p}{q}+\frac{q}{p}\right) \nonumber\\
&& + \frac{8b_{01}}{9 \cH^2 \Omega^2_{\rm m} }\left(f+ \frac{3}{2} \Omega_{\rm m}\right)- \frac{8b_{(\p\delta)^2}}{9 \cH^4 \Omega_{\rm m}^2 } p^2\,q^2  - \frac{4b_{\nabla^2\delta}}{9 \cH^4 \Omega_{\rm m}^2 } p^2\,q^2 \left(\frac{p^2}{q^2}+\frac{q^2}{p^2}\right)\bigg] p\,q \delta^K_{\ell1} +\frac{3}{2}b_{s^2}\delta^K_{\ell2}
\eea
In order to write explicitly  $ F_{\ell}^{\nabla^2 \delta_{m  {\rm CS}} (2)}(p, q;\eta),~F_{\ell}^\Theta (p,q) $ \& $F_{\ell}^\Xi (p, q)$, let us consider the following relations  (see also \cite{Tram:2016cpy})
$$(\hat \bp \cdot \hat \bq)=  \frac{1}{2}\frac{\tilde k^2}{p\,q}-\frac{1}{2}\left(\frac{p}{q}+\frac{q}{p}\right)\quad {\rm and} \quad \left[ 1+ \left(\frac{p}{q}+\frac{q}{p}\right)(\hat \bp \cdot \hat \bq)+(\hat \bp \cdot \hat \bq)^2\right]=\frac{1}{4} \frac{\tilde k^4}{p^2\,q^2}-\frac{1}{4}  \left(\frac{p}{q}-\frac{q}{p}\right)^2$$ where $\tilde k^2=p^2+q^2+2p\,q (\hat \bp \cdot \hat \bq)$\;.
Then we can recast Eqs.  (\ref{FTheta}), (\ref{FXi}) and (\ref{Fnabla^2delta_mCS}) as
 \bea\label{FTheta2}
F^\Theta (\bp, \bq, \tilde k) &=&\frac{1}{4}\left[-\frac{1}{6} - \frac{1}{3} \frac{p\,q}{ \tilde k^2} \left(\frac{p}{q}+\frac{q}{p}\right) +   \frac{1}{2} \frac{p^2\,q^2}{ \tilde k^4} \left(\frac{p}{q}-\frac{q}{p}\right)^2  \right]\;,  \\
&&\nonumber\\
\label{FXi2}
 F^\Xi (\bp, \bq, \tilde k) &=& \frac{1}{8}\left(\frac{p}{q}+\frac{q}{p}\right)-\frac{1}{4}(\hat \bp \cdot \hat \bq) +  \frac{1}{2}\frac{p^2\,q^2}{ \tilde k^2}\left[1-\frac{1}{4}\left(\frac{p}{q}+\frac{q}{p}\right)^2\right]\;,\\
  &&\nonumber\\
  F^{\nabla^2  \delta_{m  {\rm CS}} (2)}(\bp, \bq,  \tilde k; \eta) &=& - {4\over 3} \frac{1}{ \cH^2 \Omega_{\rm m}}\frac{a}{a_{\rm in}}\frac{D_{\rm in}}{D }\bigg[ \left(4 f_{\rm NL} + \frac{5}{6}\right) p\,q \,\left(p^2 + q^2\right)\, (\hat \bp \cdot \hat \bq)+  f_{\rm NL}\left(p^2 + q^2\right)^2  \nonumber\\
  && + 4  \left( f_{\rm NL} + \frac{5}{12}\right) p^2\,q^2 (\hat \bp \cdot \hat \bq)^2 \bigg]  - {4 \over 9}  \frac{1}{\cH^4 \Omega^2_{\rm m}}p^2\,q^2 \bigg[\left(1+{\F \over D^2} \right) \left(p^2 + q^2\right) + 2 p\, q \, \left(1+{\F \over D^2} \right)  (\hat \bp \cdot \hat \bq) \nonumber \\
  && +\left(1- {\F \over D^2} \right)\left(p^2 + q^2\right) (\hat \bp \cdot \hat \bq)^2+2 p\, q \, \left(1- {\F \over D^2} \right)\, (\hat \bp \cdot \hat \bq)^3  \bigg] \;,
 \eea
and consequently we find
 \bea\label{FellTheta}
F_\ell^\Theta (p, q) &=&\frac{1}{4}\left[-\frac{1}{6} \delta^K_{\ell 0} - \frac{ p\,q}{3} \left(\frac{p}{q}+\frac{q}{p}\right) \sigma_\ell^2(p, q)+   \frac{p^2\,q^2}{2}  \left(\frac{p}{q}-\frac{q}{p}\right)^2 \sigma_\ell^4(p, q) \right]\;, \nonumber \\ \\ 
\label{FellXi}
 F_\ell^\Xi(p, q) &=& \frac{1}{8}\left(\frac{p}{q}+\frac{q}{p}\right) \delta^K_{\ell 0}  -\frac{1}{4}\delta_{\ell 1}+  \frac{p^2\,q^2}{2}\left[1-\frac{1}{4}\left(\frac{p}{q}+\frac{q}{p}\right)^2\right] \sigma_\ell^2(p, q)\ \;, \\
   &&\nonumber\\
F_{\ell}^{\nabla^2 \delta_{m  {\rm CS}} (2)}(p, q;\eta) &=& - {4\over 3} \frac{1}{ \cH^2 \Omega_{\rm m}}\frac{a}{a_{\rm in}}\frac{D_{\rm in}}{D }p^2\,q^2 \Bigg\{\bigg[ f_{\rm NL} \left(\frac{p^2}{q^2}+\frac{q^2}{p^2}\right) + {10\over 3} \left( f_{\rm NL} + \frac{1}{6}\right) \bigg]\delta^K_{\ell 0}  +\left(4 f_{\rm NL} + \frac{5}{6}\right)  \,\left(\frac{p}{q}+\frac{q}{p}\right)\, \delta^K_{\ell 1} \nonumber\\
  &&+  {8 \over 3}  \left( f_{\rm NL} + \frac{5}{12}\right)  \, \delta^K_{\ell 2} \Bigg\}    - {8 \over 9}  \frac{1}{\cH^4 \Omega^2_{\rm m}}\, p^3\,q^3 \bigg[{1\over 3}\left(2+{\F \over D^2} \right) \left(\frac{p}{q}+\frac{q}{p}\right) + {2 \over 5} \left(4+{\F \over D^2} \right)  \, \delta^K_{\ell 1} \nonumber \\
  && -{1\over 3} \left(1- {\F \over D^2} \right) \left(\frac{p}{q}+\frac{q}{p}\right)  \, \delta^K_{\ell 2} + {2 \over 5} \left(1- {\F \over D^2} \right)\, \delta^K_{\ell 3 }  \bigg] \;,
 \eea
where 
$$\sigma_\ell^n(p, q)=\frac{2 \ell +1}{2}\int_{-1}^{1}  \frac{\PP_{\ell} (x)~\ud x }{\left(p^2+q^2+2p\,q \,x \right)^{n/2}}\;. $$

In Appendix \ref{NewcalM2} we will compute $\tilde F_\ell^{a}(p, q, \tilde k ; \eta)$. With this new approach we will not need to compute $\sigma_\ell^2(p, q)$.

\subsubsection{Prescription for {\rm SONG}} \label{SONG}

SONG\footnote{https://github.com/coccoinomane/song} is an open-source second-order Boltzmann code which includes all the effects of metric, CDM, baryons, photons and neutrinos, see e.g. Ref. \cite{Pettinari:2014vja}.
From this code one can compute all possible kernels in Fourier space from the radiation era, both at large and small scales; the addition of the non-Gaussianity parameter $f_{\rm NL}$ is straightforward following our prescription.

Taking into account that SONG is written in the Poisson gauge, i.e.
\bea\label{f2-b}
 f^{(2)}( \bk, \eta) &= & \int \frac{\ud^3 \bp}{(2\pi)^3} \frac{\ud^3 \bq}{(2\pi)^3} ~(2\pi)^3\delta^D \left(\bp+\bq- \bk \right)~   \K^{f (2)}_{\rm SONG}(\bp, \bq, k; \eta)   \delta_m^{(1)} (\bp, \eta) \delta_m^{(1)} (\bq, \eta) \;,
\eea
where $f^{(2)}=\{\Phi^{(2}, \Psi^{(2)}, v^{(2)}, \delta_m^{(2)}\}$, we can quickly  correlate $\K^{b (2)}$ with $F^{b (2)}$ in the following way
\be
F^{b (2)}(\bp, \bq,  k; \eta) = \K^{b (2)}_{\rm SONG}(\bp, \bq, k; \eta) \frac{\T^{\delta_{\rm m}}(\bp,\eta)}{\T^{\Phi} (\bp,\eta)} \frac{\T^{\delta_{\rm m}}(\bq,\eta)}{\T^{\Phi} (\bq,\eta)}\;.
\ee

Now, for simplicity, setting $b_e=0$,   $ \delta_{g {\rm CS}}^{(2)} \to  \delta_{m {\rm CS}}^{(2)}$, $\delta_g^{(2)}  \to \delta_m^{(2)} $ and $\delta_{g }  \to \delta_{m}^{(1)} $, from Eq.(\ref{dg2-2}) we can 
immediately obtain the gauge transformation 
 \begin{eqnarray} 
 \label{dm2}
   \delta_{m {\rm CS}}^{(2)}  &=&
\delta_m^{(2)}  + \p^i \chi^{(1)}  \p_i \delta_{m  {\rm CS}}^{(1)}  - 3\cH v^{(2)} -3\left( 3 \cH^2  - \cH'   \right) v^2 - 6\cH  v \delta^{(1)}_{m  {\rm CS}} +3  \cH v  {v}'  +2  v {\delta^{(1)}_{m {\rm CS}}}' 
  \nonumber\\
&&    - 3 \cH  \nabla^{-2}\bigg( v \nabla^2 {v}' - {v}' \nabla^2 v - 6 \p_i \Phi \p^i v - 6 \Phi \nabla^2 v \bigg)\;.
\end{eqnarray} 
Then we obtain
\bea
\label{FdeltamSC}
&&F^{  \delta_{m {\rm CS}} (2)}(\bp, \bq,  k; \eta) \T^{\Phi} (\bp,\eta) \T^{\Phi} (\bq,\eta) =  \K^{\delta_m (2)}_{\rm SONG}(\bp, \bq, k; \eta) - 3 \cH \K^{v (2)}_{\rm SONG}(\bp, \bq, k; \eta)  \nonumber\\
&&- \frac{1}{2} (\bp\cdot\bq) \left[\T^{\chi}(\bp,\eta) \T^v(\bq,\eta) +\T^{\chi}(\bq,\eta) \T^v (\bp,\eta) \right] -3\left(3 \cH^2 - \cH'\right)\T^{v}(\bq,\eta) \T^v (\bp,\eta) \nonumber\\
&&+(f-3)\cH\left[\T^{\delta_{m  {\rm CS}} }(\bp,\eta) \T^v(\bq,\eta) +\T^{\delta_{m  {\rm CS}} }(\bq,\eta) \T^v (\bp,\eta) \right] +\frac{3}{2} \cH
\left[\T^v (\bp,\eta){\T^v(\bq,\eta) }' + {\T^v (\bq,\eta)}{\T^v(\bp,\eta) }'  \right] \nonumber\\
&&+\frac{3}{2}\cH\frac{\left(k_1^2-k_2^2\right)}{k^2}\left[\T^v (\bp,\eta){\T^v(\bq,\eta) }' - {\T^v(\bp,\eta) }'  {\T^v (\bq,\eta)} \right]+9\cH\frac{\bp\cdot\bq}{k^2} \left[ \T^\Phi (\bp,\eta)\T^v(\bq,\eta)  + \T^\Phi(\bq,\eta)   \T^v (\bp,\eta)\right] \nonumber \\
&&+9\frac{\cH}{k^2} \left[ q^2\T^\Phi (\bp,\eta)\T^v(\bq,\eta)  +p^2 \T^\Phi(\bq,\eta)   \T^v (\bp,\eta)\right] \;.
\eea
Here, in Eq. (\ref{FdeltamSC}), we did not explicitly write the transfer functions because all the quantities should be numerically evaluated with SONG. 

\section{ $\M_{\ell m \ellvp \mvp \ellvq \mvq \barell \barm}^{a(2)}(k; p, q) $ for Eqs (\ref{Deltag2-loc-1}) and (\ref{Deltag2-int-3}): different  method} \label{NewcalM2}

Here below we present a different way to compute $\M_{\ell m \ellvp \mvp \ellvq \mvq \barell \barm}^{a(2)}(k; p, q) $ for  Eqs (\ref{Deltag2-loc-1}) and (\ref{Deltag2-int-3}). Finally, using this new approach, in Sec.\ \ref{F2ell-2} we will rewrite the second-order kernel at large scales.

\subsection{Eq.\ (\ref{Deltag2-loc-1})}\label{Sec:Deltag2-loc-1_2}

Starting from Eqs.\ (\ref{Deltag2-loc-1-general_relation}) and (\ref{F2-b}), let us now write the right-hand side of Eq.\ (\ref{Fourier_Deltag2-loc-1}) in a different way
\be\label{Fourier_Deltag2-loc-2}
\frac{1}{2} \Delta^{b (2)}(\bx, \eta) =  \int \frac{\ud^3\tilde \bk}{(2\pi)^3} \frac{\ud^3 \bp}{(2\pi)^3} \frac{\ud^3 \bq}{(2\pi)^3} (2\pi)^3 \delta^D \left(\bp+\bq-\tilde \bk \right)~  \frac{1}{2} F^{b (2)}(\bp, \bq, \tilde k; \eta) \T^{\Phi}(\bp, \eta)\T^{\Phi}(\bq, \eta)  \Phi_{\rm p}(\bp) \Phi_{\rm p}(\bq)\; e^{i\tilde \bk  \bx} \;.
\ee
Then, if we do not apply immediately the $\delta^D (\bp+\bq-\tilde \bk)$ constraint on $F^{b (2)}(\bp, \bq, \tilde k; \eta)$, $\tilde k$ does not depend on  $p,~q$ and $\cos(\theta_{\rm p q})$, and the second-order kernels can be expanded as
\be
\frac{1}{2} F^{b (2)}(\bp, \bq, \tilde k; \eta) =\sum_{\bar \ell} \frac{1}{2} \tilde F_{\bar \ell}^{b (2)}(p, q, \tilde k; \eta) ~\PP_{\barell} (\cos(\theta_{\rm p q}))=
\sum_{\bar \ell \bar m} \frac{4\pi}{2\bar \ell +1} ~ \frac{1}{2} \tilde F_{\bar \ell}^{b (2)}(p, q, \tilde k; \eta)    Y_{\barell \barm} (\hat \bp) Y_{\barell \barm}^*(\hat \bq)\;.
\ee
Now using
\bea
 &(2\pi)^3& \delta^D \left(\bp+\bq-\tilde \bk \right) = \int \ud^3\tilde \bx ~e^{-i\left(\bp+\bq-\tilde \bk \right)\tilde \bx} \nonumber\\
&=& \sum_{\ellvp  \mvp \ellvq  \mvq \tell \tm} \left(4 \pi\right)^3 (-1)^{\ellvp+ \ellvq+\mvp+\mvq}\; i^{\ellvp + \ellvq+ \tell} \J_{\ellvp \ellvq \tell}(p,q,\tilde k)\; \G_{-\mvp - \mvq \tm}^{\ellvp \ellvq \tell} Y_{\ellvp  \mvp}(\hat \bp) Y_{ \ellvq  \mvq}(\hat \bq) Y^*_{\tell \tm}(\hat{\tilde \bk})\;,
\eea
where
$$ \J_{\ellvp \ellvq \tell}(p,q,\tilde k)=\int_0^\infty \ud \tilde\chi \,\tilde \chi^2 j_{\ellvp} (p \tilde\chi) j_{\ellvq} (q  \tilde\chi) j_{\tell} (\tilde k  \tilde\chi)\;,$$
and after few simple algebraic manipulations, Eq.\ (\ref{Fourier_Deltag2-loc-2}) yields
\bea \label{Fourier_Deltag2-loc-3}
&&\frac{1}{2} \Delta^{b (2)}(\bx, \eta) =\sum_{ \tell \tm \ellvp \mvp \ellvq \mvq \barell \barm} (-1)^{\tm} Y_{\tell \tm} (\bn) \int \frac{\ud^3 \bp}{(2\pi)^3} \frac{\ud^3 \bq}{(2\pi)^3}  \Bigg\{ (4\pi)^3 i^{\ellvp+\ellvq} (2\bar \ell +1)^{-1} \G^{\ellvp \ellvq \tell}_{\mvp \mvq -\tm}  \int \frac{\tilde k^2 \ud \tilde k}{2 \pi^2} \nonumber\\
&&\times  \frac{1}{2} \tilde F_{\bar \ell}^{b (2)}(p, q, \tilde k; \eta)\T^{\Phi}(\bp, \eta)\T^{\Phi}(\bq, \eta)      j_{\tell} (\tilde k  \bar\chi) \J_{\ellvp \ellvq \tell}(p,q,\tilde k)  \Bigg\} Y_{\barell \barm}(\hat \bp) Y^*_{\ellvp  \mvp}(\hat \bp) Y^*_{\barell \barm}(\hat \bq)  Y^*_{ \ellvq  \mvq}(\hat \bq) \Phi_{\rm p}(\bp) \Phi_{\rm p}(\bq) 
\eea
and finally we find
\bea 
 \M_{\ell m \ellvp \mvp \ellvq \mvq \barell \barm}^{\alpha[b](2)}(k; p, q) &=& (4\pi)^3 \aleph^*_\ell(k)  (-1)^m i^{\ellvp+\ellvq} (2\bar \ell +1)^{-1} \G^{\ell \ellvp \ellvq }_{-m  \mvp \mvq} \int \ud \bar \chi~\bar \chi^2  \W (\bar \chi)\int \frac{\tilde k^2 \ud \tilde k}{2 \pi^2}
\nonumber\\ 
&\times &   \WW^{\alpha} \left(\bar \chi,  \eta, \frac{\p}{\p \bar \chi}, \frac{\p}{\p  \eta}, \right)  \left[\frac{1}{2} \tilde F_{\bar \ell}^{b (2)}(p, q, \tilde k; \eta)\T^{\Phi}(\bp, \eta)\T^{\Phi}(\bq, \eta)      j_{\ell} (\tilde k  \bar\chi) \right]  \J_{\ellvp \ellvq \ell}(p,q,\tilde k)  j_\ell (k  \bar \chi)\;.\nonumber\\
\eea

Here below we show explicitly all the terms in Eq.\ (\ref{Deltag2-loc-1}):
 \bea\label{delta_g^(2)-2}
 \M_{\ell m \ellvp \mvp \ellvq \mvq \barell \barm}^{ \delta_g^{(2)}(2)}(k; p, q) &=& (4\pi)^3 \aleph^*_\ell(k)  (-1)^m i^{\ellvp+\ellvq} (2\bar \ell +1)^{-1} \G^{\ell \ellvp \ellvq }_{-m  \mvp \mvq} \int \ud \bar \chi~\bar \chi^2  \W (\bar \chi)\int \frac{\tilde k^2 \ud \tilde k}{2 \pi^2}
\nonumber\\ 
&\times &   \left[\frac{1}{2} \tilde F_{\bar \ell}^{ \delta_g (2)}(p, q, \tilde k; \eta) \T^{\Phi}(\bp, \eta)\T^{\Phi}(\bq, \eta)  j_{\ell} (\tilde k  \bar\chi)  \J_{\ellvp \ellvq \ell}(p,q,\tilde k) \right]  j_\ell (k  \bar \chi)\;,
\eea

 \bea\label{Psi^(2)-2}
 \M_{\ell m \ellvp \mvp \ellvq \mvq \barell \barm}^{\Psi^{(2)}(2)}(k; p, q) &=& (4\pi)^3 \aleph^*_\ell(k)  (-1)^m i^{\ellvp+\ellvq} (2\bar \ell +1)^{-1} \G^{\ell \ellvp \ellvq }_{-m  \mvp \mvq} \int \ud \bar \chi~\bar \chi^2  \W (\bar \chi)\left[  -2  \left(1 - \Q\right)\right]\int \frac{\tilde k^2 \ud \tilde k}{2 \pi^2}
\nonumber\\ 
&\times &  \left[\frac{1}{2}  \tilde F_{\bar \ell}^{\Psi (2)}(p, q, \tilde k; \eta) \T^{\Phi}(\bp, \eta)\T^{\Phi}(\bq, \eta)  j_{\ell} (\tilde k  \bar\chi)  \J_{\ellvp \ellvq \ell}(p,q,\tilde k)\right]   j_\ell (k  \bar \chi)\;,
\eea

 \bea
 \M_{\ell m \ellvp \mvp \ellvq \mvq \barell \barm}^{\Phi^{(2)}(2)}(k; p, q) &=& (4\pi)^3 \aleph^*_\ell(k)  (-1)^{m+1} i^{\ellvp+\ellvq} (2\bar \ell +1)^{-1} \G^{\ell \ellvp \ellvq }_{-m  \mvp \mvq} \int \ud \bar \chi~\bar \chi^2  \W (\bar \chi) \bigg[ b_e-2 \Q -1 -   \frac{\cH'}{\cH^2} 
\nonumber\\ 
& &-  2\frac{\left(1 - \Q\right)}{\bar \chi \cH} \bigg] ~\int \frac{\tilde k^2 \ud \tilde k}{2 \pi^2} \left[\frac{1}{2}  \tilde F_{\bar \ell}^{\Phi (2)}(p, q, \tilde k; \eta) \T^{\Phi}(\bp, \eta)\T^{\Phi}(\bq, \eta)  j_{\ell} (\tilde k  \bar\chi)   \J_{\ellvp \ellvq \ell}(p,q,\tilde k) \right]   j_\ell (k  \bar \chi)\;,\nonumber\\ 
\eea

 \bea
 \M_{\ell m \ellvp \mvp \ellvq \mvq \barell \barm}^{\Psi^{(2)}{'} (2)}(k; p, q) &=& (4\pi)^3 \aleph^*_\ell(k)  (-1)^m i^{\ellvp+\ellvq} (2\bar \ell +1)^{-1} \G^{\ell \ellvp \ellvq }_{-m  \mvp \mvq} \int \ud \bar \chi~\bar \chi^2  \W (\bar \chi) \left(\frac{1}{ \cH}\right)\int \frac{\tilde k^2 \ud \tilde k}{2 \pi^2}
\nonumber\\ 
&\times & \frac{\p}{\p  \eta} \left[\frac{1}{2} \tilde F_{\bar \ell}^{\Psi (2)}(p, q, \tilde k; \eta) \T^{\Phi}(\bp, \eta)\T^{\Phi}(\bq, \eta) \right]  j_{\ell} (\tilde k  \bar\chi)  \J_{\ellvp \ellvq \ell}(p,q,\tilde k)  j_\ell (k  \bar \chi)\;,
\eea 

 \bea
 \M_{\ell m \ellvp \mvp \ellvq \mvq \barell \barm}^{\p_\| v^{(2)}  (2)}(k; p, q) &=& (4\pi)^3 \aleph^*_\ell(k)  (-1)^m i^{\ellvp+\ellvq} (2\bar \ell +1)^{-1} \G^{\ell \ellvp \ellvq }_{-m  \mvp \mvq} \int \ud \bar \chi~\bar \chi^2  \W (\bar \chi)\left[ b_e-2 \Q  -   \frac{\cH'}{\cH^2} -2  \frac{\left(1 - \Q\right)}{\bar \chi \cH} \right]
\nonumber\\ 
&\times &\int \frac{\tilde k^2 \ud \tilde k}{2 \pi^2} \frac{1}{2} \tilde F_{\bar \ell}^{v (2)}(p, q, \tilde k; \eta)  \frac{\p}{\p \bar \chi}  \left[  j_{\ell} (\tilde k  \bar\chi) \right]  ~ j_\ell (k  \bar \chi)  \J_{\ellvp \ellvq \ell}(p,q,\tilde k) ~\T^{\Phi}(\bp, \eta) \T^{\Phi}(\bq, \eta) \;,
\eea

 \bea\label{v^(2)-2}
 \M_{\ell m \ellvp \mvp \ellvq \mvq \barell \barm}^{\p_\|^2 v^{(2)} (2)}(k; p, q) &=& (4\pi)^3 \aleph^*_\ell(k)  (-1)^{m+1} i^{\ellvp+\ellvq} (2\bar \ell +1)^{-1} \G^{\ell \ellvp \ellvq }_{-m  \mvp \mvq} \int \ud \bar \chi~\bar \chi^2  \W (\bar \chi)\left(\frac{1}{ \cH }\right)\int \frac{\tilde k^2 \ud \tilde k}{2 \pi^2}
\nonumber\\ 
&\times &\frac{1}{2} \tilde F_{\bar \ell}^{v (2)}(p, q, \tilde k; \eta)  \frac{\p^2}{\p \bar \chi^2}  \left[  j_{\ell} (\tilde k  \bar\chi) \right]  \J_{\ellvp \ellvq \ell}(p,q,\tilde k)
 ~ j_\ell (k  \bar \chi) ~\T^{\Phi}(\bp, \eta)\T^{\Phi}(\bq, \eta) \;.
\eea

\subsection{Eq.\ (\ref{Deltag2-int-3})}

Starting from Eq.\ (\ref{Deltag2-int-3-klm}), [i.e. considering only the first two additive terms of Eq.\ (\ref{Deltag2-int-3})] and using Eq.\ (\ref{Fourier_Deltag2-loc-3}) with $b=\Phi^{(2)}+\Psi^{(2)}$ we find immediately
\bea 
&& \M_{\ell m \ellvp \mvp \ellvq \mvq \barell \barm}^{\alpha(2)}(k; p, q) = (4\pi)^3 \aleph^*_\ell(k)  (-1)^m i^{\ellvp+\ellvq} (2\bar \ell +1)^{-1} \G^{\ell \ellvp \ellvq }_{-m  \mvp \mvq} \int \ud \bar \chi~\bar \chi^2  \W (\bar \chi)  \int_0^{\bar \chi}\ud\tilde\chi ~ \int \frac{\tilde k^2 \ud \tilde k}{2 \pi^2}
\nonumber\\ 
&&\times   \WW^{a}\left(\bar \chi,\tilde \chi,  \eta, \tilde \eta, \frac{\p}{\p  \tilde \chi}, \frac{\p}{\p  \tilde \eta} \right) \left\{\left[\frac{1}{2} \tilde  F_{\bar \ell}^{\Phi (2)}(p, q, \tilde k; \tilde \eta)+ \frac{1}{2}  \tilde F_{\bar \ell}^{\Psi (2)}(p, q, \tilde k; \tilde \eta)\right]\T^{\Phi}(\bp, \tilde \eta)\T^{\Phi}(\bq, \tilde \eta)      j_{\ell} (\tilde k  \tilde\chi) \right\}  \J_{\ellvp \ellvq \ell}(p,q,\tilde k)  j_\ell (k  \bar \chi)\;,\nonumber\\
\eea
and find
 \bea\label{M-T^(2)-2}
 \M_{\ell m \ellvp \mvp \ellvq \mvq \barell \barm}^{T^{(2)}(2)}(k; p, q) &=& 2 (4\pi)^3 \aleph^*_\ell(k)  (-1)^{m} i^{\ellvp+\ellvq} (2\bar \ell +1)^{-1} \G^{\ell \ellvp \ellvq }_{-m  \mvp \mvq} \int \ud \bar \chi~\bar \chi  \W (\bar \chi)~(1-\Q)
 \int_0^{\bar \chi}\ud\tilde\chi ~ \int \frac{\tilde k^2 \ud \tilde k}{2 \pi^2}
\nonumber\\ 
&\times & \left[\frac{1}{2} \tilde  F_{\bar \ell}^{\Phi (2)}(p, q, \tilde k; \tilde \eta)+ \frac{1}{2}  \tilde F_{\bar \ell}^{\Psi (2)}(p, q, \tilde k; \tilde \eta)\right] \T^{\Phi}(\bp, \tilde \eta)\T^{\Phi}(\bq, \tilde \eta )      \J_{\ellvp \ellvq \ell}(p,q,\tilde k)   j_{\ell} (\tilde k  \tilde\chi)  j_\ell (k  \bar \chi)\;,  \nonumber\\
\eea
and
 \bea\label{M-I^(2)-2}
&& \M_{\ell m \ellvp \mvp \ellvq \mvq \barell \barm}^{I^{(2)}(2)}(k; p, q) =  (4\pi)^3 \aleph^*_\ell(k)  (-1)^{m+1} i^{\ellvp+\ellvq} (2\bar \ell +1)^{-1} \G^{\ell \ellvp \ellvq }_{-m  \mvp \mvq} \int \ud \bar \chi~\bar \chi^2  \W (\bar \chi)~\nonumber\\ 
&&  \times\bigg[ b_e-2 \Q  -   \frac{\cH'}{\cH^2}- 2 \frac{\left(1 - \Q\right)}{\bar \chi \cH} \bigg]
  \int_0^{\bar \chi}\ud\tilde\chi ~ \int \frac{\tilde k^2 \ud \tilde k}{2 \pi^2} ~\frac{\p}{\p  \tilde \eta} \left\{  \left[\frac{1}{2} \tilde  F_{\bar \ell}^{\Phi (2)}(p, q, \tilde k; \tilde \eta)+ \frac{1}{2}  \tilde F_{\bar \ell}^{\Psi (2)}(p, q, \tilde k; \tilde \eta)\right] \T^{\Phi}(\bp, \tilde \eta)\T^{\Phi}(\bq, \tilde \eta )  \right\}    \nonumber\\
&&  \times ~ \J_{\ellvp \ellvq \ell}(p,q,\tilde k)   j_{\ell} (\tilde k  \tilde\chi)  j_\ell (k  \bar \chi) \;.
\eea

Finally, for the last two additive terms of  Eq.\ (\ref{Deltag2-int-3}), Eq.\ (\ref{Delta_kappa-2}) becomes
\bea \label{Delta_kappa-3}
&&\frac{1}{2} \Delta^{[2\p_\perp S^{(2)} -\nabla_\perp^2 T^{(2)}] (2)}(\bx, \eta) = - (1-\Q) \int_0^{\bar \chi}\ud\tilde\chi (\bar \chi - \tilde \chi) \frac{\tilde \chi}{\bar \chi} \,   \tilde \nabla^2_\perp \left[\frac{1}{2}\Phi^{(2)}(\tilde \bx, \tilde \eta)+\frac{1}{2}\Psi^{(2)}(\tilde \bx, \tilde \eta)\right] =(1-\Q)\int_0^{\bar \chi}\ud\tilde\chi  \frac{(\bar \chi - \tilde \chi)}{\bar \chi \tilde \chi} \nonumber\\
&&\times \sum_{ \tell \tm \ellvp \mvp \ellvq \mvq \bar \ell \bar m}   (4\pi)^3  i^{\ellvp+\ellvq} (2\bar \ell +1)^{-1} (-1)^{\tm+1} ~{}^{(2)}\nabla^2 \left[  Y_{\tell \tm} (\bn)  \right]   \G^{\ellvp \ellvq \tell}_{\mvp \mvq -\tm}  \int \frac{\ud^3 \bp}{(2\pi)^3} \frac{\ud^3 \bq}{(2\pi)^3} \Bigg\{  \int \frac{\tilde k^2 \ud \tilde k}{2 \pi^2}   \nonumber\\
&&\times~ \left[\frac{1}{2}  \tilde F_{\bar \ell}^{\Phi (2)}(p, q, \tilde k;\tilde \eta)+ \frac{1}{2}  F_{\bar \ell}^{\Psi (2)}(p, q, \tilde k; \tilde \eta)\right] \T^{\Phi}(\bp, \tilde \eta)\T^{\Phi}(\bq, \tilde \eta )      \J_{\ellvp \ellvq \tilde \ell}(p,q,\tilde k)   j_{\tilde \ell} (\tilde k  \tilde\chi) \Bigg\}   Y_{\ellvp \mvp}^*(\hat \bp)  Y_{\barell \barm} (\hat \bp) Y_{\ellvq \mvq}^*(\hat \bq) Y_{\barell \barm}^*(\hat \bq)  \nonumber\\
&&\times ~ \Phi_{\rm p}(\bp) \Phi_{\rm p}(\bq)
\eea
and, using $ {}^{(2)}\nabla^2Y_{\ell m}({\bn}) =- \ell(\ell+1) Y_{\ell m}({\bn})$, we find
 \bea\label{M-kappa^(2)-2}
 &&\M_{\ell m \ellvp \mvp \ellvq \mvq \barell \barm}^{[2\p_\perp S^{(2)} -\nabla_\perp^2 T^{(2)}] (2)}(k; p, q) = (4\pi)^3 \aleph^*_\ell(k)  (-1)^{m} i^{\ellvp+\ellvq} \frac{\ell(\ell+1)}{(2\bar \ell +1)} \G^{\ell \ellvp \ellvq }_{-m  \mvp \mvq} \int \ud \bar \chi~\bar \chi  \W (\bar \chi)~(1-\Q)
 \int_0^{\bar \chi}\ud\tilde\chi \frac{(\bar \chi - \tilde \chi)}{ \tilde \chi}
\nonumber\\ 
&\times &  \int \frac{\tilde k^2 \ud \tilde k}{2 \pi^2} \left\{\left[\frac{1}{2} \tilde  F_{\bar \ell}^{\Phi (2)}(p, q, \tilde k; \tilde \eta)+ \frac{1}{2}  \tilde F_{\bar \ell}^{\Psi (2)}(p, q, \tilde k; \tilde \eta)\right]       \J_{\ellvp \ellvq \ell}(p,q,\tilde k)   j_{\ell} (\tilde k  \tilde\chi) \right\} \T^{\Phi}(\bp, \tilde \eta)\T^{\Phi}(\bq, \tilde \eta ) j_\ell (k  \bar \chi)\;. 
\eea

\subsection{$\tilde F_{\bar \ell}^{b (2)}(p, q, \tilde k; \eta)$ }\label{F2ell-2}

Now, using the results in Sec.\ \ref{LS}, we obtain immediately
  \bea
\tilde F_{\ell}^{\Phi (2)}(p, q, \tilde k;  \eta) &=& \left[3 - 2\frac{a}{a_{\rm in}}\frac{D_{\rm in}}{D } \left(f_{\rm NL} + \frac{5}{6}\right)+ \frac{2}{3} \frac{f^2}{ \Omega_{\rm m}} \right] \delta_{\ell0}^K +\frac{2}{3} \frac{1}{\Omega_{\rm m}} p\,q   \delta^K_{\ell1}  \nonumber\\
&&+ 12 \left(2 - \frac{5}{3}\frac{a}{a_{\rm in}}\frac{D_{\rm in}}{D }+ \frac{2}{3} \frac{f^2}{ \Omega_{\rm m}} \right) \tilde F_{\ell}^\Theta (p, q, \tilde k)  - \frac{4}{3}\frac{(1+\F/D^2 )}{ \Omega_{\rm m}} \tilde F_{\ell}^\Xi (p, q, \tilde k) \;, \\
 \tilde F_{\ell}^{\Psi (2)}(p, q, \tilde k;  \eta)&=& \left[ -1 - 2\frac{a}{a_{\rm in}}\frac{D_{\rm in}}{D } \left(f_{\rm NL} + \frac{5}{6}\right)+ \frac{2}{3} \frac{f^2}{ \Omega_{\rm m}}  \right] \delta_{\ell0}^K +\frac{2}{3} \frac{1}{\Omega_{\rm m}} p\,q \delta^K_{\ell1} + 12 \left(1 - \frac{5}{3}\frac{a}{a_{\rm in}}\frac{D_{\rm in}}{D } \right)  \tilde F_{\ell}^\Theta (p, q, \tilde k) \nonumber\\
&&     - \frac{4}{3}\frac{(1+\F/D^2 )}{ \Omega_{\rm m}}  \tilde F_{\ell}^\Xi (p, q, \tilde k) \;, \\
 \tilde F_{\ell}^{v (2)}(p, q, \tilde k;  \eta) &=&   \frac{2}{3}\frac{ f}{ \cH \Omega_{\rm m}}  \left[1+ 2\frac{a}{a_{\rm in}}\frac{D_{\rm in}}{D } \left(f_{\rm NL} - \frac{5}{3}\right) \right] \delta_{\ell0}^K  -\frac{4}{9} \frac{f}{\cH^3 \Omega^2_{\rm m}}  p\,q \, \delta^K_{\ell1}   - 8 \frac{ f}{ \cH \Omega_{\rm m}} \tilde F_{\ell}^\Theta (p, q, \tilde k)  \nonumber\\
 &&+ \frac{8}{9}\frac{\F' }{D^2 \cH^4 \Omega_{\rm m}^2}  \tilde F_{\ell}^\Xi (p, q, \tilde k)    \;, \\
 &&\nonumber\\
\tilde  F_{\ell}^{\delta_{m  {\rm CS}} (2)}(p, q, \tilde k;  \eta)  &=& {4\over 3} \frac{1}{ \cH^2 \Omega_{\rm m}}\frac{a}{a_{\rm in}}\frac{D_{\rm in}}{D }\left[  f_{\rm NL}\left(p^2 + q^2\right)  \delta_{\ell0}^K  +2 \left(f_{\rm NL} + \frac{5}{12}\right) p\,q \,  \delta^K_{\ell1} \right] \nonumber\\
  &&+{8 \over 27}  \frac{1}{\cH^4 \Omega^2_{\rm m}} p^2\,q^2 \left[\left(2+{\F \over D^2} \right) \delta_{\ell0}^K + \left(1- {\F \over D^2} \right) \delta^K_{\ell2} \right]\;,\\
   &&\nonumber\\
\tilde  F_{\ell}^{\nabla^2 \delta_{m  {\rm CS}} (2)}(p, q, \tilde k;  \eta) &=& - \tilde k^2 \tilde  F_{\ell}^{\delta_{m  {\rm CS}} (2)}(p, q, \tilde k;  \eta) \;, \\
 &&\nonumber\\
  \tilde F_{\ell}^{\delta_{g }(2)}(p, q, \tilde k;  \eta)  &=&  b_{10}   \tilde F_{\ell}^{\delta_{m  {\rm CS}} (2)}(p, q, \tilde k;  \eta)   - (b_e-3) \cH    \tilde F_{\ell}^{v (2)}(p, q, \tilde k;  \eta) +   b_{\nabla^2\delta} \tilde F_{\ell}^{\nabla^2 \delta_{m  {\rm CS}} (2)}(p, q, \tilde k; \eta) \nonumber\\
  &&+\Bigg\{ \frac{4b_{02}}{9 \Omega^2_{\rm m} }\left(f+ \frac{3}{2} \Omega_{\rm m}\right)^2 +   \frac{4 f^2}{9 \Omega^2_{\rm m} } \bigg[(b_e -3)^2 + \frac{\p b_e}{\p \ln a} \bigg]  + \frac{2f}{3 \Omega_{\rm m}}(b_e -3)(1-f) \nonumber\\
 &&  - \frac{8 f}{9 \cH\Omega^2_{\rm m} }\left(f+ \frac{3}{2} \Omega_{\rm m}\right)\left[b_{01}' +\cH( b_e-3)   b_{01}  \right] + \frac{4b_{20}}{9 \cH^4 \Omega^2_{\rm m} } p^2\,q^2 +\bigg[ - 2\frac{f( b_e-3)}{\Omega_{\rm m}}\frac{1}{\tilde k^2} - \frac{4f^2 b_{10}}{9 \cH^2 \Omega^2_{\rm m} } \nonumber\\
&&  -\frac{4f}{9 \cH^3 \Omega_{\rm m}^2 }  \left[b_{10}' +\cH( b_e-3)   b_{10}  \right]   + \frac{4b_{11}}{9 \cH^2 \Omega^2_{\rm m} }\left(f+ \frac{3}{2} \Omega_{\rm m}\right)  - \frac{4f}{9 \cH^3 \Omega_{\rm m}^2 } \left[b_{\nabla^2\delta}' +\cH( b_e-3)   b_{\nabla^2\delta}  \right]  \nonumber\\
&&  - \frac{4b_{\delta\nabla^2\delta}}{9 \cH^4 \Omega_{\rm m}^2 } p^2\,q^2\bigg] p\,q \left(\frac{p}{q}+\frac{q}{p}\right)  +  \frac{4b_{(\nabla^2\delta)^2} }{9 \cH^4 \Omega^2_{\rm m} } p^4\,q^4  + \frac{4f^2 b_{\nabla^2\delta}}{9 \cH^3 \Omega_{\rm m}^2 } p^2\,q^2 \left(\frac{p^2}{q^2}+\frac{q^2}{p^2}\right) \Bigg\} \delta^K_{\ell 0} \nonumber\\
&& +\bigg[-4\frac{f( b_e-3) }{ \Omega_{\rm m}}\frac{1}{\tilde k^2} + \frac{8b_{01}}{9 \cH^2 \Omega^2_{\rm m} }\left(f+ \frac{3}{2} \Omega_{\rm m}\right) + \frac{4b_{10}}{9 \cH^4 \Omega_{\rm m}^2 } p\,q \left(\frac{p}{q}+\frac{q}{p}\right) - \frac{8b_{(\p\delta)^2}}{9 \cH^4 \Omega_{\rm m}^2 } p^2\,q^2  \nonumber\\
&& - \frac{4b_{\nabla^2\delta}}{9 \cH^4 \Omega_{\rm m}^2 } p^2\,q^2 \left(\frac{p^2}{q^2}+\frac{q^2}{p^2}\right) \bigg] p\,q \delta^K_{\ell1} +\frac{3}{2}b_{s^2}\delta^K_{\ell2}\;.
\eea

Finally for $\tilde F_{\ell}^\Theta (p, q, \tilde k) $ \& $\tilde F_{\ell}^\Xi (p, q, \tilde k)$ we find
 \bea\label{FellTheta}
\tilde F_{\ell}^\Theta (p, q, \tilde k) &=&
\frac{1}{4}\left[-\frac{1}{6} - \frac{1}{3} \frac{p\,q}{ \tilde k^2} \left(\frac{p}{q}+\frac{q}{p}\right) +   \frac{1}{2} \frac{p^2\,q^2}{ \tilde k^4} \left(\frac{p}{q}-\frac{q}{p}\right)^2  \right]  \delta^K_{\ell 0} 
\;,  \\
\label{FellXi}
\tilde F_{\ell}^\Xi (p, q, \tilde k) &=&\left\{ \frac{1}{8}\left(\frac{p}{q}+\frac{q}{p}\right) +  \frac{1}{2}\frac{p^2\,q^2}{ \tilde k^2}\left[1-\frac{1}{4}\left(\frac{p}{q}+\frac{q}{p}\right)^2\right] \right\}\delta^K_{\ell 0} -\frac{1}{4}\delta^K_{\ell 1} \;.
 \eea


\end{document}